\documentclass[acmtog]{acmart}

\AtBeginDocument{%
  }

\acmJournal{TOG}
\acmVolume{45}
\acmNumber{4}
\acmArticle{1}
\acmMonth{6}

\usepackage{cancel}
\usepackage{enumitem}
\usepackage{amsmath}
\usepackage{relsize}
\usepackage{makecell}
\usepackage[symbol]{footmisc}
\usepackage{bm}
\usepackage{listings}
\usepackage{tcolorbox}
\usepackage{xcolor}
\usepackage{overpic,pbox,tikz}
\usepackage{multirow,colortbl}
\usepackage{siunitx}
\usepackage[utf8x]{inputenc}
\tcbuselibrary{listings, skins}
\usepackage[export]{adjustbox}
\usepackage[symbol]{footmisc}
\usepackage{wrapfig}
\usepackage{mathtools}
\usepackage{xspace}

\definecolor{revision_blue}{rgb}{0.2,0.2,0.7}

\definecolor{darkgreen}{rgb}{0.0, 0.5, 0.0}

\newcommand{\D}{\mathrm{d}}
\newcommand{\FLIP}{\protect\reflectbox{F}LIP\xspace}
\newcommand*{\trans}{^{\mkern-1.5mu\mathsf{T}}}

\DeclareMathOperator\erf{erf}

\makeatletter
\def\sectionautorefname{\S\@gobble}
\makeatother
\makeatletter
\def\subsectionautorefname{\S\@gobble}
\makeatother

\definecolor{lightgray}{RGB}{240,240,240}
\definecolor{darkgray}{RGB}{64,64,64}
\definecolor{darkgray}{RGB}{64,64,64}
\definecolor{solarized_base03}{RGB}{  0,      43,     54}
\definecolor{solarized_base02}{RGB}{  7,      54,     66}
\definecolor{solarized_base01}{RGB}{  88,     110,    117}
\definecolor{solarized_base00}{RGB}{  101,    123,    131}
\definecolor{solarized_base0}{RGB}{   131,    148,    150}
\definecolor{solarized_base1}{RGB}{   147,    161,    161}
\definecolor{solarized_base2}{RGB}{   238,    232,    213}
\definecolor{solarized_base3}{RGB}{   253,    246,    227}
\definecolor{solarized_yellow}{RGB}{  181,    137,    0}
\definecolor{solarized_orange}{RGB}{  203,    75,     22}
\definecolor{solarized_red}{RGB}{     220,    50,     47}
\definecolor{solarized_magenta}{RGB}{ 211,    54,     130}
\definecolor{solarized_violet}{RGB}{  108,    113,    196}
\definecolor{solarized_blue}{RGB}{    38,     139,    210}
\definecolor{solarized_cyan}{RGB}{    42,     161,    152}
\definecolor{solarized_green}{RGB}{   133,    153,    0}

\lstdefinestyle{pseudocodestyle}{
    language=Python,
    basicstyle=\ttfamily\footnotesize\color{solarized_base02},    
    commentstyle=\color{solarized_cyan},
    keywordstyle=\bfseries\color{solarized_base03},
    numberstyle=\tiny\color{solarized_base03},
    stringstyle=\color{solarized_green},    
    tabsize=2,
    breaklines=true,
    numbers=left,       
    captionpos=t,  
    escapeinside={(*@}{@*)},
}

\AtBeginDocument{%
\newtcblisting[blend into=listings]{pseudocode}[1][]{%
  arc=2pt,
  boxrule=1pt,
  colframe=solarized_base00,
  enhanced,
  listing only,
  overlay={
    \begin{tcbclipinterior}
        \fill[solarized_base00!25](frame.south west)
        rectangle ([xshift=10pt]frame.north west);
    \end{tcbclipinterior}},
  listing remove caption=false,
  listing options={numbers=left, numberstyle=\tiny,   
    basicstyle=\small\ttfamily, 
    xleftmargin=2pt, 
    aboveskip=\smallskipamount, 
    belowskip=\smallskipamount,
    columns=fullflexible,
    style=pseudocodestyle
  },
  coltitle=black,
  attach boxed title to bottom center={yshift=-0pt},
  boxed title style={enhanced jigsaw, colback=white, sharp corners, boxrule=0pt},
  #1
}
}

\begin{document}

\title{Unified Gaussian Primitives for Scene Representation and Rendering}

\author{Yang Zhou}
\authornote{
Work done as a PhD student at University of California, Santa Barbara.}
\email{yyp0502@gmail.com}
\affiliation{%
  \institution{University of California, Santa Barbara}
  \city{Santa Barbara}
  \state{California}
  \country{USA}
}
\affiliation{%
  \institution{Meta Reality Labs}
  \city{Redmond}
  \state{Washington}
  \country{USA}
}

\author{Songyin Wu}
\email{s_wu975@ucsb.edu}
\affiliation{%
  \institution{University of California, Santa Barbara}
  \city{Santa Barbara}
  \state{California}
  \country{USA}
}

\author{Lingqi Yan}
\email{lingqi.yan@mbzuai.ac.ae}
\affiliation{%
  \institution{Mohamed bin Zayed University of Artificial Intelligence}
  \city{Abu Dhabi}
  \country{United Arab Emirates}
}




\begin{teaserfigure}
	\newlength{\lenTeaser}
	\setlength{\lenTeaser}{0.5\linewidth}
	\addtolength{\tabcolsep}{-5pt}
    \center
	\begin{tabular}{cc}
		\includegraphics[width=\lenTeaser]{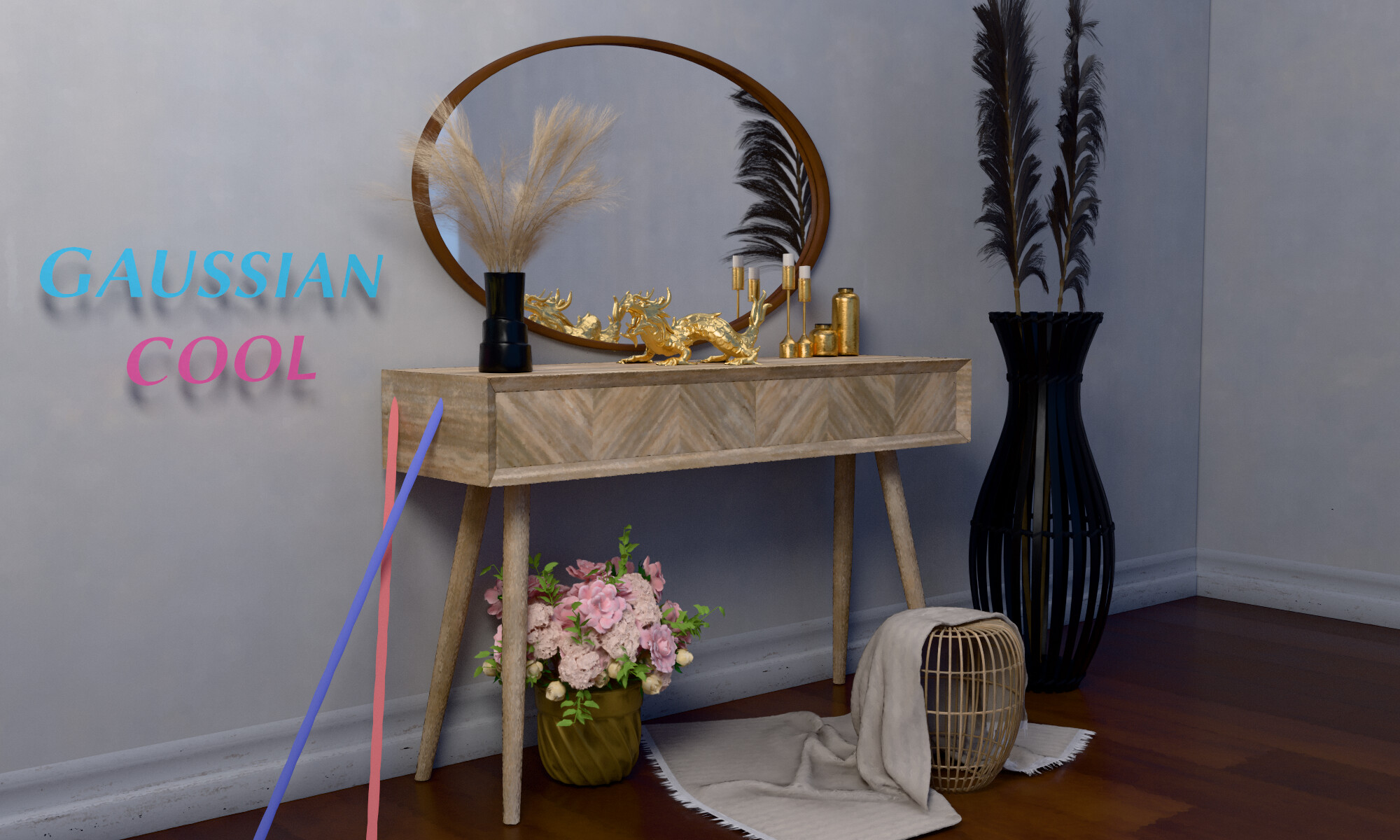}		 
		&
		\includegraphics[width=\lenTeaser]{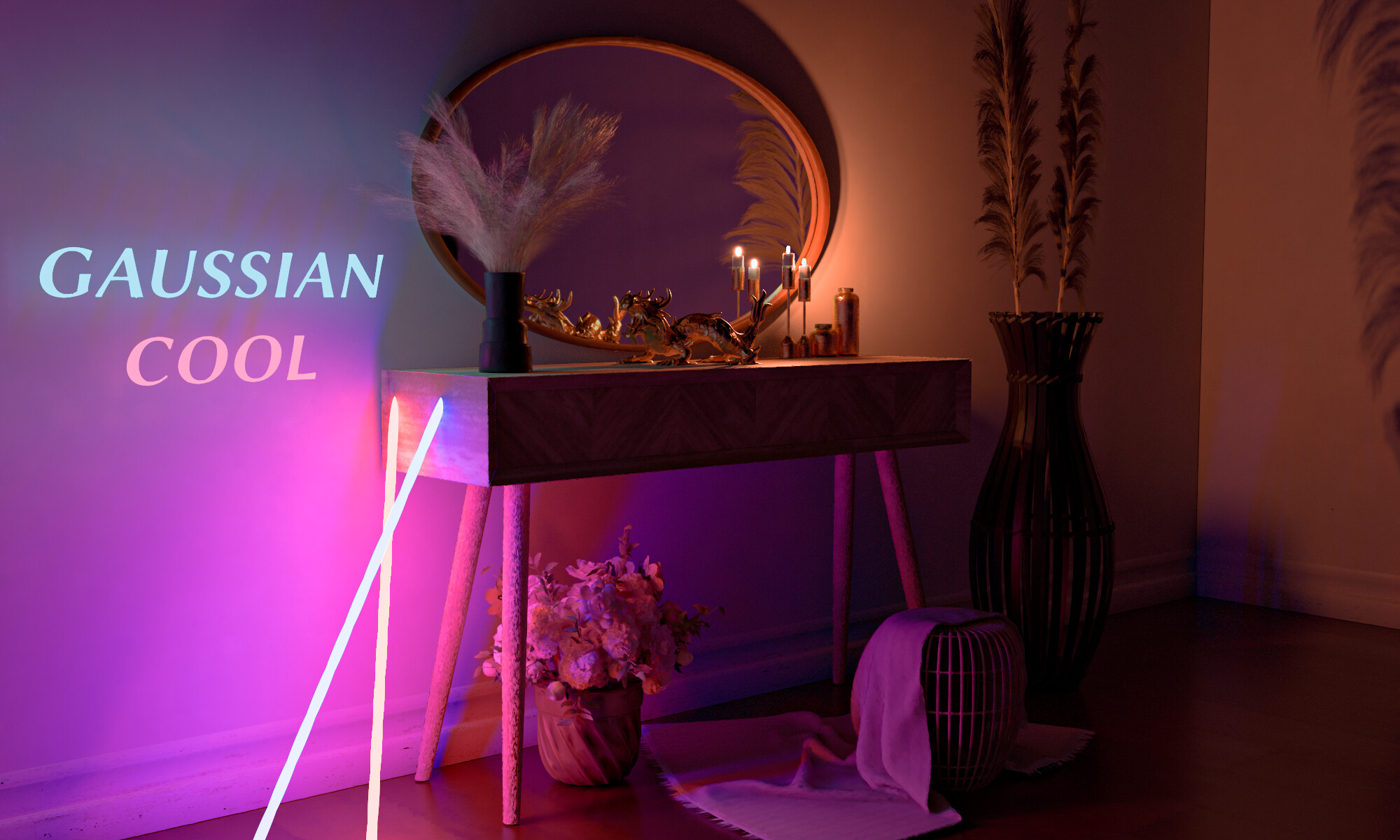}		
	\end{tabular}	
	\caption{\label{fig:teaser}
        Our Gaussian primitives are capable of representing complex scenes with global illumination support. This \emph{Dressing Table} scene is modeled
		\emph{entirely} by our primitives (\(\sim\)5.5 million primitives). It features a variety of geometric structures, ranging from hard surfaces to dense,
		unstructured aggregates, as well as different types of materials, such as near-specular, glossy, diffuse, and emissive.
		We show renders with two drastically different lighting configurations and highlight accurate effects including glossy reflections and color bleeding.
		}
\end{teaserfigure}
\begin{abstract}
Searching for a unified scene representation remains a research challenge in computer graphics. 
Traditional mesh-based representations are unsuitable for dense, fuzzy elements and introduce additional complexity for filtering and differentiable rendering. 
Conversely, voxel-based representations struggle to model hard surfaces and high-frequency details. 
We propose a general-purpose rendering primitive based on 3D Gaussian distributions for unified scene representation, 
featuring versatile appearance ranging from glossy surfaces to fuzzy elements, as well as physically based scattering to enable accurate global illumination. 
We formulate the rendering theory for the primitive based on non-exponential transport and derive efficient rendering operations to 
be compatible with Monte Carlo path tracing. 
The new representation can be converted from different sources, including meshes and 3D Gaussian splatting, and further refined via transmittance optimization 
thanks to its differentiability. We demonstrate the versatility of our representation in various rendering applications such as global illumination and appearance editing, 
while naturally supporting arbitrary lighting conditions. 
With suitable simplification, we further adapt our method to radiance field reconstruction and rendering. We conduct comprehensive comparisons of our 
representation with existing scene representations, highlighting its efficiency in capturing details and representing aggregate elements.
\end{abstract}

\maketitle

\section{Introduction} \label{sec:introduction}
Scene representation is a fundamental building block of computer graphics. It determines the types of content that can be expressed, the compatible rendering
algorithms, and ultimately influences the quality of the rendering. Throughout the development of computer graphics, numerous scene representations have been
proposed, yet achieving a unified representation that reconciles both surfaces and volumes remains a challenging problem. Polygon meshes with textures are the
most common representation, supported by mature hardware rendering acceleration for both rasterization and ray tracing. Meshes are well-suited for modeling
connected hard surfaces but struggle to represent dense, fine elements that exist in nature, such as hair, fur, grains, and foliage. Additionally, the discrete
characteristic of meshes poses challenges for essential graphics applications such as level-of-detail~\citep{vicini2021non, bako2023deep, weier2023neural, tog/ZhouHRSY25} and
differentiable rendering~\citep{zhao2020physics}. Conversely, volumes model
3D objects as fields of microscopic scatterers, excelling at representing aggregated content but are less effective for hard surfaces. 
Traditionally, volumes are discretized as (and interpolated between) voxels. They suffer from prohibitive memory costs when stored densely.
Although sparse storage~\citep{tog/Museth13, si3d/CrassinNLE09} alleviates this issue, voxels remain poorly suited for capturing high-frequency 
details due to their nature of axis-aligned spatial subdivision.
The recent success of 3D Gaussian splatting (3DGS)~\citep{kerbl20233d} suggests using anisotropic 3D Gaussian mixtures for fitting,
achieving superior reconstruction quality. However, 3DGS remains an incomplete scene representation as it only records radiance fields.

The adaptability of Gaussians to complex geometry, as demonstrated by 3DGS, inspires us to develop a Gaussian-based rendering primitive not
just for radiance fields, but for general-purpose light transport and appearance modeling. 
Crucially, we design our Gaussian primitive to be \emph{atomic}, an indivisible entity, and we do not explicitly simulate sub-primitive random walks.
A parallel can be drawn between our Gaussian primitive and a triangle as they are both the basic building block for the respective scene representations.
Each Gaussian primitive can be assigned with its own \emph{per-primitive} phase function, enabling meaningful appearance modeling and authoring.
This approach contrasts with defining an exponential volume with spatially varying extinction coefficient as a sum of Gaussians, where sub-Gaussian multiple 
scattering is required to resolve appearance. 
We further combine the Gaussians with a non-exponential, linear transmittance model~\citep{vicini2021non} to better adapt to both hard surfaces and
fuzzy, aggregated elements. 
Using our primitives, a scene becomes a novel kind of heterogeneous, non-exponential volume. We derive efficient Monte Carlo operations,
such as free-flight distribution sampling, for the volume to be rendered by Monte Carlo path tracing with full global illumination.

Our primary goal in this work is to propose a novel unified scene representation capable of expressing rich appearance and compatible with Monte Carlo path
tracing. We demonstrate the advantages of our representation via several rendering applications, including global illumination and appearance editing.
Recognizing that a new representation inherently lacks data or content compared to more mature counterparts, we provide methods to convert other popular representations
into our representation. As proof-of-concept differentiable rendering applications, we first showcase gradient-based transmittance optimization. 
With suitable simplification, we further demonstrate image-based radiance field reconstruction as a second application.
We show that our representation can be made backward compatible with pure radiance fields.
While our framework can potentially support an end-to-end inverse rendering pipeline, this is \emph{not} the focus of this work.

To summarize, our contributions include:
\begin{itemize}[leftmargin=*]
    \item A novel 3D Gaussian-based volumetric rendering primitive that can handle both hard surfaces and aggregated elements.
    \item Efficient Monte Carlo operations for our non-exponential heterogeneous scene representation that enables full path tracing.
    \item A flexible phase function that incorporates both aggregated geometric configurations and reflectance properties of each surface element.
    \item Versatile forward rendering applications including global illumination.
    \item Differentiable rendering applications including transmittance optimization and radiance field reconstruction.
\end{itemize}
\section{Related Work} \label{sec:related_work}
Scene representation is a fundamental and long-standing problem in computer graphics. Our work draws inspiration from volumetric rendering, 
point-based graphics, and the recent advancements in 3D Gaussian-based representations. In the following, we survey key related works in these fields.

\paragraph{Volumetric Light Transport}
Light transport simulation in participating media is based on the theory of radiative transfer~\citep{chandrasekhar1960radiative} and involves solving the 
radiative transfer equation (RTE). Due to its recursive nature, Monte Carlo integration is required to solve the RTE unbiasedly. Extensive studies have been 
conducted in compute graphics for efficient Monte Carlo techniques, and we refer readers to~\citet{novak2018monte} for a comprehensive review. The original RTE 
only models participating media consisting of isotropic, independently distributed microscopic scatterers. It is extended by the microflake 
theory~\citep{jakob2010radiative} to handle anisotropic scatterers, and by non-exponential transport~\citep{bitterli2018radiative,jarabo2018radiative} to 
model the spatial correlation between the scatterers. These extensions to the original RTE greatly broaden the capability of volumes to represent diverse 
objects, resulting in more versatile scene representations. 
Our method takes inspiration from both the microflake theory and the non-exponential transport study, and adapts them to establish a novel and practical 
Gaussian-based scene representation.

\paragraph{Volumetric Scene Representations}
Using volumes to represent complex geometry has been explored extensively since first introduced by \citet{kajiya1989rendering}. Volumes are traditionally 
used to approximate the rendering of dense, unstructured geometries such as fur, hair, and foliage~\citep{neyret1998modeling, decaudin2009volumetric, 
koniaris2014survey, moon2008efficient}. The microflake theory has extended volumetric representations to model fabric and cloth~\citep{zhao2011building, 
zhao2012structure, khungurn2015matching}.
Given that high-resolution volumes can be very memory-intensive, several works address the challenge of downsampling microflake volumes while preserving the 
self-shadowing effect~\citep{zhao2016downsampling,loubet2018new}. In granular material rendering, explicit grain instances are switched to volumes to achieve 
acceleration~\citep{moon2007rendering,meng15granular,muller16efficient,zhang2020multi}. Non-exponential transport has inspired studies on unified representations 
that support both opaque surfaces and volumes~\citep{vicini2021non, bako2023deep, weier2023neural}.
Our work follows the theme of abstracting complex geometry as a volume, achieving this through a novel transmittance model and phase function.

\paragraph{Neural Implicit Representations}
The seminal work of neural radiance field (NeRF)~\citep{mildenhall2020nerf} has popularized implicit neural field as an effective tool for capturing 3D 
objects~\citep{martel2021acorn, barron2022mip, muller2022instant}. Compared to traditional voxel discretization, neural fields can better reconstruct fine 
details, albeit with the added cost of training and extra inference. However, radiance fields only record the outgoing radiance under fixed illumination at 
capture time, limiting their interaction with different lighting conditions at render time. Various extensions have been proposed to predict simple material 
parameters and reflectance~\citep{bi2020neural, srinivasan2021nerv, jin2023tensoir, boss2021nerd, boss2021neural, zhang2021nerfactor, 
zheng2021neural, lyu2022neural, zhang2023nemf, zeng2023relighting}, but most are significantly constrained in simulating light transport and global illumination 
effects. Instead, we propose a general-purpose primitive for scene representation and forward rendering. 
While our work does not aim to solve the end-to-end inverse rendering problem, we demonstrate its differentiability by several meaningful differentiable rendering 
applications.

\paragraph{Point-based Graphics}
A classical family of modeling and rendering techniques uses point primitives~\citep{alexa2004point, kobbelt2004survey}. A scene is modeled by small, 
unstructured point-like primitives such as disks~\citep{pfister2000surfels} or Gaussians~\citep{zwicker2001surface, zwicker2001ewa}. Rendering of point 
primitives involves projecting them to screen space and perform proper reconstruction filtering (``splatting'') to avoid holes and aliases. More recent work
 explores the differentiability of point primitives for inverse rendering tasks \citep{yifan2019differentiable, lassner2021pulsar}. Additionally, point 
 primitives are used as proxy geometry or irradiance cache for real-time global illumination~\citep{ritschel2008imperfect, ritschel2009micro, wright2022lumen, 
 brinck2021surfels}.

\paragraph{3D Gaussian-based Representations}
Recently, \citet{kerbl20233d} develop 3D Gaussian splatting (3DGS) that extends the EWA volume splatting framework \citep{zwicker2001ewa} to be 
differentiable and uses 3D Gaussians to optimize and render radiance fields. 3DGS achieves state-of-the-art reconstruction quality and offers significantly 
faster rendering speed compared to previous NeRF approaches. Since its debut, 3DGS has inspired a number of Gaussian-based representations with different 
focuses, such as for level-of-detail rendering~\citep{kerbl2024hierarchical}, mesh reconstruction~\citep{Huang2DGS2024, guedon2023sugar}, avatar 
rendering~\citep{saito2023rgca}, inverse rendering~\citep{gao2023relightable}, and high-dimensional function fitting~\citep{diolatzis2024n}. While not modeling 
the full light transport, 3DGS demonstrates the effectiveness of anisotropic Gaussians in adapting to complex geometries, especially thin structures.
3DGS relies on an approximate global depth sort operation that introduces inaccuracies such as popping. Recent works explore more accurate rendering methods 
than splatting, such as hierarchical rasterization~\citep{radl2024stopthepop} or ray tracing~\citep{moenne20243d}. However, they are either limited to 
primary intersection or Whitted-style ray tracing. Our method supports Monte Carlo path tracing with arbitrary bounces and non-specular appearance. 
\citet{tog/CondorSBBGDJ25} propose to ray trace Gaussian and other particle primitives for rendering scattering and emissive media. Their work is 
similar to ours in the sense of supporting volumetric light transport. However, their method is limited to cloud (smoke)-like appearance with simple phase functions.
\citet{tog/JiangSLWLR25} combine Gaussian surfels and radiosity theory for differentiable light transport, but their method inherits the limitations of radiosity in 
modeling high-frequency appearance and light transport effects.
\section{Preliminaries}

\subsection{3D Gaussian-based Representations}
A scaled 3D Gaussian distribution is defined as
\begin{equation}
    G(x; c, \mu, \Sigma) = \frac{c}{(2\pi)^{\frac{3}{2}} |\Sigma|^{\frac{1}{2}}} \exp \Big(-\frac{1}{2}(x-\mu)\trans \Sigma^{-1} (x-\mu) \Big),
\end{equation}
where $\mu$ is the mean, $\Sigma$ is the covariance matrix, and $c$ is the magnitude. The covariance matrix can be decomposed into a rotation matrix $R$ and a 
scale matrix $S$:
\begin{equation} \label{eq:covariance_decomposition}
    \Sigma = (RS) (RS)\trans.
\end{equation}
Intuitively, $\mu$, $R$, and $S$ form an affine transform that transforms an isotropic Gaussian distribution centered at the origin to an anisotropic one 
centered at $\mu$. 

\subsection{Volumetric Light Transport} \label{subsec:vol_light_transport}
In its integral form, the radiative transfer equation (RTE) \citep{chandrasekhar1960radiative} defines the outgoing radiance as 
a recursive integral over the distance a ray traveled within the volume
\begin{alignat}{3} \label{eq:rte}
    &L(x,\omega)    \,&&= &&\,\, \int_0^b \mathcal{P}(x, x_t) L_s(x_t, \omega) \, \D t + \mathcal{T}(x, x_b)L_b(x_b, \omega), \notag\\
    &L_s(x, \omega) \,&&= &&\,\, L_e(x, \omega) + \int_{\mathbb{S}^2} f_p(\omega_i, \omega; x) L(x, \omega_i) \D\omega_i,
\end{alignat}
where $x_t = x + t\omega$, $b$ is the distance to the closest external boundary surface or infinity if none,
$\mathcal{P}$ is the free-flight distribution, $\mathcal{T}$ is the transmittance function, 
$L_b$ is the external emission from either the boundary surface or free space, and 
$L_s$ is the source term. $\mathcal{P}$ and $\mathcal{T}$ are interdependent as the former is a probability distribution function (PDF), and the latter 
is one minus the corresponding cumulative distribution function (CDF):
\begin{equation*}
    \mathcal{T}(x, x_t) = 1 - \int_0^t \mathcal{P}(x_{t'}) \D{t'} \Leftrightarrow \mathcal{P}(x, x_t) = -\frac{\partial}{\partial t} \mathcal{T}(x, x_t).
\end{equation*}
The source term $L_s$ consists of self-emission $L_e$ and the in-scattering term, which is the inner product of the phase function $f_p$ and the (recursive) 
incident radiance. Note that we have factored absorption into the phase function, similar to the formulation by \citet{zhao2016downsampling}.

Traditional volumetric representations are modeled as microscopic scatterers that are independently distributed in 3D space, leading to an exponential 
free-flight distribution and transmittance function:
\begin{align} \label{eq:exp_ff}
    \mathcal{T}(x, x_t) &= \exp \Big(-\int_0^t \sigma_{\mathrm{t}} (x_{t'}) \D{t'} \Big), \\ 
    \mathcal{P}(x, x_t) &= -\frac{\partial}{\partial t} \mathcal{T}(x, x_t) = \sigma_{\mathrm{t}}(x_t) \exp \Big(-\int_0^t \sigma_{\mathrm{t}} (x_{t'}) \D{t'} \Big),
\end{align} 
where $\sigma_{\mathrm{t}}$ is the spatially varying extinction coefficient, which intuitively controls the density of the volume. 
Note that $\mathcal{T}$ and $\mathcal{P}$ only differ by $\sigma_{\mathrm{t}}$. This is coincidentally due to the unique property of the exponential function 
being invariant under differentiation. 

\subsection{Non-exponential Transport}
Non-exponential transport has been introduced to model the spatial correlation in participating media and thus enhance the expressiveness of volumetric 
representations \citep{jarabo2018radiative, bitterli2018radiative, vicini2021non}. In non-exponential transport, $\mathcal{T}$ and $\mathcal{P}$ are no longer 
required to be exponential, and thus do not share a similar form.
In particular, \citet{vicini2021non} propose a 
transmittance function that interpolates between exponential and linear transmittance
\vspace{-0.41cm}
\begin{align}
    \mathcal{T}(x, x_t) &= \gamma \exp(-\tau) + (1-\gamma) \max(0, 1 - \frac{1}{2}\tau), \\
    \tau &= \int_0^t \sigma_{\mathrm{t}}(x_t') \D{t'},
\end{align}
where $\gamma$ is the interpolation weight. The $1/2$ factor is applied to ensure that two modes have the same mean free path. \citet{vicini2021non} have 
performed extensive experiments to demonstrate that the linear component reflects the negative correlation exhibited by hard surfaces. This, in turn, helps a 
volumetric representation to better model surface-like objects and reduce artifacts such as leaking.
\section{Linear Transmittance Gaussian Primitives} \label{sec:primitives}
Our goal is to define a general-purpose volumetric rendering primitive based on 3D Gaussian distribution. We begin by analyzing the feasibility and requirements
for defining such a primitive. 3DGS has convincingly demonstrated the advantages of anisotropic 3D Gaussians for adapting to complex shapes. However, to be truly usable in light transport, we need to define how these Gaussians interact with light. This includes the attenuation of light, controlled by the free-flight distribution or transmittance, and the scattering (or absorption) of light, controlled by the appearance or phase function.
For the primitive to be practically valuable in modeling and rendering applications, the following properties are desirable:

\textbf{Prioritization of Opaque Surfaces and Elements.}
Instead of true participating media such as clouds and smoke, we intend to use the primitive to represent the wider range of content consisting of opaque
geometries. Objects may exhibit a spectrum of geometric characteristics from continuous surfaces to dense elements (which in aggregation produce a fuzzy look,
but are nonetheless opaque individually). Exponential transport is not suitable in most of these cases, as the geometries are usually
not independently distributed. Only when the elements are sufficiently decorrelated, the free-flight statistics may approach exponential in the far-field
limit. However, this is also rarely the case: For a scene to reach high fidelity with sharp details, typically a single primitive only represents a surface
patch or a small cluster of oriented elements.
To address this issue, we draw inspiration from previous literature~\citep{bitterli2018radiative, vicini2021non} and incorporate the linear transmittance model to
handle the negative correlation for opaque surfaces. We validate this design choice in \autoref{fig:linear_vs_exp}.

\begin{figure}[tb]
    \newlength{\lenPerPrimitiveIllus}
	\setlength{\lenPerPrimitiveIllus}{\linewidth}
    \centering
    \includegraphics[width=\lenPerPrimitiveIllus]{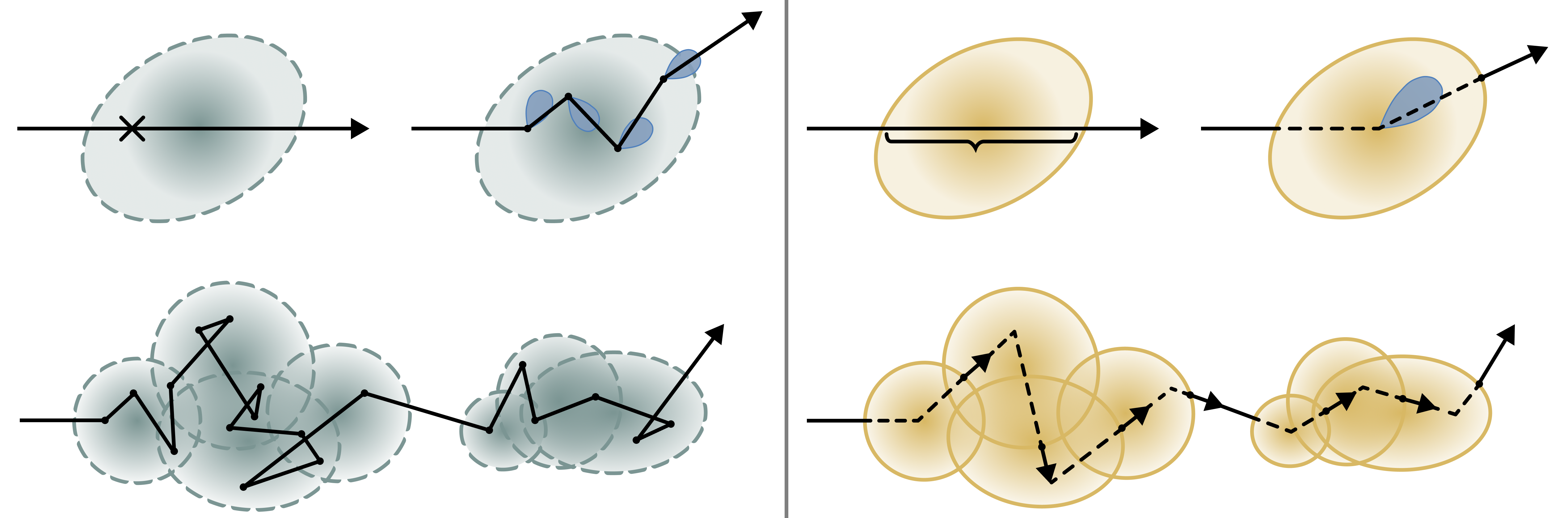}
    \caption{\label{fig:per_prim_illus}
        In conventional volume path tracing, a collision point is sampled at each time and the apparent appearance of a primitive depends on multiple scattering
        (left). Our method samples the entire primitive and uses the \emph{per-primitive} phase function to scatter light once, essentially treating it as an
        indivisible entity.
    }
\end{figure}

\textbf{Intuitive Appearance Definition.}
It is important to define the appearance of the primitive in a way that accommodates intuitive authoring and editing.
Traditionally, the appearance of a volume is specified by the (single-scattering) phase function at each point in space, which governs how light is scattered at
each scattering event.
However, light transport in the volume undergoes multiple scattering, which contributes significantly to the final rendered image. This typically results in more
saturated color and a softer look\footnote[2]{The relationship between the phase function and the apparent color of the volume is non-obvious and
sometimes even ambiguous~\citep{zhao2014high, wrenninge2017path}.}. Multiple scattering is also costly as it requires explicit simulation of random walks inside
each primitive. We argue that such \emph{per-point} appearance definition is unnecessarily complex. In contrast, we propose to define appearance at a
\emph{per-primitive} level: the phase function describes the aggregated behavior for an entire primitive and abstracts away all sub-primitive scattering
interactions. This also necessitates a new kind of discrete free-flight sampling where a primitive is sampled each time, in contrast to the common
continuous free-flight sampling where a collision point is sampled each time. Once a primitive is selected, light is scattered once according to the \emph{per-primitive}
phase function and then exits. Further self-intersections are ignored during this event.
Conceptually, the role of our Gaussian primitive resembles that of a triangle in mesh-based representations: a basic, \emph{atomic} entity for scene authoring
with probabilistic scattering replacing deterministic intersection tests. Multiple scattering between primitives should still be tracked for global illumination.
The conceptual differences between conventional volume path tracing and our formulation are illustrated in \autoref{fig:per_prim_illus}.

\begin{figure}[tb]
    \newlength{\lenFFIllustration}
	\setlength{\lenFFIllustration}{\linewidth}
    \centering
    \includegraphics[width=\lenFFIllustration]{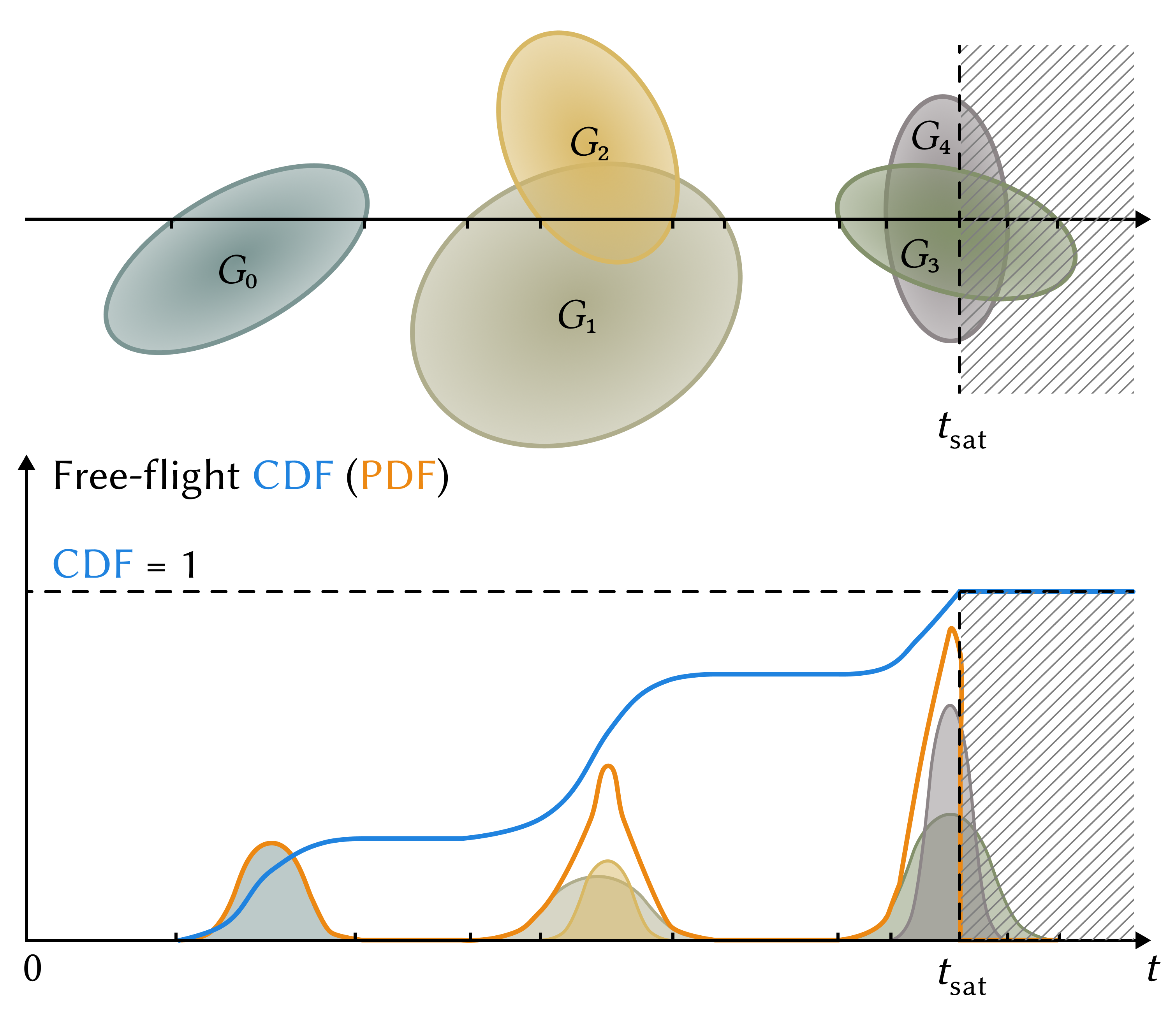}
    \caption{\label{fig:ff_illustration}
    2D flatland illustration of a ray traveling through a heterogeneous volume consisting of 5 Gaussian primitives (top) and the corresponding free-flight CDF
    and PDF (bottom). The CDF is accumulated from $0$ to $1$, where it reaches the saturating distance $t_{\mathrm{sat}}$. The ray is effectively ``blocked''
    at this point, and the PDF is 0 afterwards.
    }
\end{figure}

To meet these properties, we propose a novel kind of heterogeneous, non-exponential volume by combining the merits of Gaussians and the linear transmittance
function from \citet{vicini2021non}. The transmittance and the corresponding free-flight distribution are defined as follows:
\begin{align}
    \mathcal{T}(x, x_t) &= \max \Big(0, 1 - \frac{1}{2}\int_0^t \Big( \sum_k G_k(x_{t'}) \Big)  \D{t'} \Big), \label{eq:transmittance} \\
    \mathcal{P}(x, x_t) &= -\frac{\partial}{\partial t} \mathcal{T}(x, x_t) =
        \begin{cases}
            \frac{1}{2}\sum_k G_k(x_t), & \text{if } t \leq t_{\mathrm{sat}}, \\
            0, & \text{otherwise},
        \end{cases} \label{eq:free_flight_pdf}
\end{align}
where the \emph{saturating distance} $t_{\mathrm{sat}}$ is the ray travel distance such that $\mathcal{T}(x, x_{t_{\mathrm{sat}}}) = 0$.
\autoref{fig:ff_illustration} illustrates the above definitions in 2D flatland.
It is clear from \autoref{eq:free_flight_pdf} that the free-flight PDF is additive: the total free-flight PDF is simply the sum of the PDFs of individual primitives.
This implies that when integrating \autoref{eq:rte}, 
we can sample an individual primitive and compute its contribution separately from other primitives, a desirable property that simplifies the integration process\footnote[1]{
    It is possible to achieve similar behavior for exponential volumes via analog decomposition tracking (\autoref{sec:decomp_track}).}.
The linear transmittance improves the ability to model negatively correlated opaque geometries~\citep{vicini2021non}.
On the other hand, it does not compromise the ability to model objects with more stochastic features such as foliage. Indeed, as we will demonstrate in
\autoref{subsec:transmittance_optim}, with suitable optimization, our final heterogeneous transmittance can fit these objects well. In \autoref{fig:linear_vs_exp},
we will also validate that the linear transmittance model is overall superior to the exponential model for different variants of opaque geometries.

\subsection{Primitive Operations}
3D Gaussians support several key operations for volumetric rendering that are either in closed forms or can be computed efficiently. These operations serve as
the building blocks of our scene rendering algorithms, introduced in \autoref{sec:scene_traveral}. We further define the corresponding appearance for a
Gaussian primitive in \autoref{sec:appearance}.

\paragraph{Ray Integral}
Given a ray $x_t = x + t\omega, t \in [t_0, t_1]$ and a Gaussian primitive $G(x)$, we aim to compute the probability of the ray being scattered by the primitive.
With the linear transmittance model, this is essentially the integral of $G(x)$ along $r$ from $t_0$ to $t_1$:
\begin{equation} \label{eq:ray_integral}
    I(t_0, t_1) = \frac{c}{(2\pi)^{\frac{3}{2}} |\Sigma|^{\frac{1}{2}}}  \int_{t_0}^{t_1} \exp \Big(-\frac{1}{2}(x_t-\mu)\trans \Sigma^{-1} (x_t-\mu) \Big) \,\D{t}.
\end{equation}
This integral can be solved in closed form (utilizing the error function $\mathtt{erf}$). Detailed derivation and the final expression are provided in~\autoref{sec:ray_integral_derivation}.

\paragraph{Ray Sampling}
To sample along a ray that intersects a single Gaussian primitive, we can simply invert \autoref{eq:ray_integral}. The inverse
error function $\mathtt{erfinv}$ is standard in mainstream numerical libraries.
We then consider the case when a ray intersects multiple overlapping primitives and solve it again by CDF inversion.
Given a random number $u \in [0, 1)$ and a ray, we seek to find a root $t \in [t_0, t_1]$ for
\begin{equation} \label{eq:sampling_overlaps}
    u = F(t; t_0) = \frac{1}{2}\sum_k I_k(t_0, t).
\end{equation}
Although $F(t)$ cannot be analytically inverted, we observe that $F(t)$ is non-decreasing, making the Newton-Raphson method suitable for solving
\autoref{eq:sampling_overlaps}.
The derivative $\D F / \D t$ is
simply the sum of the evaluations of all primitives. As will be discussed in \autoref{sec:scene_traveral}, our full algorithm can prune away most cases, making such explicit inversion rarely needed. When it is indeed required, we can always guarantee the existence of a unique solution,
and provide a fairly tight initial bracket $[t_0, t_1]$, such that usually only a few iterations are required for convergence.
Alternatively, one could explore other analytic sampling techniques even when the CDF cannot be analytically inverted \citep{heitz2020can}.

\paragraph{Bounding Shapes}
A Gaussian distribution has an infinite support in 3D space. In practice, we would like to truncate its contribution at a certain extent to give it a finite size,
thereby accelerating intersection tests. We first determine the ellipsoidal isosurface where the distribution evaluates to less than a threshold of the peak:
\begin{equation} \label{eq:cutoff_threshold}
    \frac{G(x; c, \mu, \Sigma)}{G(\mu; c, \mu, \Sigma)} = \epsilon \Leftrightarrow  \Big\lVert \sqrt{\frac{-1}{2 \ln \epsilon}}(RS)^{-1}(x-\mu) \Big\rVert = 1,
\end{equation}
where we utilize \autoref{eq:covariance_decomposition}. Here, $\epsilon$ is usually set to $0.01$, and any contribution outside the ellipsoid is discarded.
We can further calculate the bounding box of the ellipsoid to be used by intersection acceleration structures.

\section{Scene Traversal and Rendering Operations} \label{sec:scene_traveral}

\begin{table*}[tb]
	\centering
	\caption{ \label{tab:highlevel}
        A side-by-side pseudocode comparison of the high-level rendering algorithm for a mesh scene (left) and for 
        our representation (right).
    }
	\begin{tabular}{p{.48\textwidth}p{.48\textwidth}}
    \multicolumn{1}{c}{\small{\textsf{Standard Surface Path Tracer}}} & \multicolumn{1}{c}{\small{\textsf{Our Volumetric Path Tracer}}} \\
      \begin{pseudocode}[label={list:pseudocode_mesh_pt}]
L, beta, pdf_prev = 0, 1, 0
for bounce in range(max_bounces + 1):
    hit = (*@\textbf{\textcolor{solarized_orange}{intersect}}@*)(ray)
    L += beta * unidirectional_Le(hit, ray, pdf_prev)
    if not hit: break
    light_dir, Le, pdf_light = sample_light(hit.p)
    v   = (*@\textbf{\textcolor{solarized_orange}{visibility}}@*)(hit.p, light_dir)
    w = mis(pdf_light, hit.(*@\textbf{\textcolor{solarized_orange}{bsdf}}@*).pdf(-ray.dir, light_dir))
    L  += beta * hit.(*@\textbf{\textcolor{solarized_orange}{bsdf}}@*).eval(-ray.dir, light_dir) * v * w * Le / pdf_light
    wi, f_over_pdf, pdf_prev = hit.(*@\textbf{\textcolor{solarized_orange}{bsdf}}@*).sample(-ray.dir)
    beta *= f_over_pdf;  ray = spawn(hit.p, wi)        
      \end{pseudocode}
      & 
      \begin{pseudocode}[label={list:pseudocode_gsr_pt}]
L, beta, pdf_prev = 0, 1, 0
for bounce in range(max_bounces + 1):
    hit = (*@\textbf{\textcolor{solarized_blue}{sample\_free\_flight}}@*)(ray)
    L += beta * unidirectional_Le(hit, ray, pdf_prev)
    if not hit: break
    light_dir, Le, pdf_light = sample_light(hit.p)
    tr  = (*@\textbf{\textcolor{solarized_blue}{transmittance}}@*)(hit.p, light_dir)
    w = mis(pdf_light, hit.(*@\textbf{\textcolor{solarized_blue}{phase}}@*).pdf(-ray.dir, light_dir))
    L  += beta * hit.(*@\textbf{\textcolor{solarized_blue}{phase}}@*).eval(-ray.dir, light_dir) * tr * w * Le / pdf_light
    wi, f_over_pdf, pdf_prev = hit.(*@\textbf{\textcolor{solarized_blue}{phase}}@*).sample(-ray.dir)
    beta *= f_over_pdf;  ray = spawn(hit.p, wi)
      \end{pseudocode}
	\end{tabular}
	\label{table:pipeline_overview} 
\end{table*}

\subsection{Overview}


Our representation is designed to integrate seamlessly into Monte Carlo rendering pipelines. Just as our Gaussian 
primitive resembles a triangle, we can also draw parallels between our volumetric path tracer and a standard surface path 
tracer, as shown in \autoref{tab:highlevel}. Both follow the standard path-tracing skeleton with next 
event estimation (NEE); only the highlighted operations differ.

The first set of differences is to replace deterministic visibility sampling/tests with their stochastic variants. As prefaced 
in \autoref{subsec:vol_light_transport}, the two necessary operations are
\emph{free-flight distribution sampling} and \emph{transmittance evaluation}. 
As both operations involve 
traversing the scene along a ray, we utilize the bounding shapes of primitives to build a \emph{kd-tree} acceleration structure. In the following, we describe 
efficient techniques for both operations using the kd-tree. Specifically, our sampling technique relies on the non-overlapping spatial subdivision by the 
kd-tree, which is why we do not use a bounding volume hierarchy that can produce overlapping nodes.

\subsection{Free-flight Distribution Sampling} \label{subsec:free_flight_sampling}
Given a ray $x_t = x + t\omega, t \in [t_0, t_1]$, and a random number $u \in [0, 1)$, traditional Monte Carlo volumetric rendering usually performs free-flight 
distance sampling, where the goal is to generate random samples of ray travel distances $t$ such that $t \nolinebreak \sim \nolinebreak \mathcal{P}(x, x_t)$. 
Distinctively, we aim at sampling discrete primitives $k$ such that $k \nolinebreak \sim \nolinebreak I_k(t_0, t_1)$. We do not need to determine the exact 
scattering locations inside the primitives.

\begin{figure}[tb]
  \newlength{\lenSampleFF}
  \setlength{\lenSampleFF}{0.98\linewidth}
  \addtolength{\tabcolsep}{-4pt}
  \renewcommand{\arraystretch}{0.5}
  \centering
  \begin{tabular}{c}
  \includegraphics[width=\lenSampleFF]{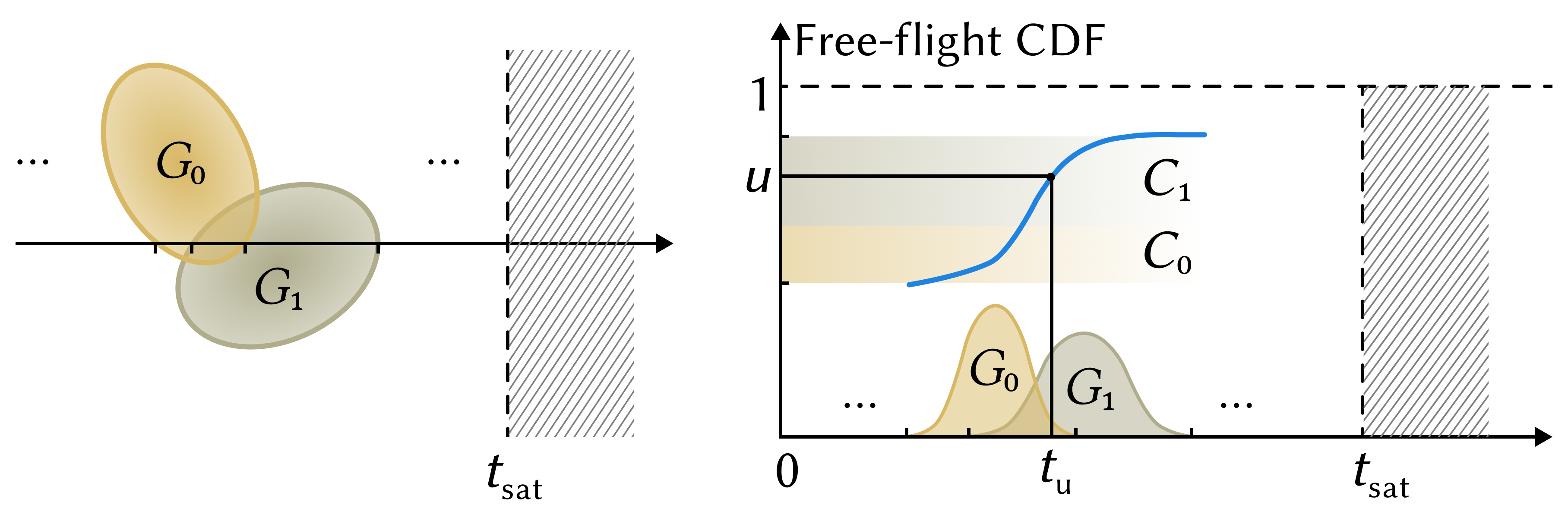} \\ \small{\textsf{(a)}} \\
  \includegraphics[width=\lenSampleFF]{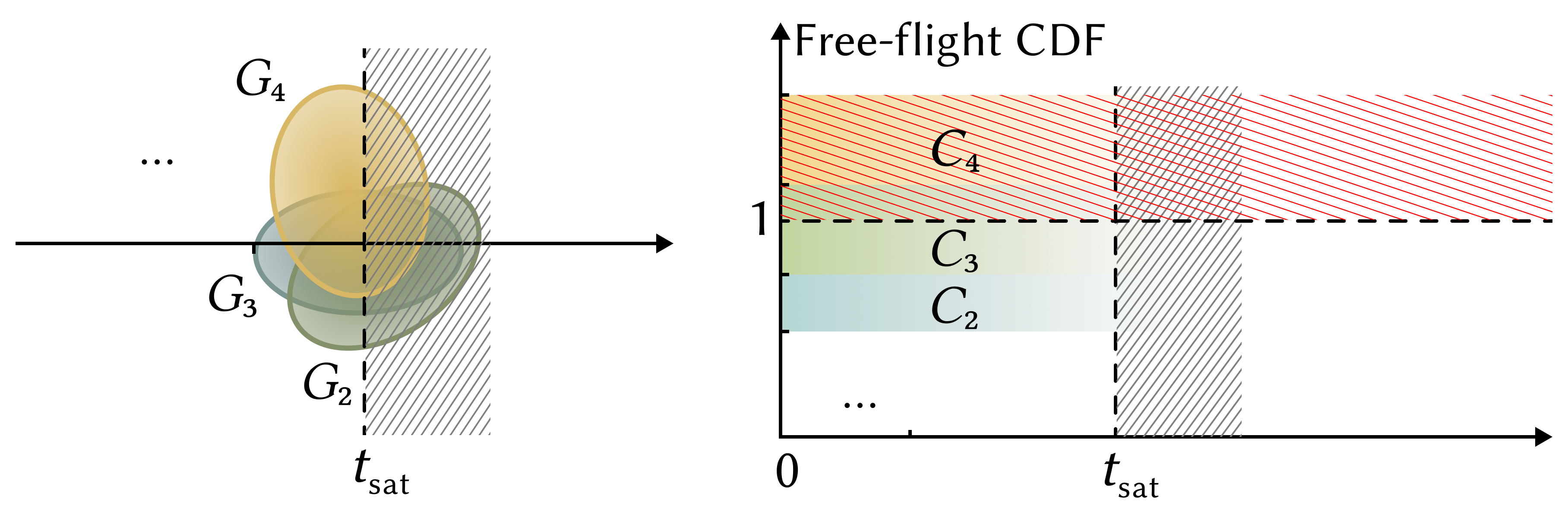} \\ \small{\textsf{(b)}} \\
  \includegraphics[width=\lenSampleFF]{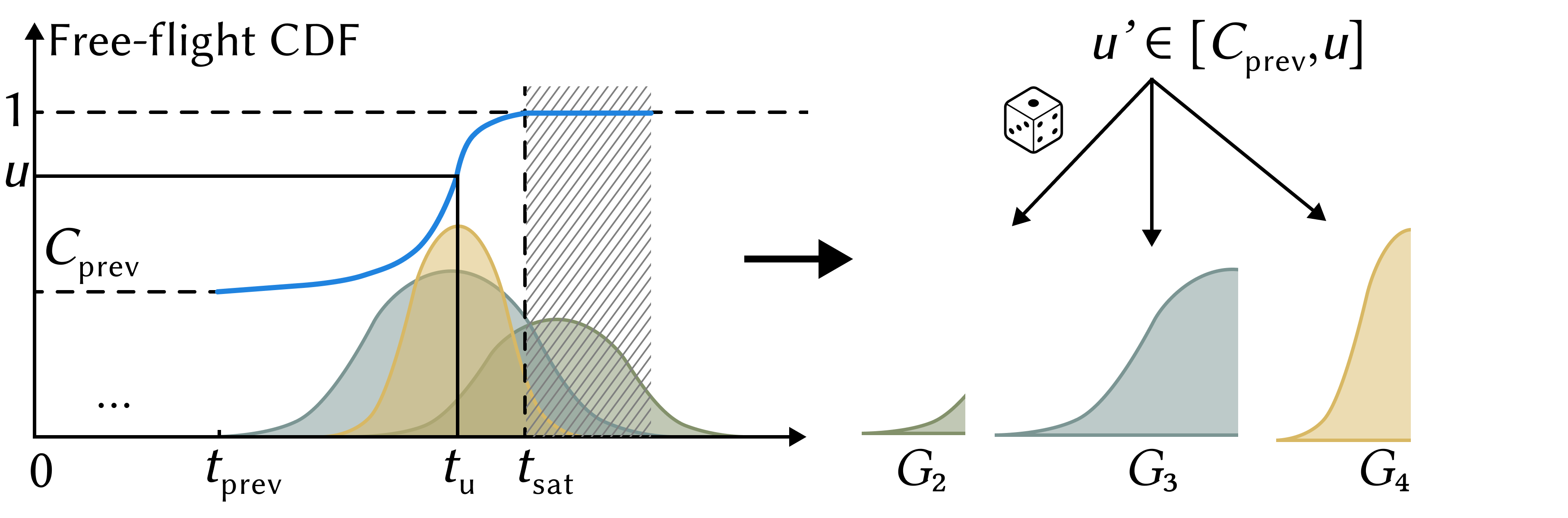} \\ \small{\textsf{(c)}}
  \end{tabular}
  \caption{\label{fig:sample_ff}
    Illustration for free-flight distribution sampling.
    \textbf{(a)} If the primitives do not touch $t = t_{\mathrm{sat}}$, it is safe to invert the free-flight CDF by accumulating the per-primitive contributions 
    and checking the interval $u$ falls within. Overlapped primitives are automatically handled.
    \textbf{(b)} However, when primitives are ambiguous, doing so will skew the free-flight distribution. Here, $G_4$ will never be sampled.  
    \textbf{(c)} Instead, we need to perform a disambiguation step by first solving for $t_u$ and then re-sampling the primitives based on the clamped distributions.
  }
\end{figure}

We achieve this by inverting the heterogeneous CDF of \autoref{eq:free_flight_pdf}, as illustrated in \autoref{fig:sample_ff}a. Thanks to the 
linear transmittance model, this is straightforward for the most part because $\mathcal{P}$ is ``almost'' a linear sum of all involved primitives. We traverse the 
scene along the ray, accumulate the CDF contributed by each visited primitive, and check if the sum reaches $u$. If so, we return the last visited primitive as 
the sample. If the CDF never reaches $u$, the ray reaches free space, and thus we sample the background. 
In fact, \autoref{eq:free_flight_pdf} 
implies that it is not necessary to traverse the primitives in any specific order as long as the traversal does not exceed the saturating distance.

However, there is a catch that lies in the nonlinearity of $\mathcal{P}$ caused by the clamping at the saturating distance $t_{\mathrm{sat}}$. If multiple primitives 
touch the saturating boundary $t = t_{\mathrm{sat}}$, we call them \emph{ambiguous}. This situation is illustrated in \autoref{fig:sample_ff}b. In this case, 
inverting the free-flight CDF by accumulating those primitives one by one results in bias. 
To understand this situation, let $M \coloneq \{ m_1, ..., m_q \}$ be the set of ambiguous primitives in the order of traversal.
Let $C_{\mathrm{prev}}$ be the accumulated CDF prior to visiting $M$, and $t_{\mathrm{prev}}$ be the travel distance so far. 
There exists a particular primitive $m_i$ such that
\begin{equation*}
    C_{\mathrm{prev}} + \frac{1}{2}\sum_{j=1}^{i-1} I_{m_j} (t_{\mathrm{prev}}, t_{\mathrm{sat}}) < 
    1 \leq 
    C_{\mathrm{prev}} + \frac{1}{2}\sum_{j=1}^i I_{m_j} (t_{\mathrm{prev}}, t_{\mathrm{sat}}).
\end{equation*}
It is clear that CDF inversion by sequential accumulation will only consider $\{m_1, ..., m_i\}$ and discard $\{m_{i+1}, ..., m_q\}$, 
thus incorrectly skewing the free-flight distribution.
The correct way to \emph{disambiguate} them, illustrated in \autoref{fig:sample_ff}c, is to first perform ray sampling by \autoref{eq:sampling_overlaps} 
to find the exact distance $t_u$ such that
\begin{equation} \label{eq:disambiguation_solve}
C_{\mathrm{prev}} + \int_{t_{\mathrm{prev}}}^{t_u} \mathcal{P}(x, x_t) \D{t} = u.
\end{equation}
Then, we sample the $j$-th primitive in $M$ proportional to $I_j(t_{\mathrm{prev}}, t_u)$. 
The disambiguation step is not always required for every sampling operation, as it may have already finished before visiting any ambiguous primitive. 
In fact, it is required at most once for each sampling operation.

\begin{figure}[h]
  \newlength{\lenSampleFFLineSweep}
  \setlength{\lenSampleFFLineSweep}{0.82\linewidth}
  \addtolength{\tabcolsep}{-4pt}
  \renewcommand{\arraystretch}{0.5}
  \centering
  \begin{tabular}{c}
  \includegraphics[width=\lenSampleFFLineSweep]{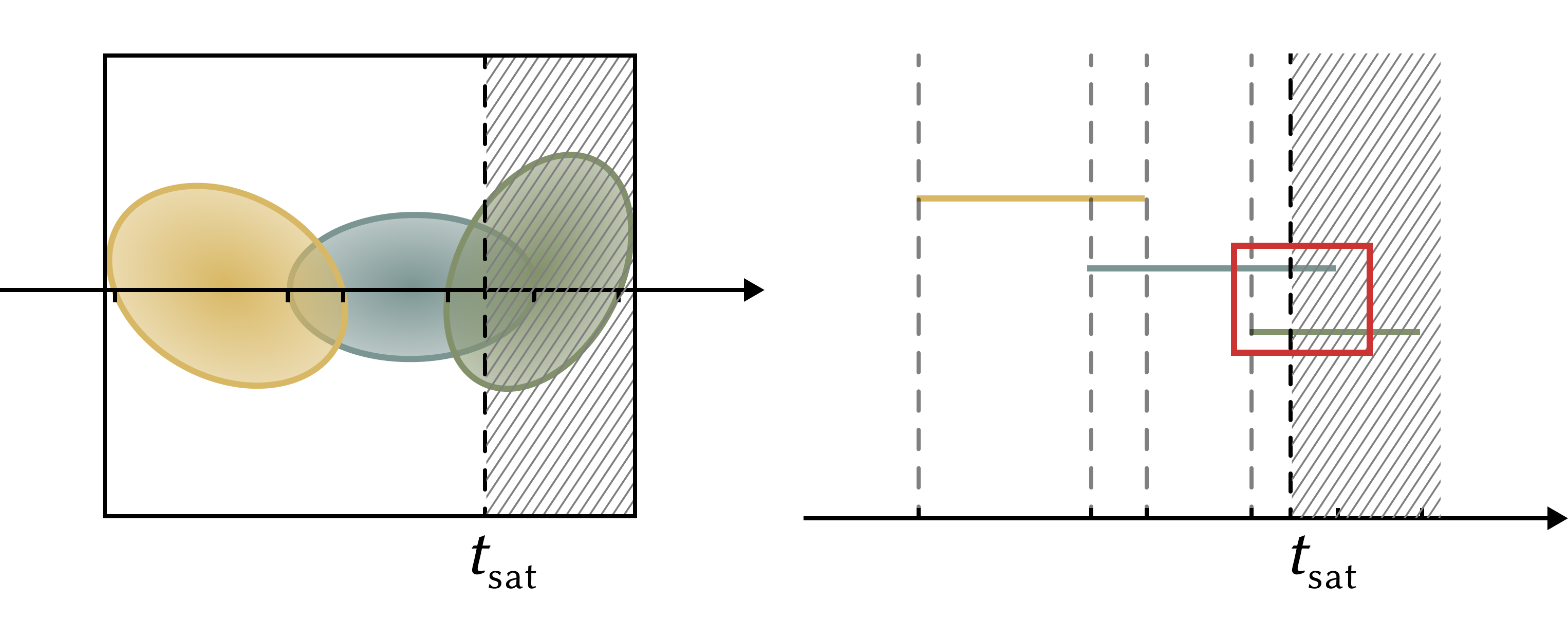}
  \end{tabular}
  \caption{\label{fig:sample_ff_linesweep}
    Even when there are multiple primitives in a leaf node, it is possible that not all of them are ambiguous. We use line sweeping to find the exact ambiguous 
    segment (highlighted by the red box).
  }
\end{figure}

In the full sampling algorithm, a ray traverses the scene using the kd-tree to prune non-intersecting primitives while keeping track of the accumulated CDF. 
At each interior node, we recursively visit its front and back children. This establishes an implicit front-to-back order without explicit sorting, which is 
necessary even if our free-flight PDF is order-independent because we need to avoid tracing behind $t_{\mathrm{sat}}$.
At each leaf node, there are several possible cases:
\begin{enumerate}[leftmargin=*]
    \item There is only one primitive. In this case, it does not matter whether it is ambiguous or not, and we simply perform per-primitive CDF accumulation.
    \item There are multiple non-ambiguous primitives. We perform per-primitive CDF accumulation for each primitive.
    \item There are multiple ambiguous primitives. We need to perform a disambiguation step.
\end{enumerate}
Case (3) can be further optimized by partitioning the leaf node into sub-node segments that consist of different subsets of the primitives in the node, 
as illustrated in \autoref{fig:sample_ff_linesweep}. We can then repeat the above classification on a per-segment level and only an ambiguous segment requires a disambiguation step. 
This further simplifies the convergence of ray sampling.
Here, our strategy depends on the implementation platform. In our CPU implementation, the partition uses the Bentley-Ottmann line sweeping algorithm \citep{o1998computational}.
In our GPU implementation, however, we switch to a simpler binning strategy by partitioning the ray-node intersection interval into a fixed number of bins to avoid dynamic memory allocation. 
Please refer to \autoref{table:fwd_timings} for performance measurements on different platforms.

\begin{figure}[h]
	\newlength{\lenRandomSceneSample}
	\setlength{\lenRandomSceneSample}{0.24\linewidth}
    \addtolength{\tabcolsep}{-4pt}
    \renewcommand{\arraystretch}{0.5}
    \centering
    \begin{tabular}{cccc}
      \frame{\includegraphics[width=\lenRandomSceneSample]{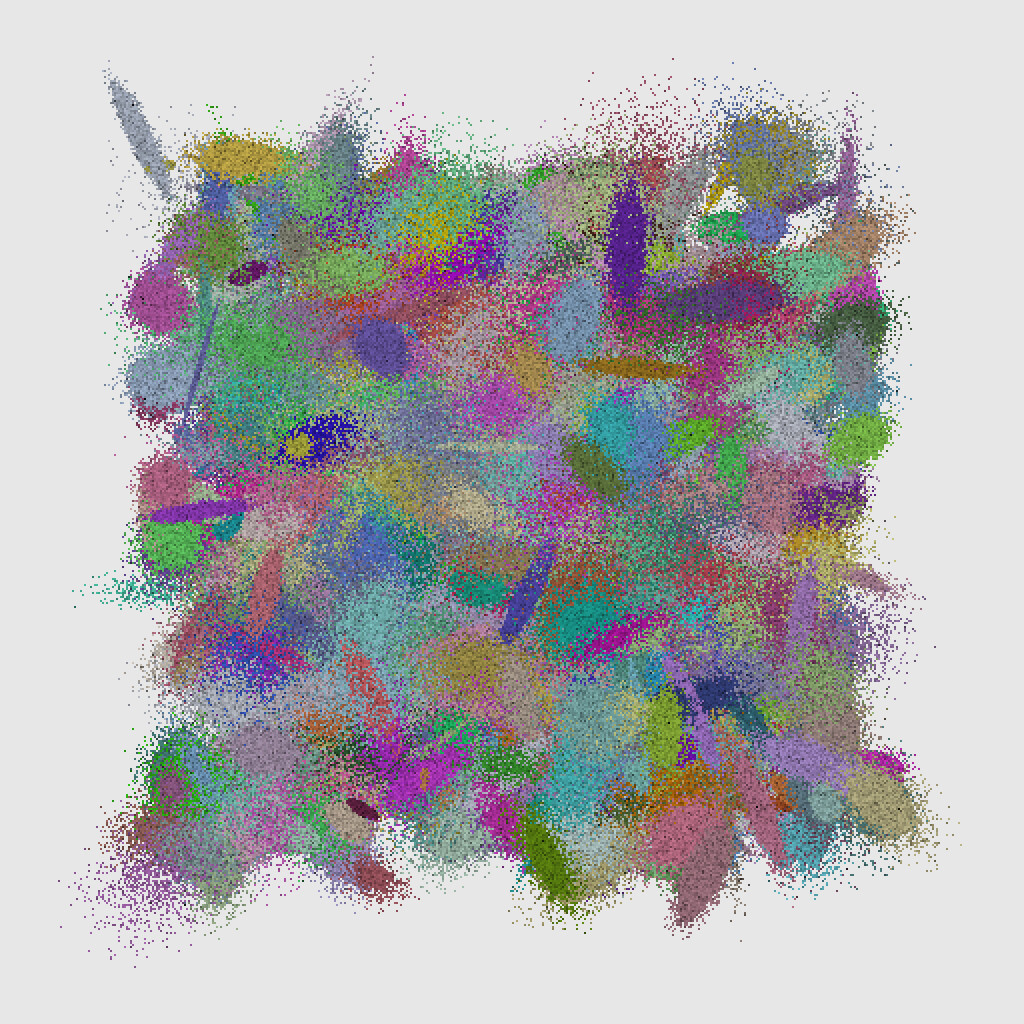}}
      &
      \frame{\includegraphics[width=\lenRandomSceneSample]{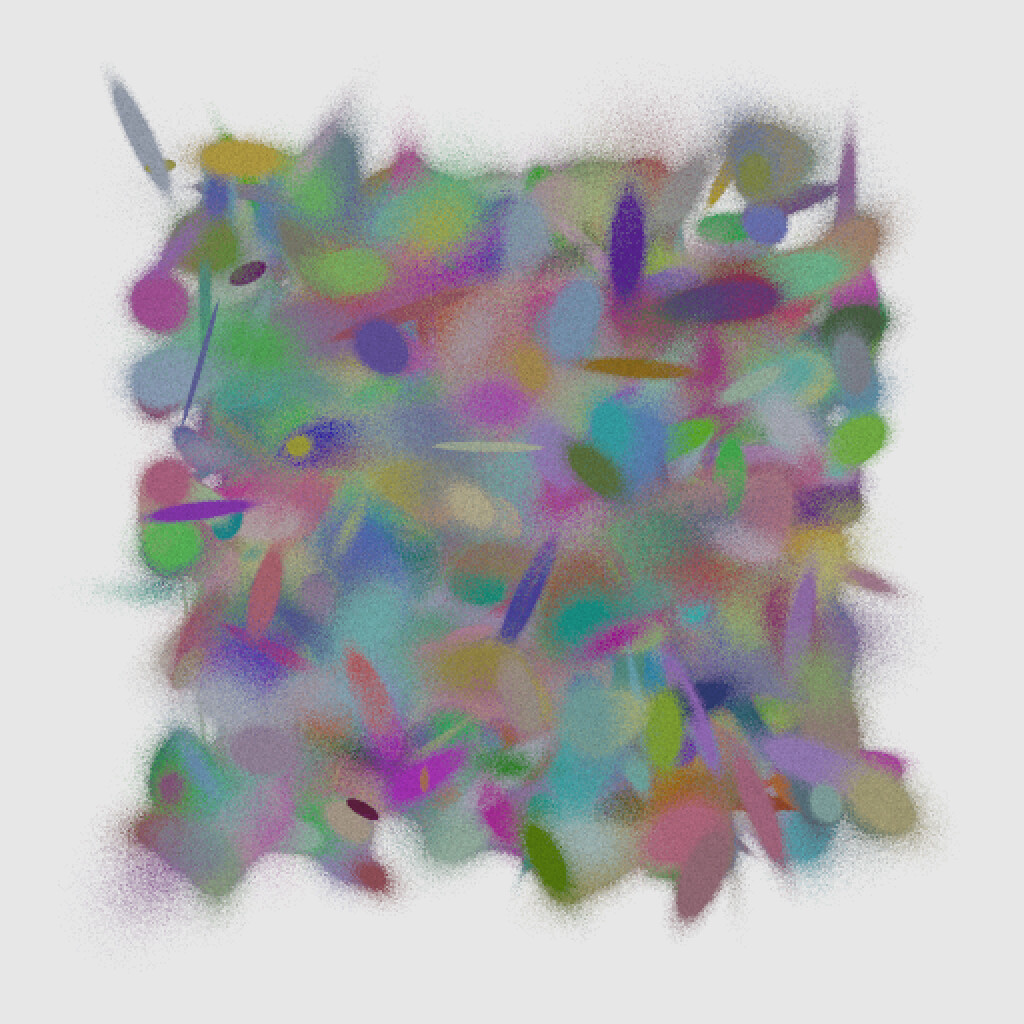}}
      &
      \frame{\includegraphics[width=\lenRandomSceneSample]{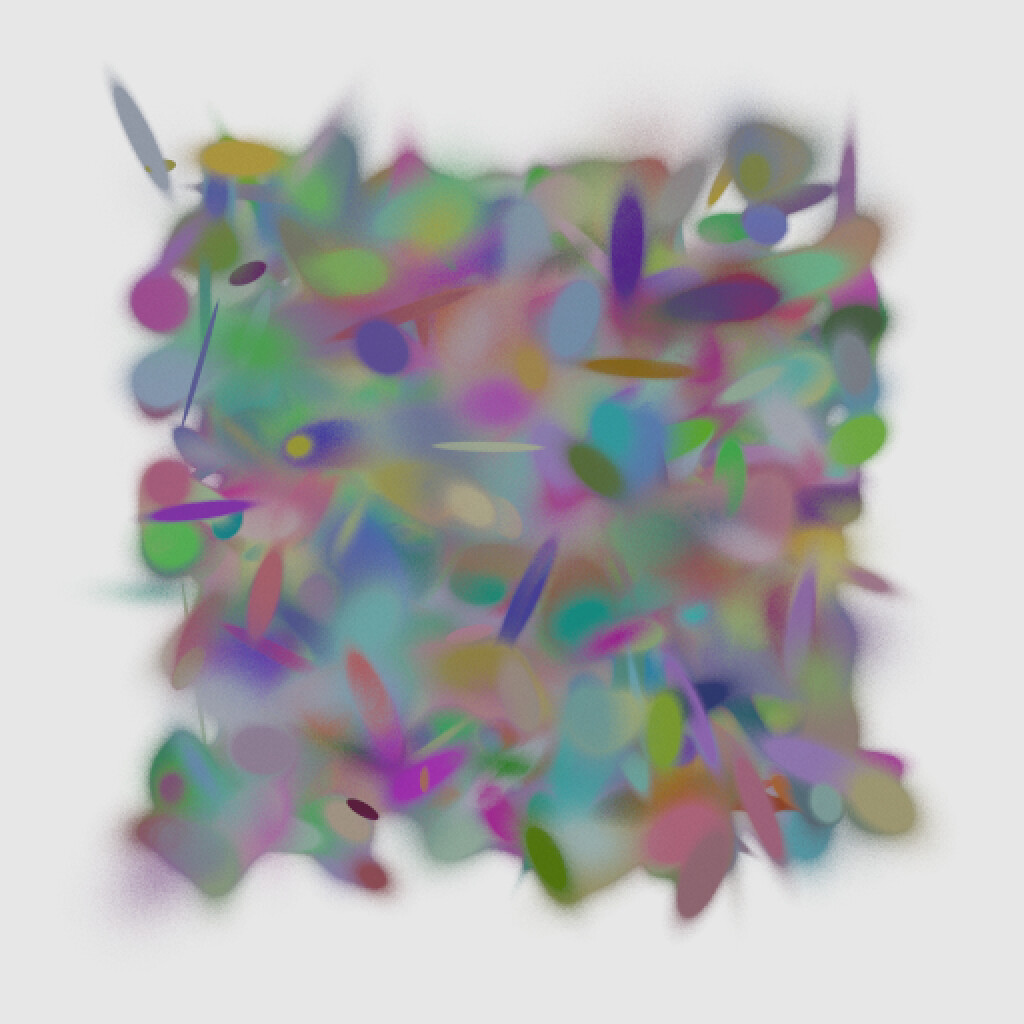}}
      &
      \frame{\includegraphics[width=\lenRandomSceneSample]{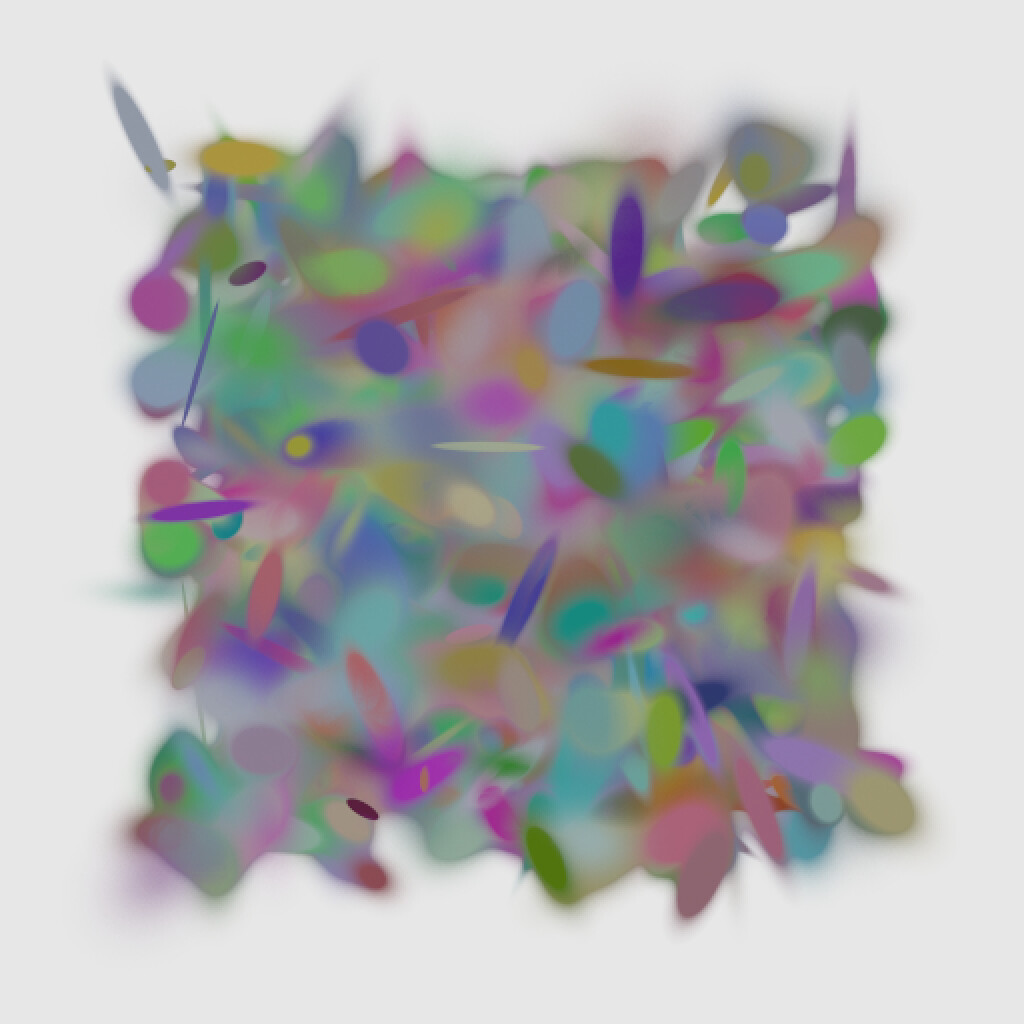}}
      \\
      \frame{\includegraphics[width=\lenRandomSceneSample]{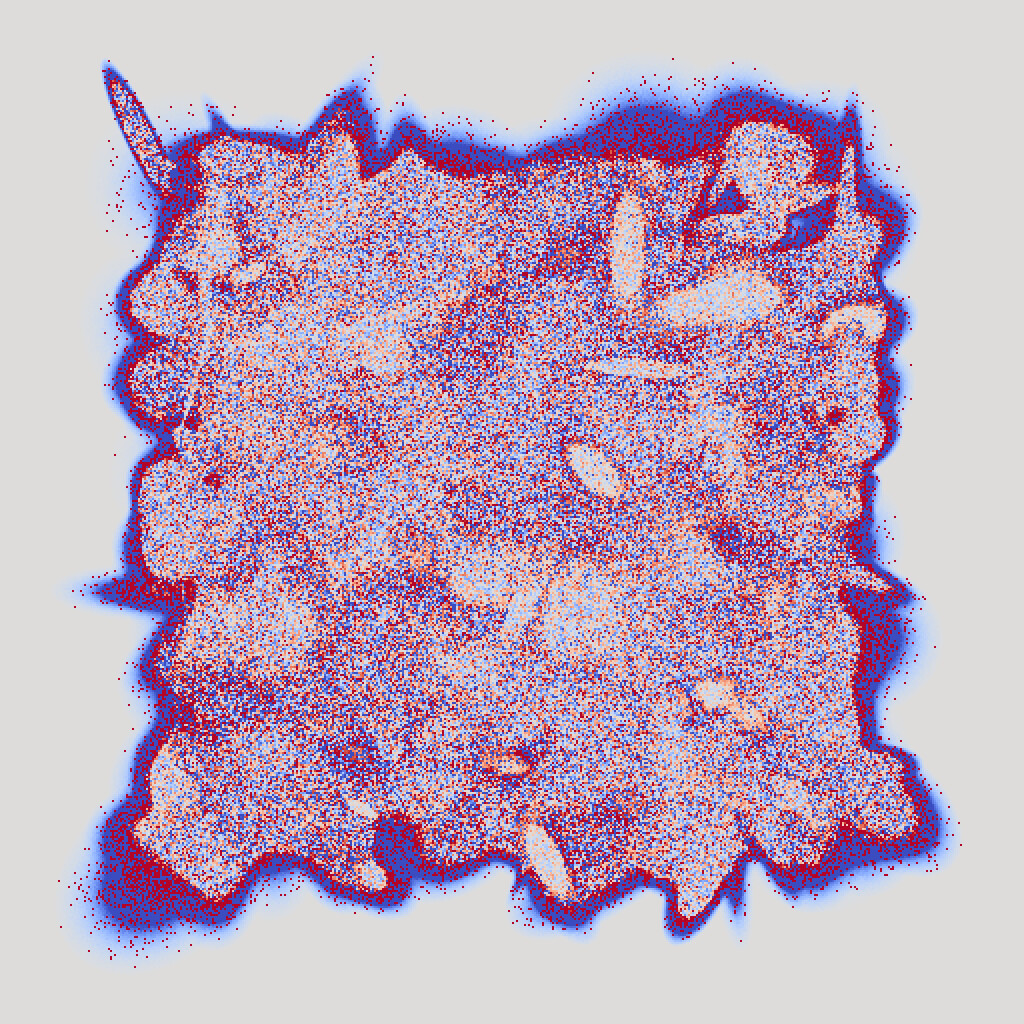}}
      &
      \frame{\includegraphics[width=\lenRandomSceneSample]{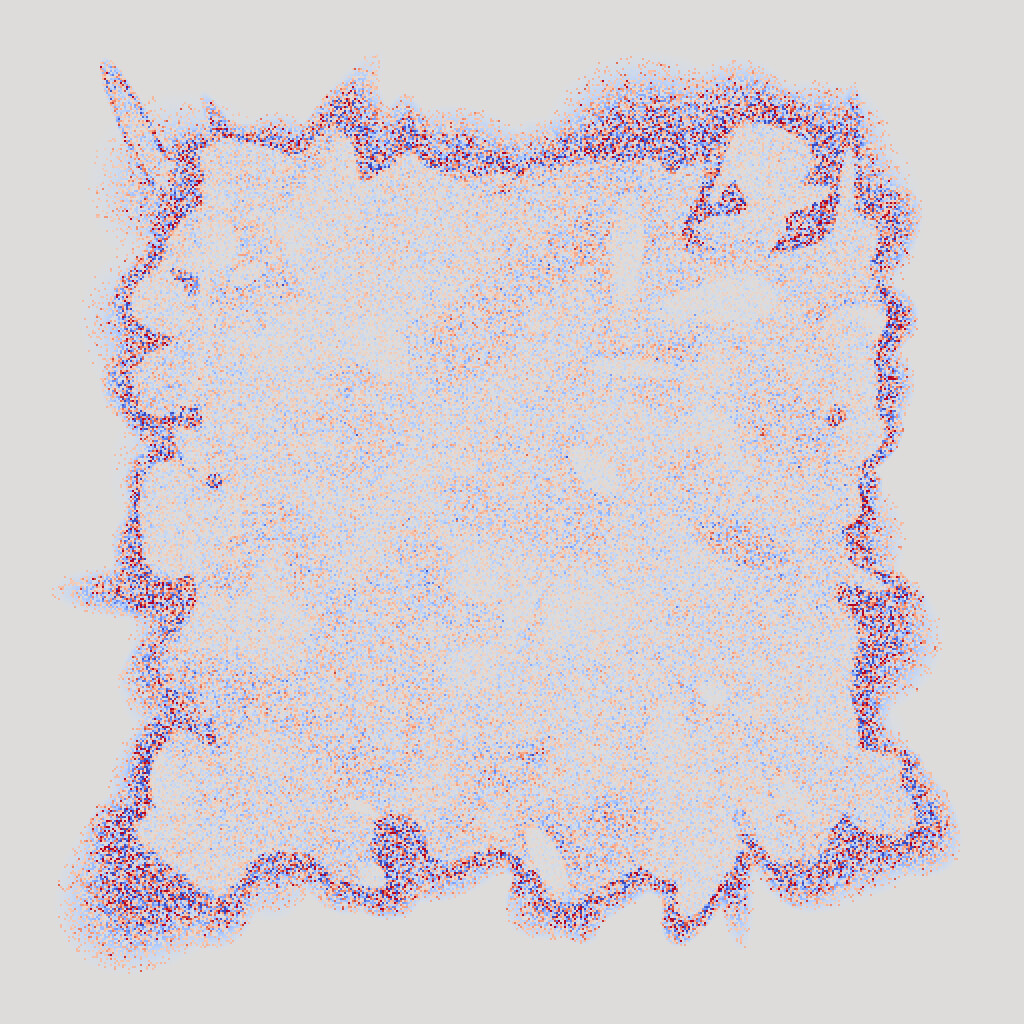}}
      &
      \frame{\includegraphics[width=\lenRandomSceneSample]{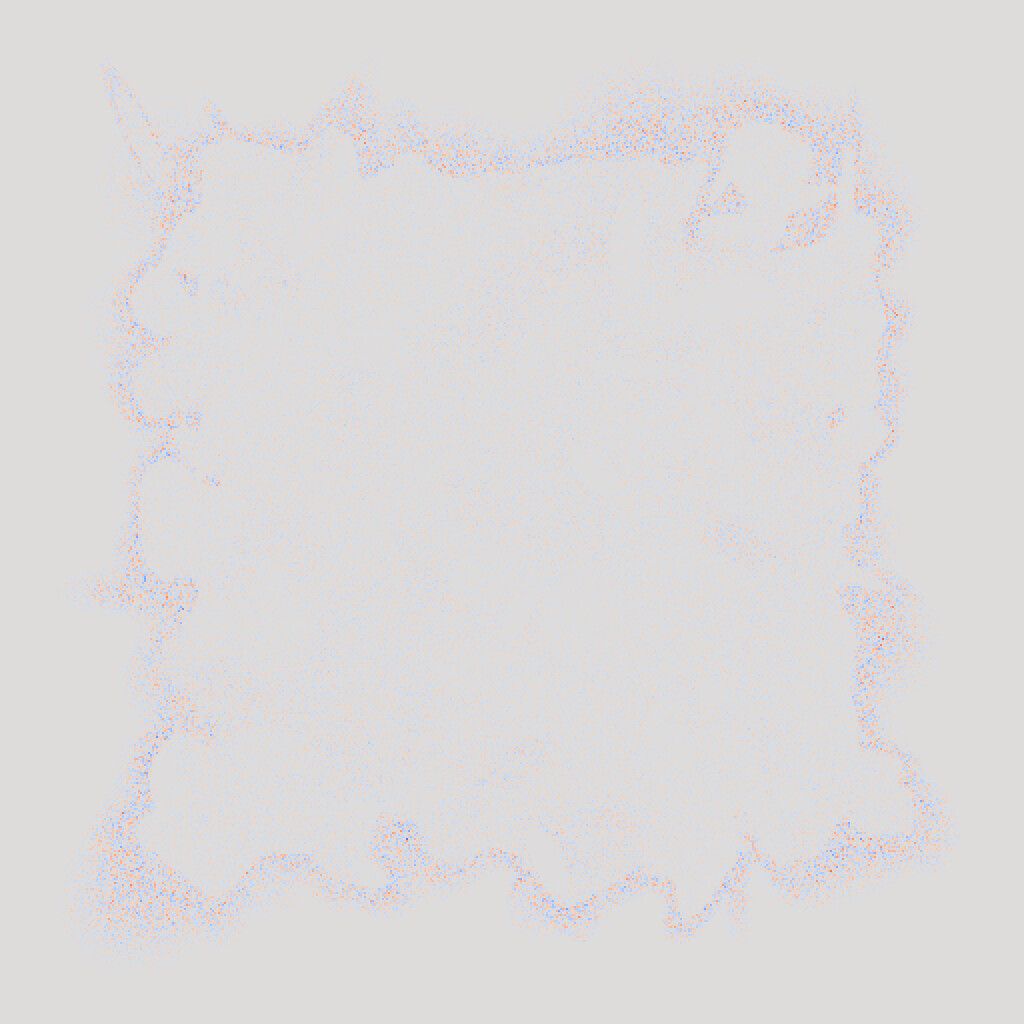}}
      &
      \multicolumn{1}{l}{\begin{overpic}[height=\lenRandomSceneSample,unit=1mm, frame]{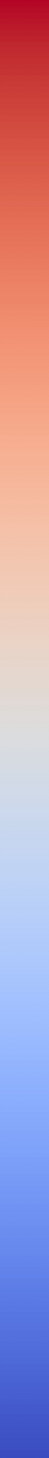}
        \put(5, 91){\small 0.1}
        \put(5, 1){\small -0.1}
      \end{overpic}}            
      \\
      \small{\textsf{1 spp}} & \small{\textsf{16 spp}} & \small{\textsf{256 spp}} & \small{\textsf{Reference}}      
    \end{tabular}
    \caption{\label{fig:random_scene_sample}
        Rendering 1K random Gaussians assigned with random colors using our free-flight sampling algorithm with different samples per pixel (spp). 
        Difference images against the reference are provided.
    }
\end{figure}

We provide pseudocode for our free-flight distribution sampling in 
\autoref{list:sample_free_flight}. The algorithm only requires 2 random numbers and is thus friendly to stratification. \autoref{fig:random_scene_sample} 
validates the convergence of rendering using the algorithm.

\begin{figure}[tb]
\begin{pseudocode}[label={list:sample_free_flight}, title={Pseudocode for free-flight distribution sampling.}]
def (*@\textbf{\textcolor{solarized_blue}{sample\_free\_flight}}@*)(ray):
  u = rnd() # draw a random number
  return sample(kdtree.root, ray, u, 0)

def (*@\textbf{\textcolor{solarized_blue}{sample}}@*)(node, ray, u, cdf):
  if not node.is_leaf():
    # traverse front-to-back  
    children = [node.front(ray), node.back(ray)]
    for c in children:
      t0, t1 = ray.intersect(c.bound)
      r = Ray(ray.o, ray.d, t0, t1)
      G = sample(c, r, u, cdf) 
      if G: return G
  else:
    if node.n_prim == 1 or cdf + 0.5 * (*@$\sum_i$@*)node.prim[i].I(t0, t1) < 1:
      # case (1) and (2)
      for p in node.prims:
        cdf += 0.5 * p.I(t0, t1) # (*@\textcolor{solarized_cyan}{\autoref{eq:ray_integral}}@*)
        if u < cdf: return p                      
    else:
      # case (3)
      # partition by line sweeping; note in actual implementation this and the iteration over segments are done in one pass
      segs = node.partition(ray, t0, t1)
      for s in segs:
        if s.n_prim == 1 or not s.ambiguous():
          # same way to determine an ambiguous node
          for p in s.prims:
            cdf += 0.5 * p.I(s.t_start, s.t_end)
            if u < cdf: return p 
        else:
          # disambiguation ((*@\textcolor{solarized_cyan}{\autoref{eq:disambiguation_solve}}@*))
          t_u = solve(s.prims, s.t_start, u - cdf)
          u_seg = lerp(cdf, u, rnd())
          for p in s.prims:
              cdf += 0.5 * p.I(s.t_start, t_u) 
              if u_seg < cdf: return p
  # the ray does not scatter in this node                   
  return null
\end{pseudocode}
\end{figure}

\subsection{Transmittance Evaluation} \label{subsec:transmittance_eval}
The transmittance evaluation is much more straightforward compared to free-flight distribution sampling. We simply traverse the scene and decrease the 
transmittance by each visited primitive until it either reaches 0 or we exit the traversal. We also employ Russian 
roulette for efficiency trade-off.
The pseudocode for transmittance evaluation is in \autoref{list:eval_transmittance}.

\begin{figure}[tb]
\begin{pseudocode}[label={list:eval_transmittance}, title={Pseudocode for transmittance evaluation.}]
def (*@\textbf{\textcolor{solarized_blue}{eval\_transmittance}}@*)(ray):
    u = rnd()    
    return eval(kdtree.root, ray, ray.t0, ray.t1)

def (*@\textbf{\textcolor{solarized_blue}{eval}}@*)(node, ray, t0, t1):
    T = 1
    while node:
    if not node.is_leaf():
        for c in node.children:
          t0, t1 = ray.intersect(c.bound)
          T -= eval(c, ray, t0, t1) 
          if T <= 0: return 0
        else:
          for p in node.prims:
            T -= 0.5 * p.I(t0, t1) # (*@\textcolor{solarized_cyan}{\autoref{eq:ray_integral}}@*)
            # Russian roulette with terminating probability q
            if T < epsilon:                
              if rnd() < q: T = 0:
              else: T /= (1 - q)
            if T <= 0: return 0                         
    return T
\end{pseudocode}
\end{figure}
\section{Appearance} \label{sec:appearance}
The free-flight sampling in \autoref{subsec:free_flight_sampling} allows us to probabilistically select a Gaussian primitive to scatter the light. 
The next ingredient is the primitive appearance that governs how light is scattered into different directions.
As motivated in \autoref{sec:primitives}, we define appearance models at the primitive level and abstract away all sub-primitive scattering interactions. 

\begin{figure}[tb]
    \newlength{\lenPhaseFunction}
    \setlength{\lenPhaseFunction}{0.75\linewidth}
    \addtolength{\tabcolsep}{-4pt}
    \renewcommand{\arraystretch}{0.5}
    \centering
    \begin{tabular}{c}
    \includegraphics[width=\lenPhaseFunction]{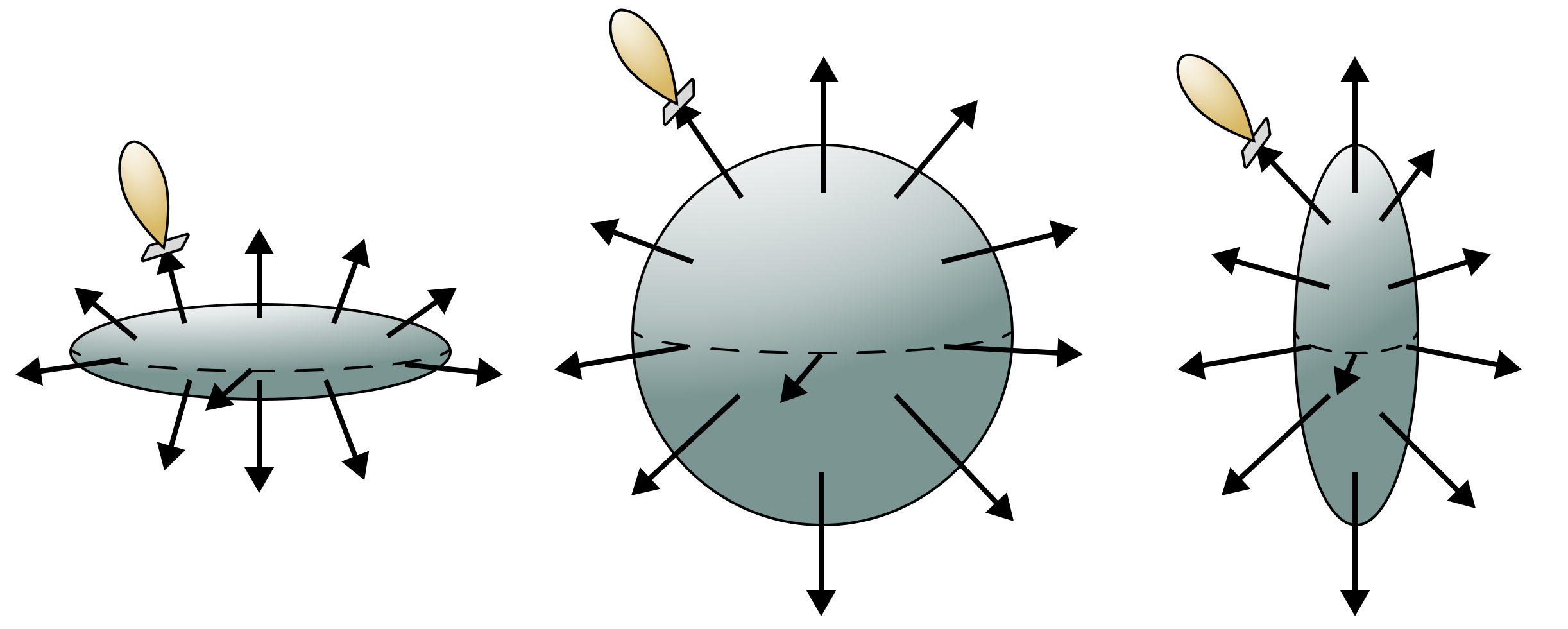}
    \end{tabular}
    \caption{\label{fig:phase_function}
        The phase function incorporates the underlying geometric configuration by modulating the base BSDF (yellow) with the (V)NDF (green). 
        It can thus represent a collection of oriented surface elements that overall behaves like a surface (left), isotropic scatterer (middle), or fiber (right).
    }
\end{figure}

A Gaussian primitive can represent a simple surface patch or a collection of small surface elements with different orientations. Therefore, a useful appearance model for the primitive should incorporate this aggregated effect while still being compatible with simple flat scenarios.
Inspired by the microflake theory \citep{jakob2010radiative, heitz2015sggx}, we describe the orientations by a normal distribution function (NDF) 
$D(n)$ and its visible normal distribution function (VNDF) $D_{\omega_o}(n)$ when conditioned by a viewing direction $\omega_o$:
\begin{align} \label{eq:vndf}
    \begin{split}
    D_{\omega_o}(n) &= \frac{1}{\sigma(\omega_o)} D(n) \langle n \cdot \omega_o \rangle, \\
    \sigma(\omega_o) &= \int_{\mathbb{S}^2}  D(n) \langle n \cdot \omega_o \rangle \,\D{n},
    \end{split}
\end{align}
where $\sigma(\omega)$ is the projected area that serves as the normalization term for $D_{\omega_o}(n)$, and $\langle \cdot \rangle$ is the clamped dot product. 
The appearance of a Gaussian primitive is affected by both the VNDF and the (cosine-weighted) base surface BSDF $f(\omega_i, \omega_o; n)$ that describes each oriented surface 
element. Formally, it is defined as a phase function which is an inner product of the VNDF and the base surface BSDF:
\begin{equation} \label{eq:phase_function}
    f_p(\omega_i, \omega_o) = \int_{\mathbb{S}^2}  D_{\omega_o}(n) f(\omega_i, \omega_o; n) \,\D{n}.
\end{equation}
In the case when the primitive only models a single flat surface, $D(n)$ becomes a delta distribution, and $f_p$ falls back
to the usual surface BSDF multiplied by the foreshortening term. Even if the base BSDF is strictly defined for a surface, the overall phase function could 
represent flexible geometric configurations from being surface-like to fiber-like, as illustrated in \autoref{fig:phase_function}.    
We use the SGGX distribution for the (V)NDF due to its expressiveness and simplicity to evaluate, importance sample, and fit 
\citep{heitz2015sggx}.

The base surface BSDF can be an arbitrary valid BSDF with standard evaluation and importance sampling procedures. In this work, we primarily 
work with the Disney principled BSDF~\citep{burley2015extending} due to its capability to model a wide range of materials. The Disney principled BSDF mainly 
consists of a microfacet specular component and a diffuse component with empirical retroflection. We refer readers to \citet{burley2015extending} for the full definition. 
In \autoref{fig:leaf}, we show an example using an alternative thin-surface translucent base BSDF.
We will also make the assumption that scattering in a primitive can be modeled accurately without an exitant positional distribution. 
See \autoref{sec:conclusion} for further discussion. 

For our phase function to be compatible with a Monte Carlo renderer, it must support several operations, namely evaluation 
(evaluate \autoref{eq:phase_function} given $\omega_i$ and $\omega_o$), importance sampling (sample a suitable $\omega_i$ given $\omega_o$), and 
preferably PDF evaluation (of the importance sampling procedure) for multiple importance sampling (MIS). 
In the following, we describe those operations in detail.

\paragraph{Evaluation}
The integral in \autoref{eq:phase_function} cannot be evaluated in closed form for most non-trivial base BSDFs (one exception is perfect specular reflection). 
For Lambertian base BSDF, \citet{heitz2015sggx} suggest a simple Monte Carlo estimator by sampling the VNDF. We propose an improvement by utilizing the existing 
sampling routine for the base BSDF and applying an \emph{internal} multiple importance sampling (MIS) between VNDF sampling and base BSDF sampling. Specifically for the Disney BSDF 
which is a linear combination of multiple components, we estimate each component by MIS to achieve even greater variance reduction. For its specular component, 
we reverse the input and output of the microfacet distribution sampling \citep{walter2007microfacet}, generating normal $n$ given half vector $\omega_h$. 
For its diffuse component, we again similarly use cosine-weighted hemisphere sampling to generate normal $n$ given $\omega_i$. The pseudocode for our improved
stochastic evaluation is in \autoref{list:eval_phase}.

\begin{figure}[tb]
\begin{pseudocode}[label={list:eval_phase}, title={Pseudocode for improved stochastic phase function evaluation.}]
def (*@\textbf{\textcolor{solarized_blue}{eval\_phase\_stochastic}}@*)((*@$\omega_i$@*), (*@$\omega_o$@*)):
    # SGGX VNDF sampling
    n1, pdf_n1 = sample_Dvis((*@$\omega_o$@*), rnd_2d())
    # "reversed" base BSDF sampling for specular and MIS
    (*@$\omega_h$@*) = ((*@$\omega_i$@*) + (*@$\omega_o$@*)).normalized()
    n2, pdf_n2 = sample_micro((*@$\omega_h$@*), rnd_2d())
    # f_s and f_d are cosine-weighted
    # mis(f, g) gives the 1-sample balance heuristic weight
    fp_s = f_s(n1, (*@$\omega_i$@*), (*@$\omega_o$@*)) * mis(pdf_n1, pdf_micro(n1))
    fp_s += Dvis((*@$\omega_o$@*), n2) * f_s(n2, (*@$\omega_i$@*), (*@$\omega_o$@*)) / pdf_n2 * mis(pdf_n2, Dvis(n2))    
    # "reversed" base BSDF sampling for diffuse and MIS
    n3, pdf_n3 = sample_cos((*@$\omega_i$@*), rnd_2d())
    fp_d = f_d(n1, (*@$\omega_i$@*), (*@$\omega_o$@*)) * mis(pdf_n1, pdf_cos(n1, (*@$\omega_i$@*)))
    fp_d += Dvis((*@$\omega_o$@*), n3) * f_d(n3, (*@$\omega_i$@*), (*@$\omega_o$@*)) / pdf_n3 * mis(pdf_n3, Dvis(n3))
    # return sum of estimates for each component
    return fp_s + fp_d;

\end{pseudocode}
\end{figure}

\paragraph{Importance Sampling}
We follow the original importance sampling strategy by \citep{heitz2015sggx}. Given a view direction $\omega_o$, we first sample the VNDF to generate a sample $n$.
We then sample an incident direction $\omega_i$ using the existing strategy of the base BSDF. Since the VNDF sampling for SGGX is perfect, the sample weight is 
simply that of the base BSDF sampling. In practice, we spawn an indirect ray from the center of intersection 
segment and disable self-intersection. This is similar to common practice of sampling a \emph{Bidirectional Curve Scattering Distribution Function} (BCSDF) in 
hair rendering.

\paragraph{Multiple Importance Sampling}
It is desirable to be able to compute the PDF from the importance sampling so that the renderer can apply useful variance reduction techniques, such 
as MIS between light sampling and phase function sampling for next-event estimation (not to be confused with the internal MIS for the stochastic evaluation).
However, the PDF follows a similar form to \autoref{eq:phase_function} and also cannot be computed in closed form:
\begin{equation} \label{eq:phase_function_pdf}
    \mathrm{pdf}_{f_p}(\omega_i | \omega_o) = \int_{\mathbb{S}^2}  D_{\omega_o}(n) \mathrm{pdf}_{f}(\omega_i | \omega_o; n) \,\D{n},
\end{equation}
where $\mathrm{pdf}_{f}$ is the PDF from the base BSDF sampling technique. Moreover, a stochastic estimator for the PDF is not useful because the MIS weight 
requires the reciprocal of the PDF, and expectation \emph{does not commute with} division \citep{heitz2016multiple, zeltner2020specular, qin2015unbiased}.
Fortunately, MIS does not require an exact PDF to work correctly,
thus we provide a simple approximation to \autoref{eq:phase_function_pdf} for this purpose. The approximate PDF uses a roughened SGGX for the specular 
component and a cosine-weighted hemisphere for the diffuse component, both parameterized by half vector. Please refer to \autoref{sec:approx_mis_pdf} 
for details. 

\begin{figure*}[tb]
    \newlength{\lenMaterialSphere}
	\setlength{\lenMaterialSphere}{0.70in}
    \addtolength{\tabcolsep}{-4pt}
    \renewcommand{\arraystretch}{0.5}
    \centering
    \begin{tabular}{cccccccccc}
        & & \small{\emph{Smooth Gold}} & & & \small{\emph{Rough Copper}} & & & \small{\emph{Plastic}} & \\
        \\
        \raisebox{5pt}{\rotatebox{90}{\footnotesize{\textsf{Delta Surface}}}}
        &
        \frame{\includegraphics[height=\lenMaterialSphere]{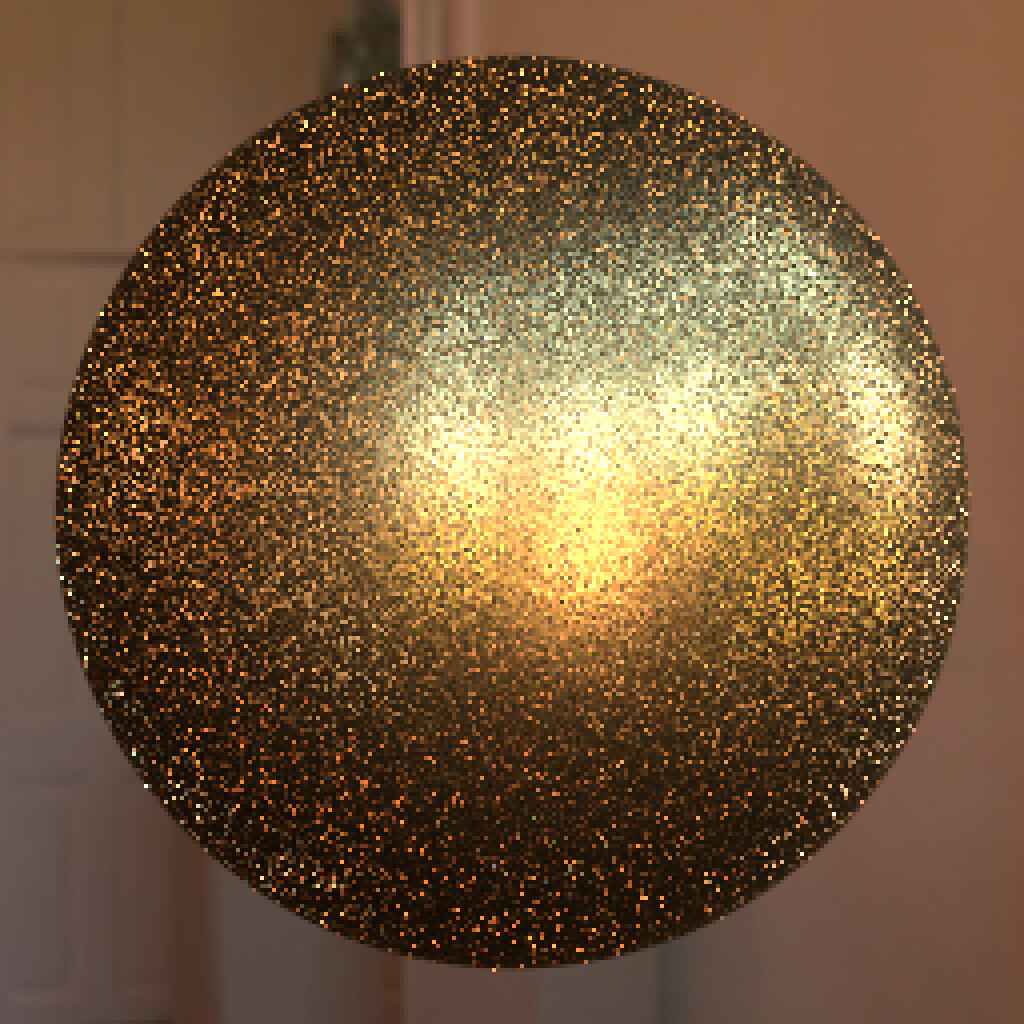}}
        &
        \frame{\includegraphics[height=\lenMaterialSphere]{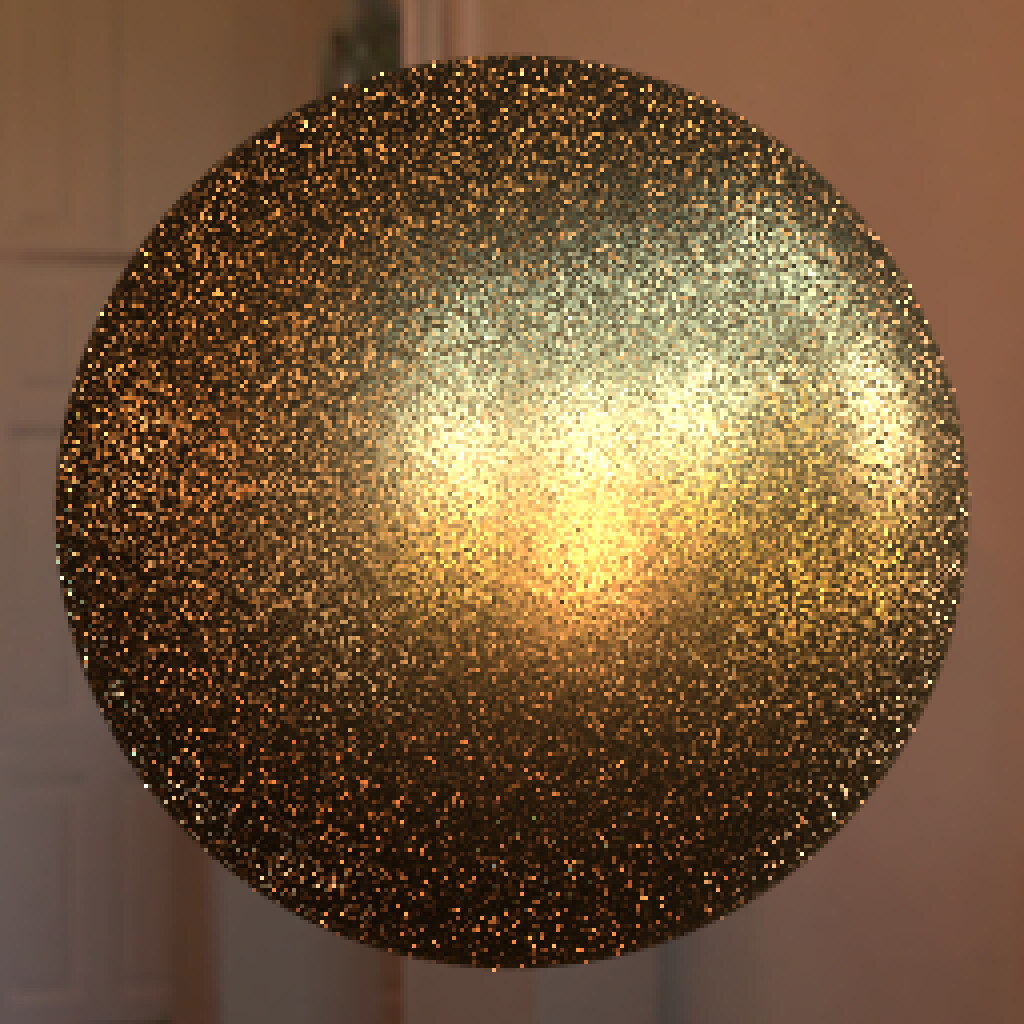}}
        &
        \frame{\includegraphics[height=\lenMaterialSphere]{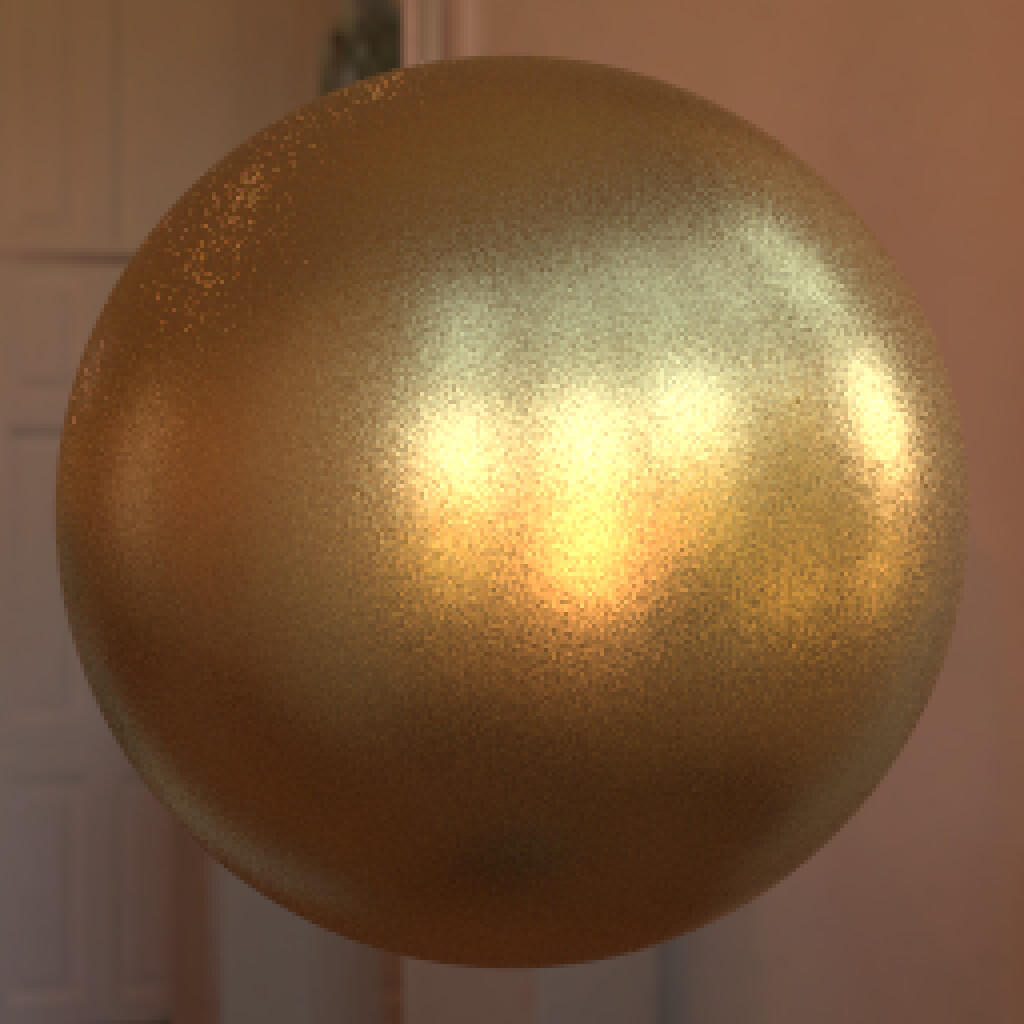}}
        &        
        \frame{\includegraphics[height=\lenMaterialSphere]{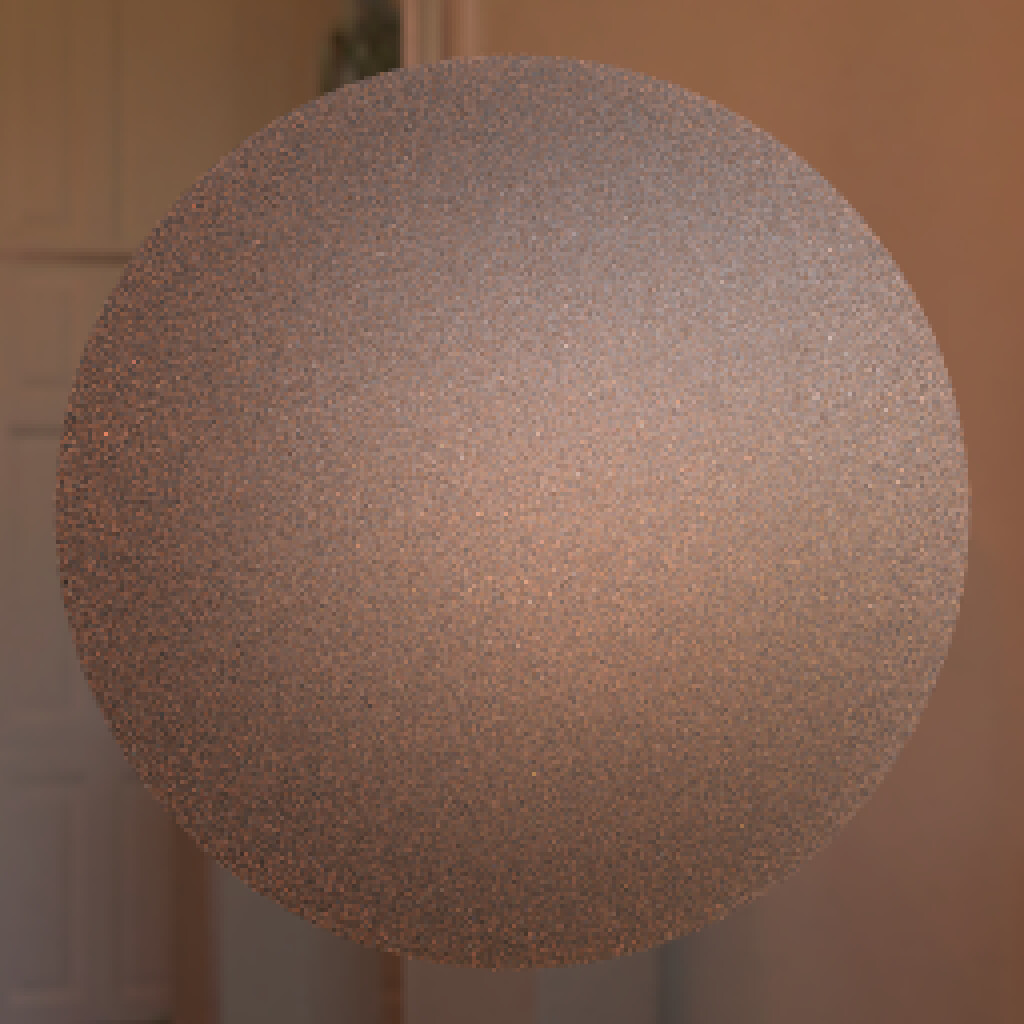}}
        &
        \frame{\includegraphics[height=\lenMaterialSphere]{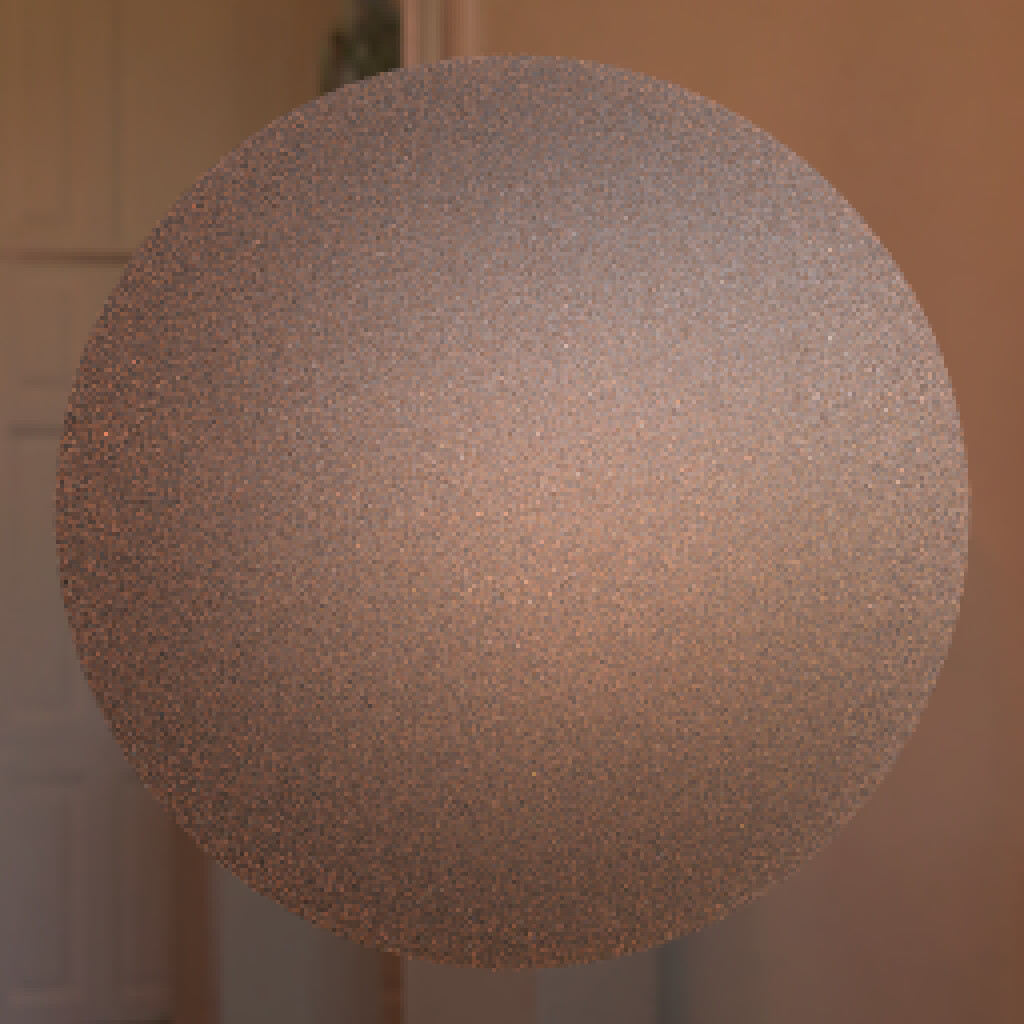}}
        &
        \frame{\includegraphics[height=\lenMaterialSphere]{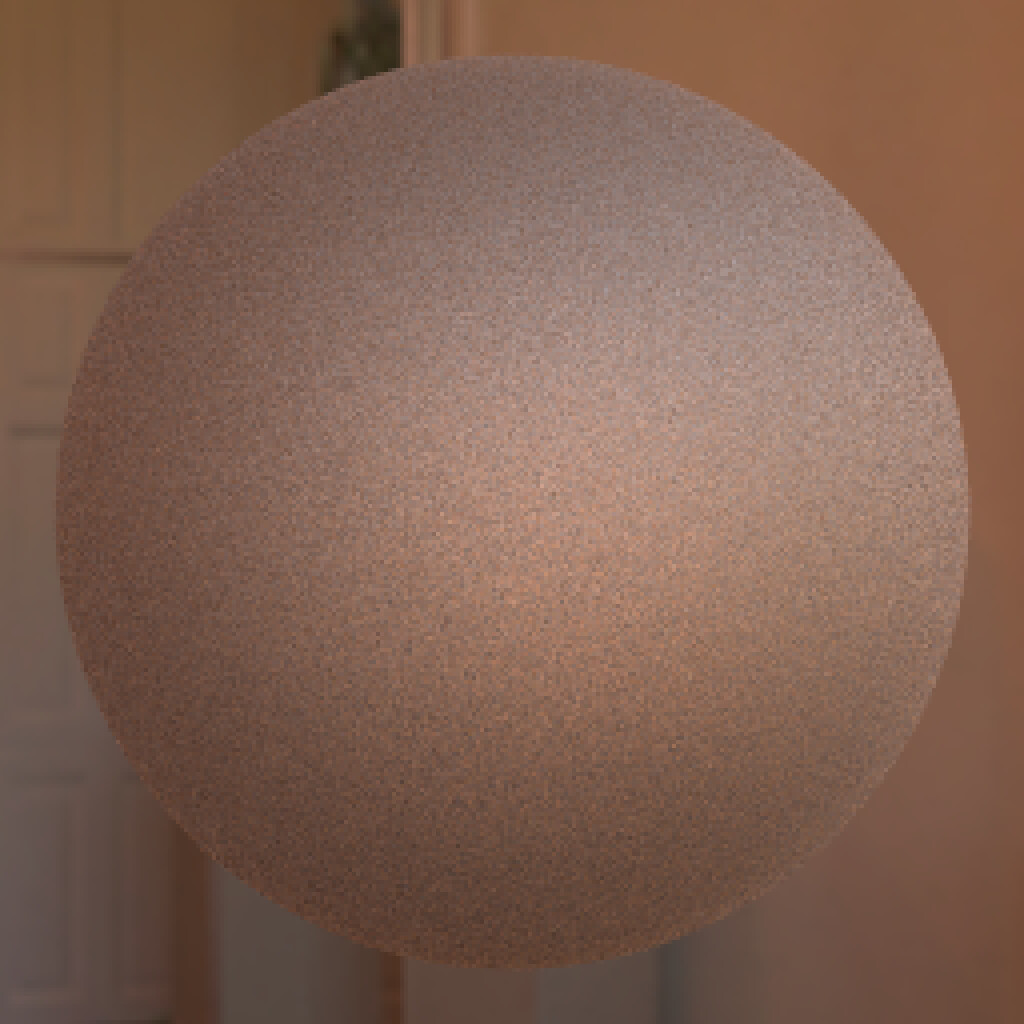}}
        &
        \frame{\includegraphics[height=\lenMaterialSphere]{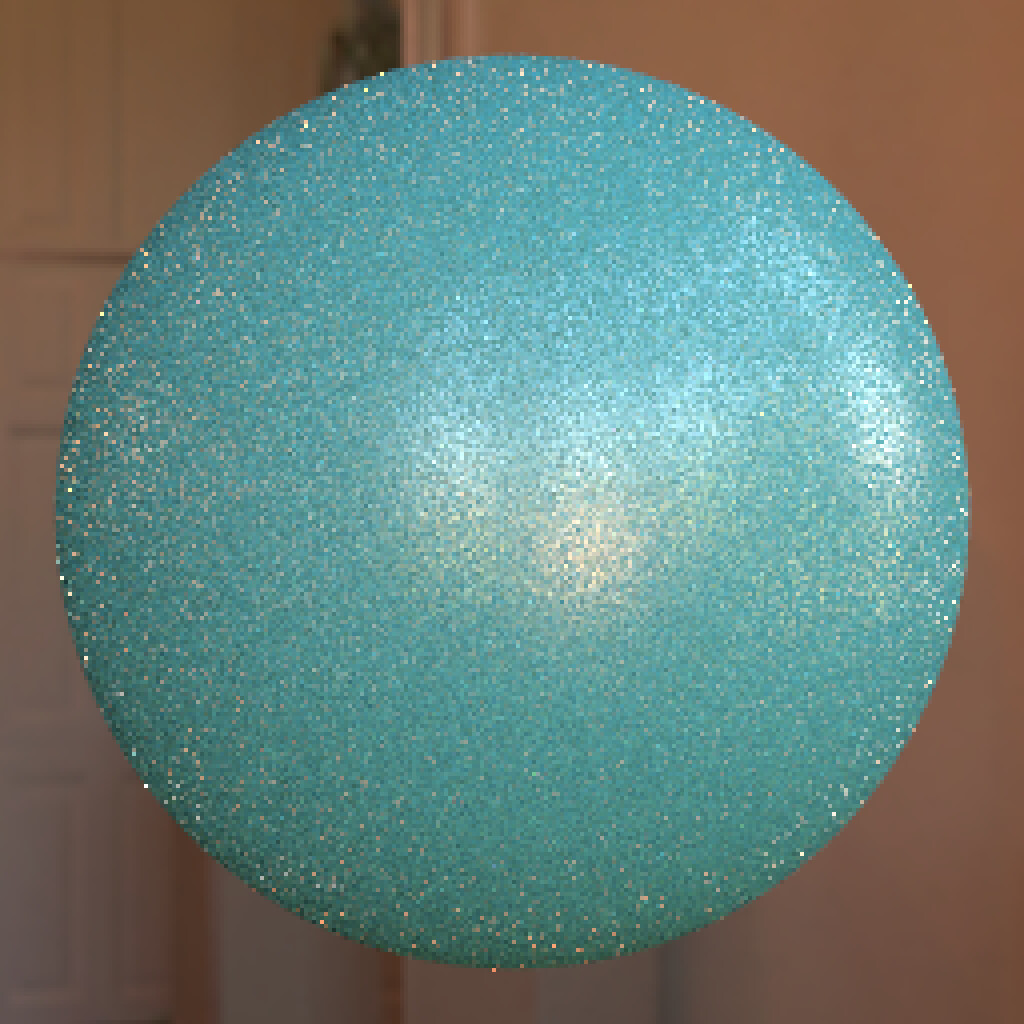}}
        &
        \frame{\includegraphics[height=\lenMaterialSphere]{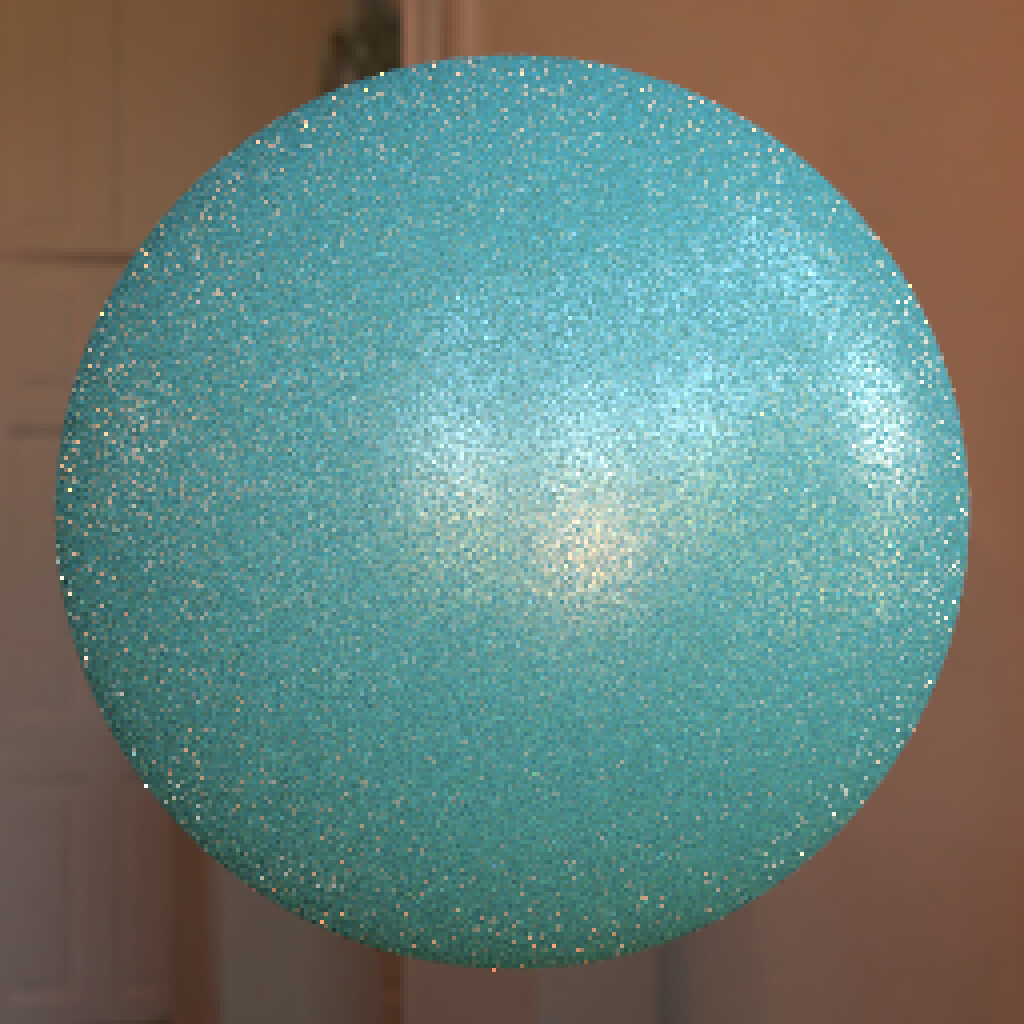}}
        &
        \frame{\includegraphics[height=\lenMaterialSphere]{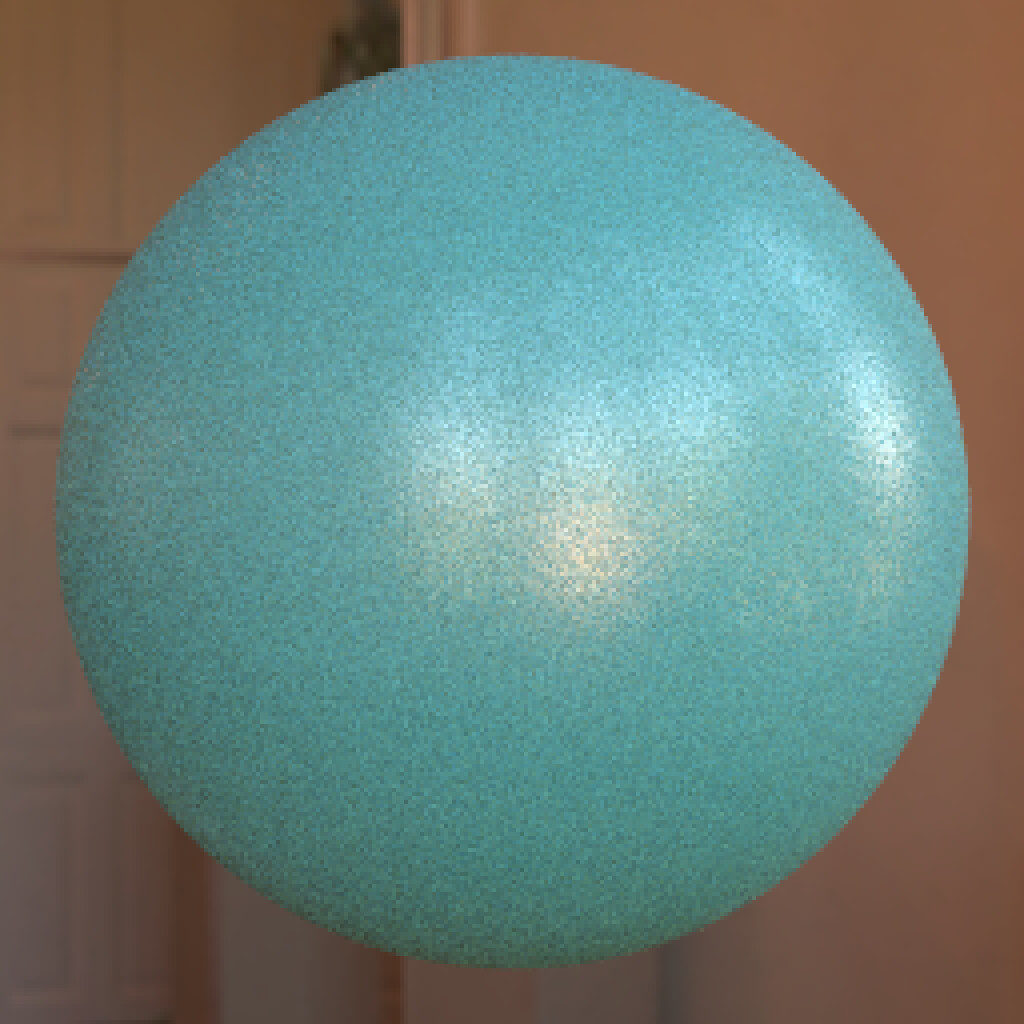}}
        \\   
        \raisebox{5pt}{\rotatebox{90}{\footnotesize{\textsf{Surface-like}}}}
        &
        \frame{\includegraphics[height=\lenMaterialSphere]{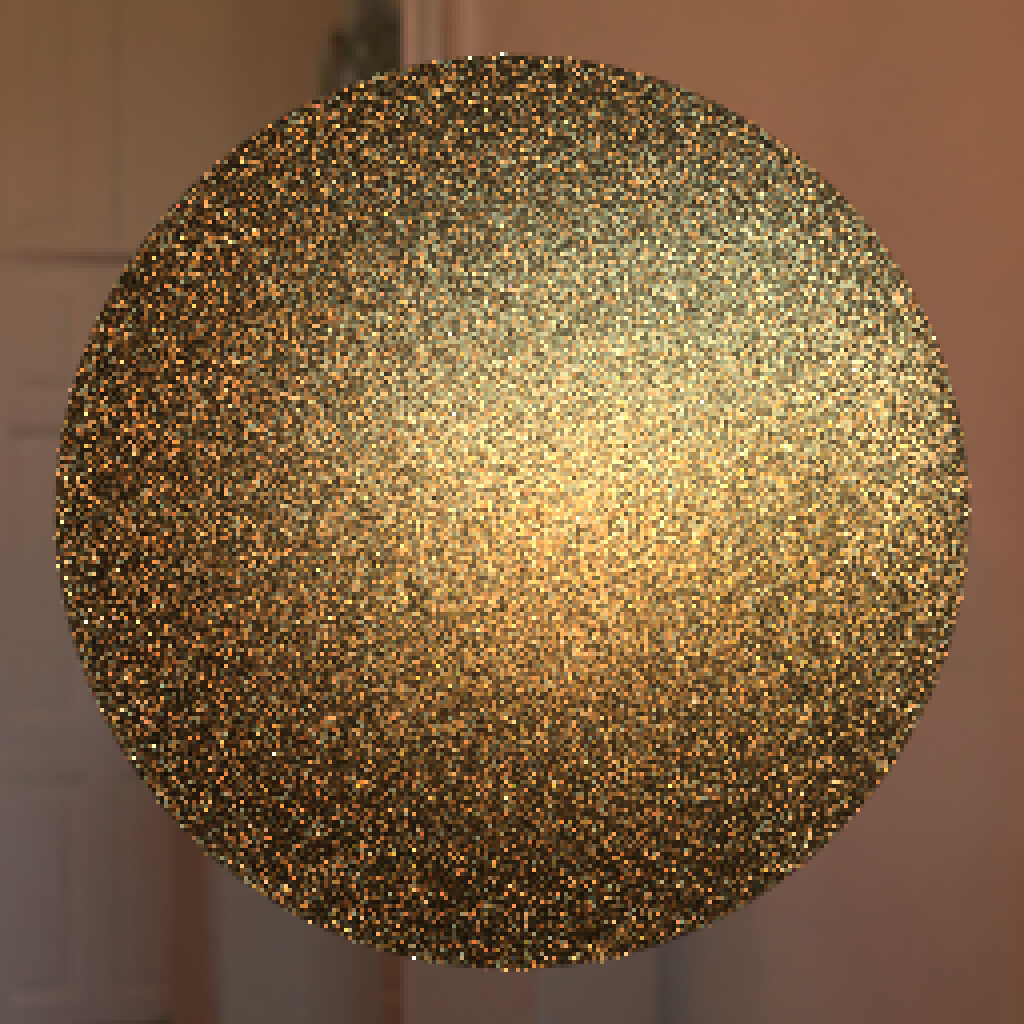}}
        &
        \frame{\includegraphics[height=\lenMaterialSphere]{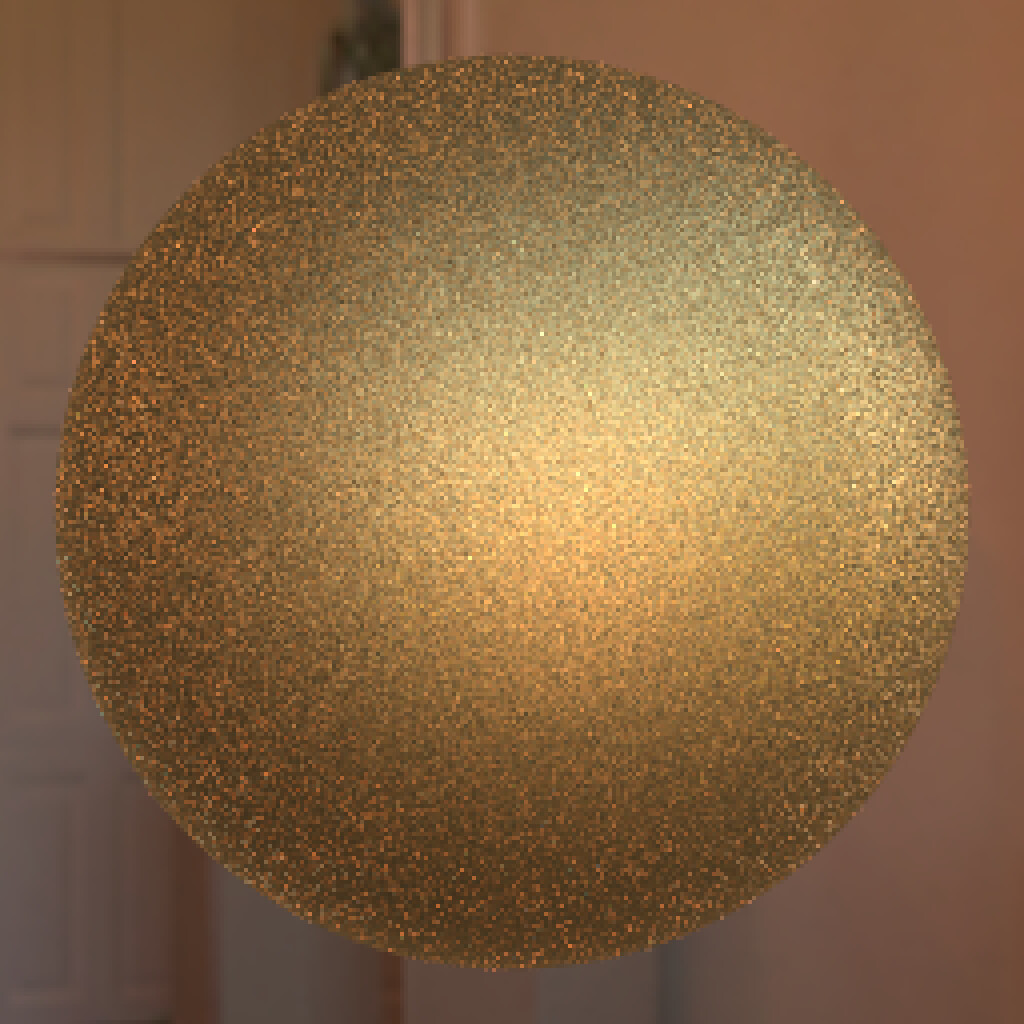}}
        &
        \frame{\includegraphics[height=\lenMaterialSphere]{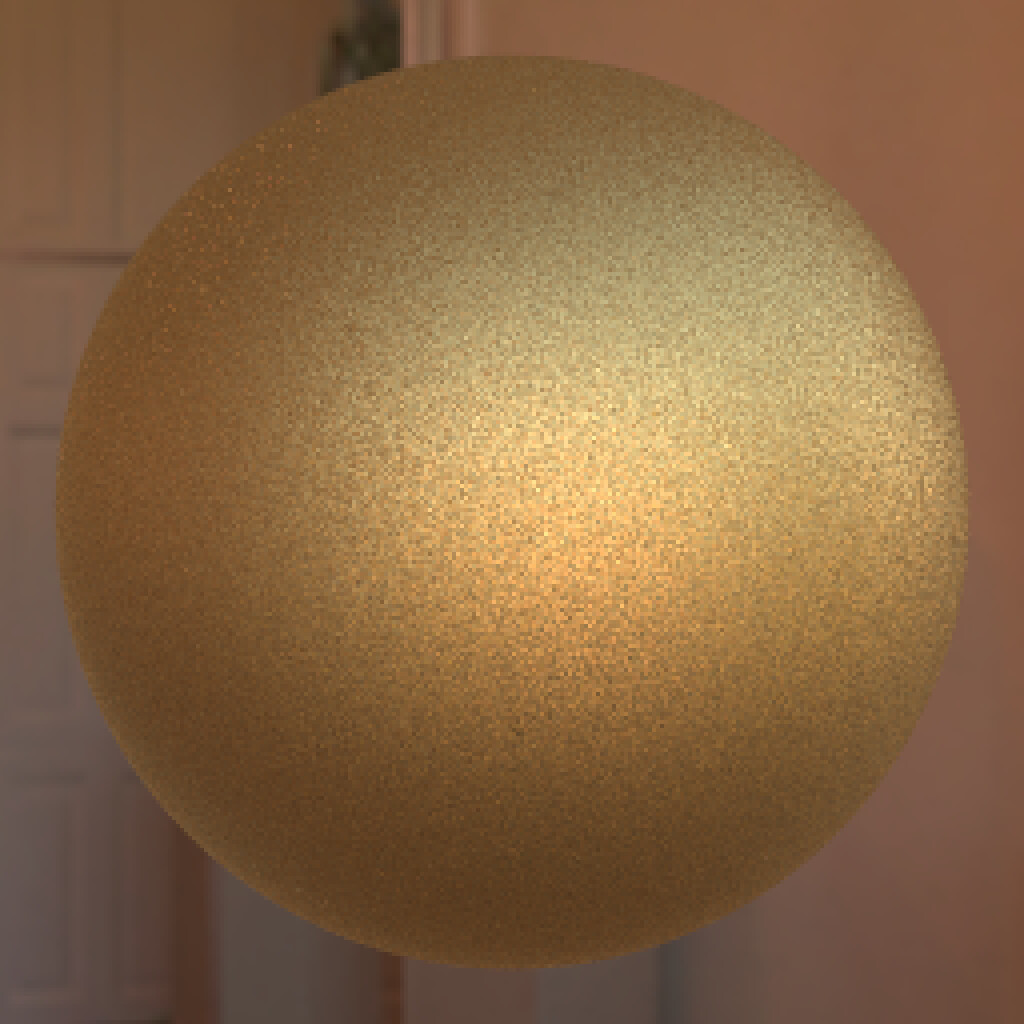}}
        &        
        \frame{\includegraphics[height=\lenMaterialSphere]{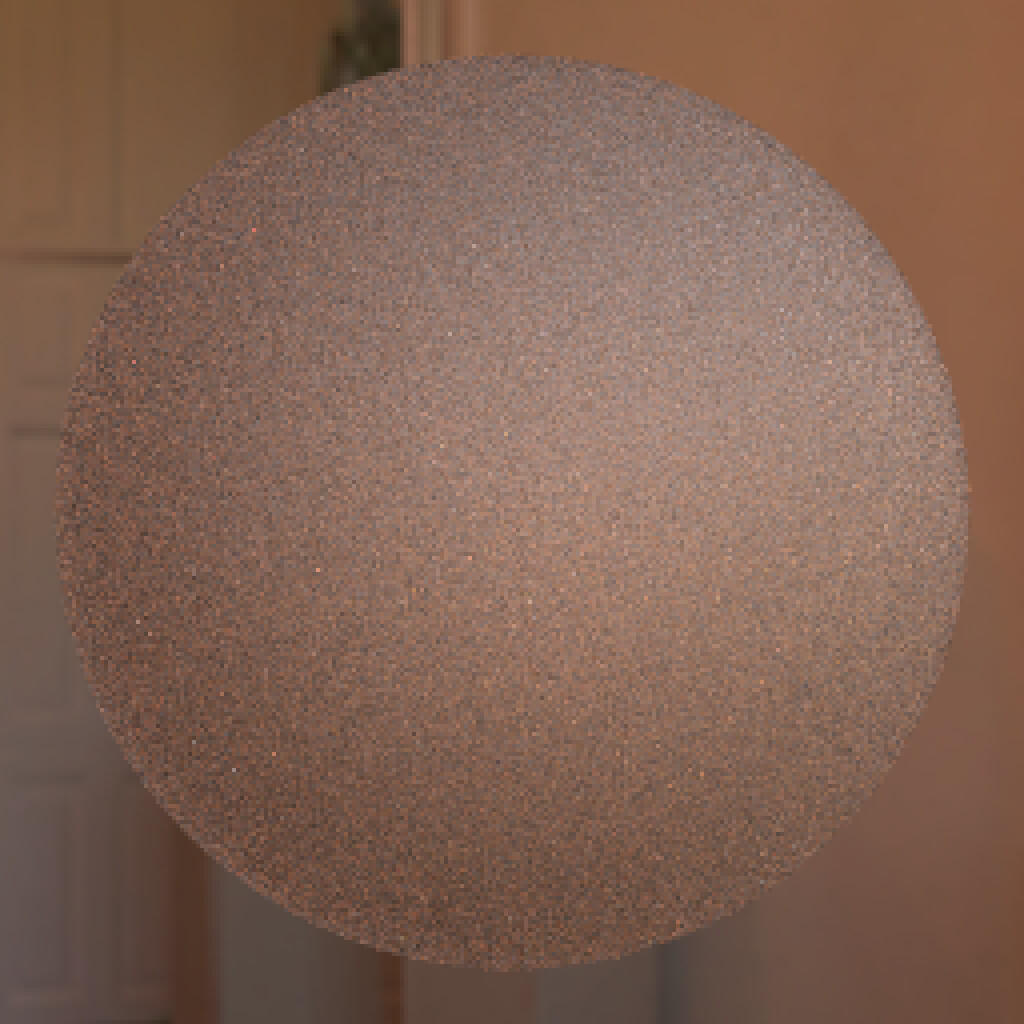}}
        &
        \frame{\includegraphics[height=\lenMaterialSphere]{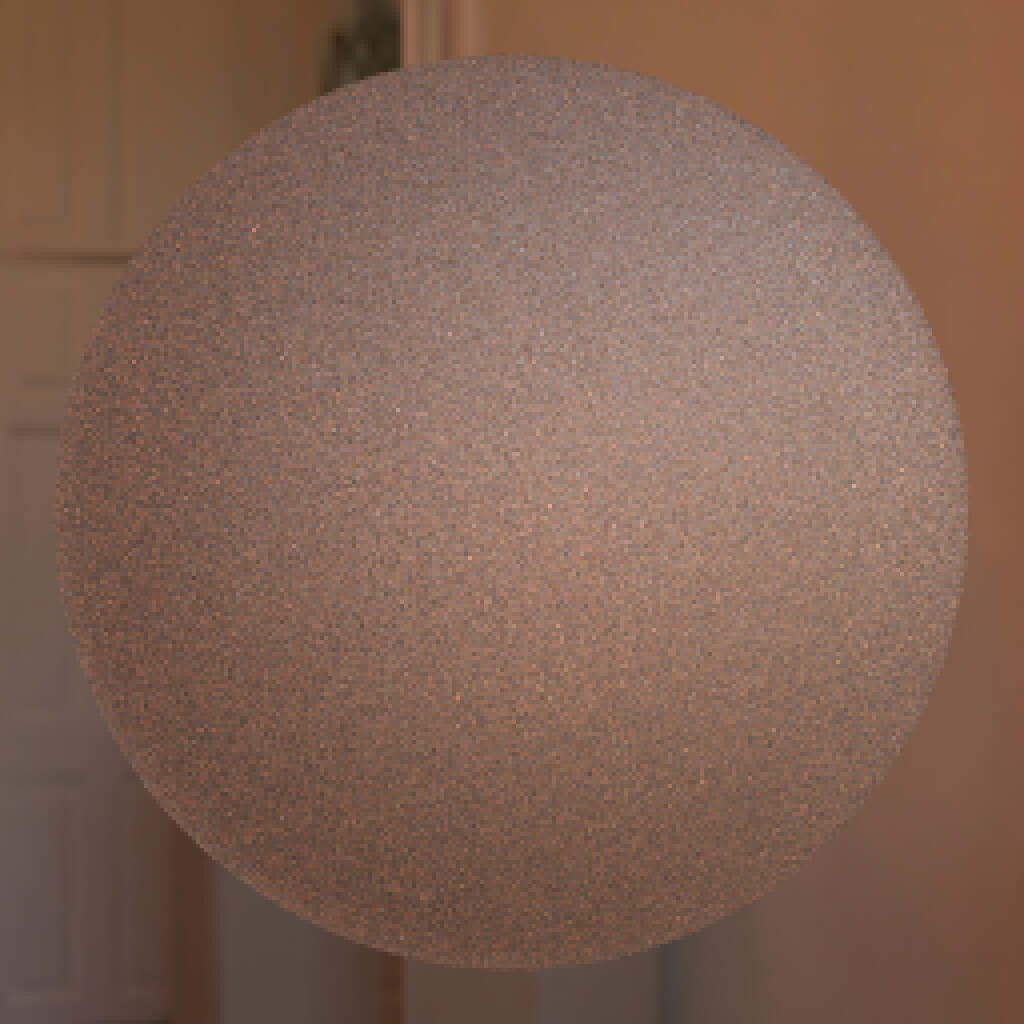}}
        &
        \frame{\includegraphics[height=\lenMaterialSphere]{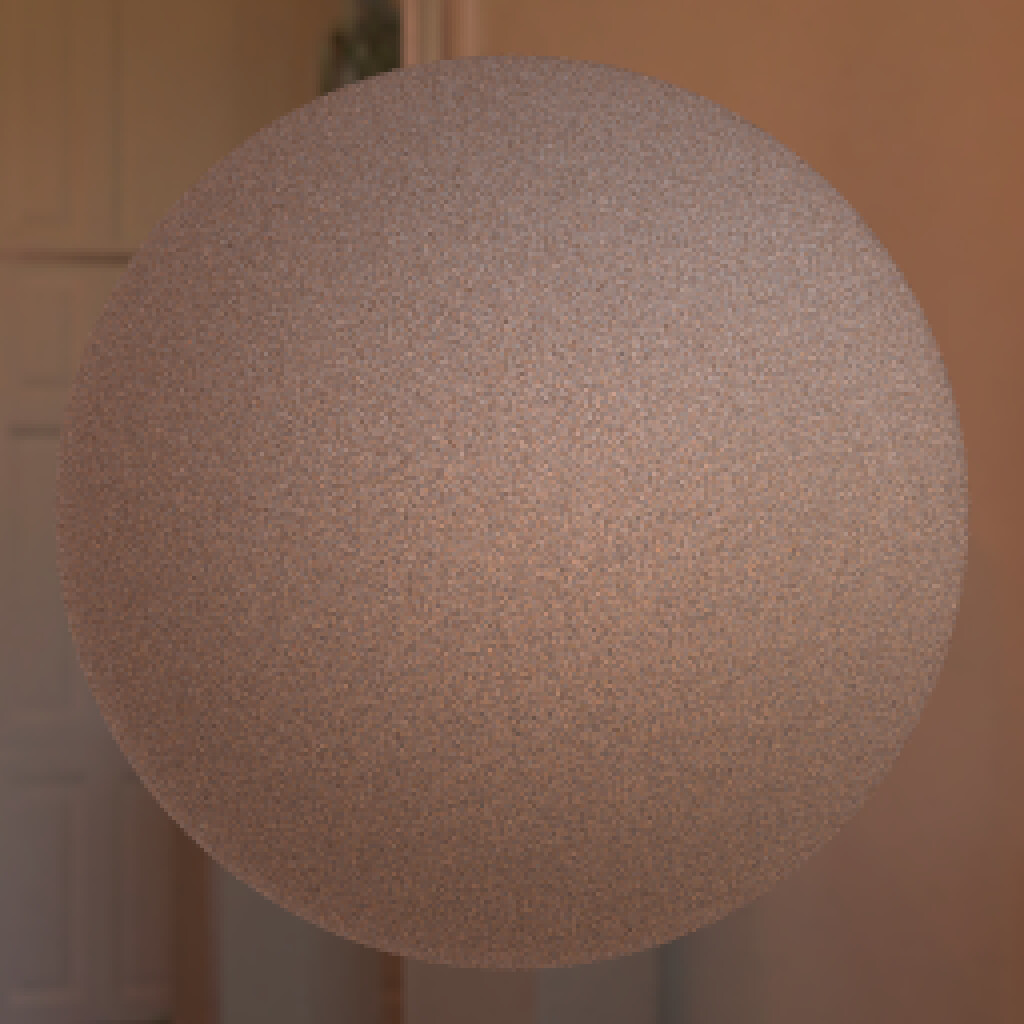}}
        &
        \frame{\includegraphics[height=\lenMaterialSphere]{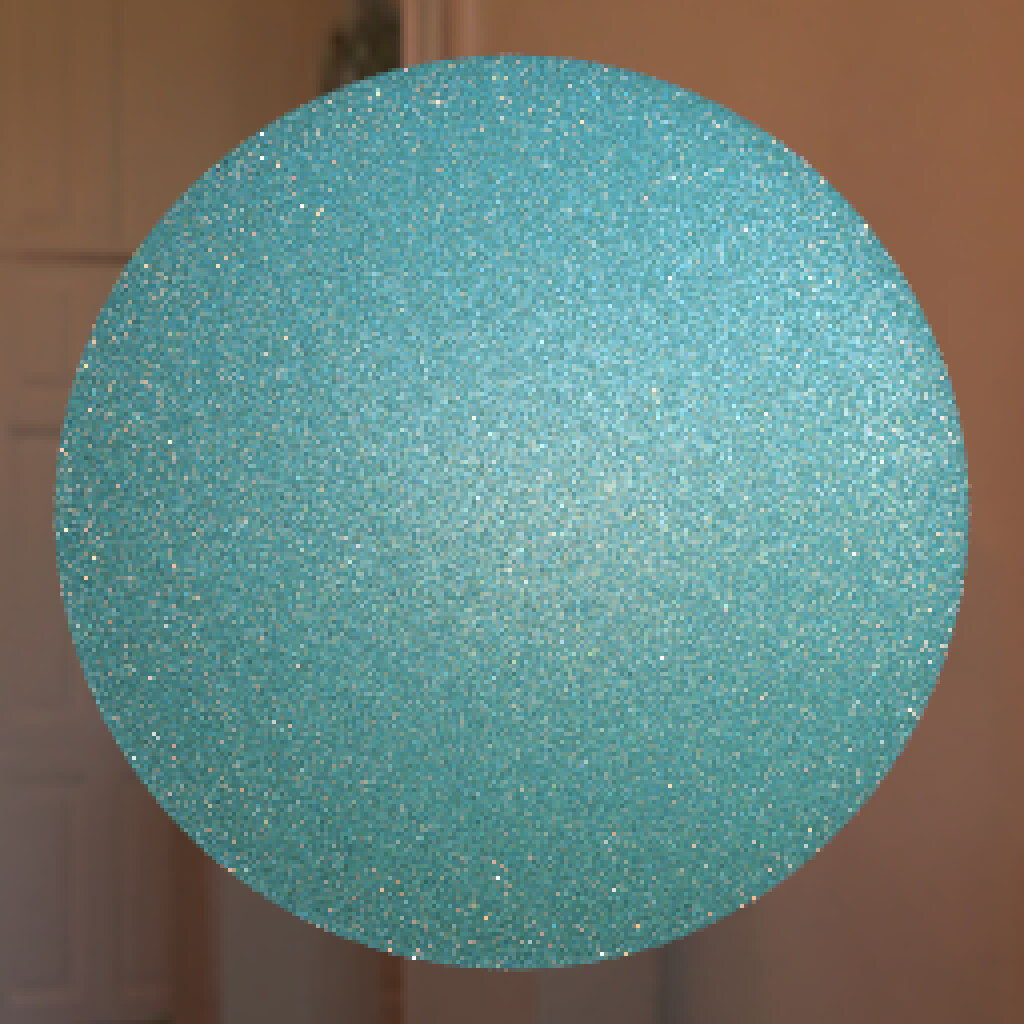}}
        &
        \frame{\includegraphics[height=\lenMaterialSphere]{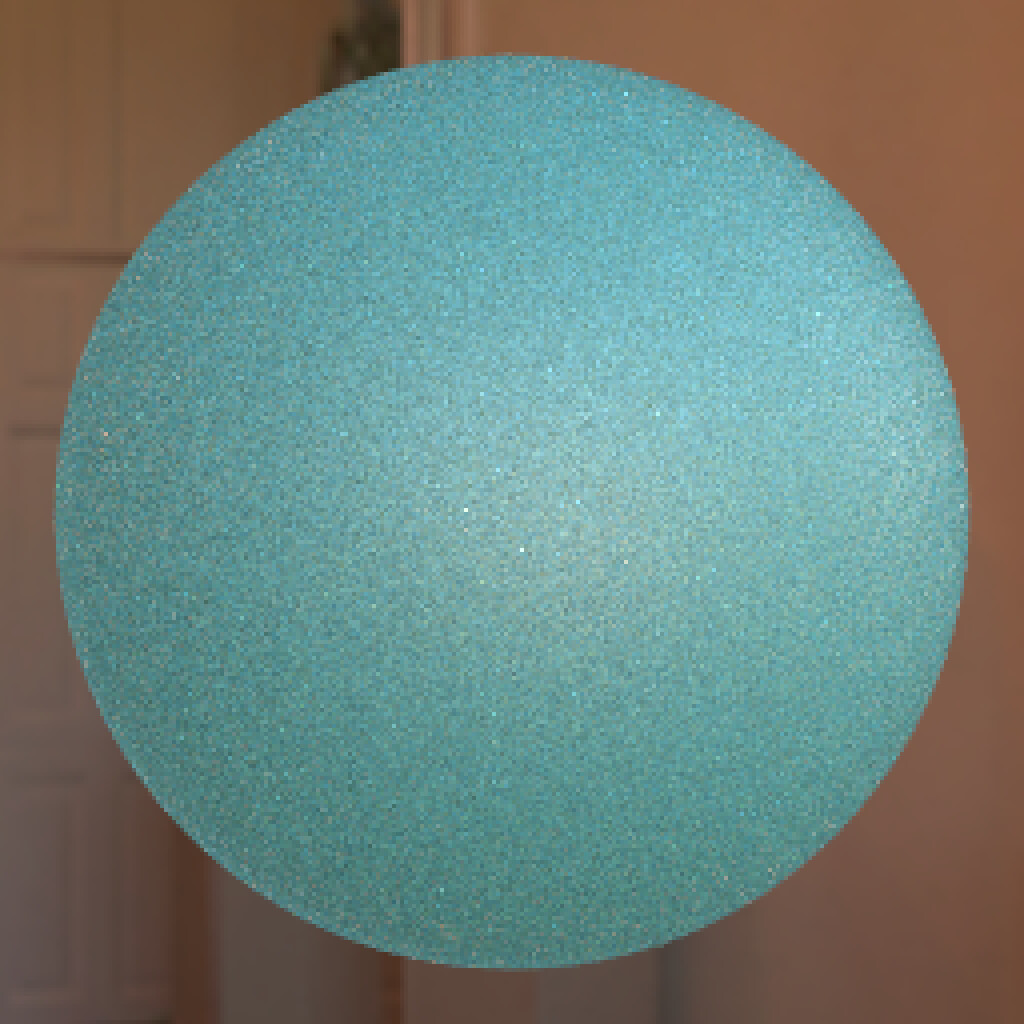}}
        &
        \frame{\includegraphics[height=\lenMaterialSphere]{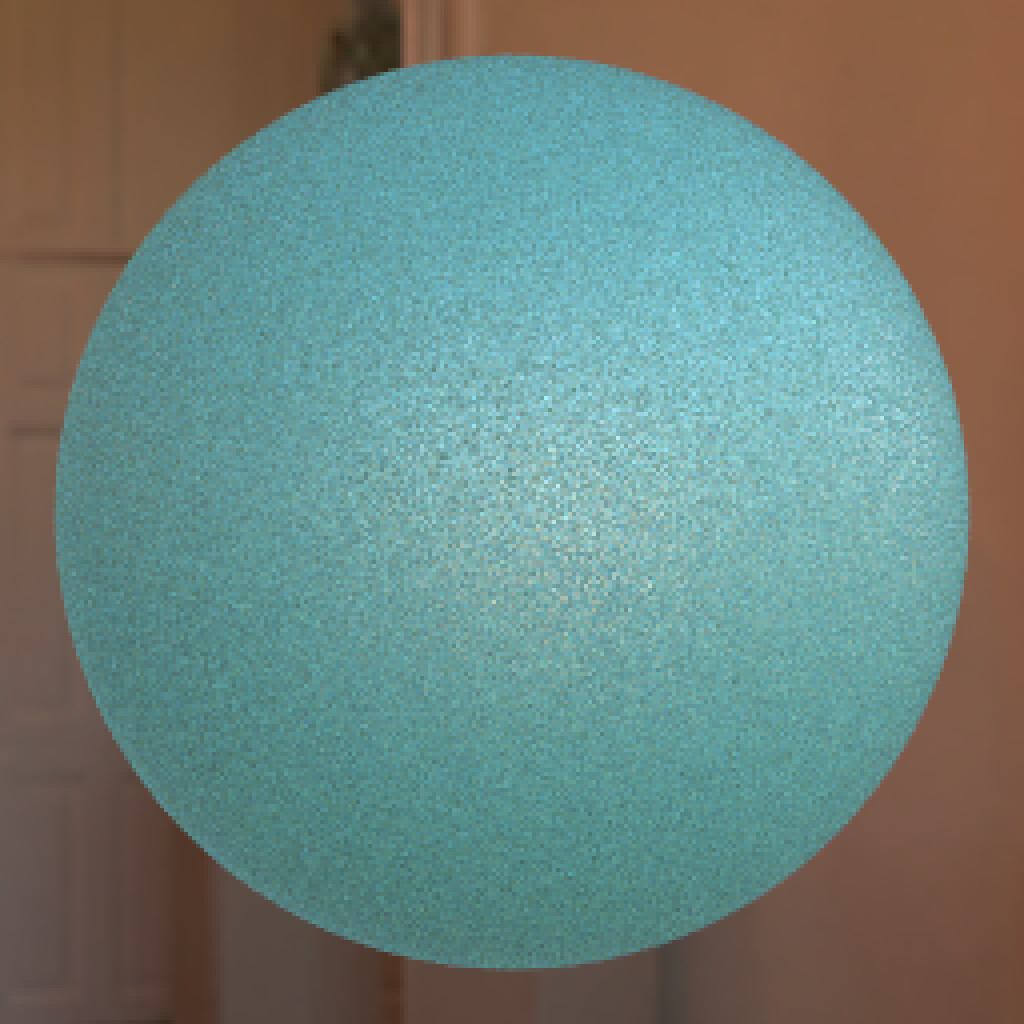}}
        \\   
        \raisebox{12pt}{\rotatebox{90}{\footnotesize{\textsf{Isotropic}}}}
        &
        \frame{\includegraphics[height=\lenMaterialSphere]{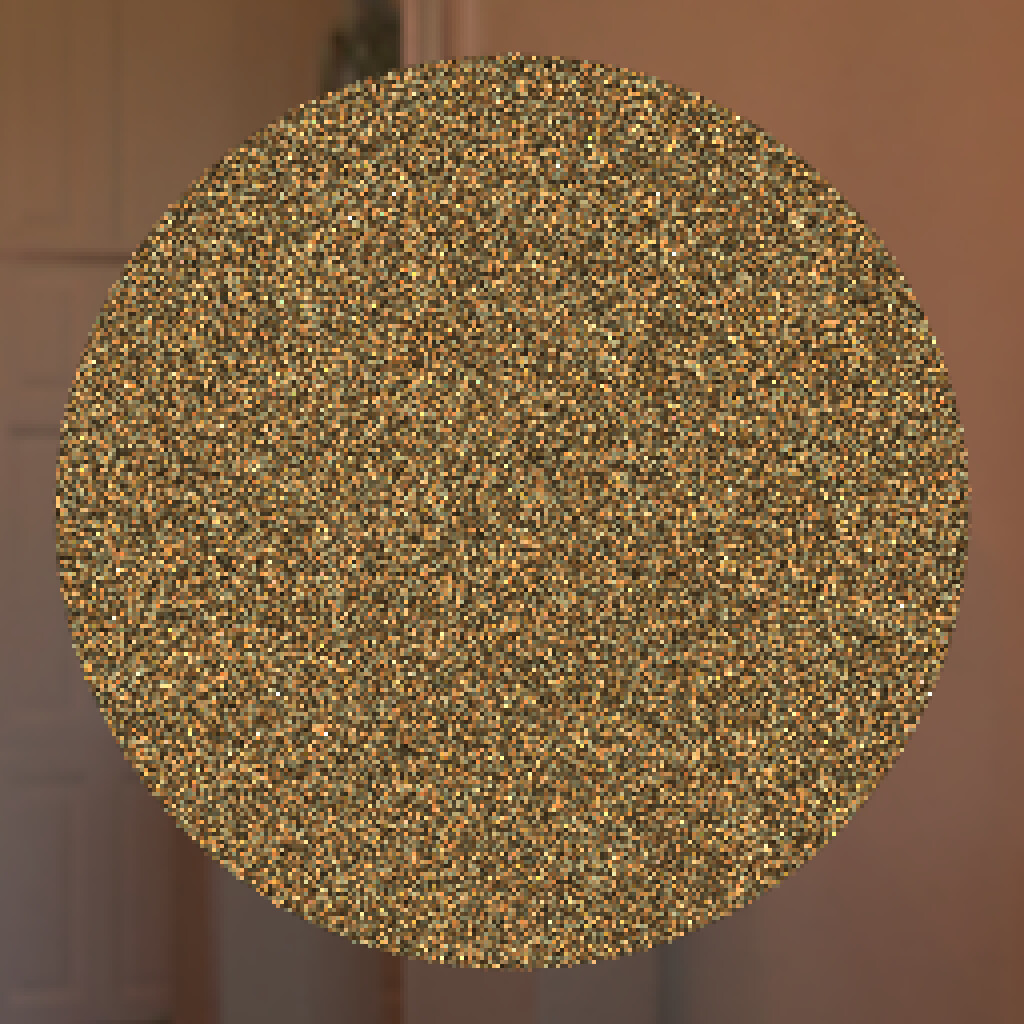}}
        &
        \frame{\includegraphics[height=\lenMaterialSphere]{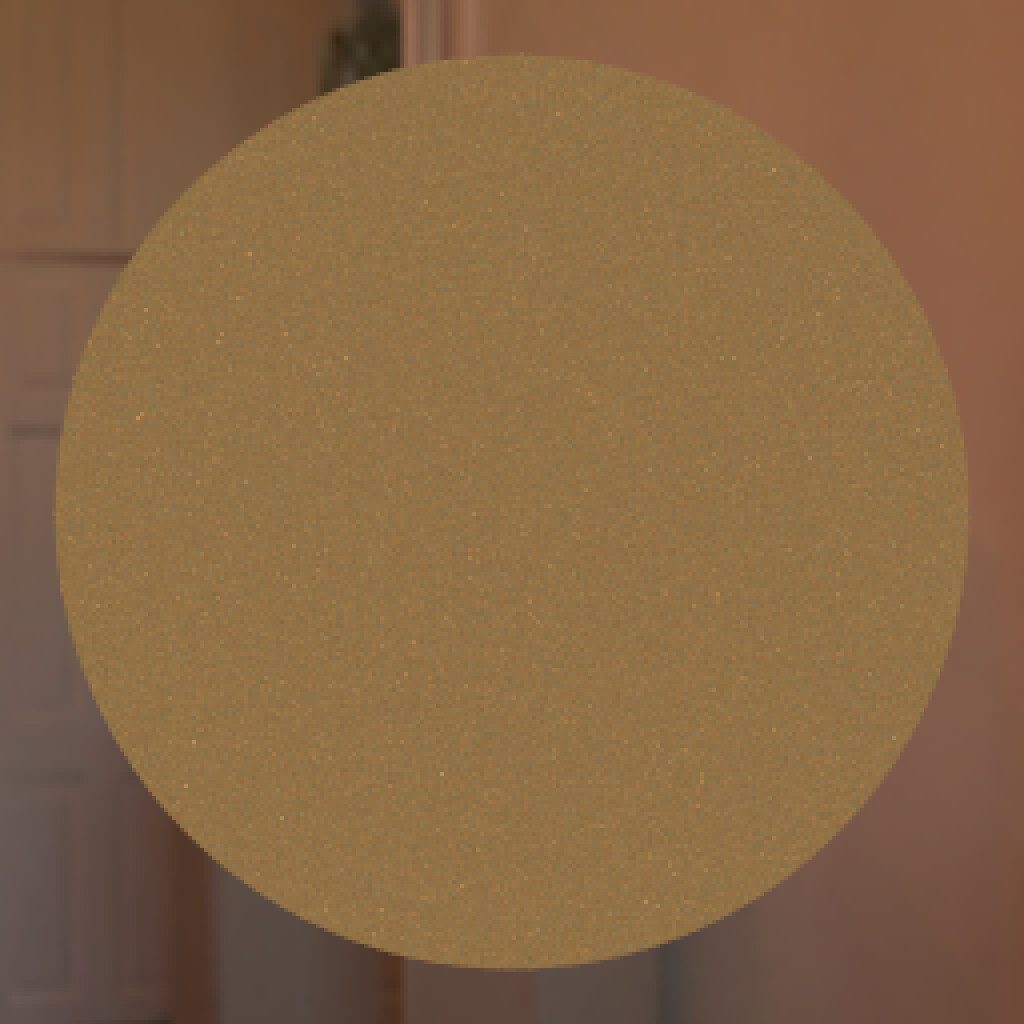}}
        &
        \frame{\includegraphics[height=\lenMaterialSphere]{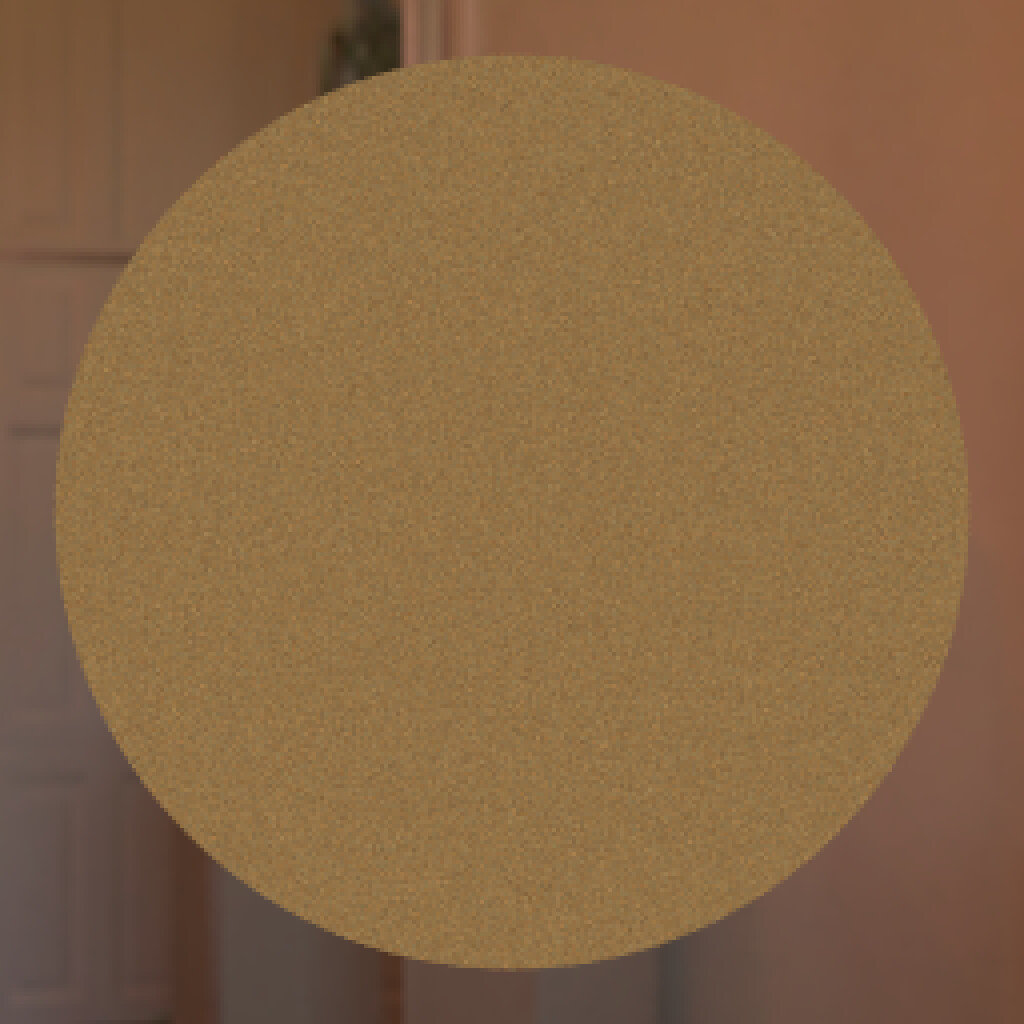}}
        &        
        \frame{\includegraphics[height=\lenMaterialSphere]{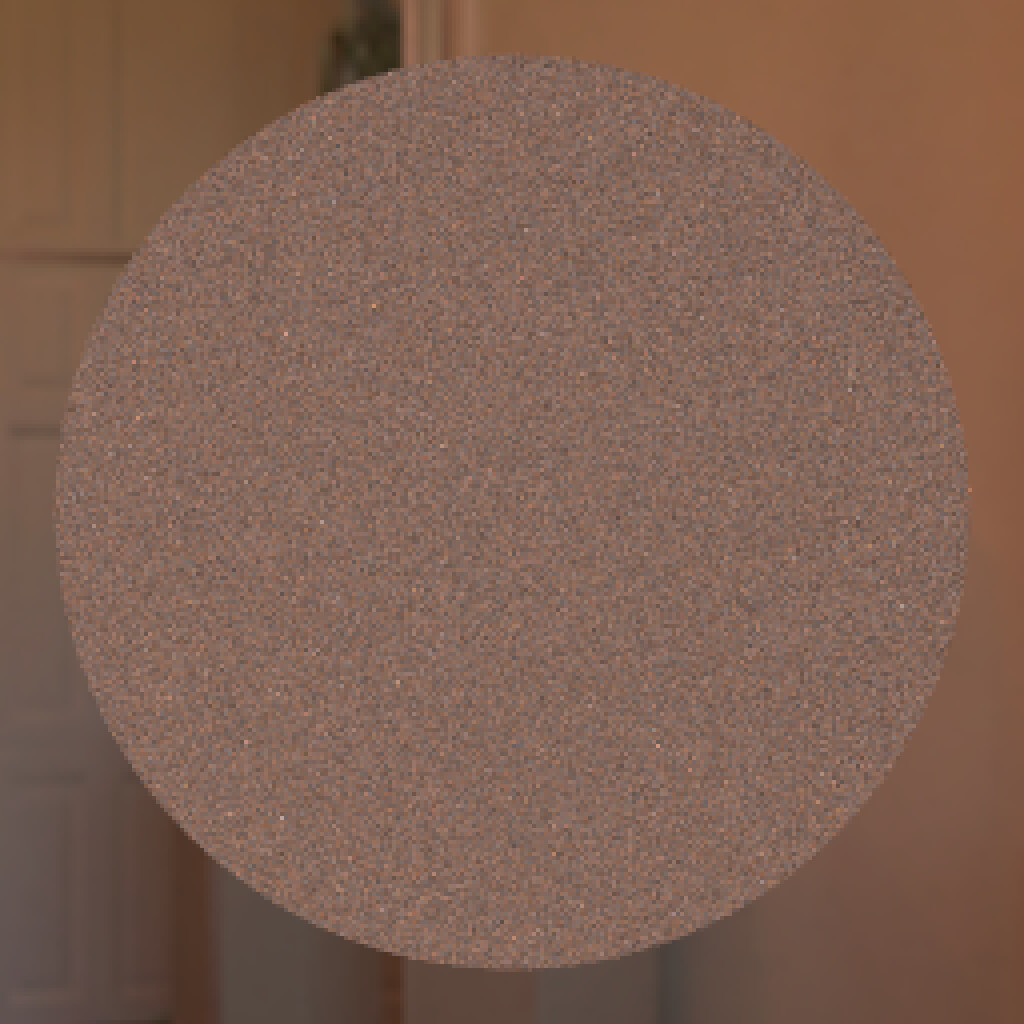}}
        &
        \frame{\includegraphics[height=\lenMaterialSphere]{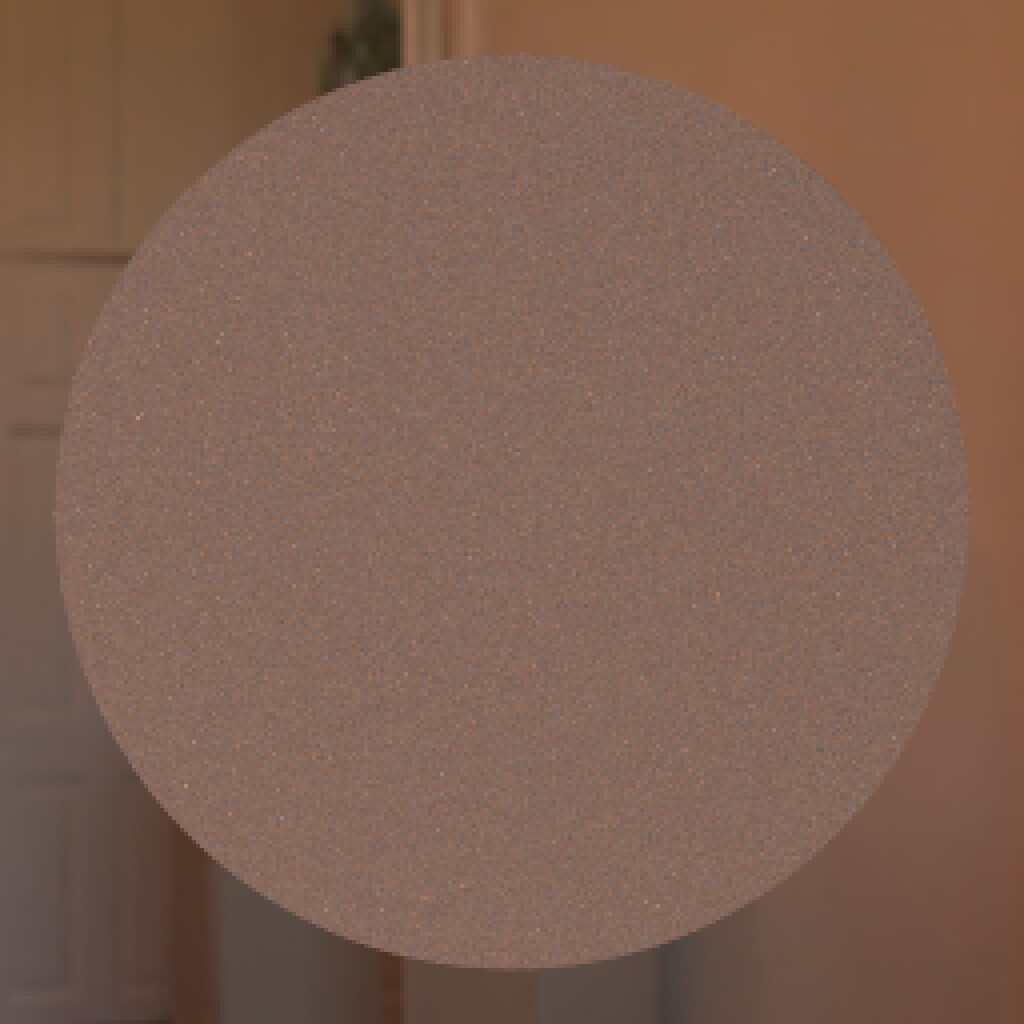}}
        &
        \frame{\includegraphics[height=\lenMaterialSphere]{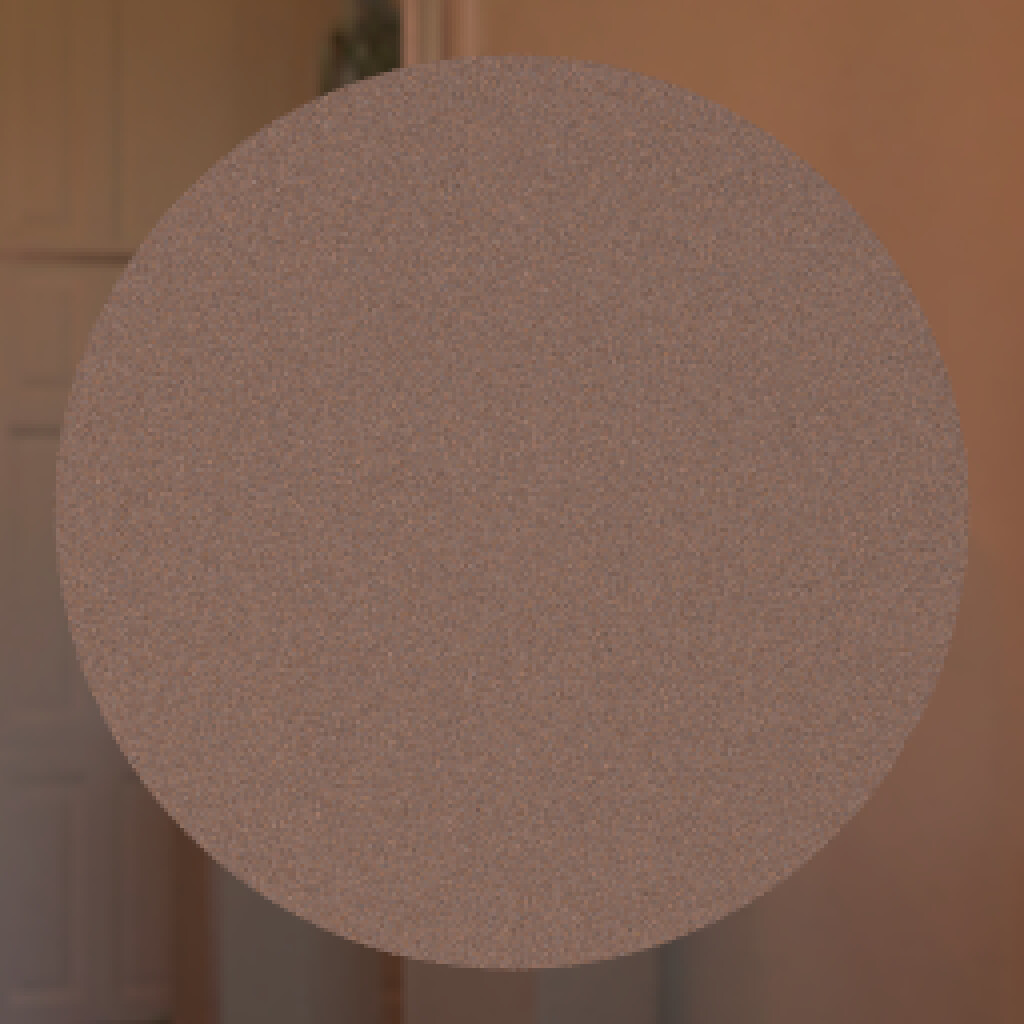}}
        &
        \frame{\includegraphics[height=\lenMaterialSphere]{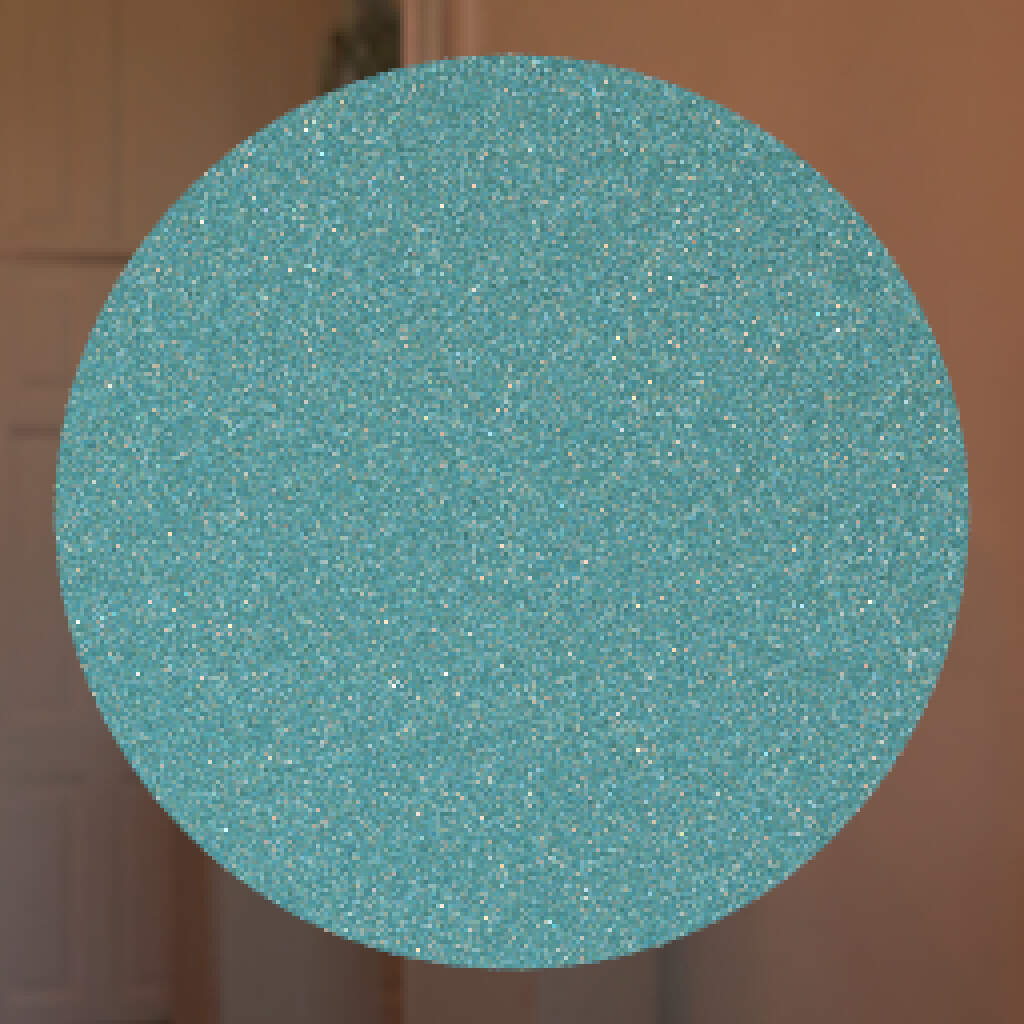}}
        &
        \frame{\includegraphics[height=\lenMaterialSphere]{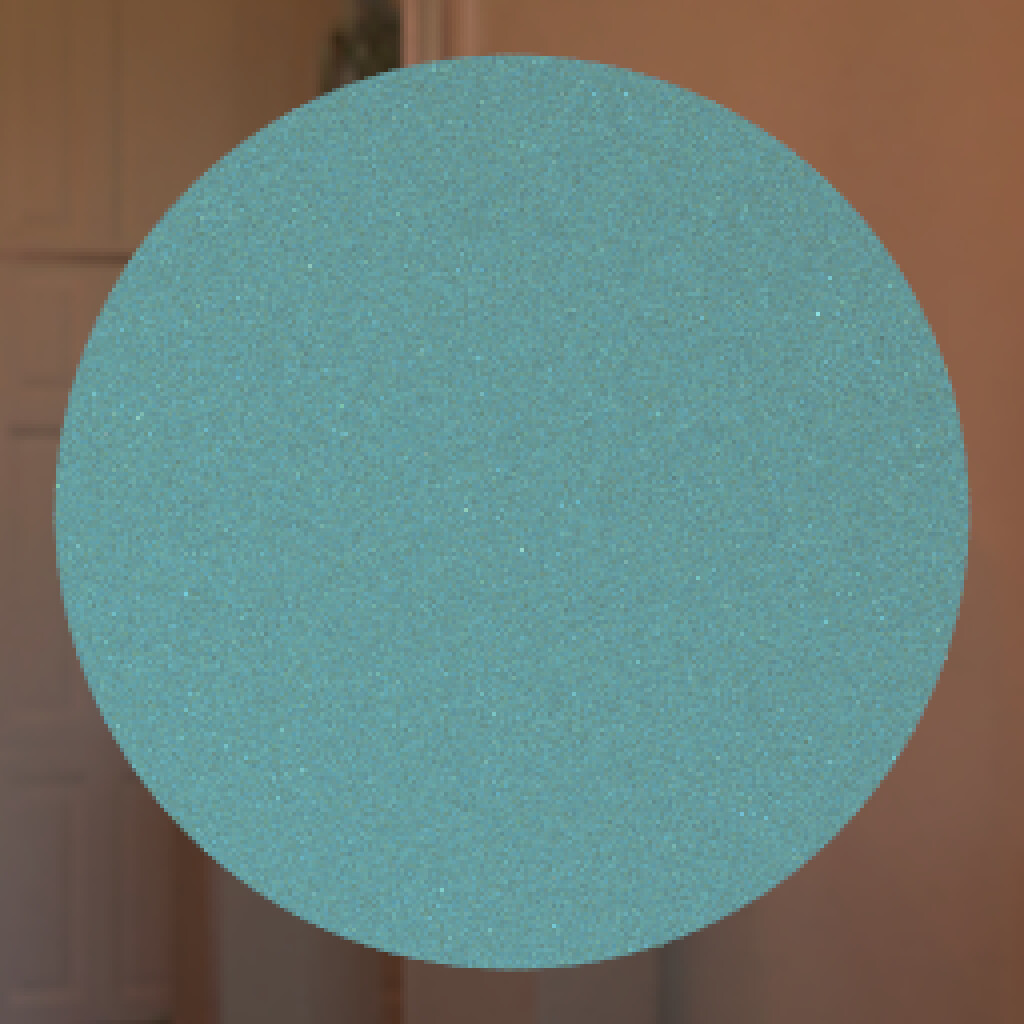}}
        &
        \frame{\includegraphics[height=\lenMaterialSphere]{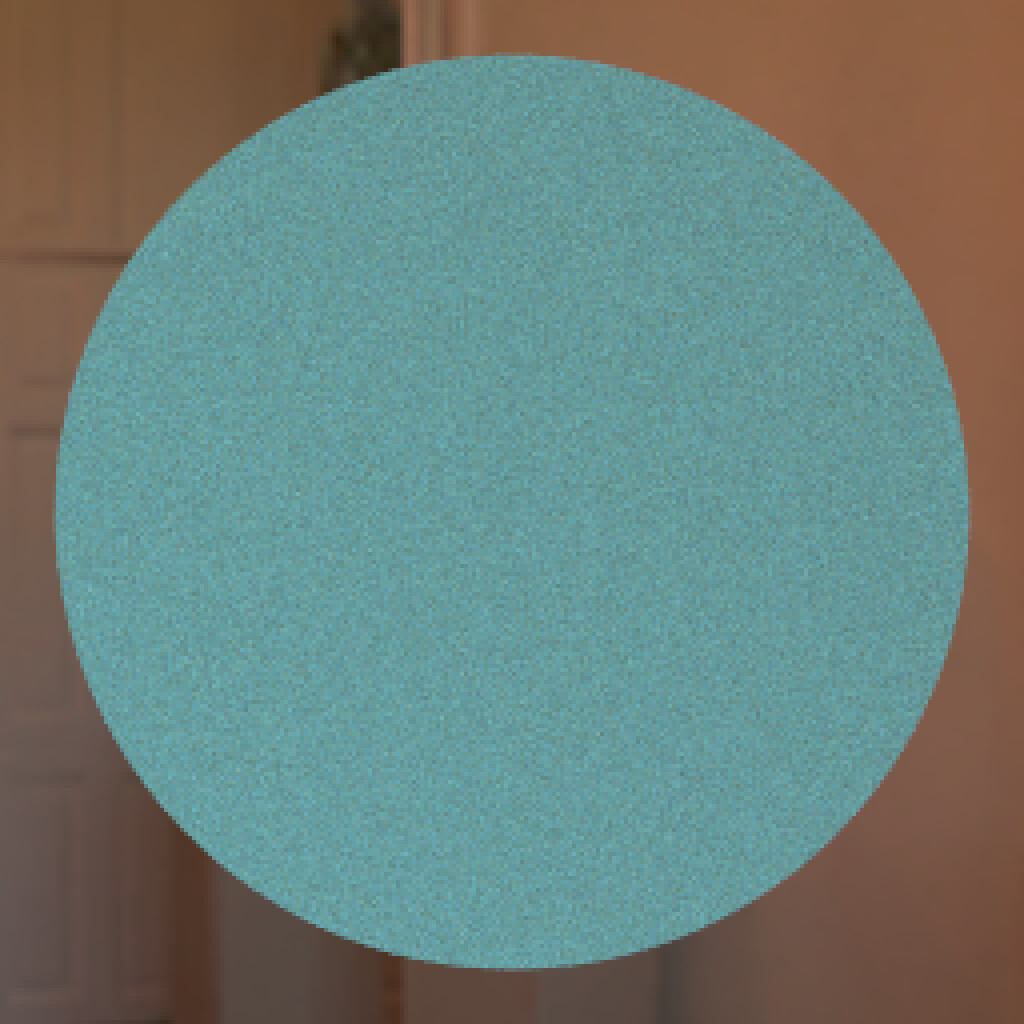}}
        \\   
        \raisebox{12pt}{\rotatebox{90}{\footnotesize{\textsf{Fiber-like}}}}
        &
        \frame{\includegraphics[height=\lenMaterialSphere]{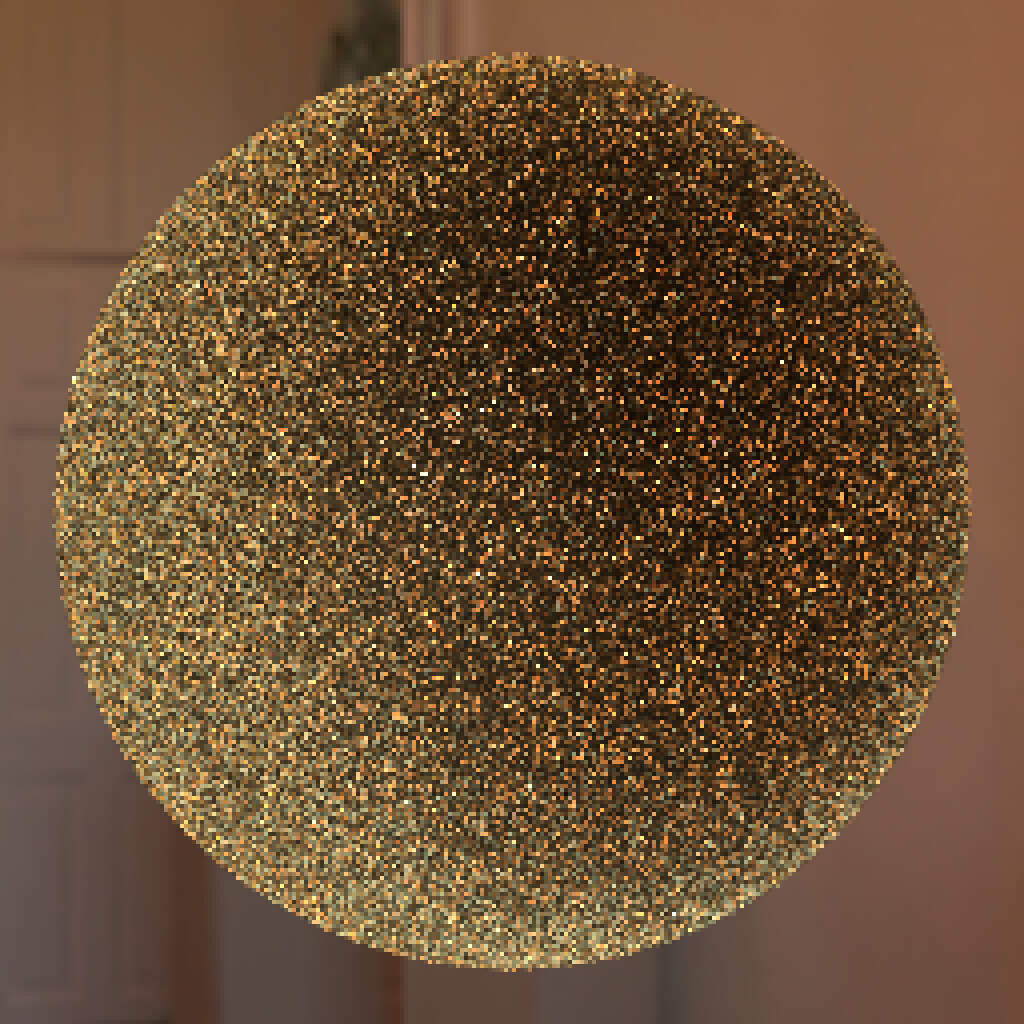}}
        &
        \frame{\includegraphics[height=\lenMaterialSphere]{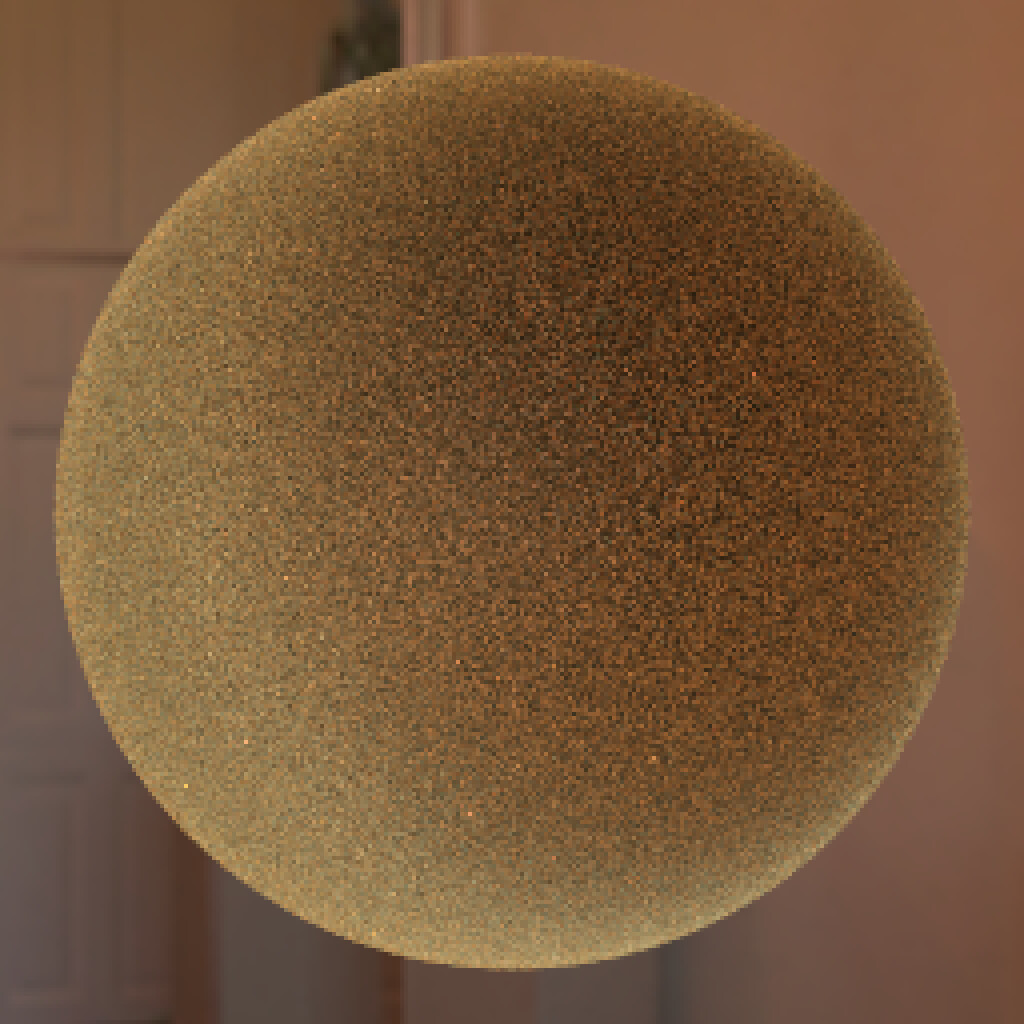}}
        &
        \frame{\includegraphics[height=\lenMaterialSphere]{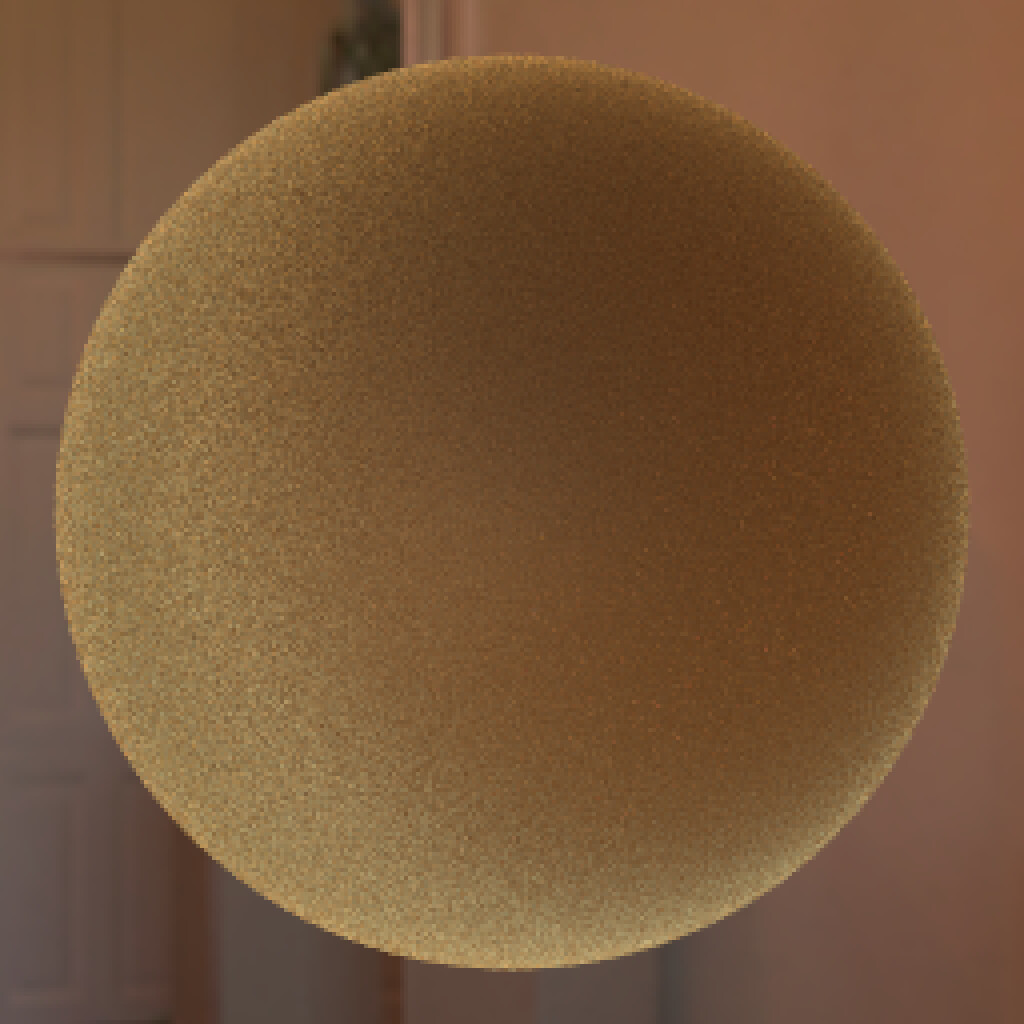}}
        &        
        \frame{\includegraphics[height=\lenMaterialSphere]{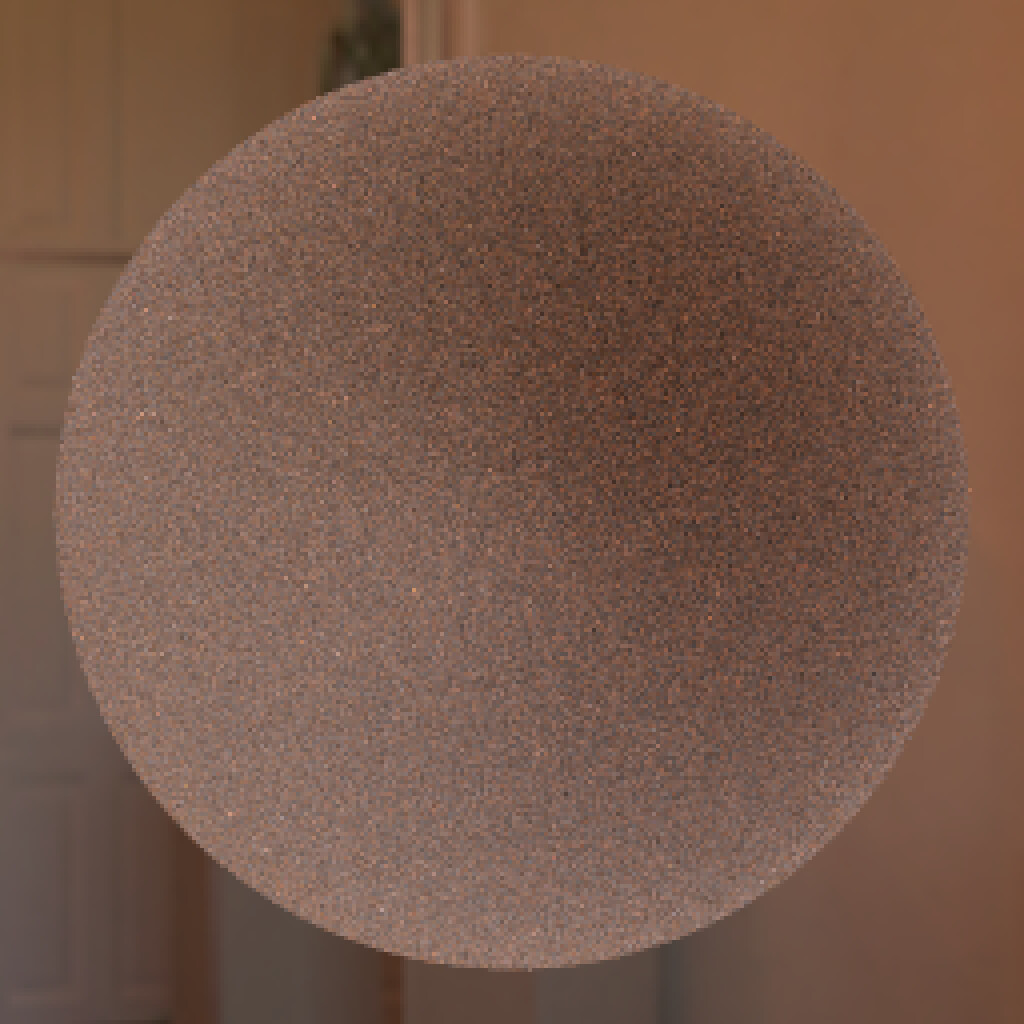}}
        &
        \frame{\includegraphics[height=\lenMaterialSphere]{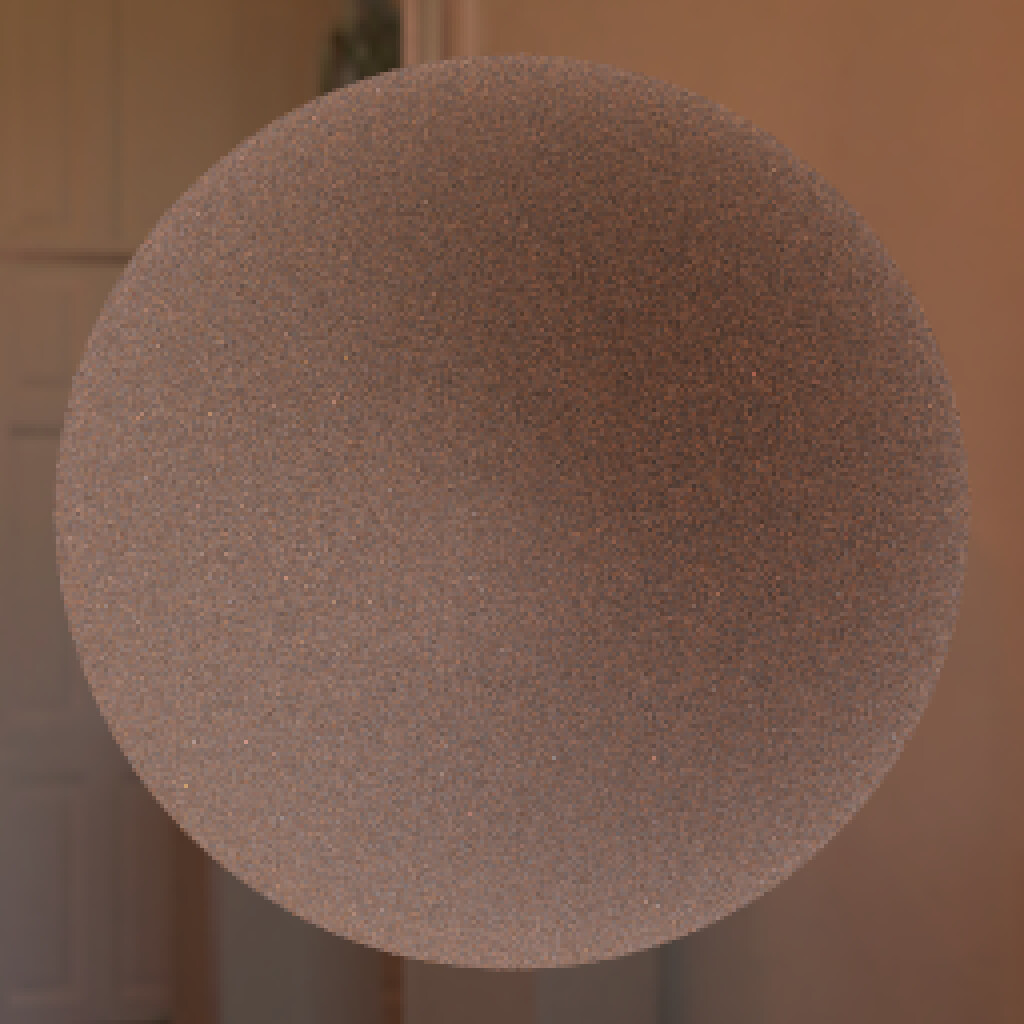}}
        &
        \frame{\includegraphics[height=\lenMaterialSphere]{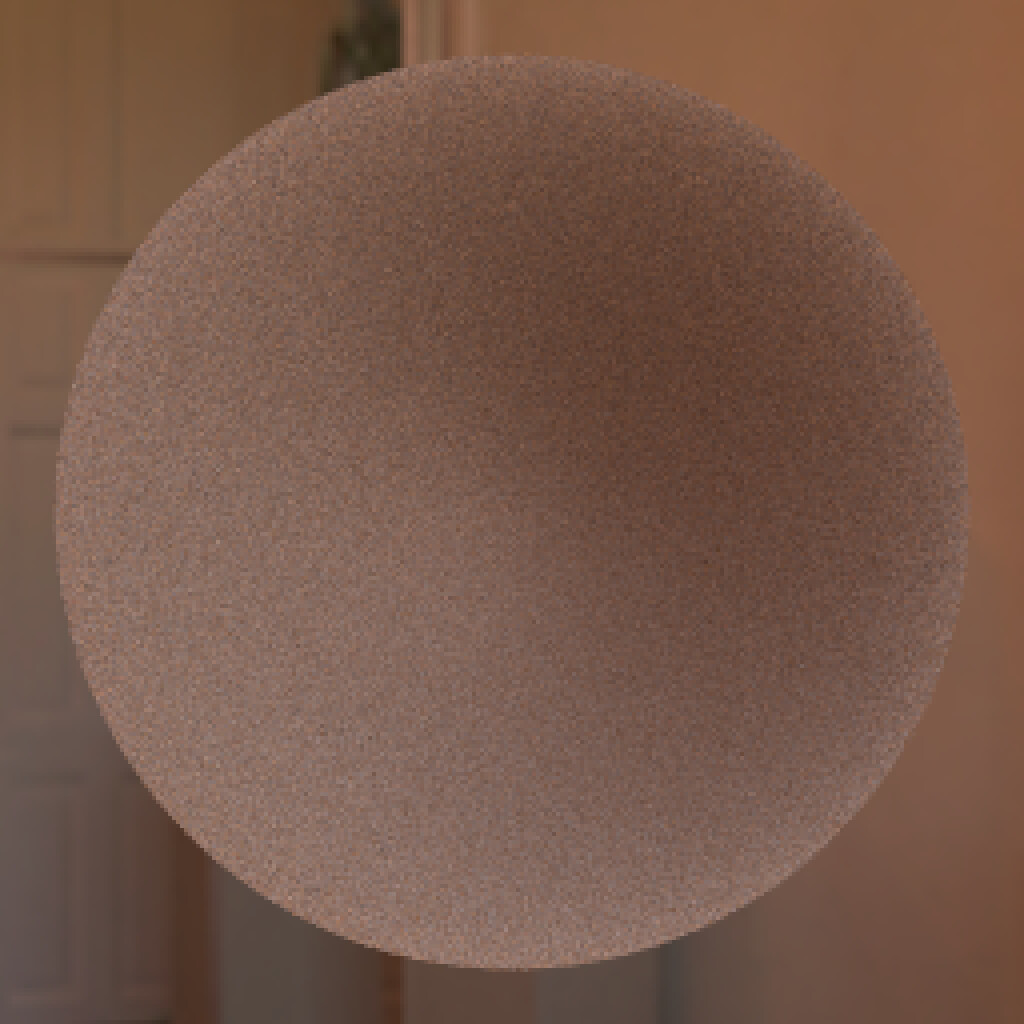}}
        &
        \frame{\includegraphics[height=\lenMaterialSphere]{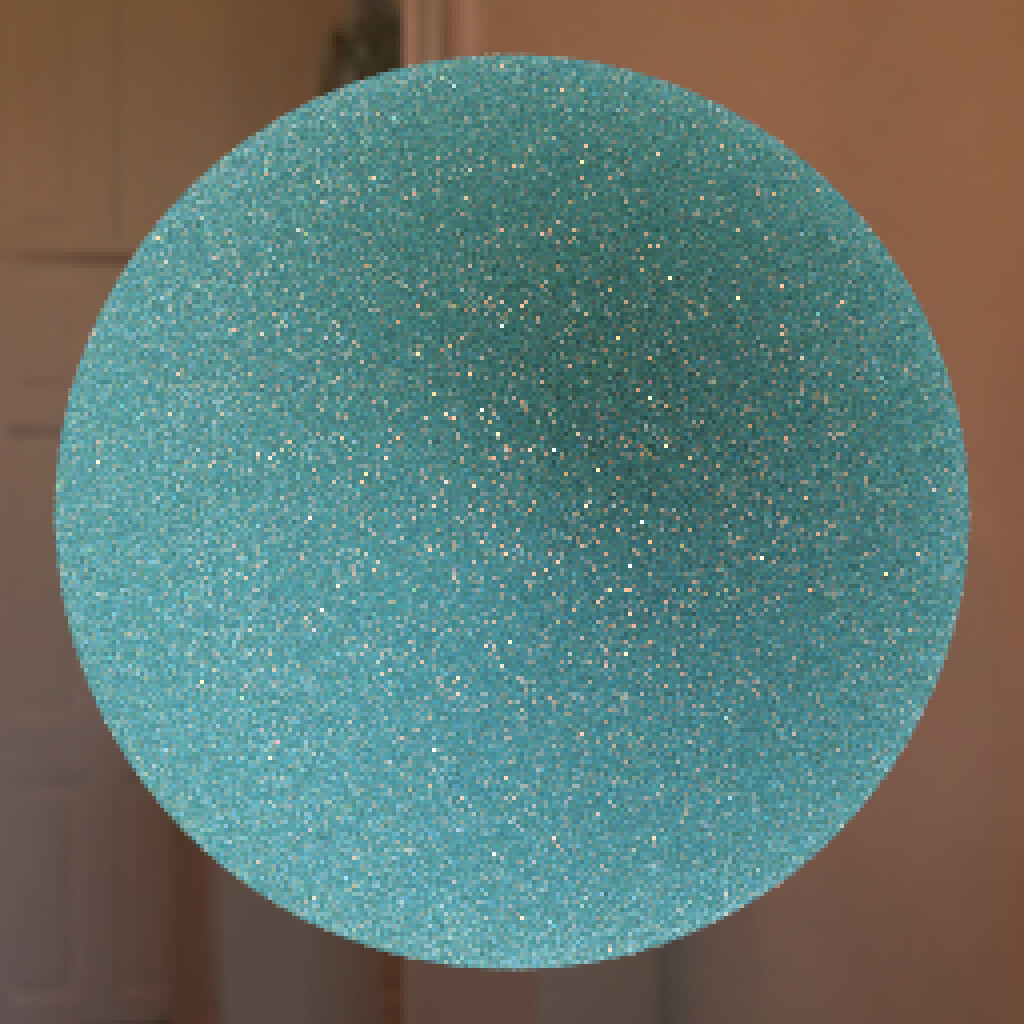}}
        &
        \frame{\includegraphics[height=\lenMaterialSphere]{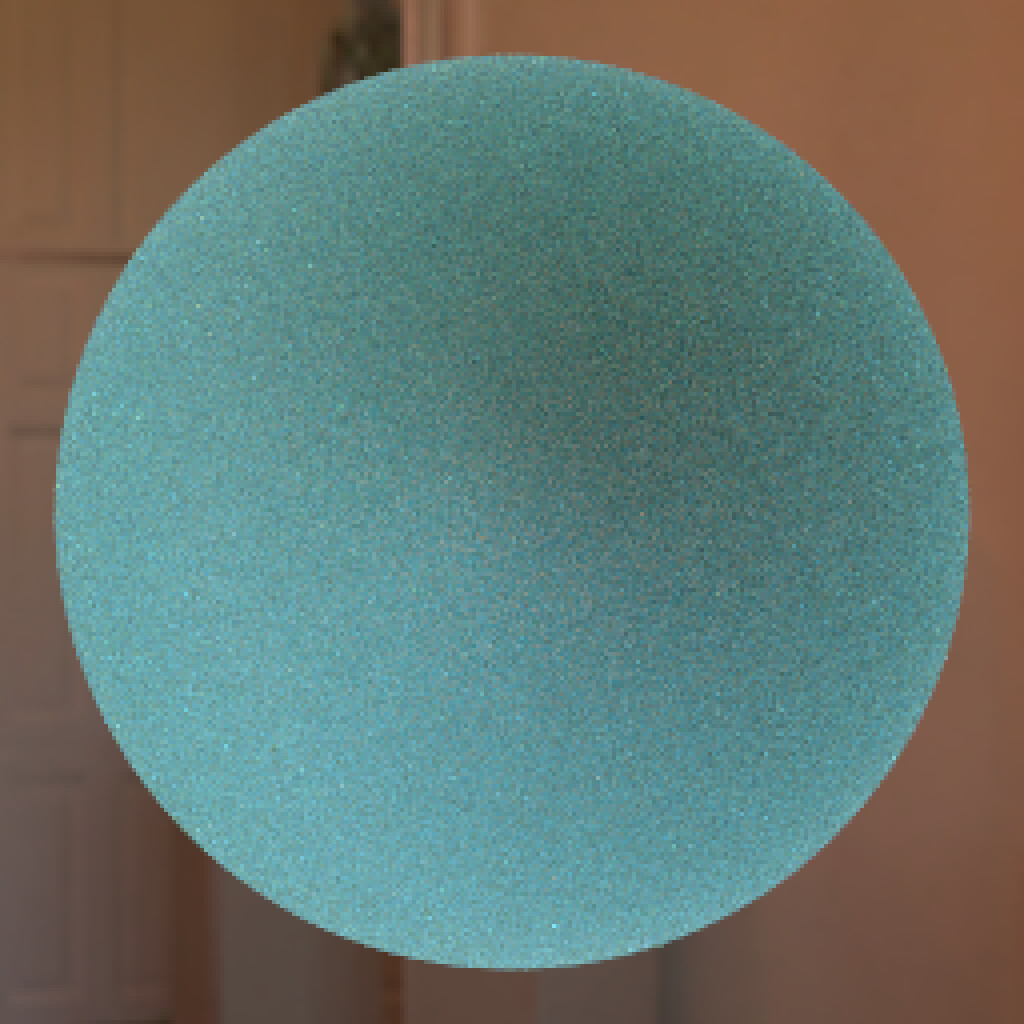}}
        &
        \frame{\includegraphics[height=\lenMaterialSphere]{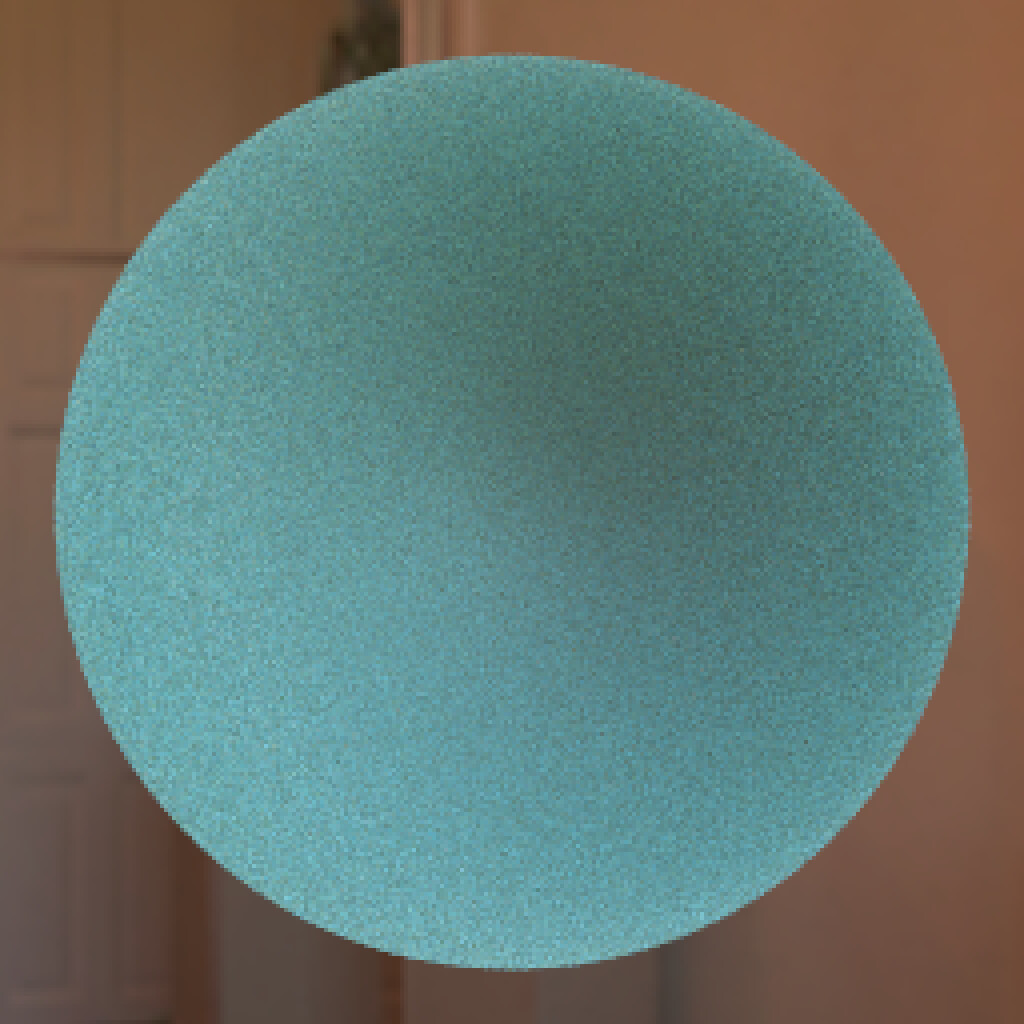}}
        \\
        &
        \footnotesize{\textsf{Na\"ive}} & \footnotesize{\textsf{Improved}} & \footnotesize{\textsf{Improved + MIS}}
        &
        \footnotesize{\textsf{Na\"ive}} & \footnotesize{\textsf{Improved}} & \footnotesize{\textsf{Improved + MIS}}
        &
        \footnotesize{\textsf{Na\"ive}} & \footnotesize{\textsf{Improved}} & \footnotesize{\textsf{Improved + MIS}}                      
    \end{tabular}
    \caption{\label{fig:material_sphere}
        Equal-sample rendering comparison of different phase function configurations with varying NDFs (rows) and base BSDFs (columns). 
        Our improved stochastic evaluation (``Improved'') is much more efficient than the simple stochastic evaluation based on SGGX VNDF sampling only (``Na\"ive''). 
        Combining it with MIS (``Improved + MIS'') provides further variance reduction. All images are rendered using 64 samples per pixel.
    }                    
\end{figure*}
\section{Data Acquisition} \label{sec:data_acquisition}
In this section, we describe methods to acquire data from other synthetic source data for our representations. 
As previously mentioned, our work focuses on defining a unified scene representation with our Gaussian primitive and does not attempt to solve the full 
inverse rendering problem. Therefore, in order to acquire full scene data for rendering, we propose conversion processes from meshes and 3DGS data, 
two widely adopted \emph{existing} representations. In other words, the input of these processes assumes 3D ground truth, and the output supports light 
transport simulation. 
Alternatively, for applications that do not require light transport simulation, our representation also supports image-based radiance field reconstruction, 
a more powerful differentiable rendering technique (\autoref{sec:radiance_field}).

\subsection{Conversion from Existing Representations} \label{subsec:conversion}
\paragraph{Conversion from Meshes} 
We provide a heuristic method to convert a mesh to a set of Gaussian primitives. This method generates flat, opaque ellipse-like primitives to cover the 
original surface. We uniformly sample points on the mesh to initialize flat Gaussians aligned to the mesh surface. 
Let $A$ be the surface area of the mesh, $N$ be the number of Gaussians, and $(s_x, s_y, s_z)$ be the diagonal elements of the scale matrix $S$ 
\autoref{eq:covariance_decomposition}. We set $s_x$ and $s_y$ to $\epsilon k A / n$, where $\epsilon$ is the cutoff threshold \autoref{eq:cutoff_threshold}, 
and $k$ is an adjustable parameter set to $16$ for our experiments. $s_z$ is then set to $0.1 s_x$. 

After initializing the shape parameters, we determine the appearance counterpart for each Gaussian by assigning those parameters from the source mesh.
We consider the spatial neighborhood of each Gaussian while heuristically rejecting outliers to better maintain the original silhouette and texture details 
(if textured).
We generate $2048$ samples $\{p_i\}$ 
according to the Gaussian distribution and project them to the plane defined by the center $\mu$ and normal. For each $p_i$, we query the nearest point on the 
mesh and gather its BSDF parameter vector $\phi_i$, normal $n_i$, and the distance $d_i$ between $p_i$ and the returned query. We examine the similarity between a sample and the center 
point $\mu$ based on those attributes, following the heuristic formula
\begin{equation}
    o_i = \left(\bigvee\limits_{k} \|\phi_{i,k} - \phi_{\mu, k}\| > \epsilon_k \right) \lor \left(n_i\cdot n_{\mu} < 1 - \epsilon_n\right) \lor \left(d_i > \epsilon_d\right),
\end{equation}
where $\phi_{*,k}$ is each BSDF parameter and $\epsilon_*$ are the thresholds for different attributes. If $o_i$ is true, the sample is rejected as an outlier.
We use the distance between the closest outlier $p'$ and $\mu$ to be $s_x$, and set the corresponding eigenvector $R_x$ as the direction from $\mu$ to $p'$. The 
last eigenvector follows as $R_y = n_\mu \times \ R_x$. $s_y$ is clamped to the maximum distance such that the bounding ellipsoid of the Gaussian does 
not contain any outlier sample points. Finally, we average $\phi_i$ across the accepted samples to obtain the BSDF parameters for the primitive.
The Gaussian primitives produced from this method are different from those by the concurrent work by \citet{Huang2DGS2024}, which are pure 2D surfel-like 
\emph{surface-only} representations. Their method focuses on reconstruction and does not support full appearance (weakly direction-dependent SH colors only) 
or further light transport. Our representation is not limited to flat Gaussians, as can be seen in the following.

\begin{figure}[h]
	\newlength{\lenMeshConvert}
	\setlength{\lenMeshConvert}{0.48\linewidth}
    \addtolength{\tabcolsep}{-4pt}
    \renewcommand{\arraystretch}{0.5}
    \centering
    \begin{tabular}{cc}
      \frame{\includegraphics[width=\lenMeshConvert]{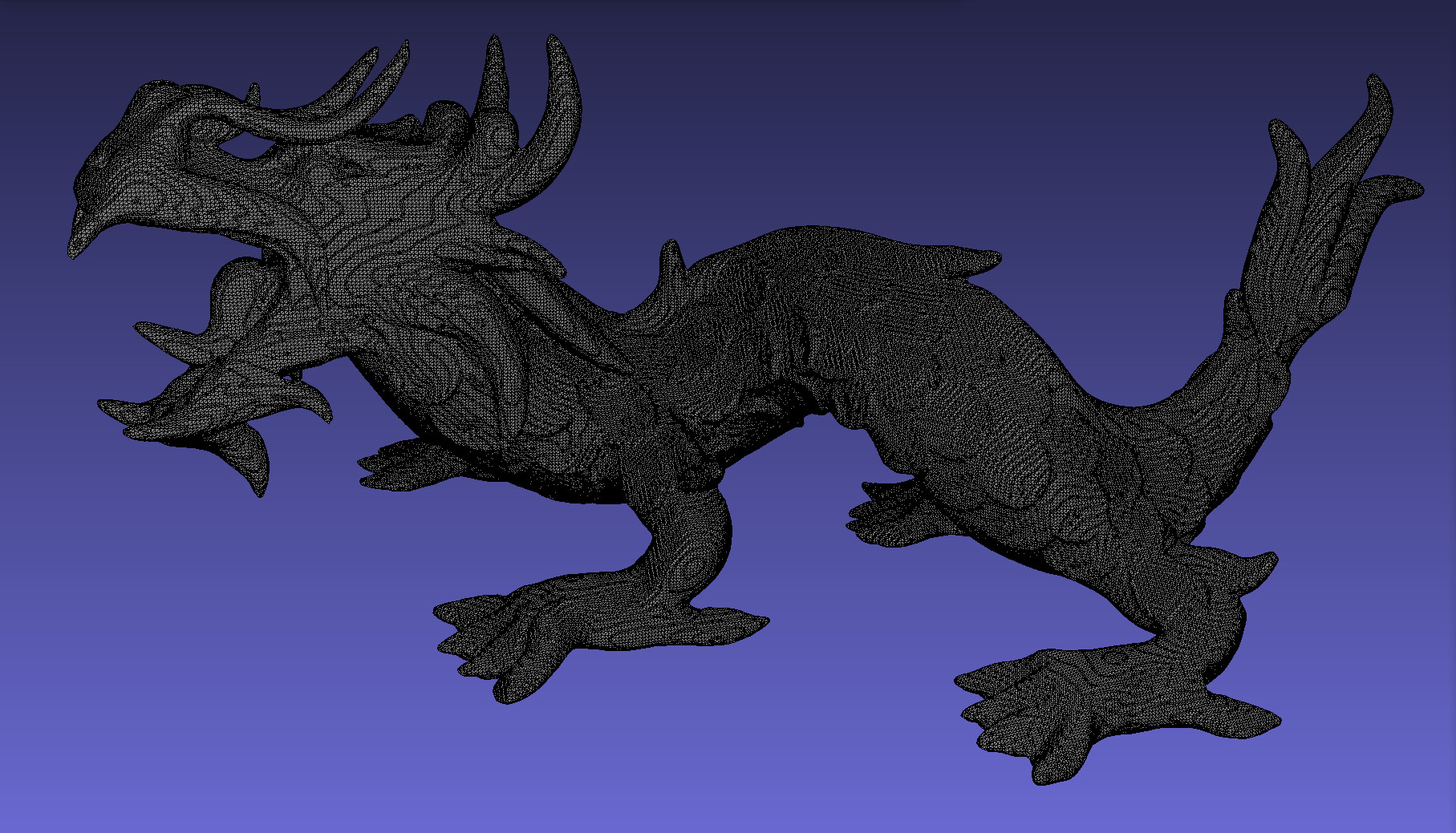}}
      &
      \frame{\includegraphics[width=\lenMeshConvert]{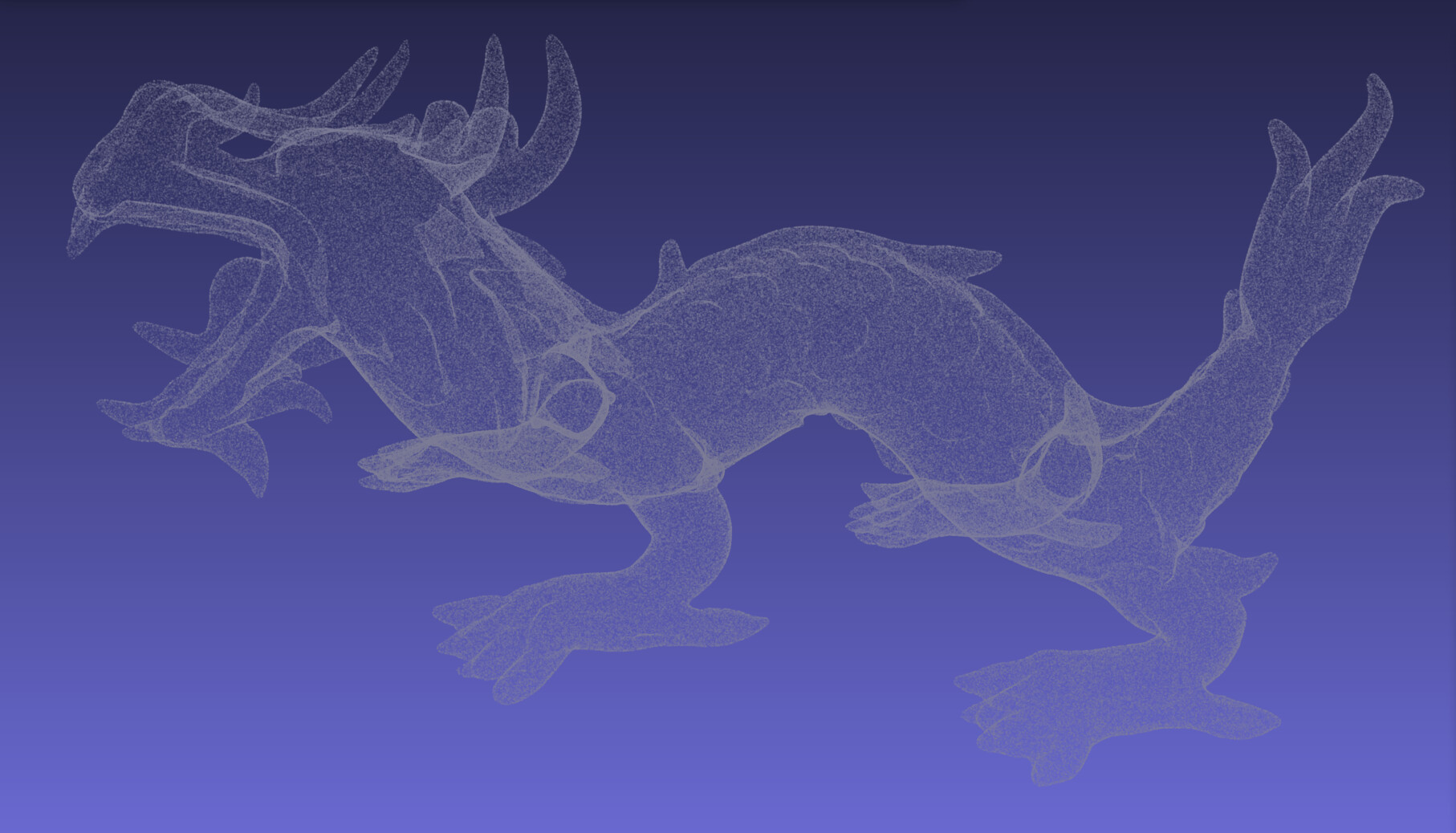}}
      \\
      \small{\textsf{(a) Mesh wireframe}} & \small{\textsf{(b) Point samples}} \\
      \frame{\includegraphics[width=\lenMeshConvert]{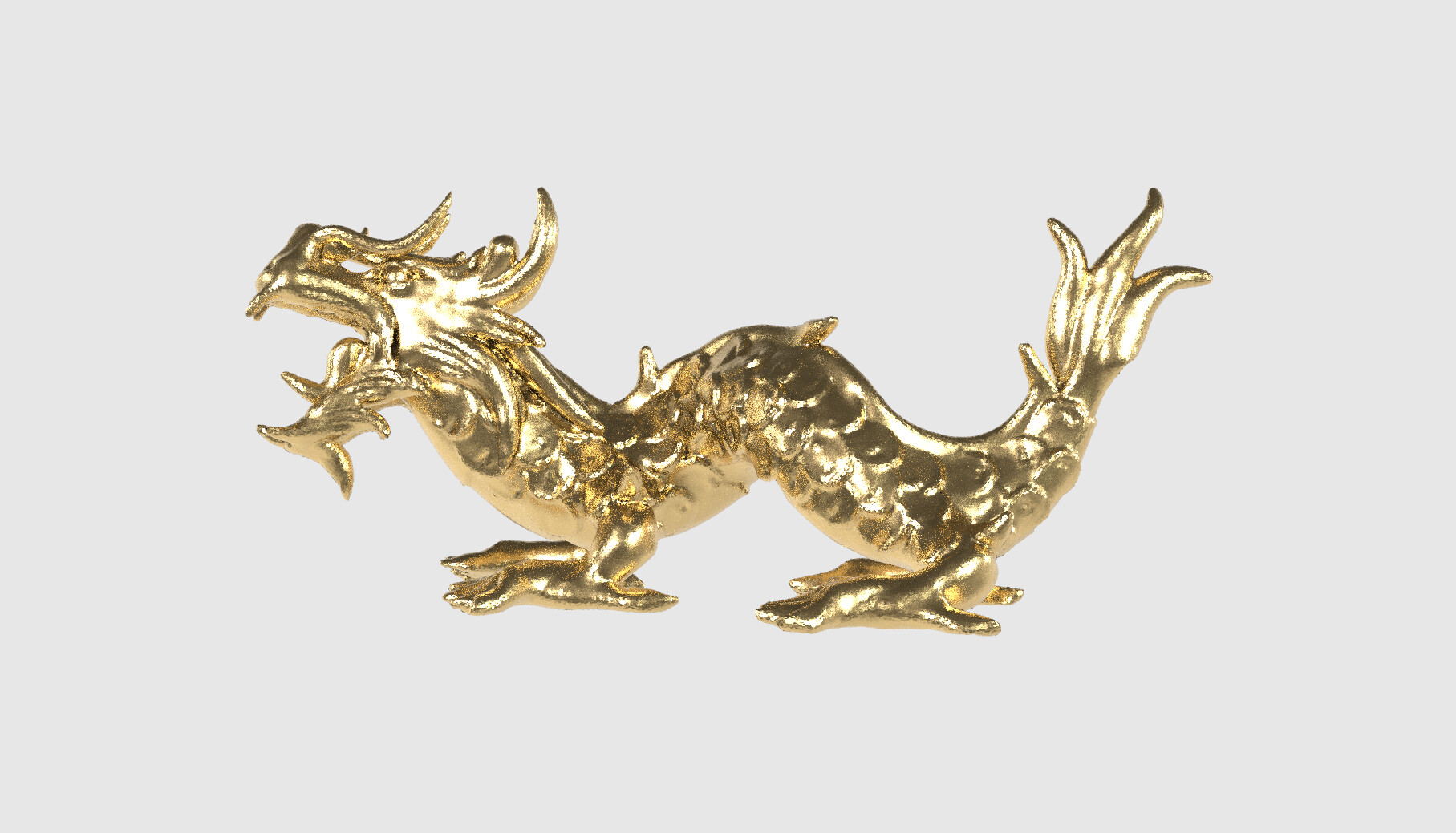}}    
      &
      \frame{\includegraphics[width=\lenMeshConvert]{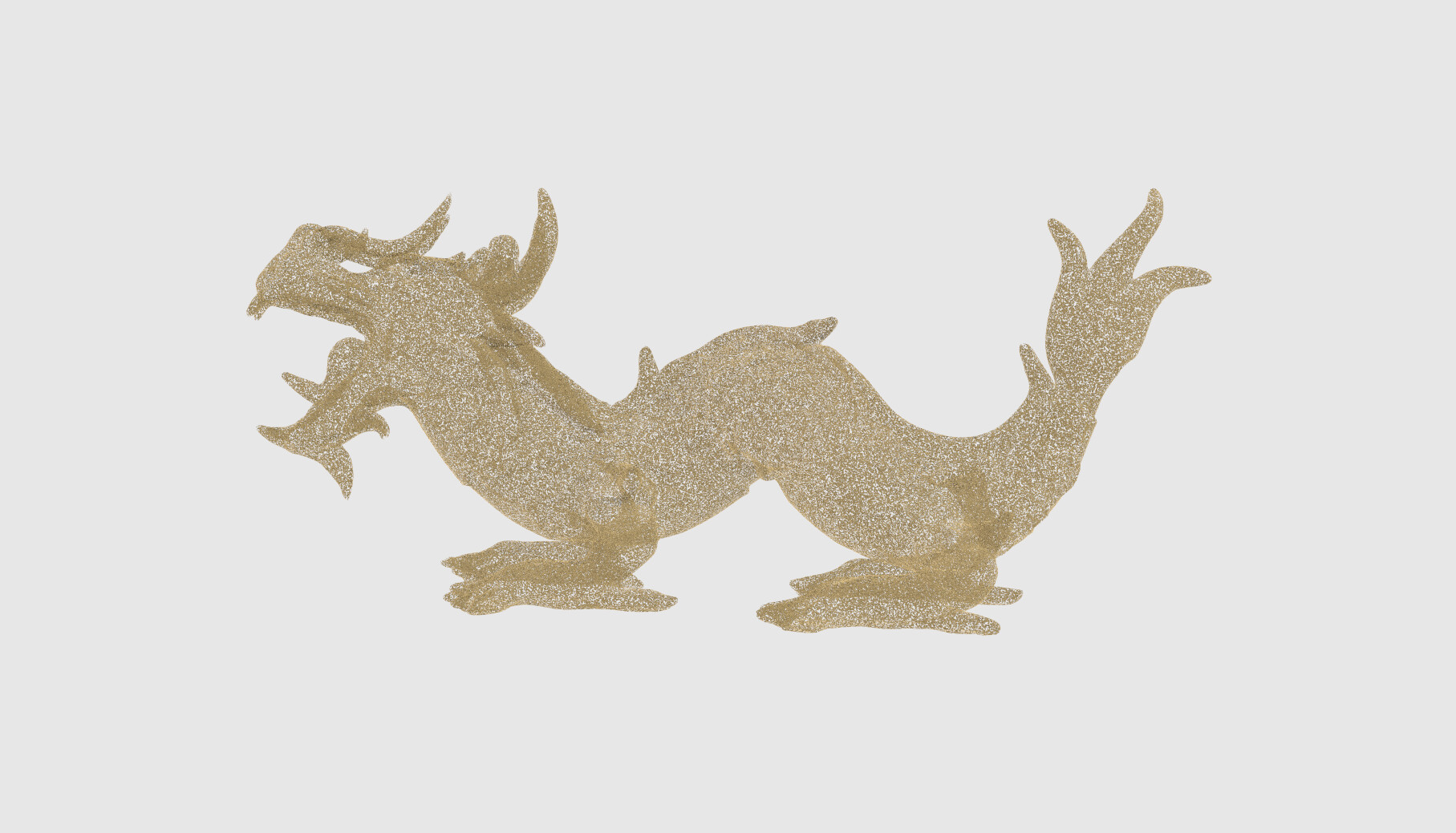}}
      \\
      \small{\textsf{(c) Render}} & \small{\textsf{(d) Shrunken Gaussians}}
    \end{tabular}
    \caption{\label{fig:mesh_convert}
        Our heuristic method samples points from the original mesh and generates flat, opaque Gaussians to cover the mesh surface. We visualize the generated 
        Gaussians by shrinking them (d).
    }
\end{figure}

\paragraph{Conversion from 3DGS}
3DGS can optimize radiance fields using anisotropic Gaussians that adapt to complex shapes. While radiance fields cannot be directly used as input for our 
framework, we find 3DGS a viable tool for the initial conversion of synthetic models that include dense and thin elements, where the previous heuristic 
conversion method is not suitable. We generate a synthetic dataset for the model by rendering it from multiple views, which is used as the input for 3DGS. 
We then extract the shape parameters $\mu, R$, and $S$ for our primitives. Interestingly, there exists a mapping between the opacity $\alpha$ in 3DGS and our 
magnitude $c$, which we detail in \autoref{sec:3dgs_conversion}. We use this remapping to initialize our magnitudes, and further optimize transmittance as 
detailed in \autoref{subsec:transmittance_optim}. 

For each primitive, we then perform similar sampling according to the Gaussian distribution. 
Each sample is projected to the closest point on the original model to query normal and BSDF parameters. 
We obtain the phase function by averaging the BSDF parameters and fitting the SGGX NDF according to the procedure by \citet{heitz2015sggx}.
While a full inverse rendering pipeline 
is out of the scope of this work, we demonstrate the differentiation capability of our representation in \autoref{subsec:transmittance_optim}, and later in \autoref{sec:radiance_field}. 
Additionally, developing digital content creation tools for our representation, similar to those for mesh modeling and signed distance field sculpting \citep{substance3dmodeler}, 
would be a desirable future direction. 

\subsection{Transmittance Optimization} \label{subsec:transmittance_optim}
For differentiable rendering, our Gaussian primitive benefits from not requiring dedicated techniques to handle the discontinuities in the rendering 
integral \citep{zhao2020physics}, as it maintains continuity similar to other volumetric representations. To demonstrate the usefulness of our 
primitive's differentiability, we develop a proof-of-concept system that optimizes the transmittance of our models. Thanks to its simple formulation, 
differentiating our linear transmittance model is considerably simpler than the traditional exponential model or the hybrid model by \citet{vicini2021non}.
Given a ray, the accumulated free-flight CDF (one minus transmittance) is a simple sum of the free-flight probabilities from each primitive:
\begin{equation} \label{eq:accum_cdf}
    C = \sum_{k=1}^{n} I_k(t^-_k, \min(t^+_k, t_{\mathrm{sat}})),
\end{equation}
where each primitive intersects the ray from $t^-_k$ to $t^+_k$ (or clamped at the saturating distance $t_{\mathrm{sat}}$). 
Let $\phi_k$ be a parameter of the $k$-th primitive, the partial derivative of $C$ w.r.t. 
$\phi_k$ is trivially
\begin{equation*}
    \frac{\partial C}{\partial \phi_k} = \frac{\partial I_k (t^-_k, \min(t^+_k, t_{\mathrm{sat}}))}{\partial \phi_k}.
\end{equation*}
The right term can be effectively computed by applying automatic differentiation (AD) to \autoref{eq:ray_integral}. Strictly speaking, $C$ should be less than or 
equal to 1, but we relax its definition during the optimization. The case when $C > 1$ is analogous to a ray penetrating multiple layers of surfaces. 
We currently do not apply any special handling to this case, but it could be interesting to consider improvements to better handle objects with complex 
interior structure.

A model is initially converted by either method discussed above, and its transmittance is then refined to better match that of the source asset. We render 
transmittance images and compute the L2 loss across multiple views. We trace one ray for each pixel with jittered sub-pixel offsets to compute and 
differentiate \autoref{eq:accum_cdf}. We optimize this loss over 
all parameters of all Gaussian primitives using the Adam optimizer \citep{kingma2014adam} with a learning rate of \num{2e-5} for rotation and \num{1e-4} for all 
other parameters. After each iteration, scales and magnitudes are clamped to be positive, and rotations are re-normalized.

\begin{figure*}[tb]
	\newlength{\lenDiff}
	\setlength{\lenDiff}{0.15\linewidth}
    \addtolength{\tabcolsep}{-4pt}
    \renewcommand{\arraystretch}{0.5}
    \centering
    \begin{tabular}{cccccccc}
        && \emph{Plant} & && \emph{Tall Vase} &
        \\
        \raisebox{25pt}{\rotatebox{90}{\small{\textsf{Render}}}}
        &
        \frame{\includegraphics[width=\lenDiff]{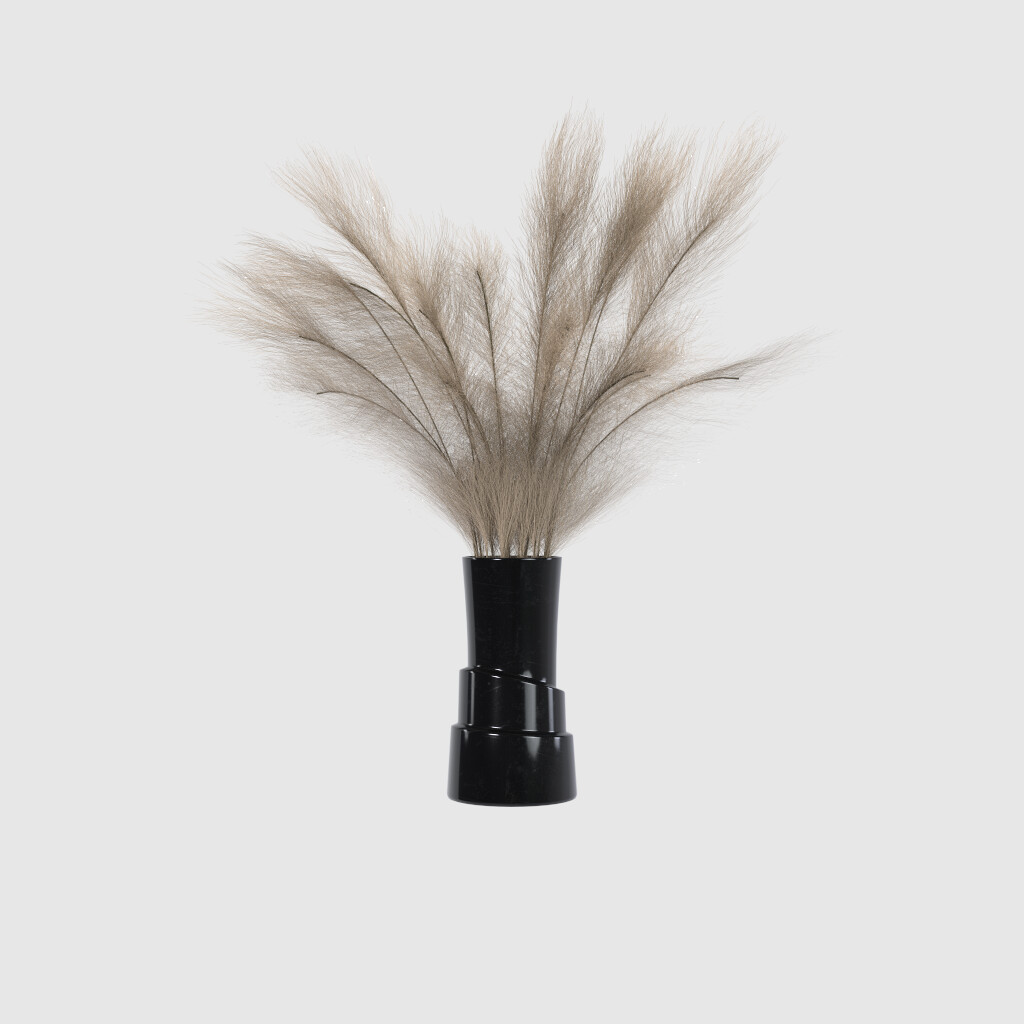}}
        &
        \frame{\includegraphics[width=\lenDiff]{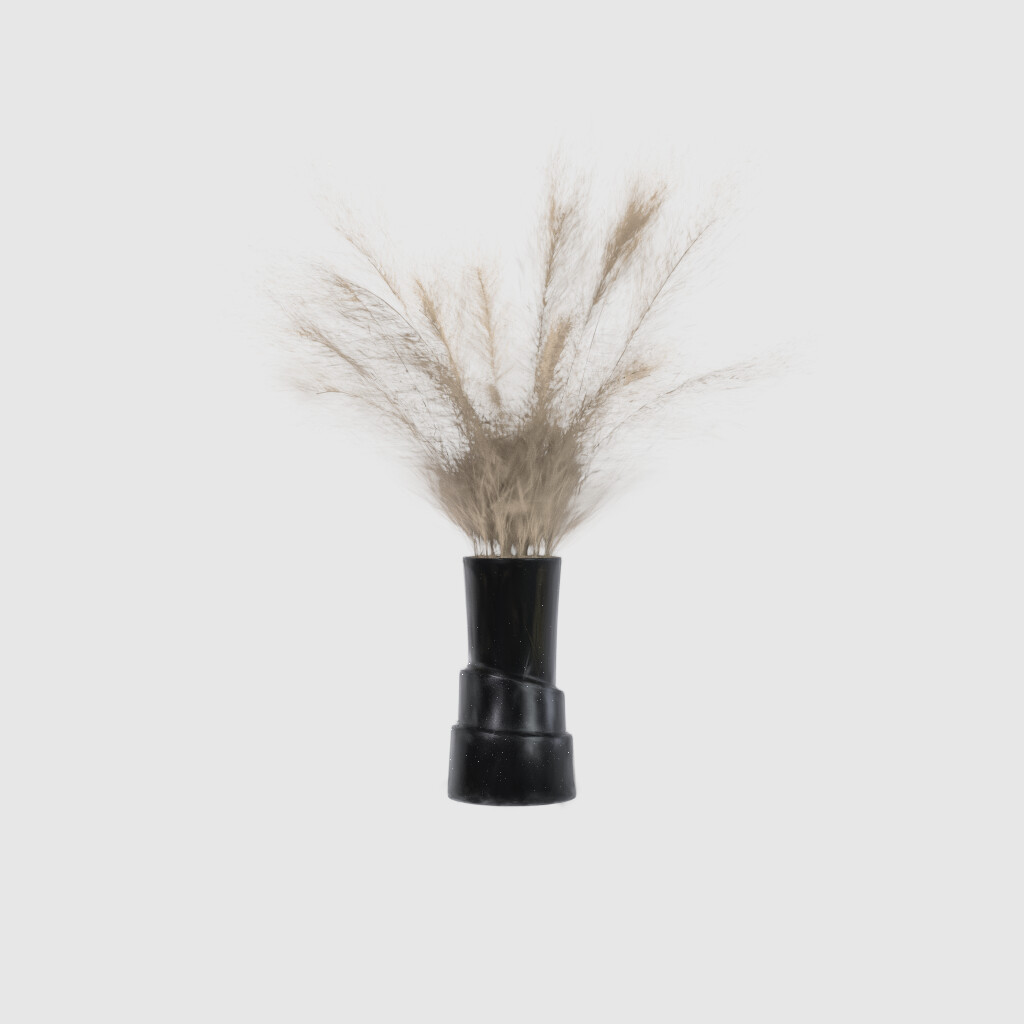}}
        &
        \frame{\includegraphics[width=\lenDiff]{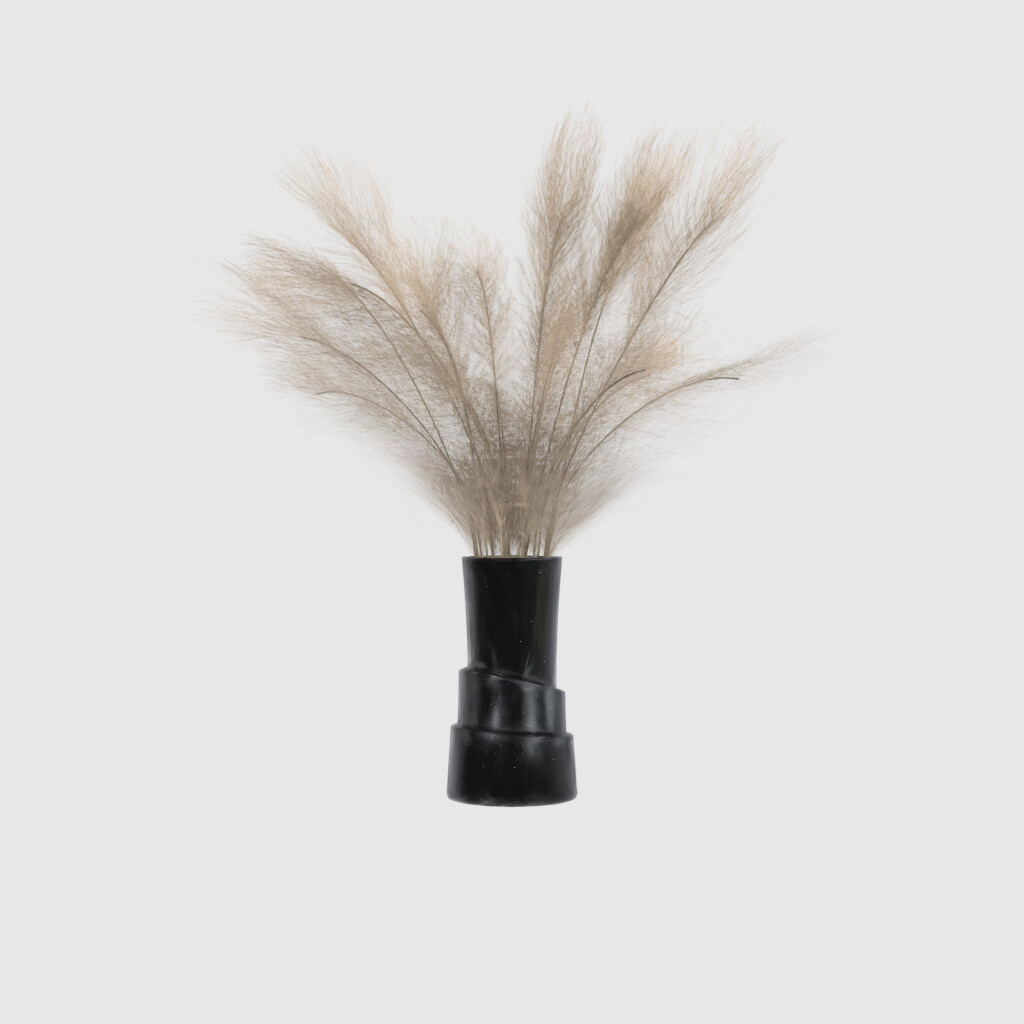}}        
        &        
        \frame{\includegraphics[width=\lenDiff]{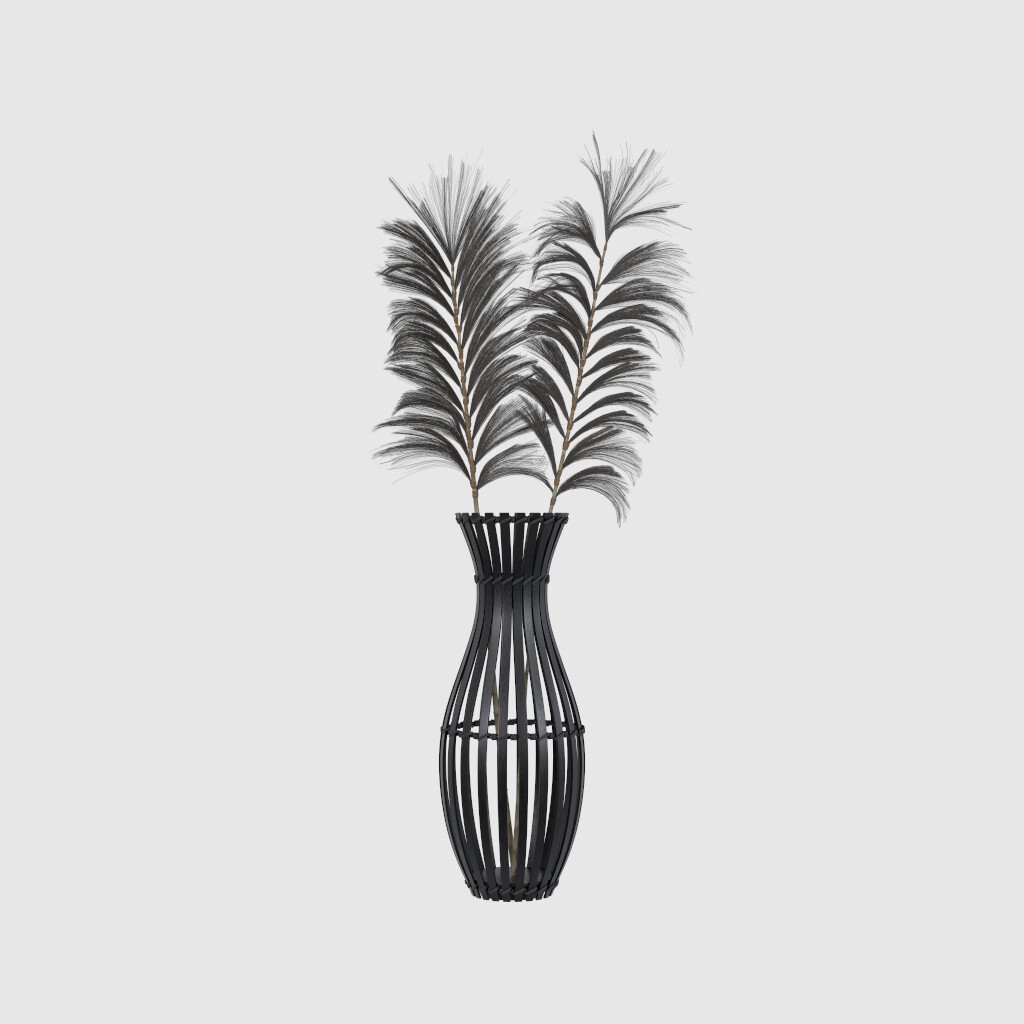}}
        &
        \frame{\includegraphics[width=\lenDiff]{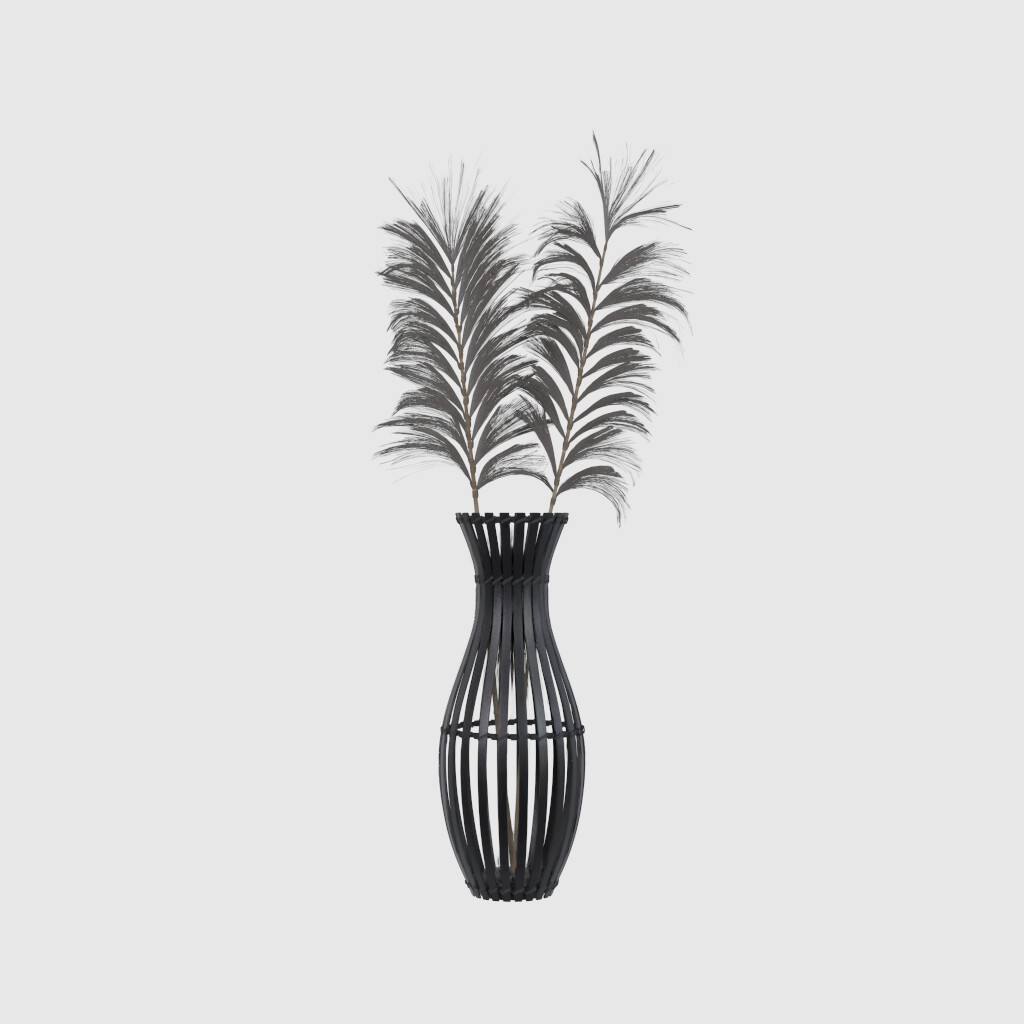}}
        &
        \frame{\includegraphics[width=\lenDiff]{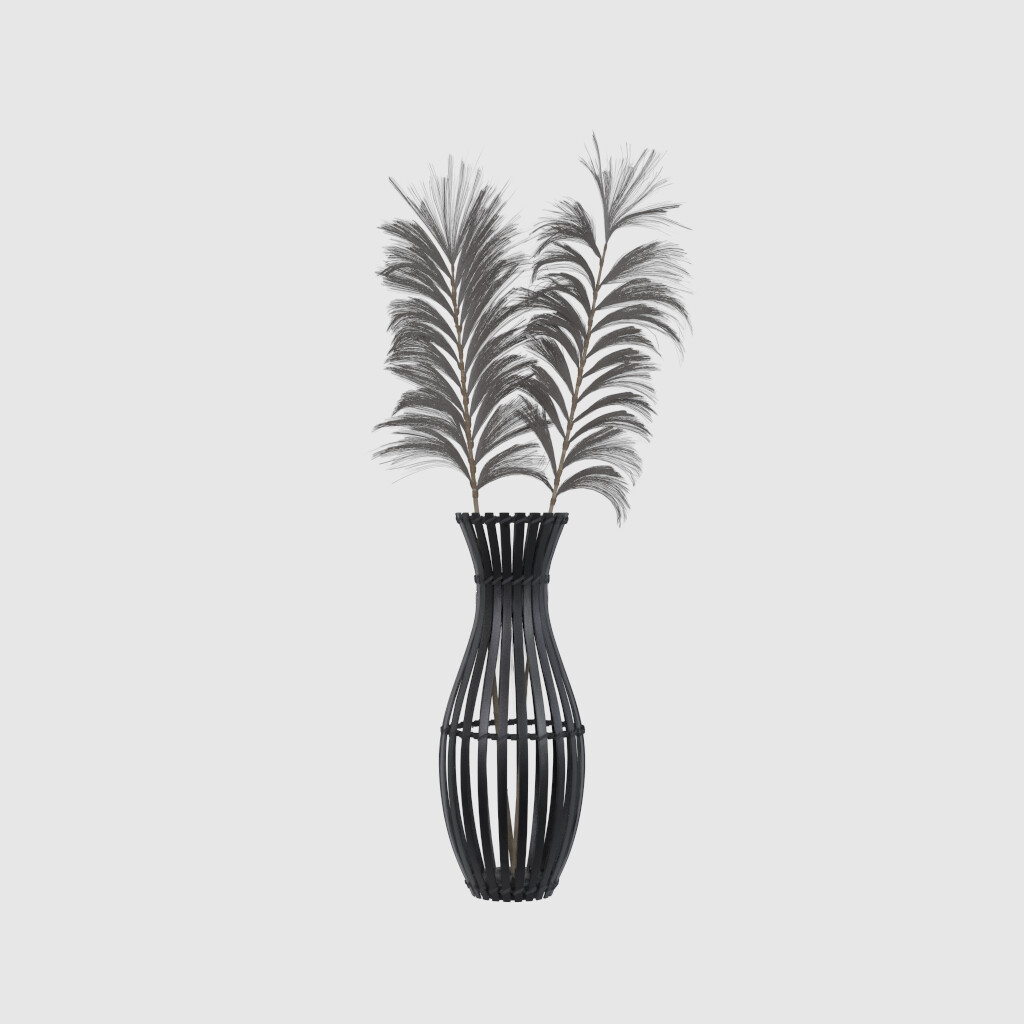}}
        &    
        \\        
        \raisebox{12pt}{\rotatebox{90}{\small{\textsf{Transmittance}}}}
        &
        \frame{\includegraphics[width=\lenDiff]{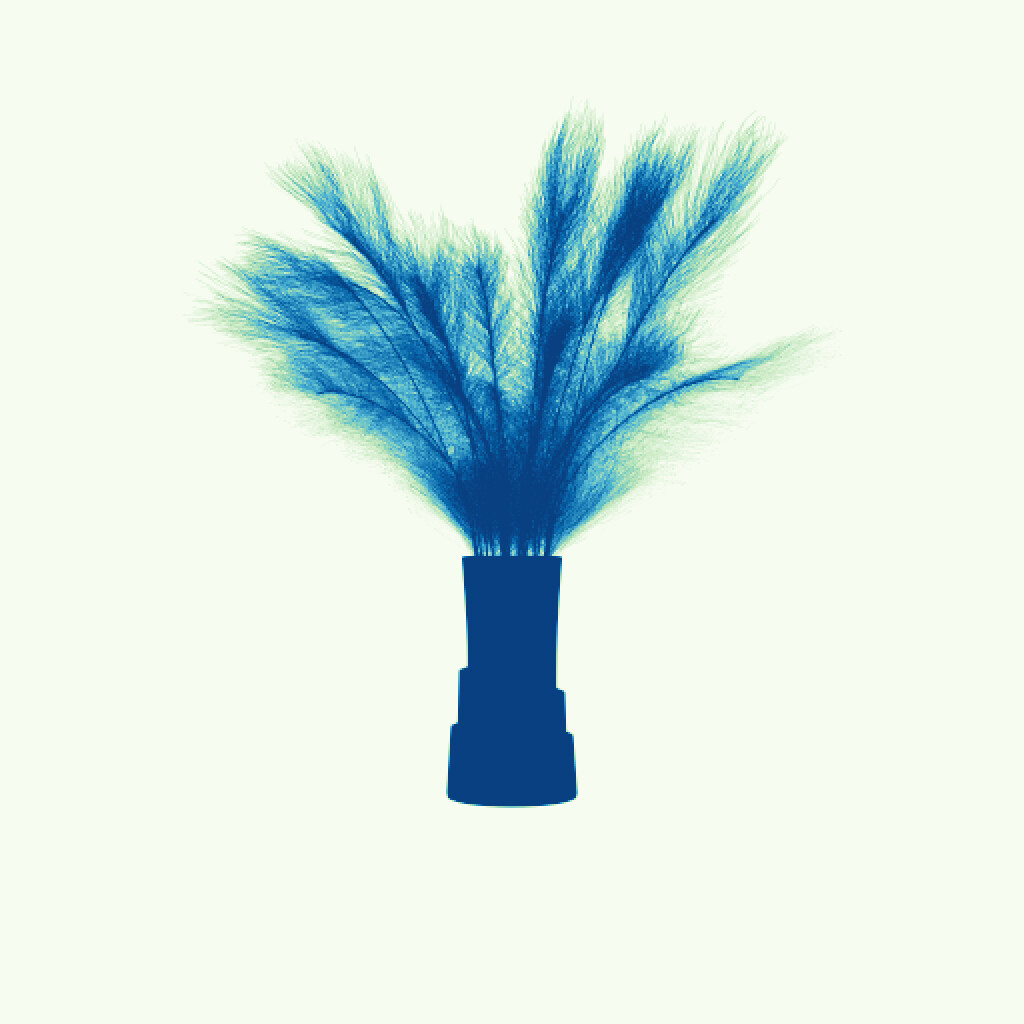}}
        &        
        \frame{\includegraphics[width=\lenDiff]{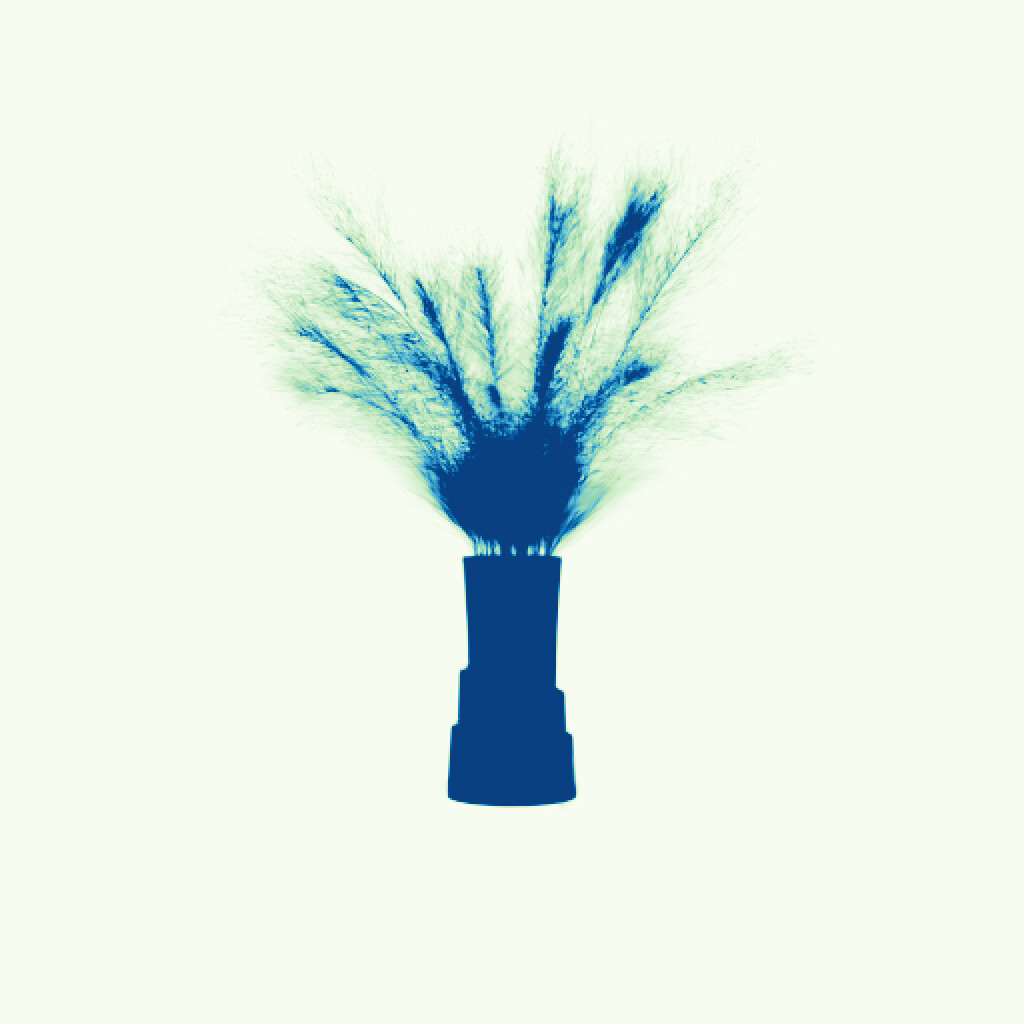}}
        &
        \frame{\includegraphics[width=\lenDiff]{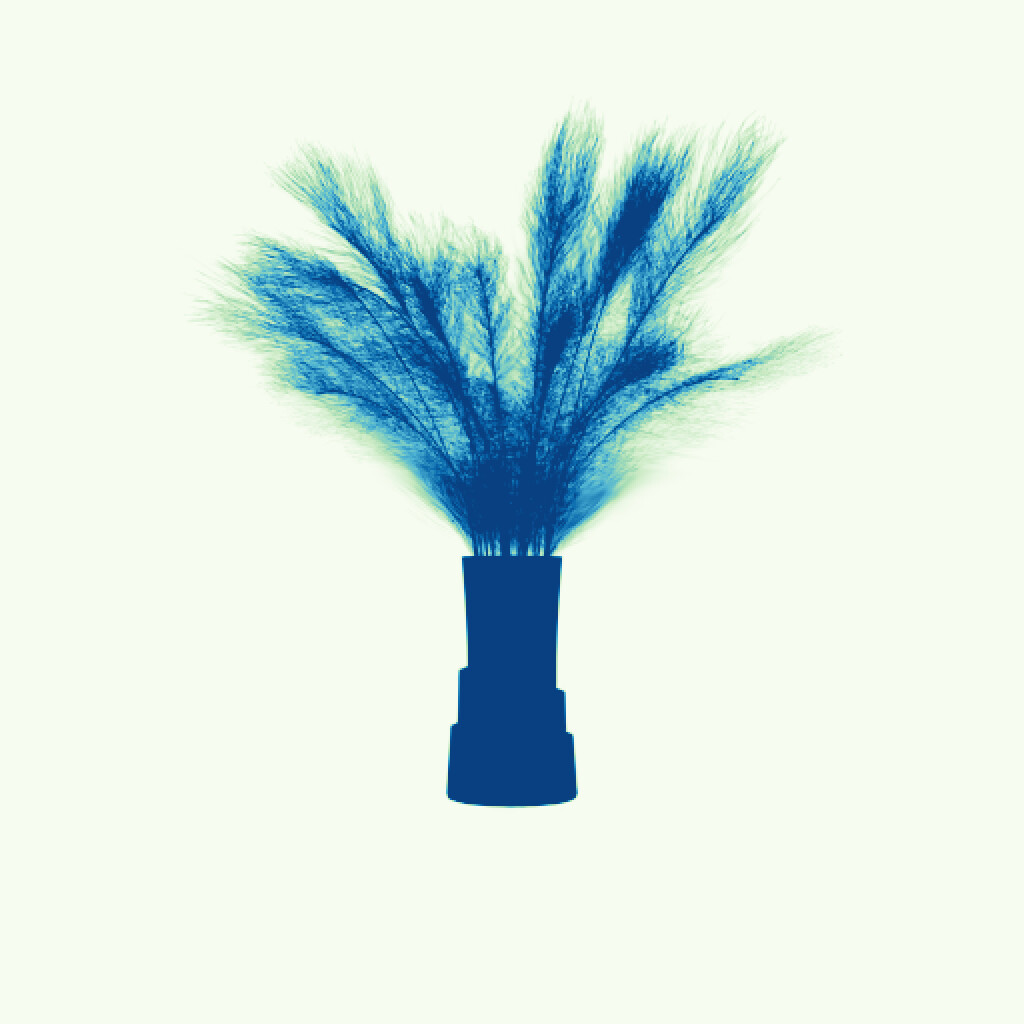}}
        &
        \frame{\includegraphics[width=\lenDiff]{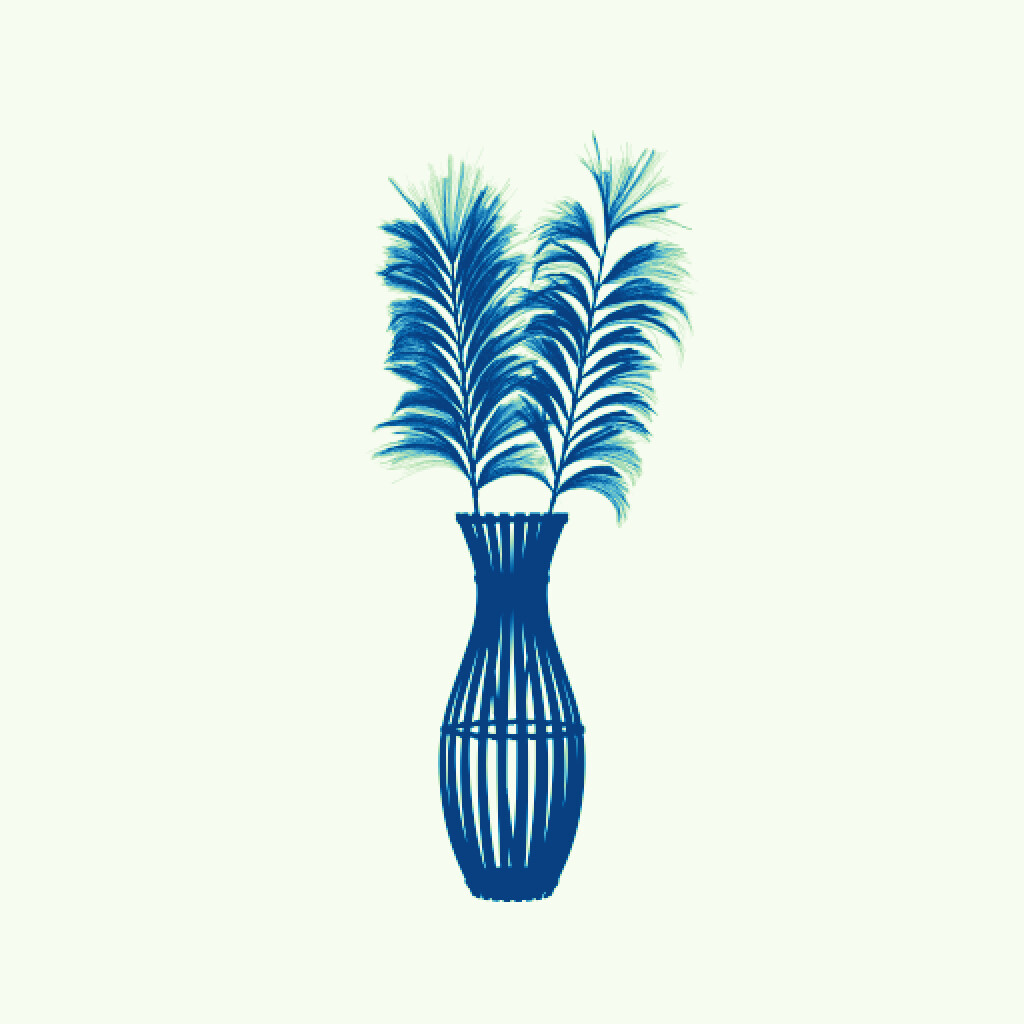}}
        &        
        \frame{\includegraphics[width=\lenDiff]{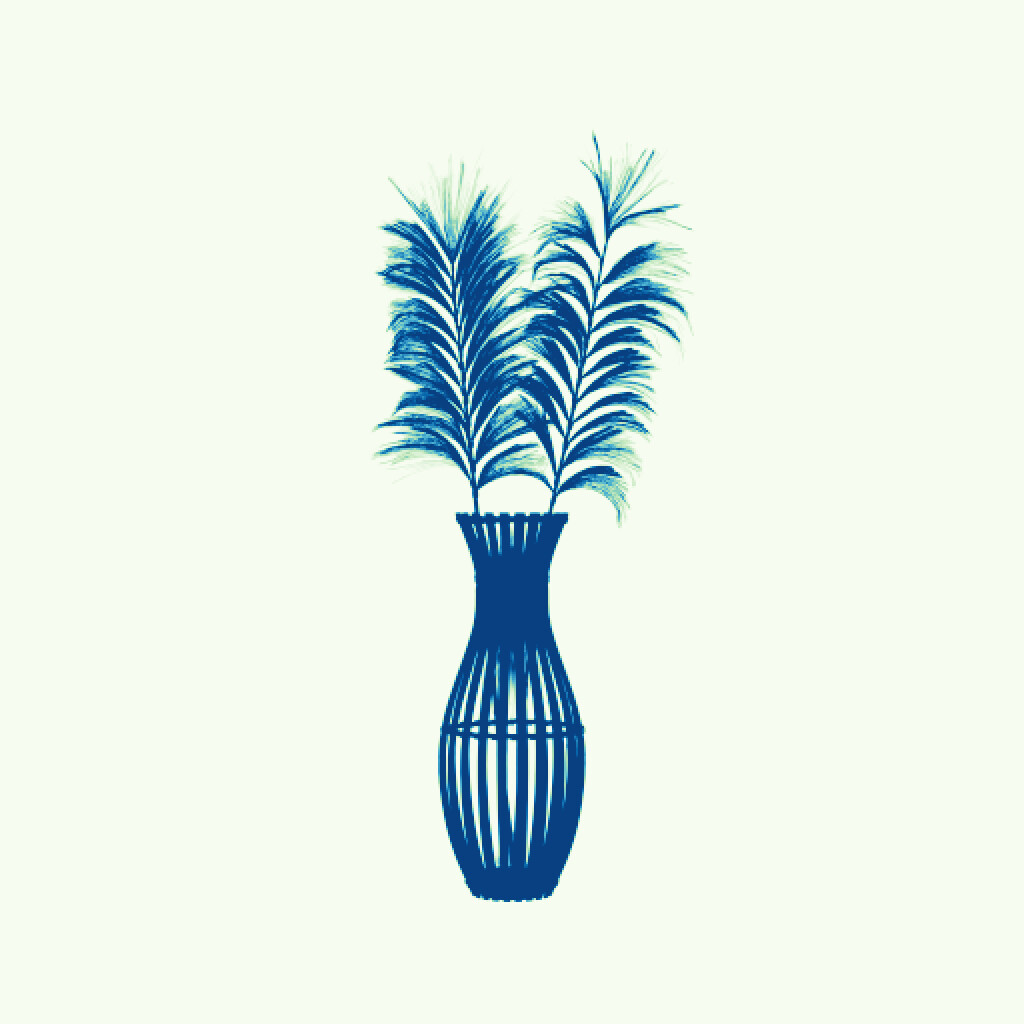}}
        &
        \frame{\includegraphics[width=\lenDiff]{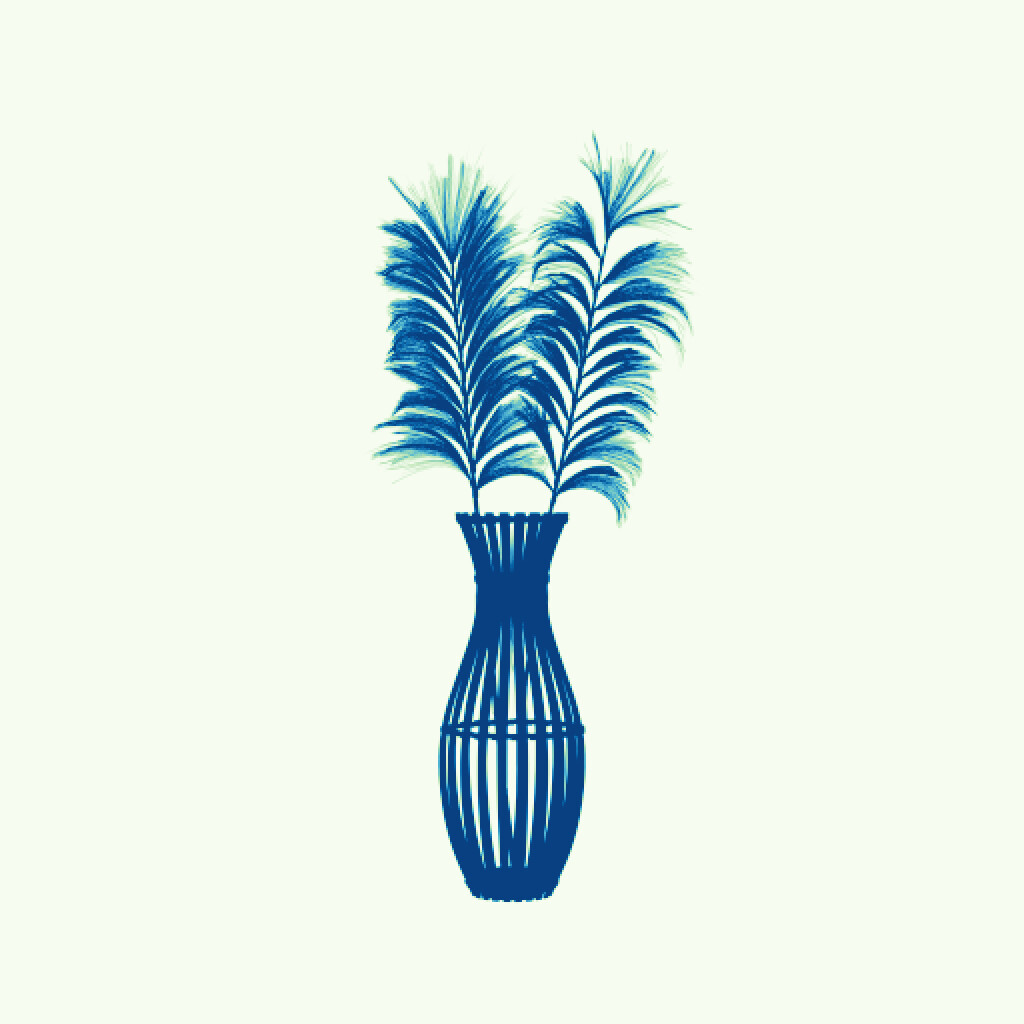}}        
        &
		\multicolumn{1}{l}{\begin{overpic}[height=\lenDiff,unit=1mm, frame]{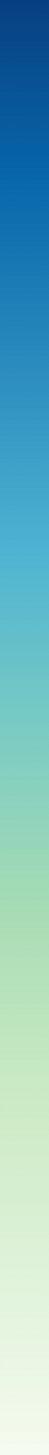}
			\put(5, 91){\small 1.0}
			\put(5, 1){\small 0.0}
		\end{overpic}}
        \\
        \raisebox{25pt}{\rotatebox{90}{\small{\textsf{Tr. Loss}}}}
        &
        \includegraphics[width=\lenDiff]{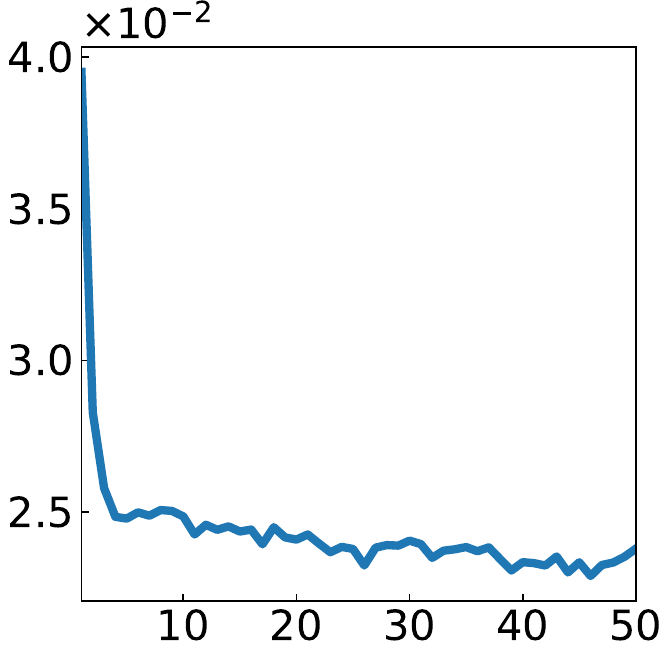}
        &
        \begin{overpic}[width=\lenDiff, frame]{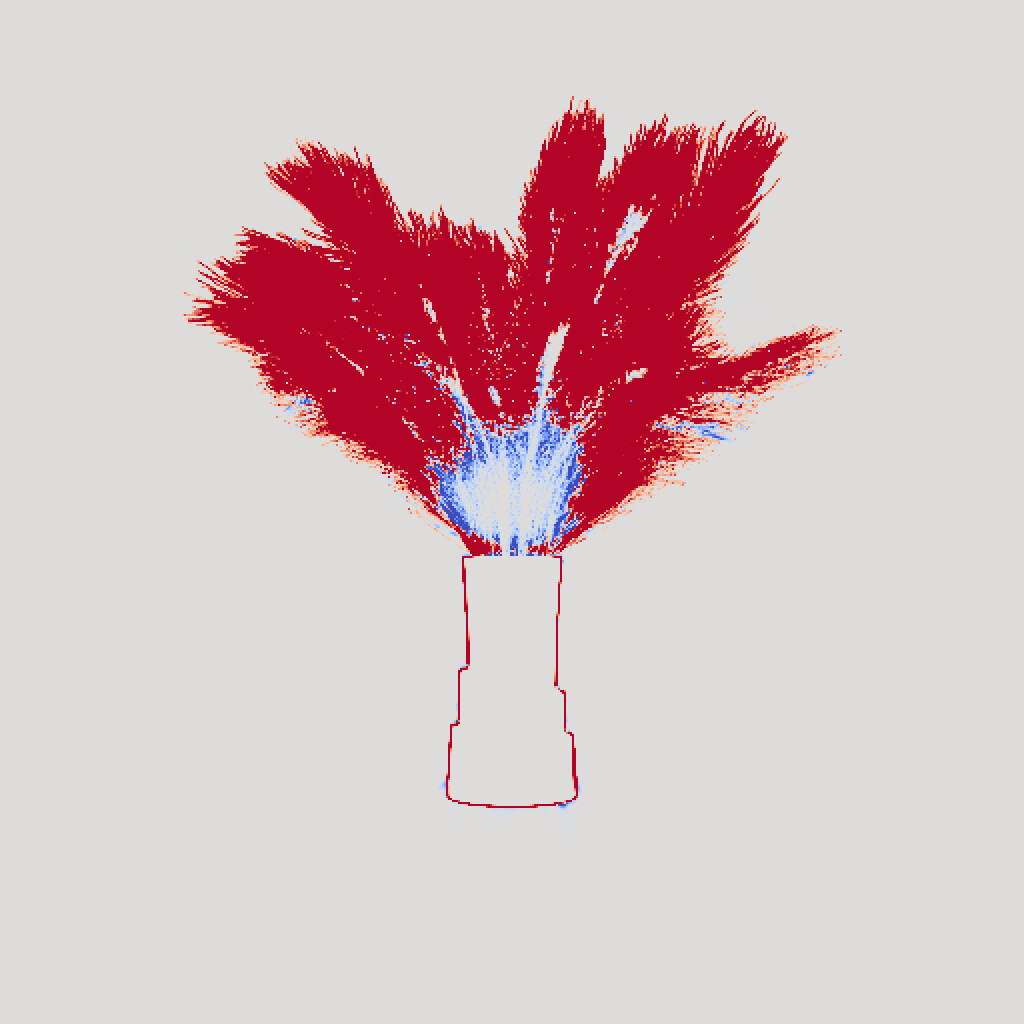}
			\put(60, 5){\small{\num{3.95e-2}}}
        \end{overpic}
        &
        \begin{overpic}[width=\lenDiff, frame]{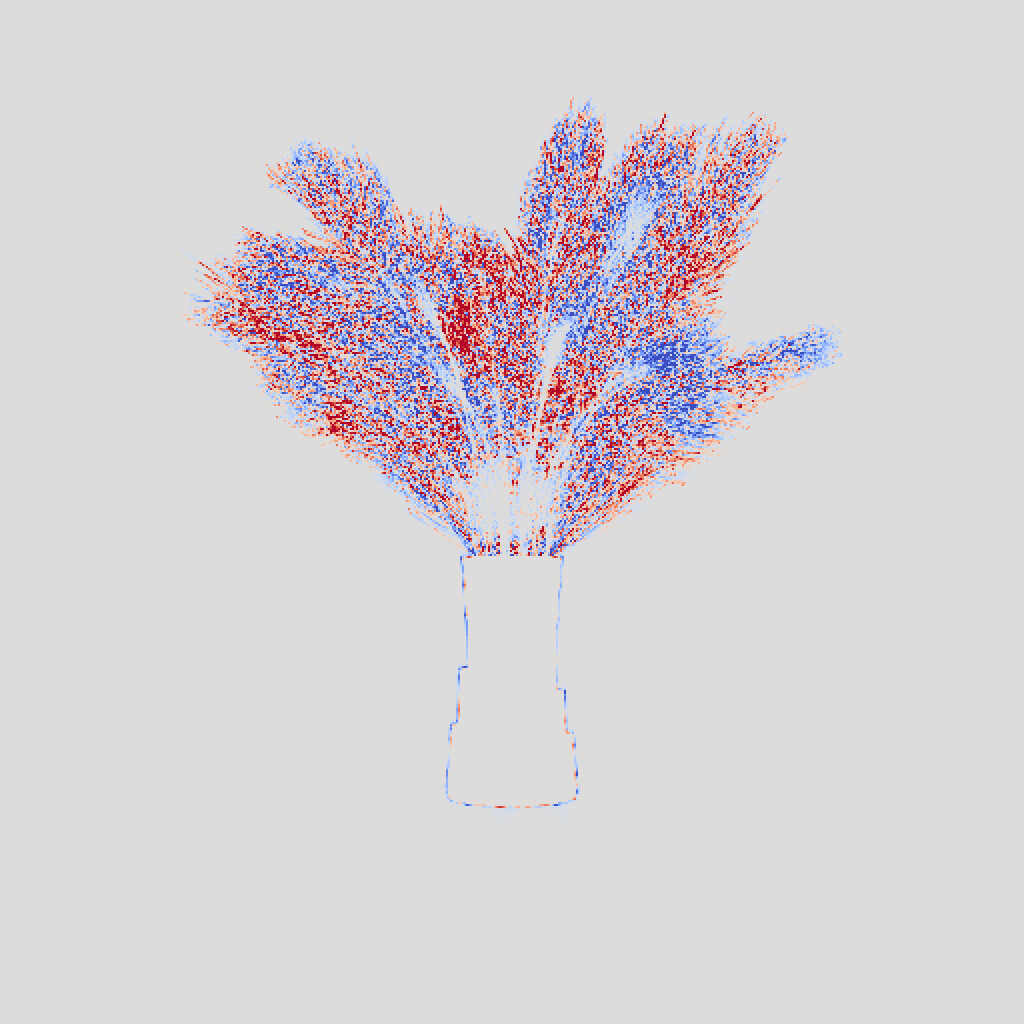}
            \put(60, 5){\small{\num{2.38e-2}}}
        \end{overpic}
        &
        \includegraphics[width=\lenDiff]{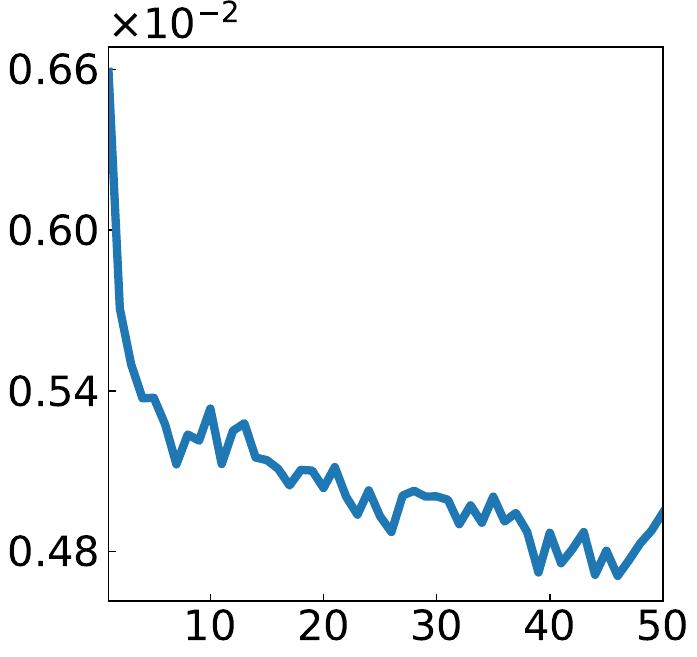}
        &     
        \begin{overpic}[width=\lenDiff, frame]{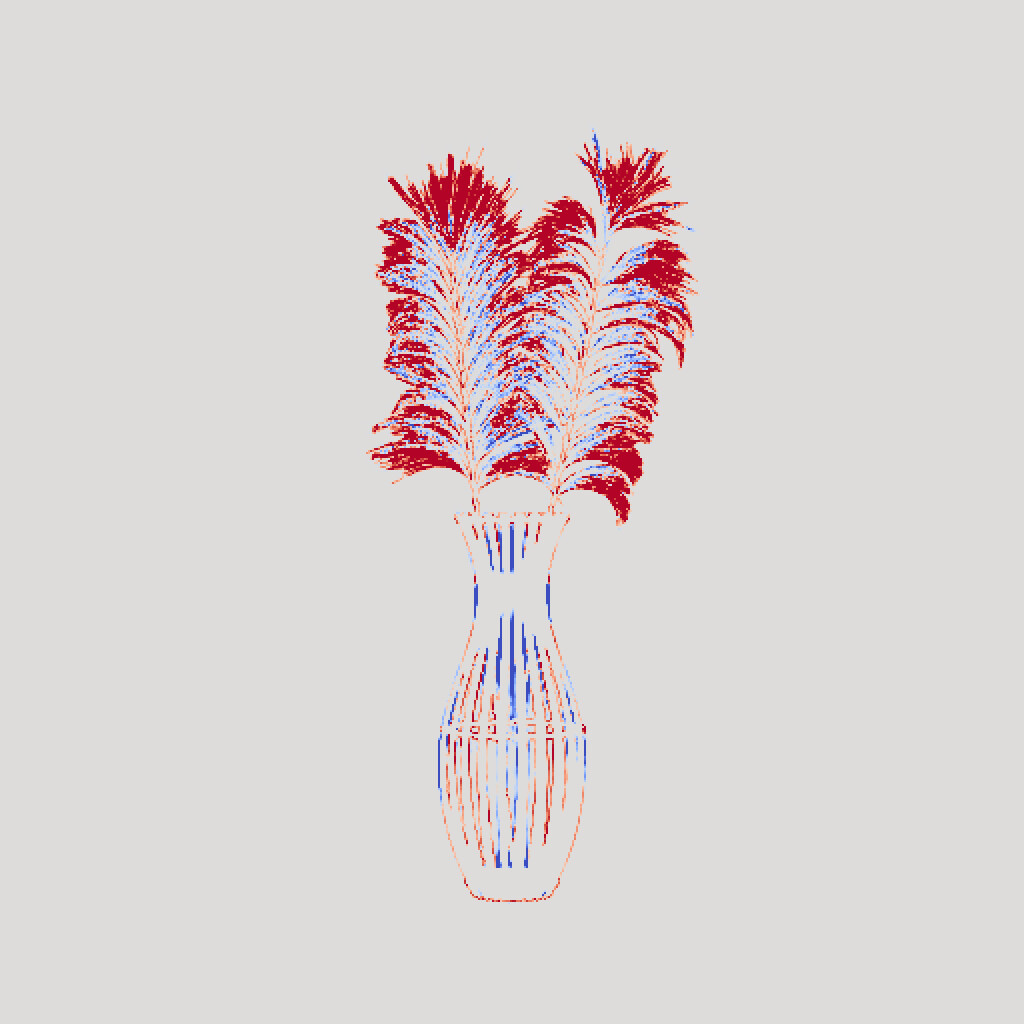}
            \put(60, 5){\small{\num{6.69e-3}}}
        \end{overpic}
        &
        \begin{overpic}[width=\lenDiff, frame]{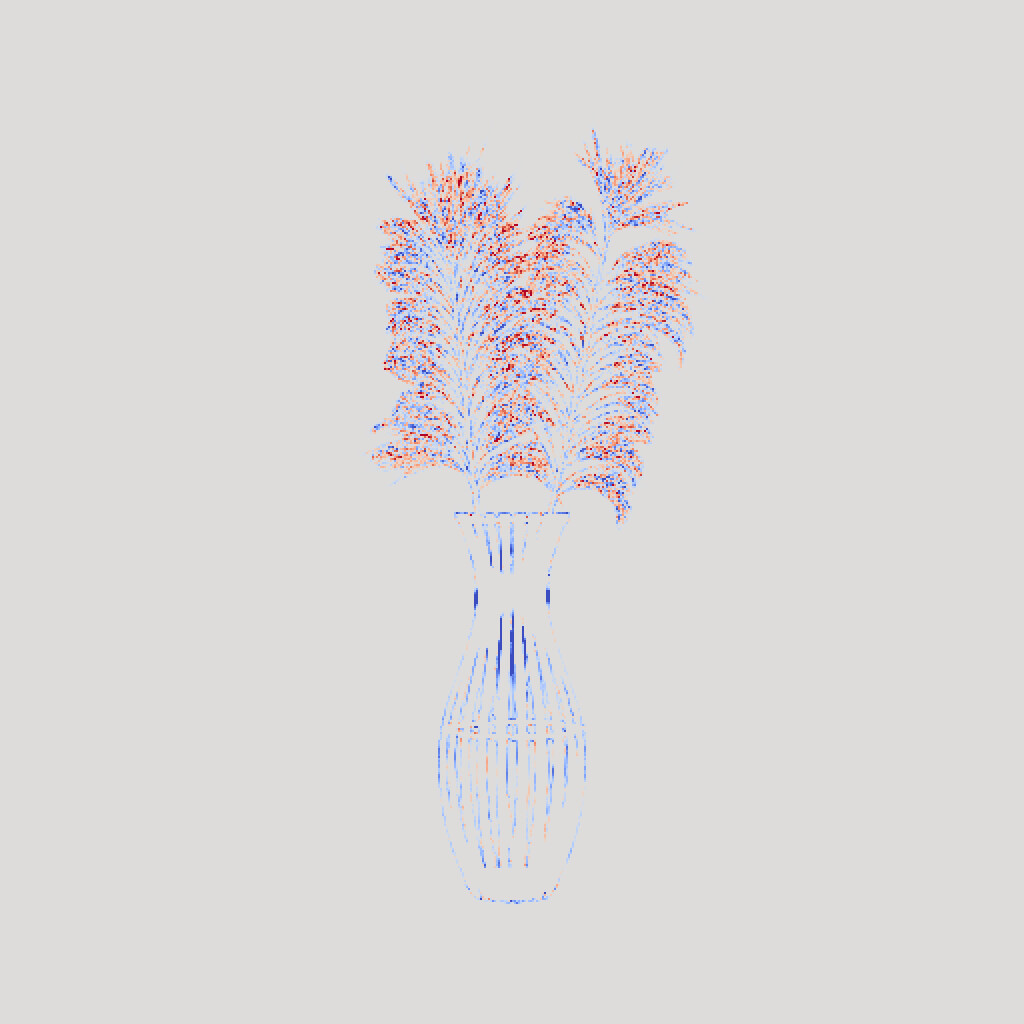}
            \put(60, 5){\small{\num{4.94e-3}}}
        \end{overpic}
        &
		\multicolumn{1}{l}{\begin{overpic}[height=\lenDiff,unit=1mm, frame]{resources/coolwarm_v.jpg}
			\put(5, 91){\small 0.1}
			\put(5, 1){\small -0.1}
		\end{overpic}}
        \\
        & \footnotesize{\textsf{Original / loss curve}} & \small{\textsf{Initial}} & \small{\textsf{Iter. 50}} & \footnotesize{\textsf{Original / loss curve}} & \small{\textsf{Initial}} & \small{\textsf{Iter. 50}} \\    
    \end{tabular}
    \caption{\label{fig:diff}
        Applying transmittance optimization to Gaussian models initialized from 3DGS. We visualize the transmittance (from a selected view) of each model 
        before and after the optimization, and compare it to that of the original model. Transmittance losses and loss curves are provided. 
        We encourage readers to zoom in to better identify the differences.
    }                    
\end{figure*}

In \autoref{fig:diff}, we apply the transmittance optimization to two models initialized by 3DGS. The \emph{Plant} model has 220K primitives, while the 
\emph{Tall Vase} model has 799K primitives. Both models are optimized for 50 iterations using 8 views at 
512$\times$512 resolution. Both models consist of long, thin branches that are not captured by 3DGS faithfully, but can be recovered quickly by our post-optimization. In addition, our optimization improves the silhouette of the spokes of the \emph{Tall Vase}. Overall improvement in transmittance can be 
verified by the loss curves. 
The process takes 16 and 26 minutes on CPU, respectively (see \autoref{sec:results} for machine specifications).

\section{Radiance Field Reconstruction} \label{sec:radiance_field}
Our formulation encompasses the general volumetric light transport (\autoref{eq:rte}), including effects caused by multiple scattering. 
Within this framework, radiance fields essentially belong to a special case with no scattering, but only emission and absorption. 
Specifically, the source term in \autoref{eq:rte} becomes simply $L_s(x, \omega) = L_e(x, \omega)$.
Given the prevalence of radiance fields in 3D reconstruction and image-based rendering~\citep{mildenhall2021nerf, muller2022instant, kerbl20233d}, 
we choose to use image-based radiance field reconstruction (and rendering) as a more powerful differentiable rendering application to 
demonstrate the versatility of our method.

Compared with our full formulation, the rendering of a radiance field is much simpler as \autoref{eq:rte} collapses to
\begin{align} \label{eq:rf}
    L(x,\omega) &= \sum_{k=1}^n \int_0^{t_{\mathrm{sat}}} \mathcal{P}(x, x_t) L_{e,k}(\omega) \, \D t \notag\\
    &= \sum_{k=1}^n I_k(t_k^-, \min{}(t_k^+, t_{\mathrm{sat}}))L_{e,k}(\omega),
\end{align}
where $k$ enumerates the intersecting primitives and $L_{e,k}(\omega)$ is the emission of the $k$-th primitive.
As such, an analytic approach exists and there is no need for Monte Carlo. 
As each ray traverses a radiance field scene, we accumulate $L_{e,k}$ up to the saturating distance $t_{\mathrm{sat}}$. The ambiguity 
situation can still arise if there are multiple primitives touching the saturating boundary $t = t_{\mathrm{sat}}$. This can be solved by the same 
disambiguation step (\autoref{subsec:free_flight_sampling}) by setting $u = 1$ to exhaust all contributing primitives instead of drawing a 
sample in the middle. 

A common choice to represent the view-dependent emission $L_{e,k}(\omega)$ is to use the spherical harmonics basis functions. However, we choose 
to omit the view dependency in favor of simplicity by letting $L_{e,k}(\omega) = L_{e,k}$. 
Our framework is agnostic to this simplification, and view-dependent effects can be easily introduced back for improved reconstruction of highlights 
and glossy reflections.

Image-based radiance field reconstruction requires gradients of pixel intensities w.r.t. the parameters of all relevant primitives $\{\phi_k\}$.
Because \autoref{eq:rf} is a linear combination, the gradient w.r.t. the $k$-th primitive is again relatively straightforward to 
obtain by differentiating the equation
\begin{equation}
    \frac{\partial L(x, \omega)}{\partial \phi_k} = \frac{\partial \, I_k (t^-_k, \min(t^+_k, t_{\mathrm{sat}})) L_{e,k}}{\partial \phi_k}.
\end{equation}

\begin{figure}[h]
	\newlength{\lenRFProblem}
	\setlength{\lenRFProblem}{0.7\linewidth}
    \centering
    \includegraphics[width=\lenRFProblem]{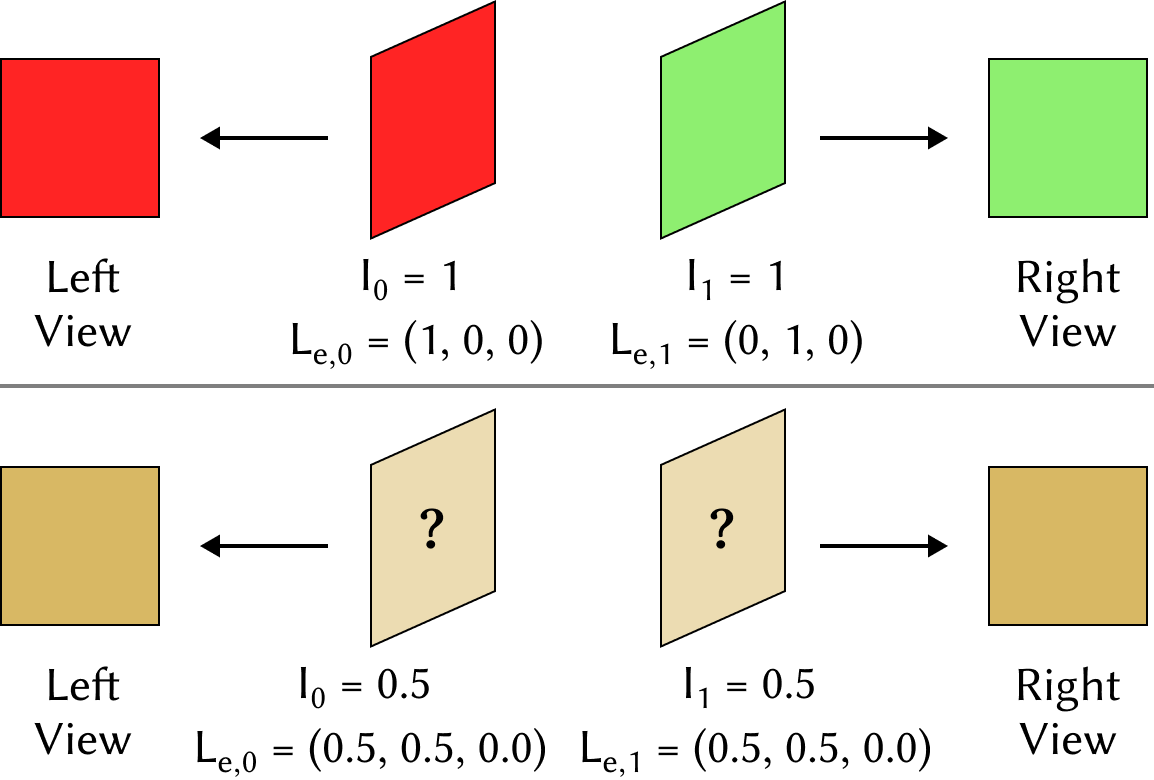}
    \caption{\label{fig:rf_problem}
        A toy failure example of radiance field optimization using linear transmittance primitives and vanilla image loss.
    }
\end{figure}

While the gradient computation is simple thanks to our linear transmittance model, it unfortunately introduces a unique difficulty to the 
overarching optimization problem. Unlike the exponential transmittance model, the linear transmittance model itself does not imply a notion of occlusion. 
This can also be reflected via the fact that the gradient of one primitive does not depend on others in front of it (unless it is partially ``saturated'').
This is problematic because the optimization may not be able to progress towards a meaningful direction using common image losses.
Consider the following illustrative example in 1D flatland in \autoref{fig:rf_problem}, where the goal is to reconstruct two opaque planes 
from observation from two sides (top). 
Assuming any image loss, it is possible that at a given state (bottom), the two primitives receive exactly the same gradient, causing the optimization to stall.
In practice, this is more likely to happen at the early stage of the optimization because as with typical optimization schemes for particle-based radiance fields, 
we start by setting all primitives to be uniformly transparent.

To overcome this difficulty, we introduce the Radiance Surface loss~\citep{zhang2025loss} into our optimization process, extending its usage from 
neural fields to our primitive-based representation. The Radiance Surface loss encourages a surface-biased reconstruction. While this may be less 
suitable for primitives based on exponential transmittance~\citep{zhang2025loss}, it instead has natural synergy with our linear transmittance model that prioritizes 
opaque surfaces. From our experiments, 
we find that it helps the optimization to quickly concentrate on primitives in non-empty regions and prune away those in empty regions. This allows the 
optimization to largely avoid the failure case discussed above. Therefore, our full loss combines both the Radiance Surface loss $\mathcal{L}_{\mathrm{rs}}$ and the image loss terms, 
which is a linear combination of $\mathcal{L}_1$ loss and a D-SSIM term:
\begin{equation} \label{eq:rf_full_loss}
    \mathcal{L} = (1-\beta)\mathcal{L}_{\mathrm{rs}} + \beta\big((1 - \lambda)\mathcal{L}_1 + \lambda\mathcal{L}_{\mathrm{D-SSIM}}\big),
\end{equation}
where $\lambda$ is set to $0.2$ following common practice~\citep{kerbl20233d}. We gradually transition $\beta$ from $0$ to $1$ during the optimization to achieve the 
aforementioned quick surface formation behavior before switching to image losses for fine-tuning. In addition, we adopt the improved initialization method of 
EDGS~\citep{kotovenko2025edgs}. See \autoref{sec:results} for more details.

Our adaptation of the Radiance Surface loss is heuristic because the loss internally assumes an exponential free-flight background distribution~\citep{zhang2025loss}. 
This does not match our linear formulation but we have not observed issues in practice, partially because the optimization is incentivized to push primitives to be near-opaque 
with near-delta free-flight distributions anyway. It may be possible to further extend the loss with alternative free-flight background distributions and we leave it to future work.




\section{Results and Discussion} \label{sec:results}


In this section, we present the results of various rendering and differentiable rendering applications using our Gaussian primitives. 
We also provide a supplemental video that includes rendering sequences with camera and light animations. 
We implement our method in a custom hybrid framework that includes both CPU and GPU code paths. 
Specifically, our forward path tracer is available on both CPU and GPU, where the GPU version follows a straightforward megakernel style with 
the binning strategy described in \autoref{subsec:free_flight_sampling}. 
The kd-tree construction is shared between CPU and GPU and parallelized across lower subtrees.
For differentiable rendering applications, the transmittance optimization (\autoref{subsec:transmittance_optim}) is implemented only on CPU, 
and the radiance field optimization (\autoref{sec:radiance_field}) is implemented only on GPU. 
We use the Slang shading language~\citep{tog/BangaruWLMBRDLH23} for automatic differentiation.
All measurements have been conducted on a desktop computer with an Intel® Core™ i9-13900K CPU and an NVIDIA® GeForce RTX™ 5090 GPU.

\paragraph{Complex Scene Rendering with Global Illumination}
In \autoref{fig:teaser}, we demonstrate the versatility of our Gaussian primitive to represent objects with a wide range of geometric and material characteristics.
The \emph{Dressing Table} scene is modeled \emph{entirely} by our primitives and contains parts that are acquired in different ways. The room, table, dragon,
and logo are converted from meshes, while the plants, candle set, and blanket are converted from 3DGS reconstruction. Additionally, several objects, such as the 
mirror and neon light bars, are modeled analytically. Our Gaussian primitives can adapt to different shapes, including flat surfaces and thin fibers, thanks
to their anisotropic definition. The volumetric formulation naturally handles the fuzzy appearance from dense, stochastic details.

The scene also features a variety of materials that include near-specular, glossy, and diffuse, demonstrating the expressiveness of our appearance definition.
The phase function of our primitive incorporates the effect of base BSDF and NDF, thus allowing it to aggregate the appearance caused by many differently
oriented small elements, such as the plants. Meanwhile, for flat surfaces like the floor and table, it naturally reverts to the familiar surface BSDF formulation. Thanks to our improved stochastic evaluation scheme and the approximated PDF for MIS (\autoref{fig:material_sphere}), the variance
contributed by phase functions is low and  diminishes quickly as more samples are used by the path tracing integrator.

\begin{figure*}[tb]
	\newlength{\heightMeshComp}
	\setlength{\heightMeshComp}{0.119\linewidth}
    \addtolength{\tabcolsep}{-4pt}
    \renewcommand{\arraystretch}{0.5}
    \centering
    \begin{tabular}{cccccccc}
        \small{\textsf{Mesh Reference}} &
        \small{\textsf{Mesh 50$\%$ \#Tri.}} & 
        \small{\textsf{Mesh 10$\%$ \#Tri.}} & 
        \small{\textsf{Mesh 1$\%$ \#Tri.}} & 
        \small{\textsf{Ours 100$\%$ \#Prim.}} & 
        \small{\textsf{Ours 50$\%$ \#Prim.}} & 
        \small{\textsf{Ours 10$\%$ \#Prim.}} & 
        \small{\textsf{Ours 1$\%$ \#Prim.}}
        \\
        \includegraphics[height=\heightMeshComp, frame]{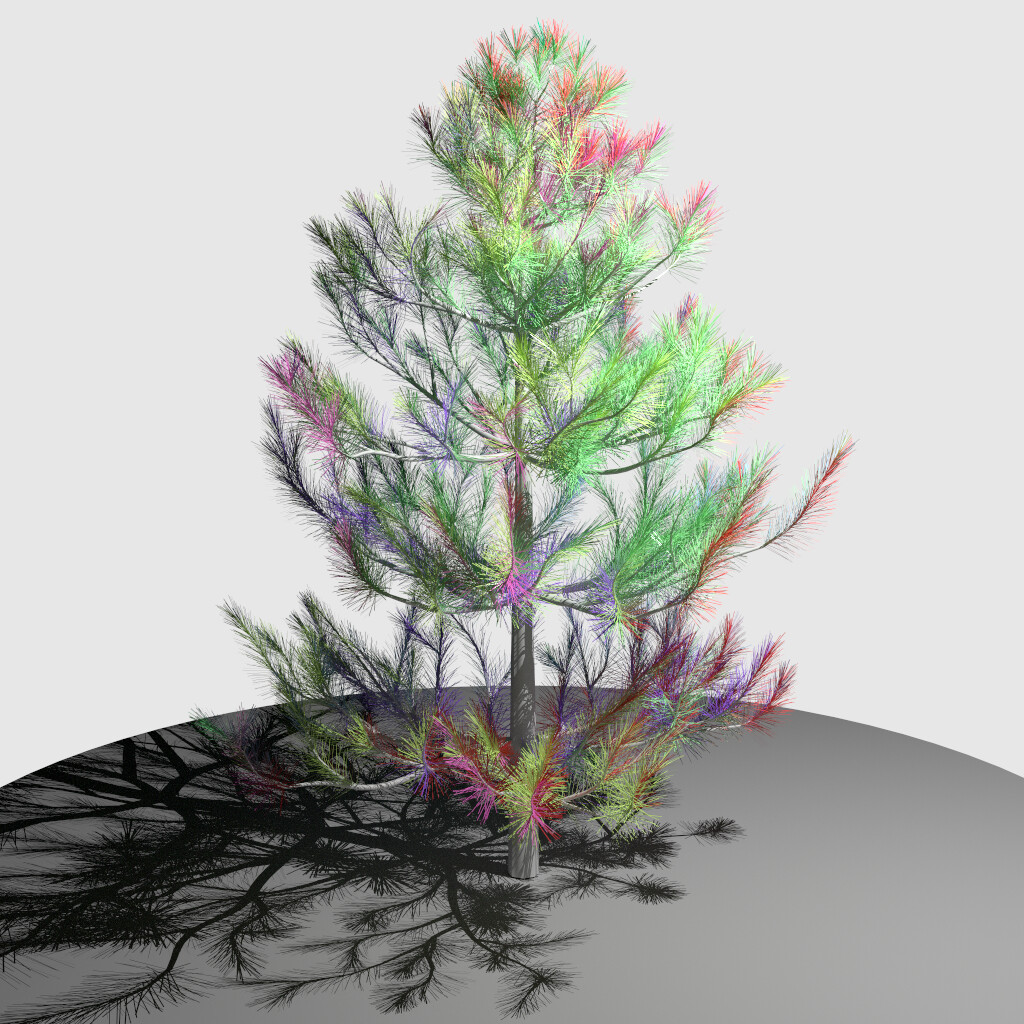}
        &
        \includegraphics[height=\heightMeshComp, frame]{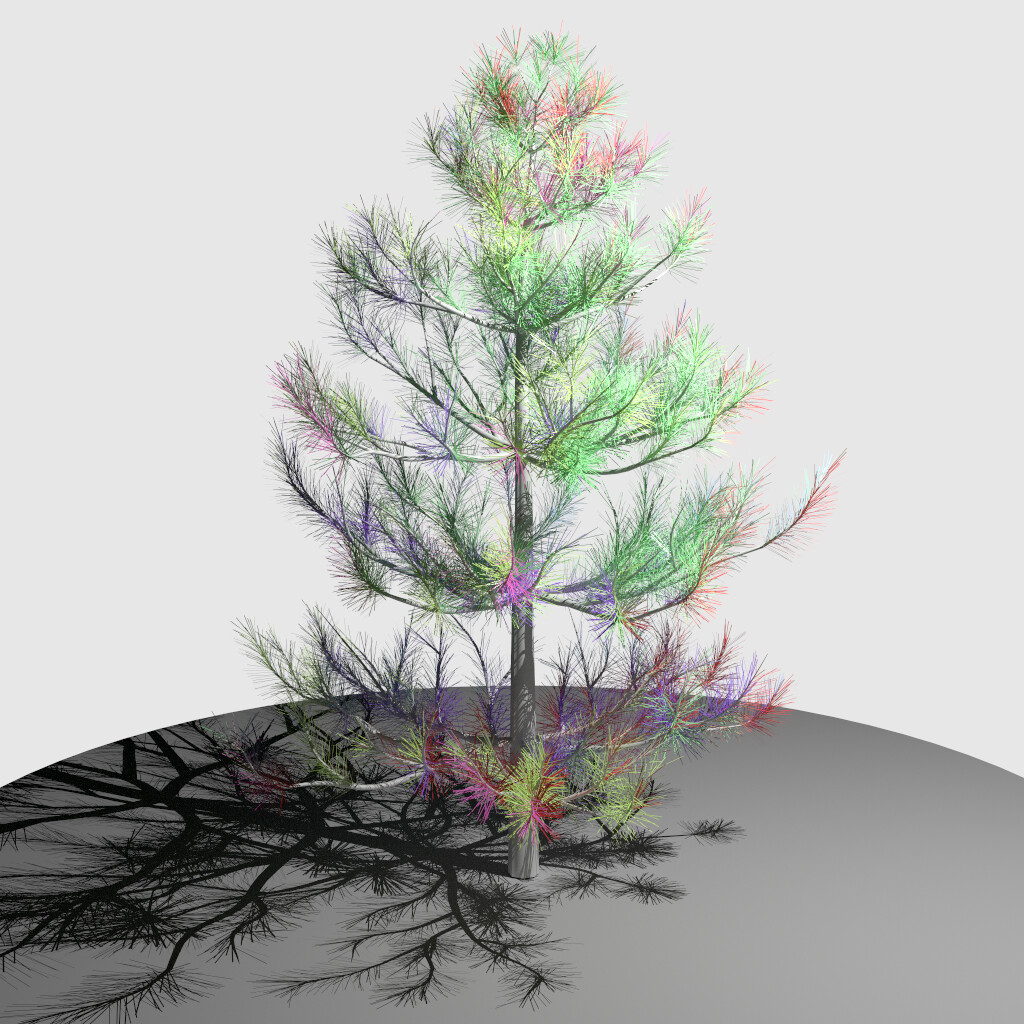}
        &
        \includegraphics[height=\heightMeshComp, frame]{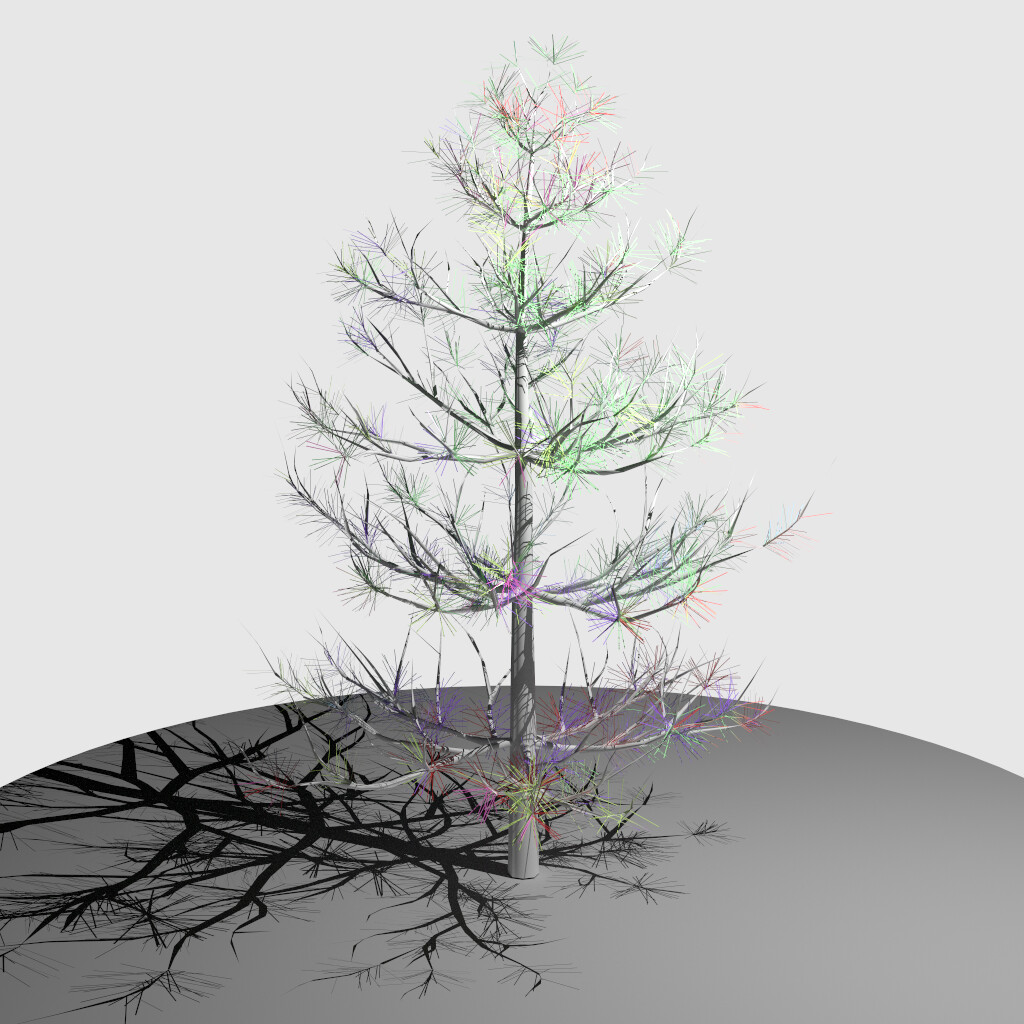}
        &
        \includegraphics[height=\heightMeshComp, frame]{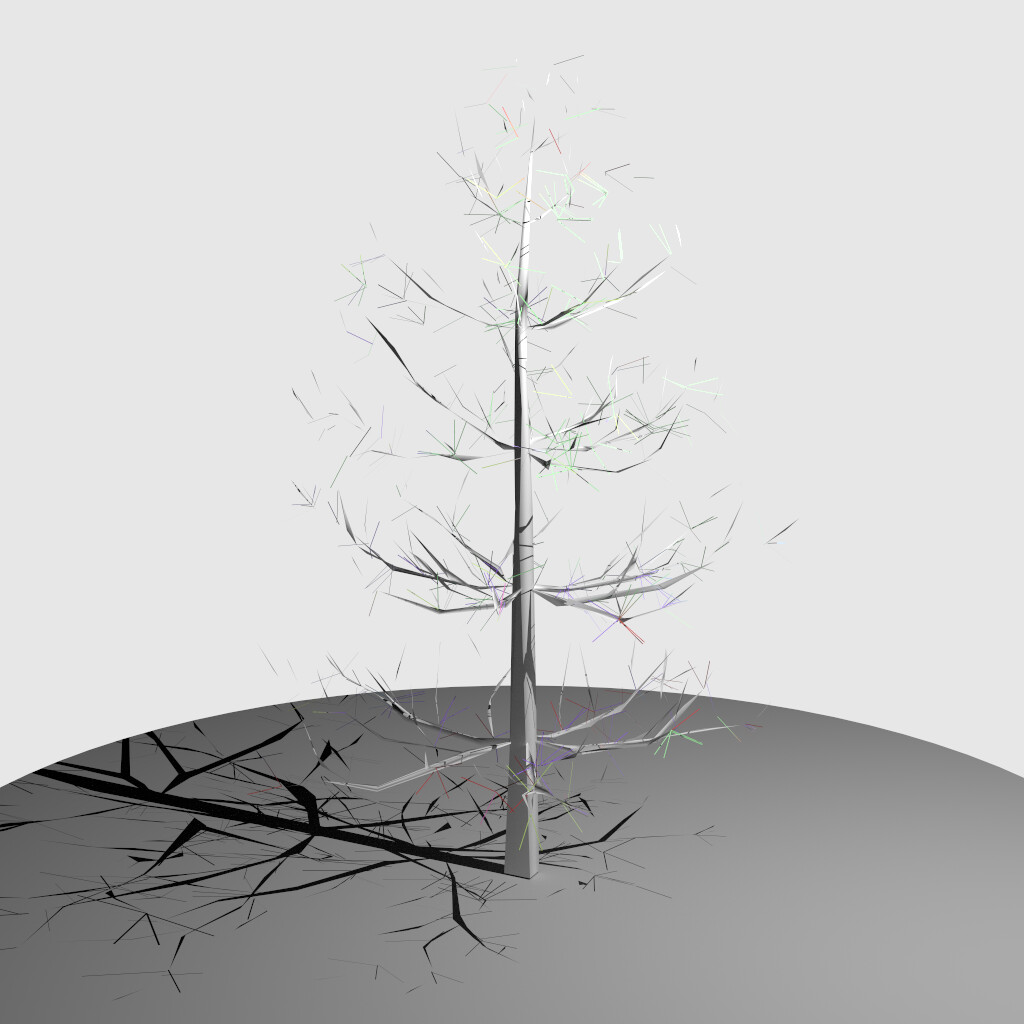}
        &
        \includegraphics[height=\heightMeshComp, frame]{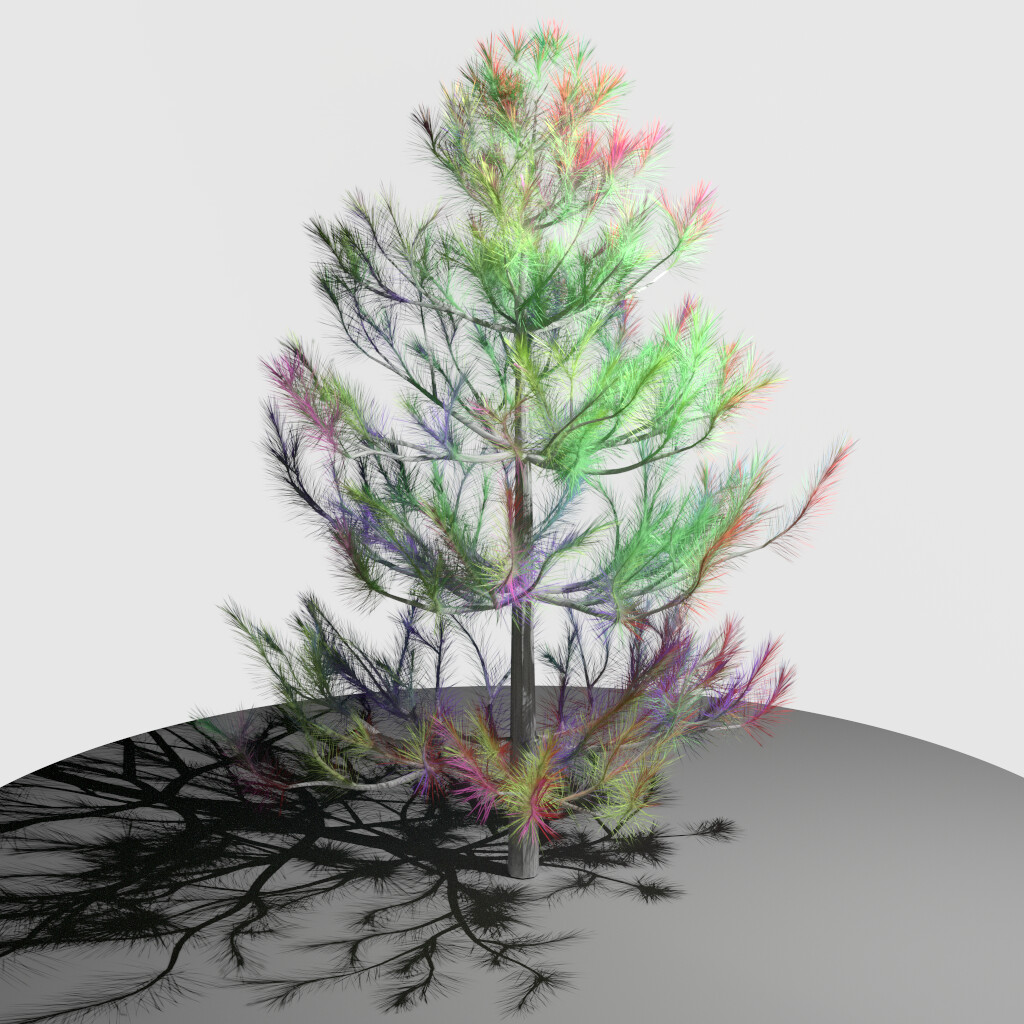}
        &
        \includegraphics[height=\heightMeshComp, frame]{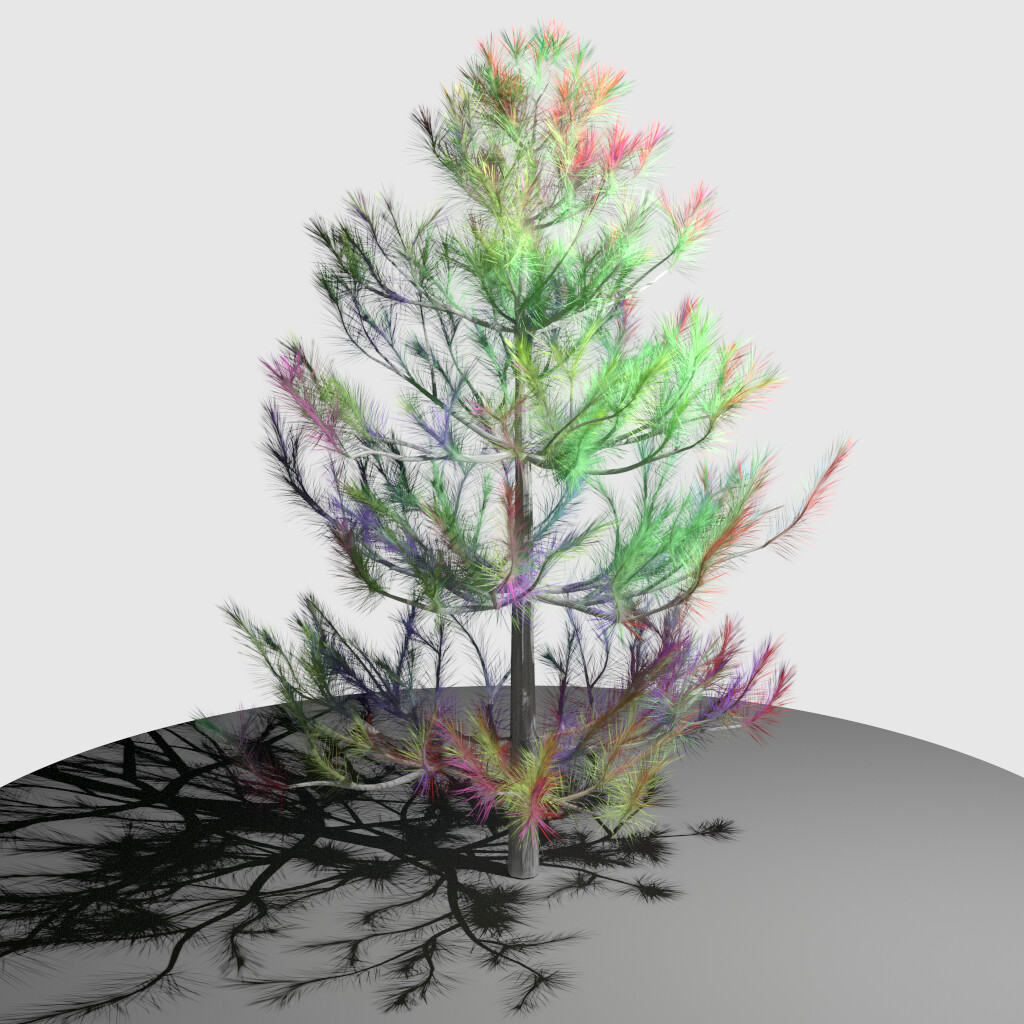}
        &
        \includegraphics[height=\heightMeshComp, frame]{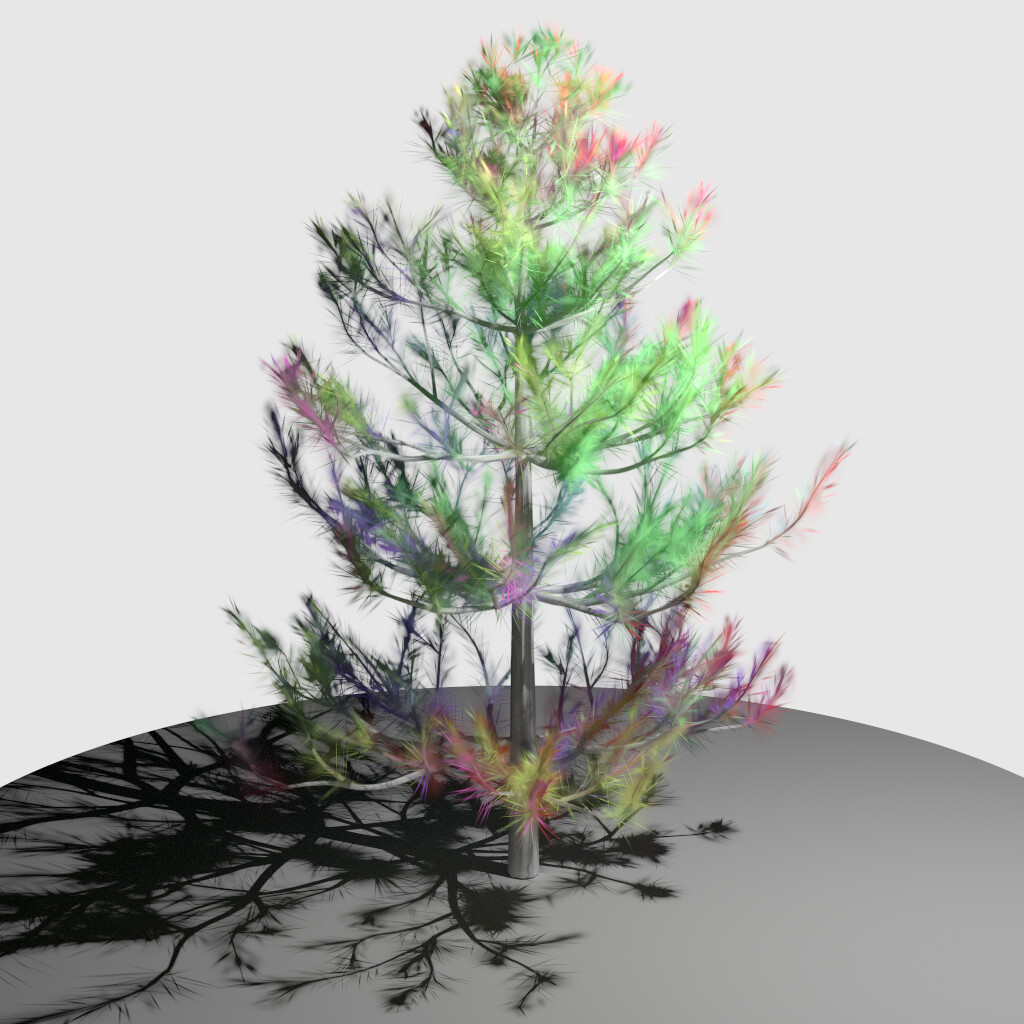}   
        &
        \includegraphics[height=\heightMeshComp, frame]{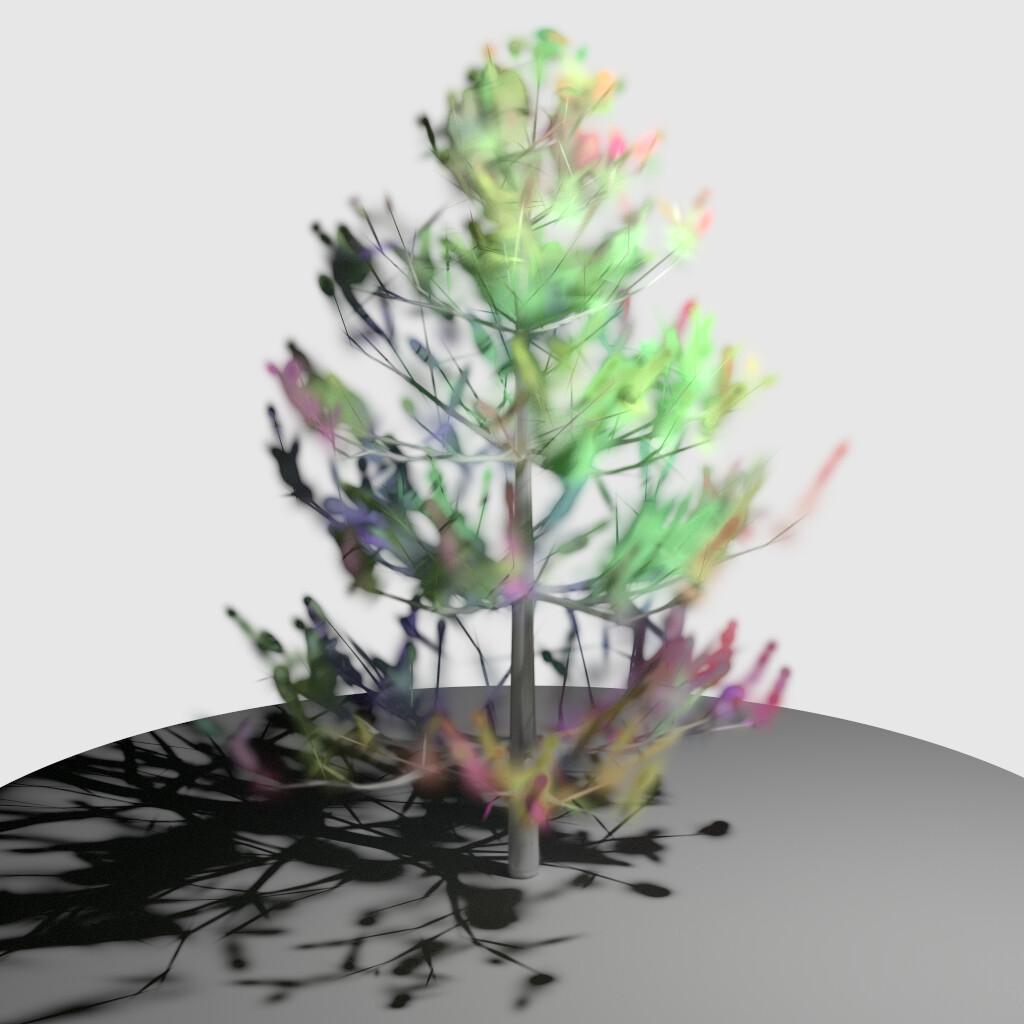}
        \\
		\multicolumn{1}{r}{\begin{overpic}[height=\heightMeshComp, frame,unit=1mm, frame]{resources/coolwarm_v.jpg}
			\put(-18, 91){\small 0.5}
			\put(-23, 1){\small -0.5}      
            \put(-75, 90){\small \emph{Color Tree}}                  
		\end{overpic}}
        &
        \includegraphics[height=\heightMeshComp, frame]{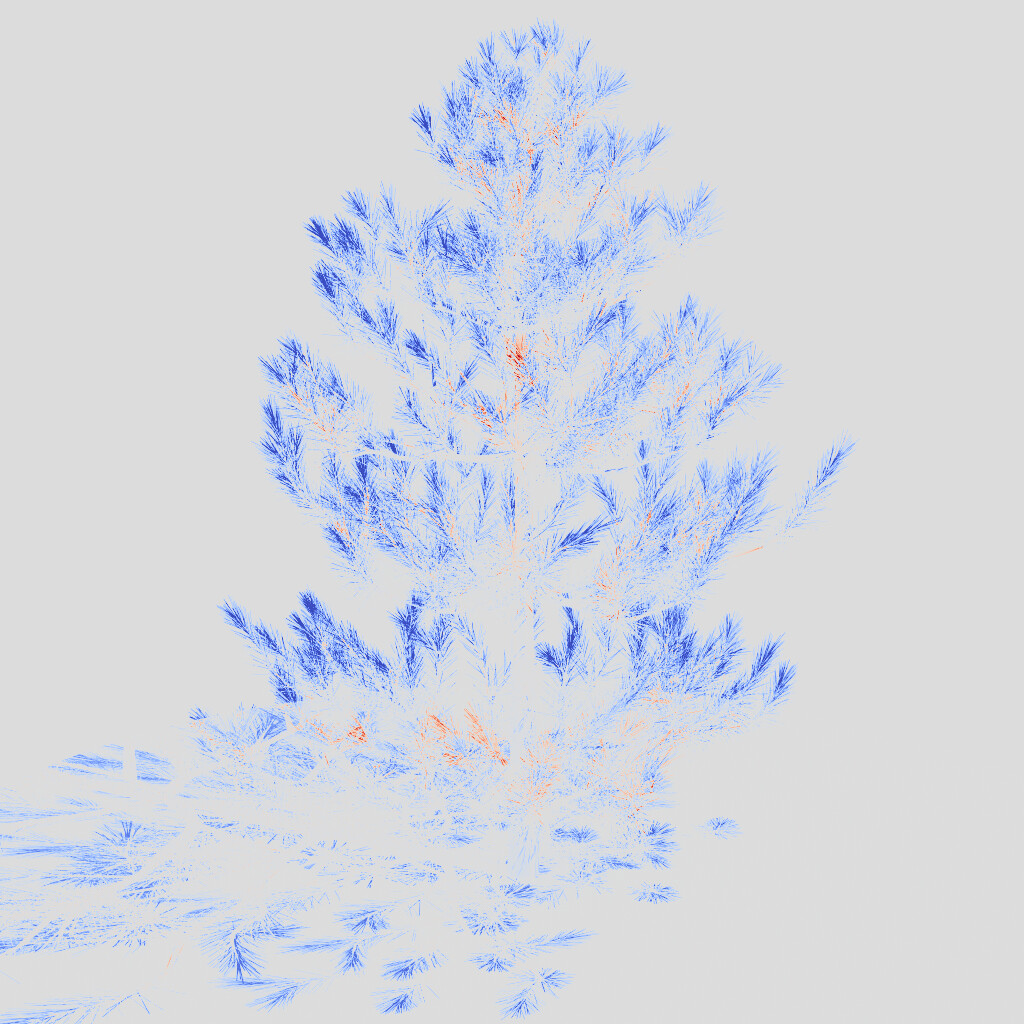}
        &
        \includegraphics[height=\heightMeshComp, frame]{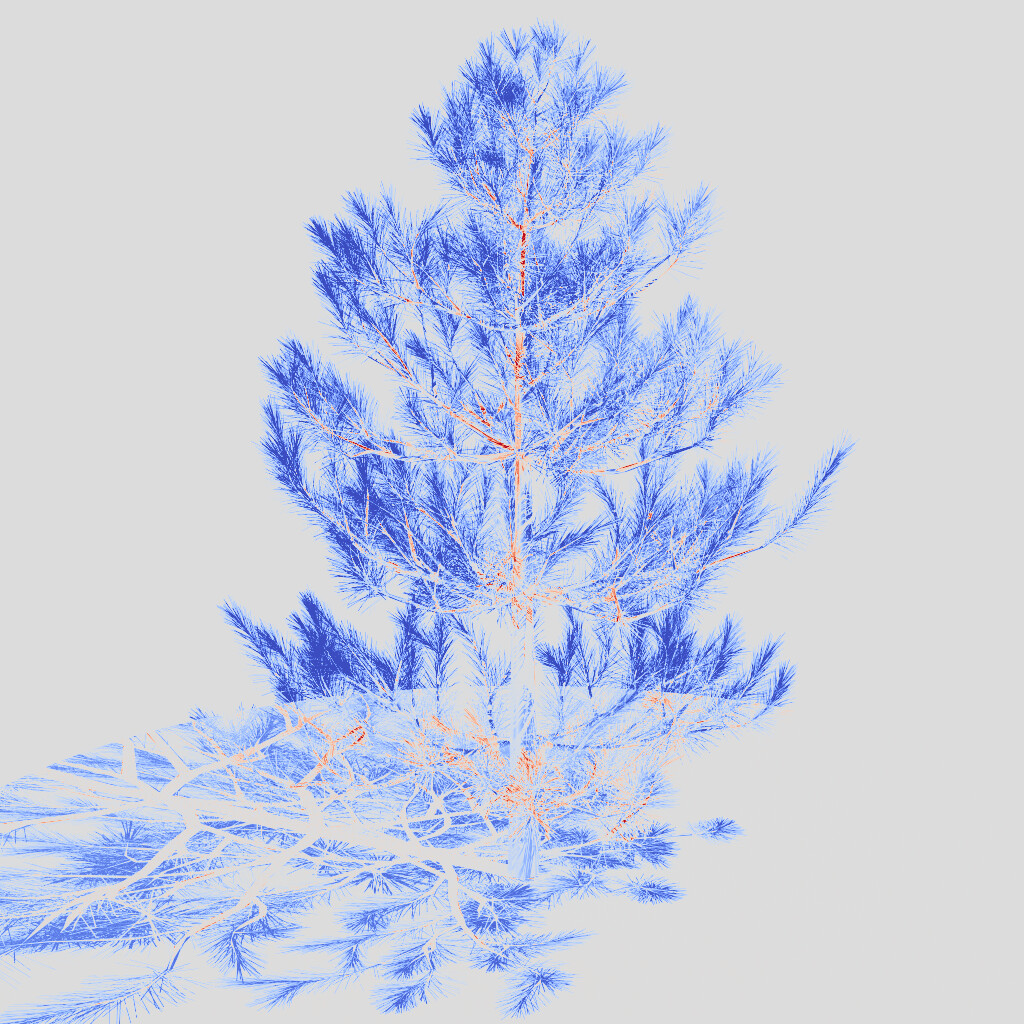}
        &
        \includegraphics[height=\heightMeshComp, frame]{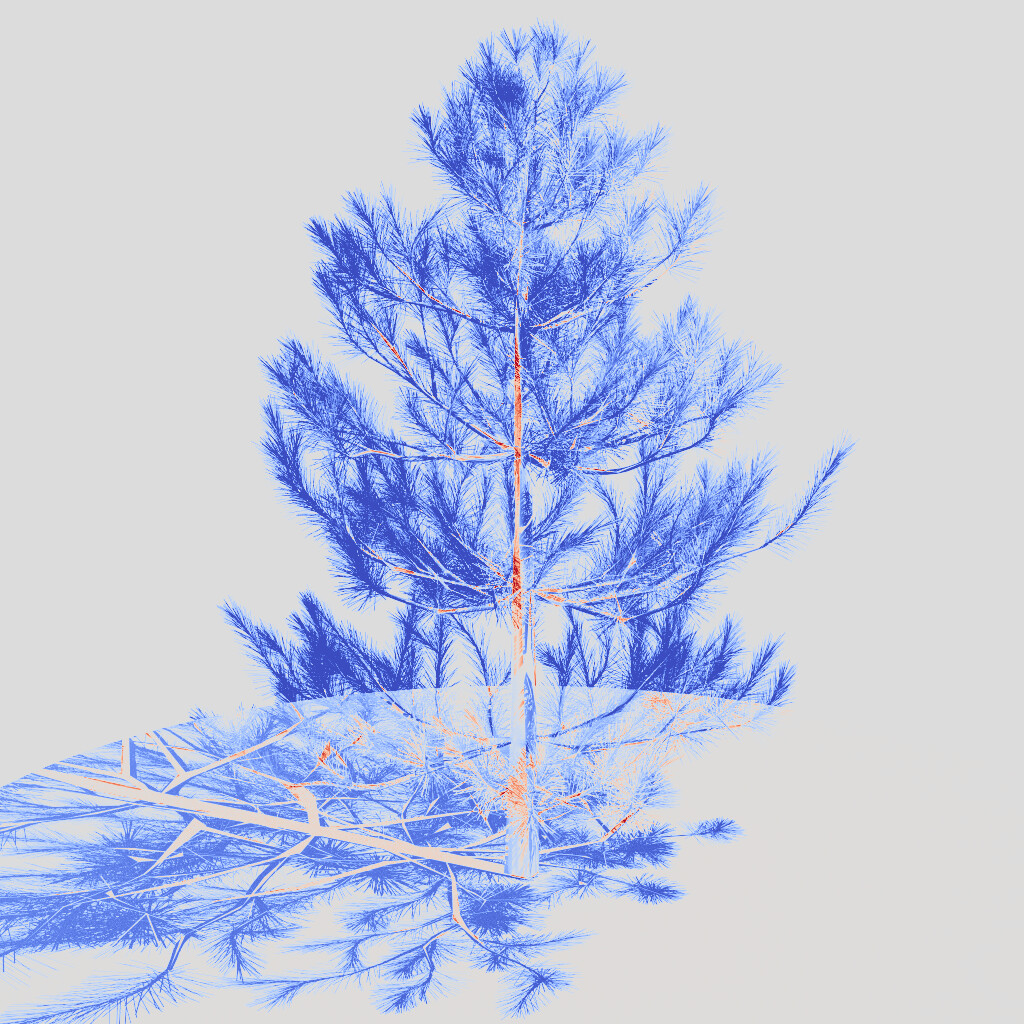}
        &
        \includegraphics[height=\heightMeshComp, frame]{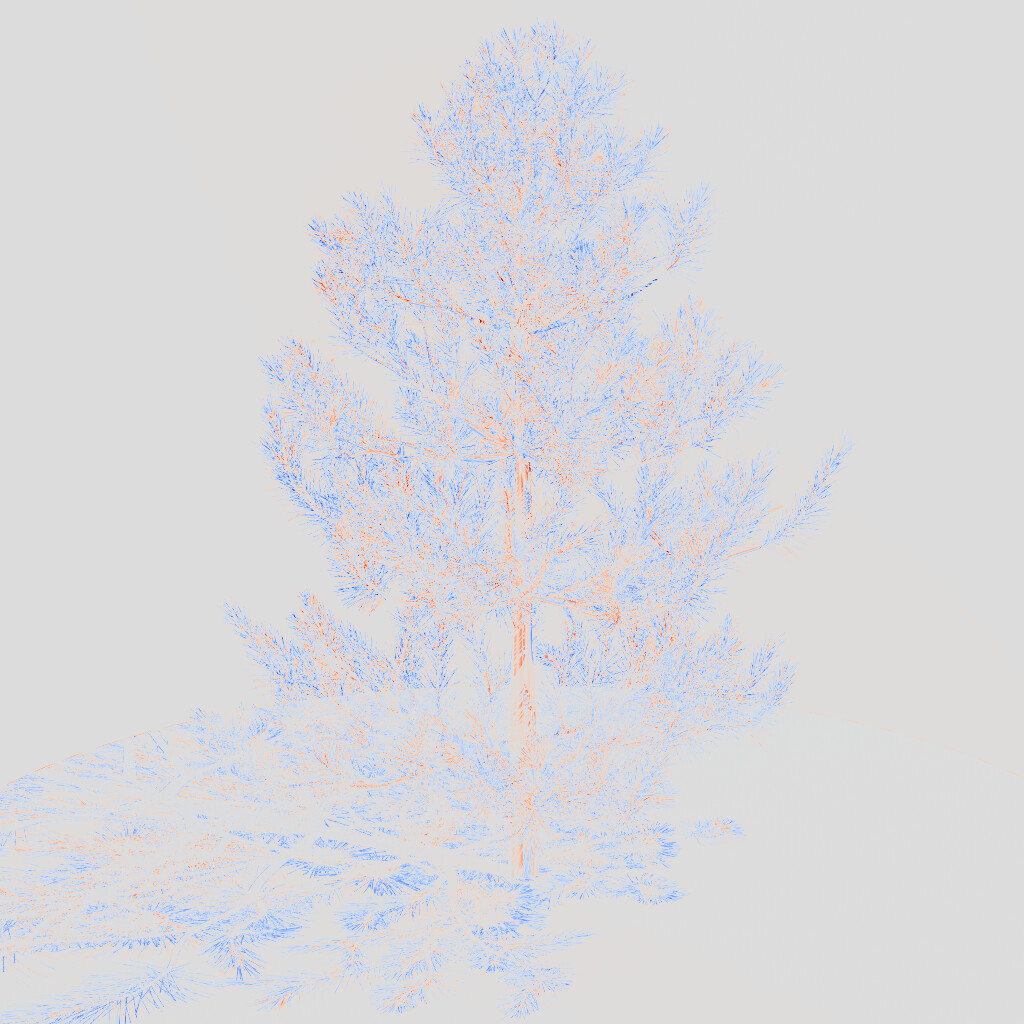}
        &
        \includegraphics[height=\heightMeshComp, frame]{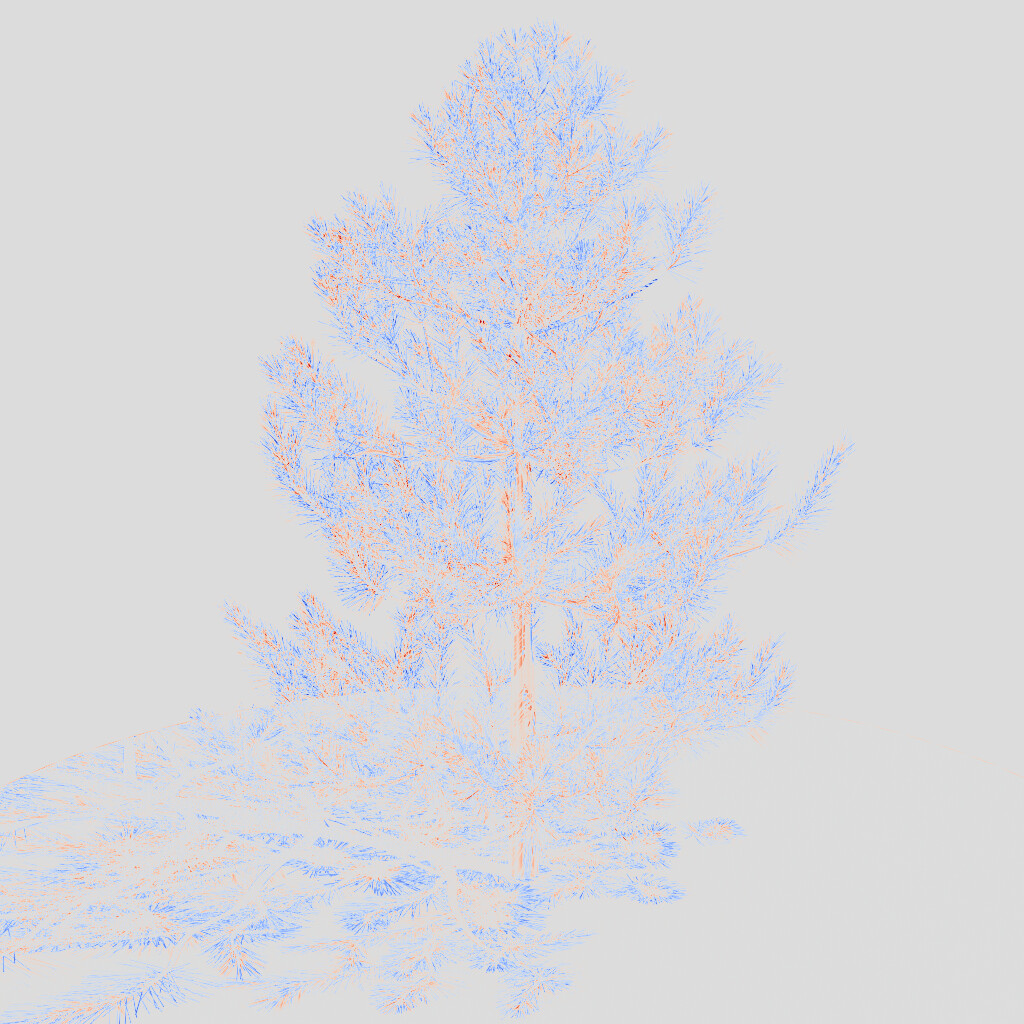}
        &
        \includegraphics[height=\heightMeshComp, frame]{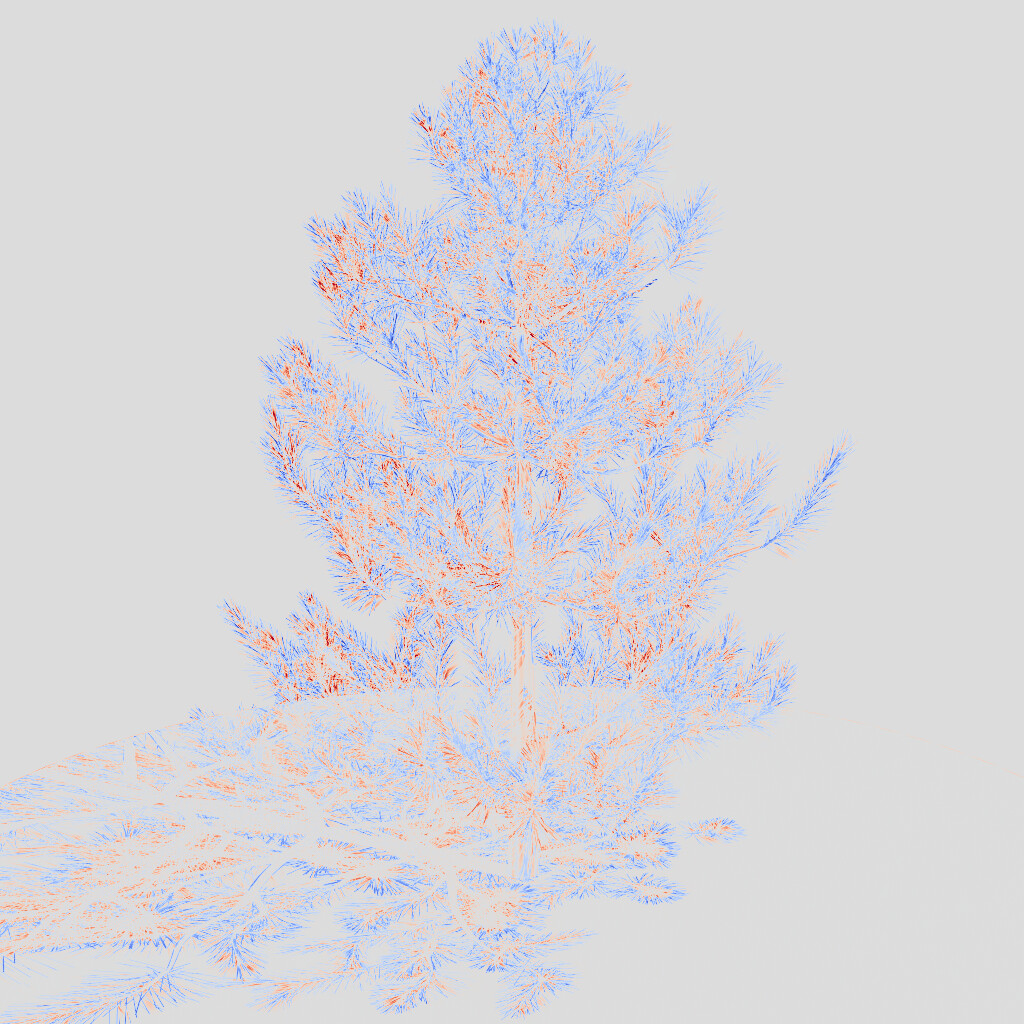}   
        &
        \includegraphics[height=\heightMeshComp, frame]{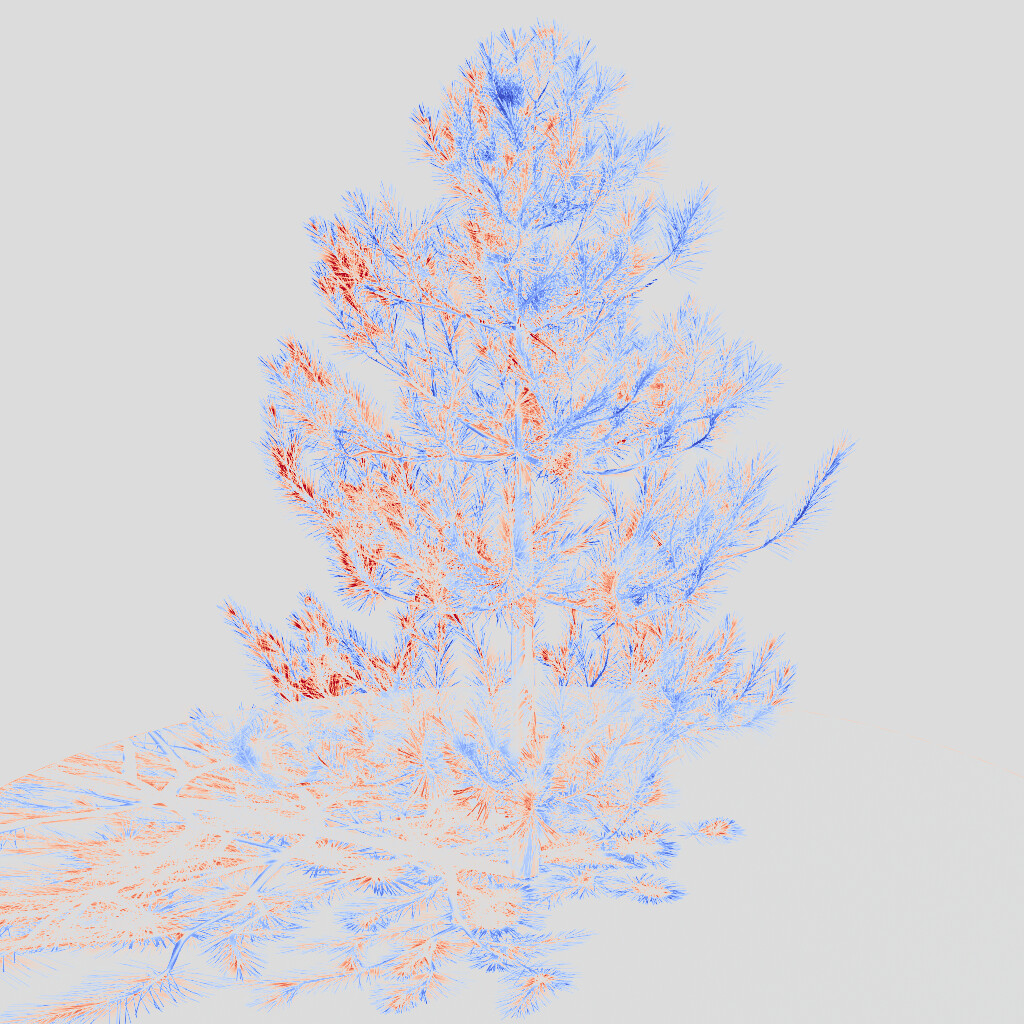}  
        \\
        \multicolumn{1}{l}{{\small{\textsf{PSNR$\uparrow$:}}}} & 
        {\small{\textsf{21.38}}} & 
        {\small{\textsf{16.54}}} & 
        {\small{\textsf{14.67}}} & 
        {\small{\textsf{24.80}}} & 
        {\small{\textsf{24.16}}} & 
        {\small{\textsf{22.25}}} & 
        {\small{\textsf{20.94}}} \\
        \multicolumn{1}{l}{{\small{\textsf{SSIM$\uparrow$:}}}} & 
        {\small{\textsf{0.864}}} & 
        {\small{\textsf{0.677}}} & 
        {\small{\textsf{0.619}}} & 
        {\small{\textsf{0.861}}} & 
        {\small{\textsf{0.841}}} & 
        {\small{\textsf{0.741}}} & 
        {\small{\textsf{0.706}}} \\
        \multicolumn{1}{l}{{\small{\textsf{LPIPS$\downarrow$:}}}} & 
        {\small{\textsf{0.111}}} & 
        {\small{\textsf{0.364}}} & 
        {\small{\textsf{0.399}}} & 
        {\small{\textsf{0.111}}} & 
        {\small{\textsf{0.121}}} & 
        {\small{\textsf{0.203}}} & 
        {\small{\textsf{0.271}}} \\
        \multicolumn{1}{l}{{\small{\textsf{\FLIP$\downarrow$:}}}} & 
        {\small{\textsf{0.128}}} & 
        {\small{\textsf{0.222}}} & 
        {\small{\textsf{0.263}}} & 
        {\small{\textsf{0.102}}} & 
        {\small{\textsf{0.102}}} & 
        {\small{\textsf{0.123}}} & 
        {\small{\textsf{0.150}}} \\          
        \multicolumn{1}{l}{{\small{\textsf{CPU Time$\downarrow$: 98.4}}}} & 
        {\small{\textsf{81.2}}} & 
        {\small{\textsf{57.3}}} & 
        {\small{\textsf{43.8}}} & 
        {\small{\textsf{262.8}}} & 
        {\small{\textsf{137.6}}} & 
        {\small{\textsf{69.4}}} & 
        {\small{\textsf{55.3}}} \\

        \includegraphics[height=\heightMeshComp, frame]{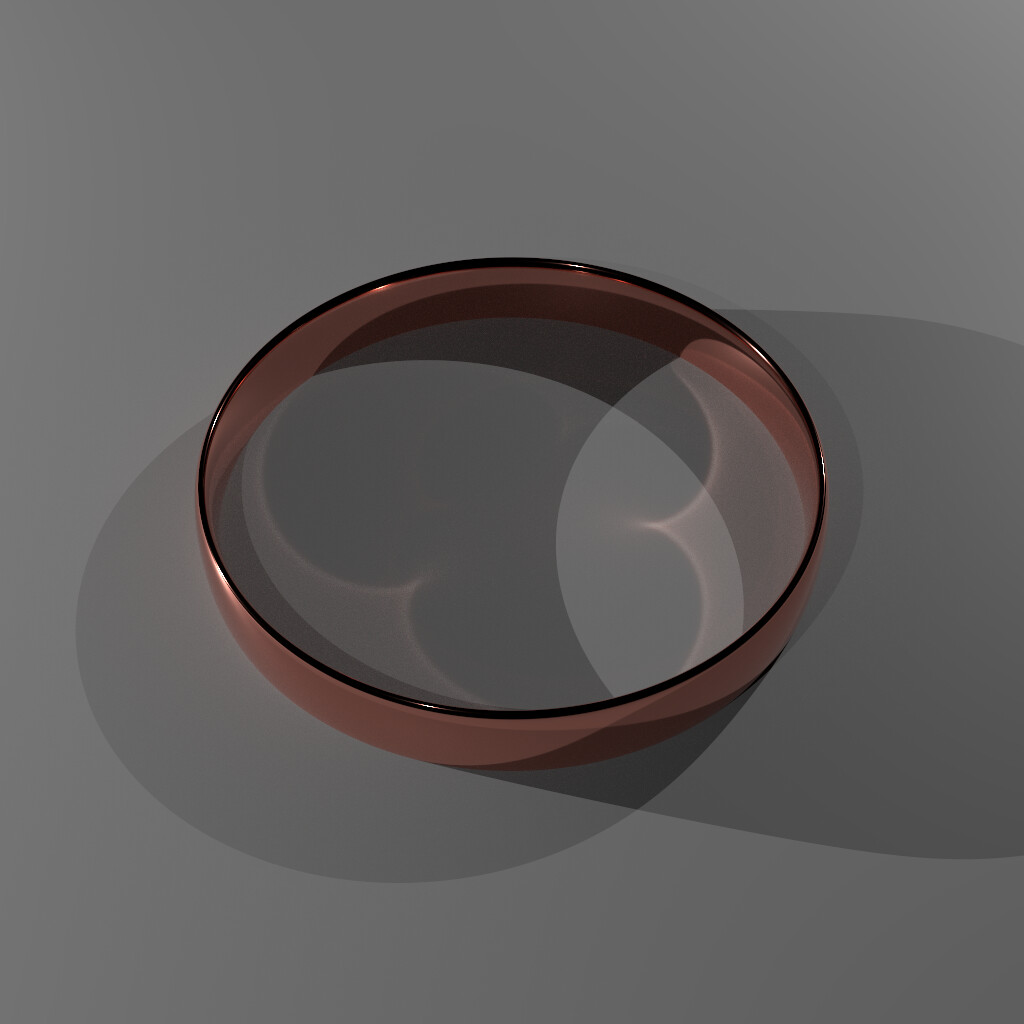}
        &
        \includegraphics[height=\heightMeshComp, frame]{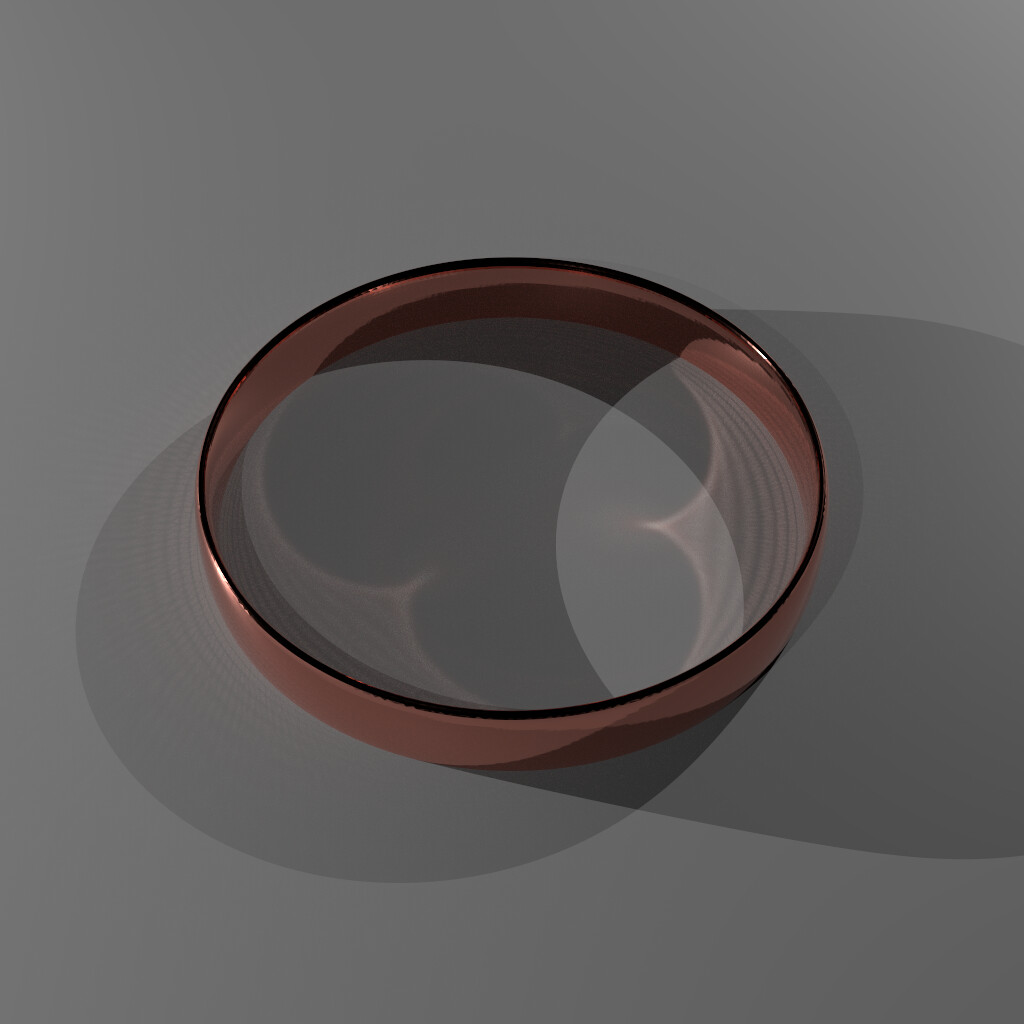}
        &
        \includegraphics[height=\heightMeshComp, frame]{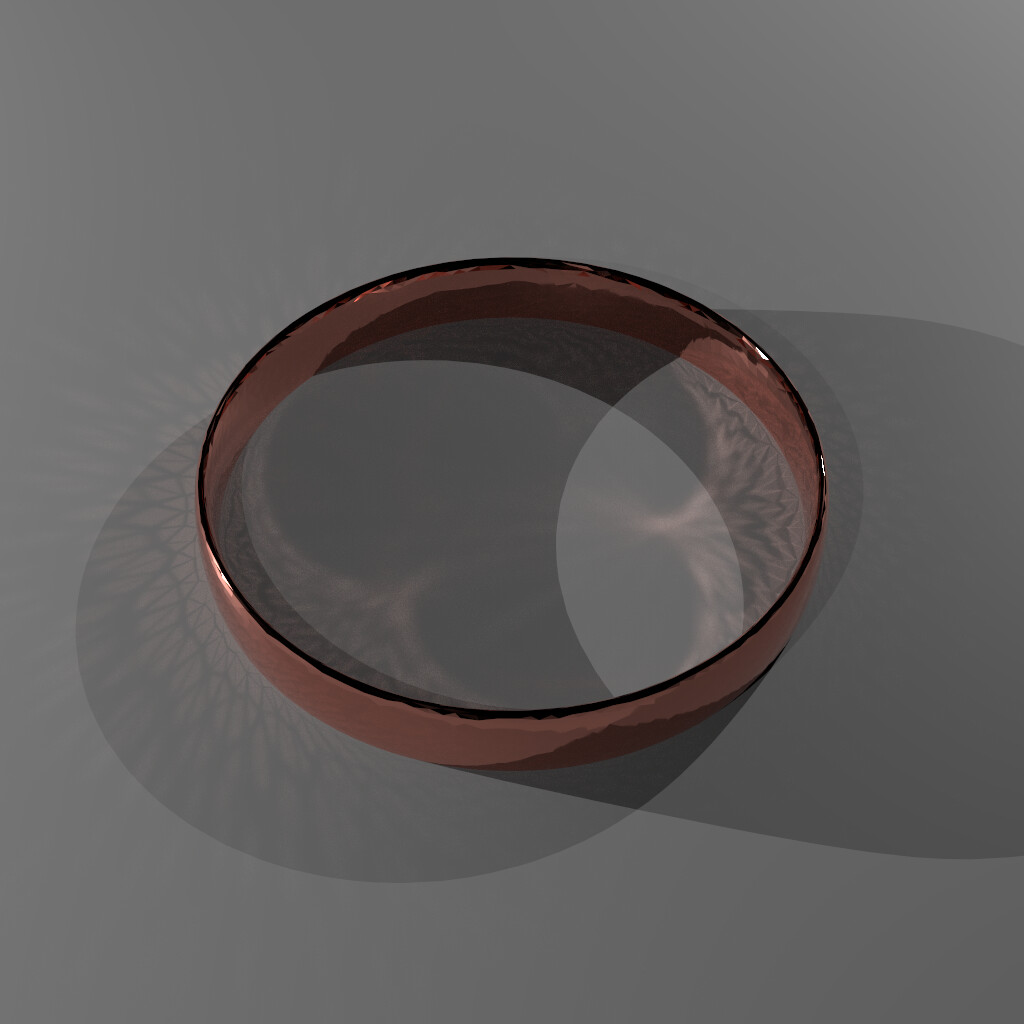}
        &
        \includegraphics[height=\heightMeshComp, frame]{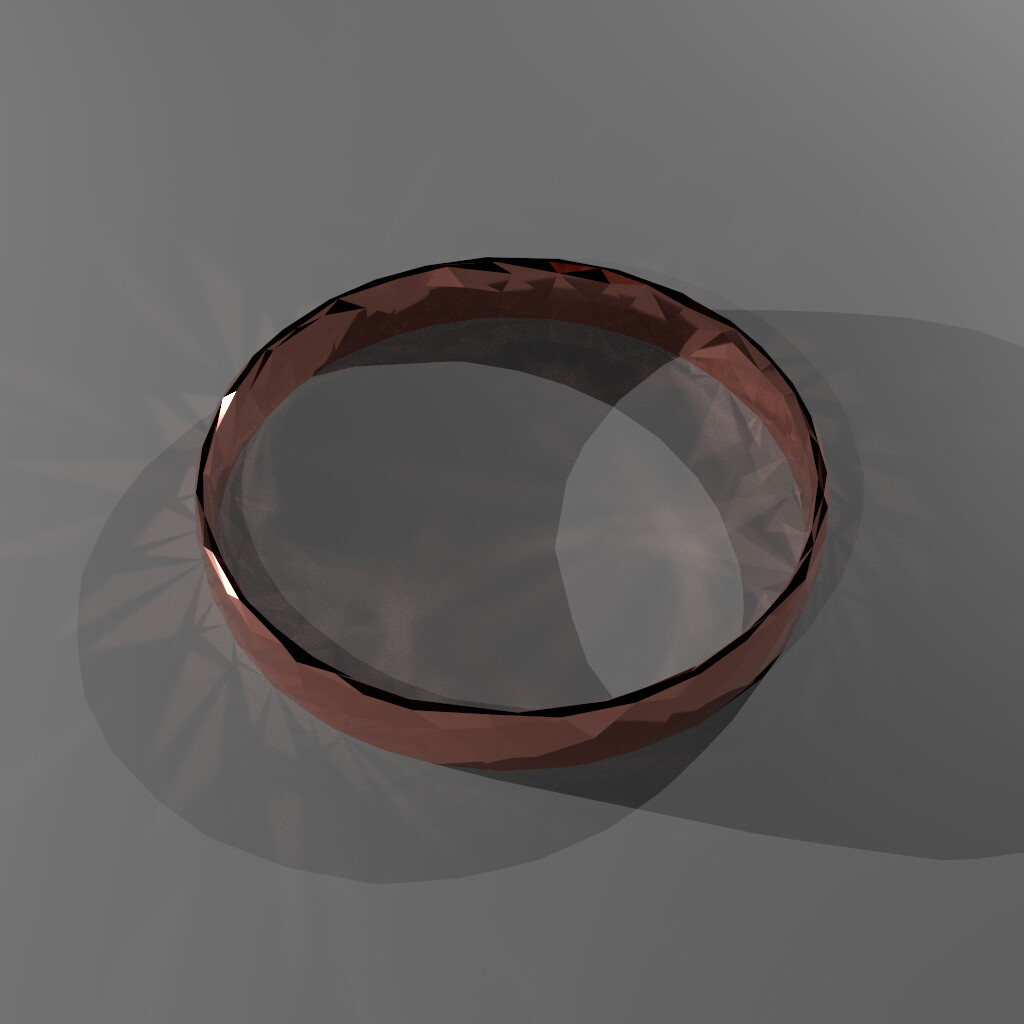}
        &
        \includegraphics[height=\heightMeshComp, frame]{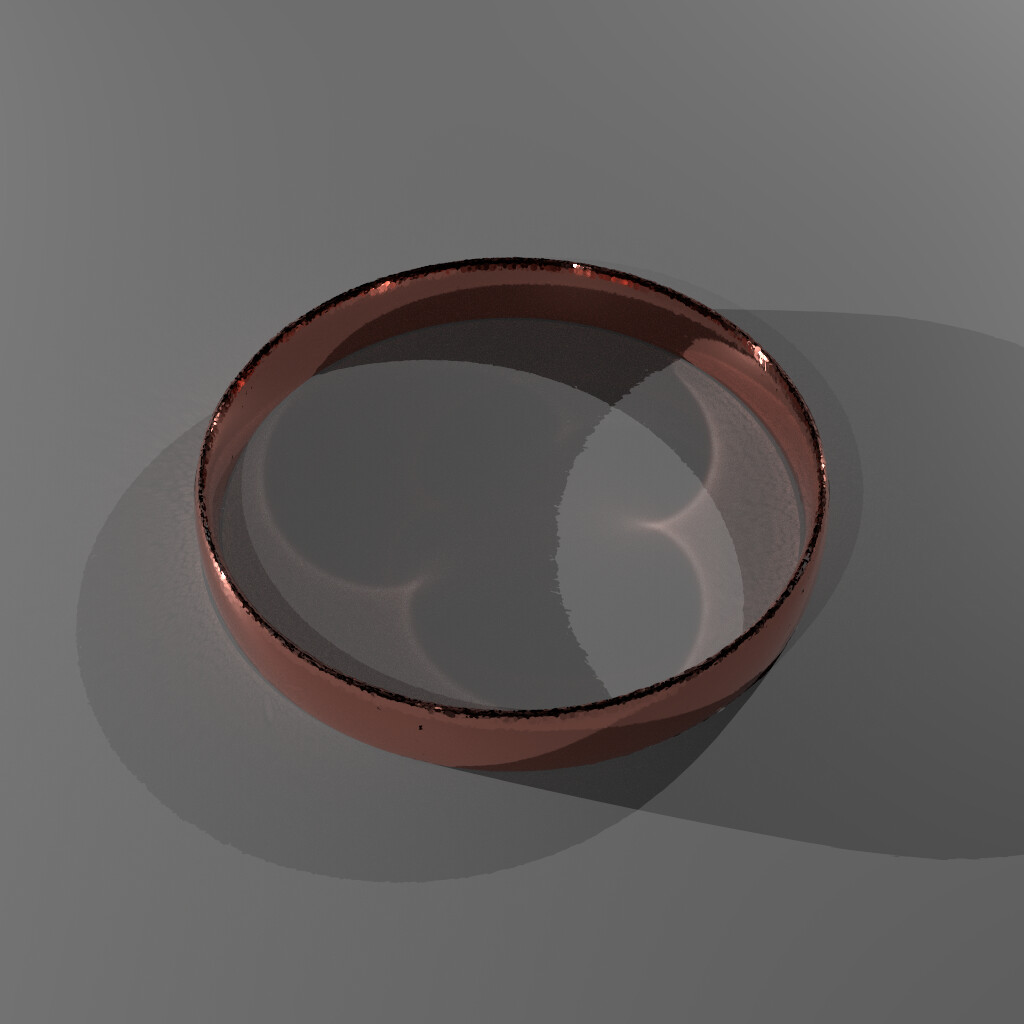}
        &
        \includegraphics[height=\heightMeshComp, frame]{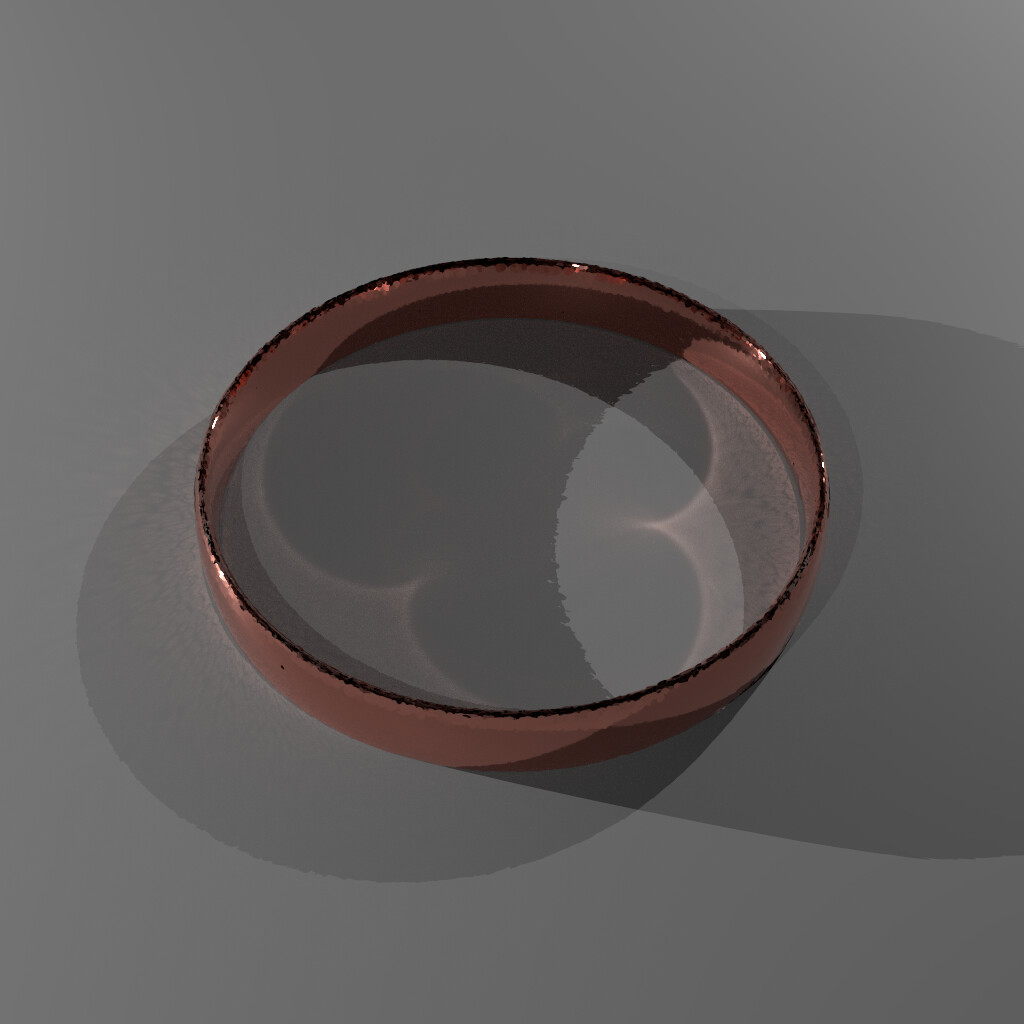}
        &
        \includegraphics[height=\heightMeshComp, frame]{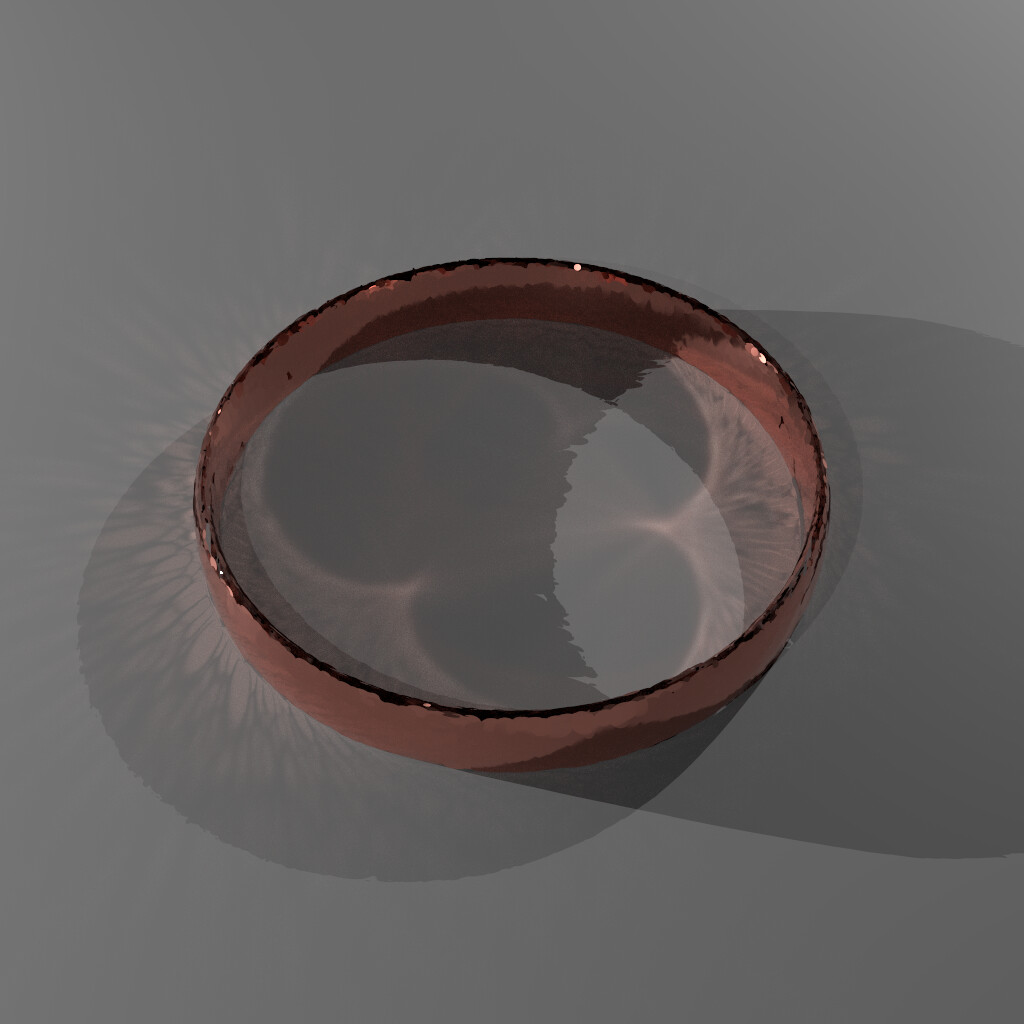}   
        &
        \includegraphics[height=\heightMeshComp, frame]{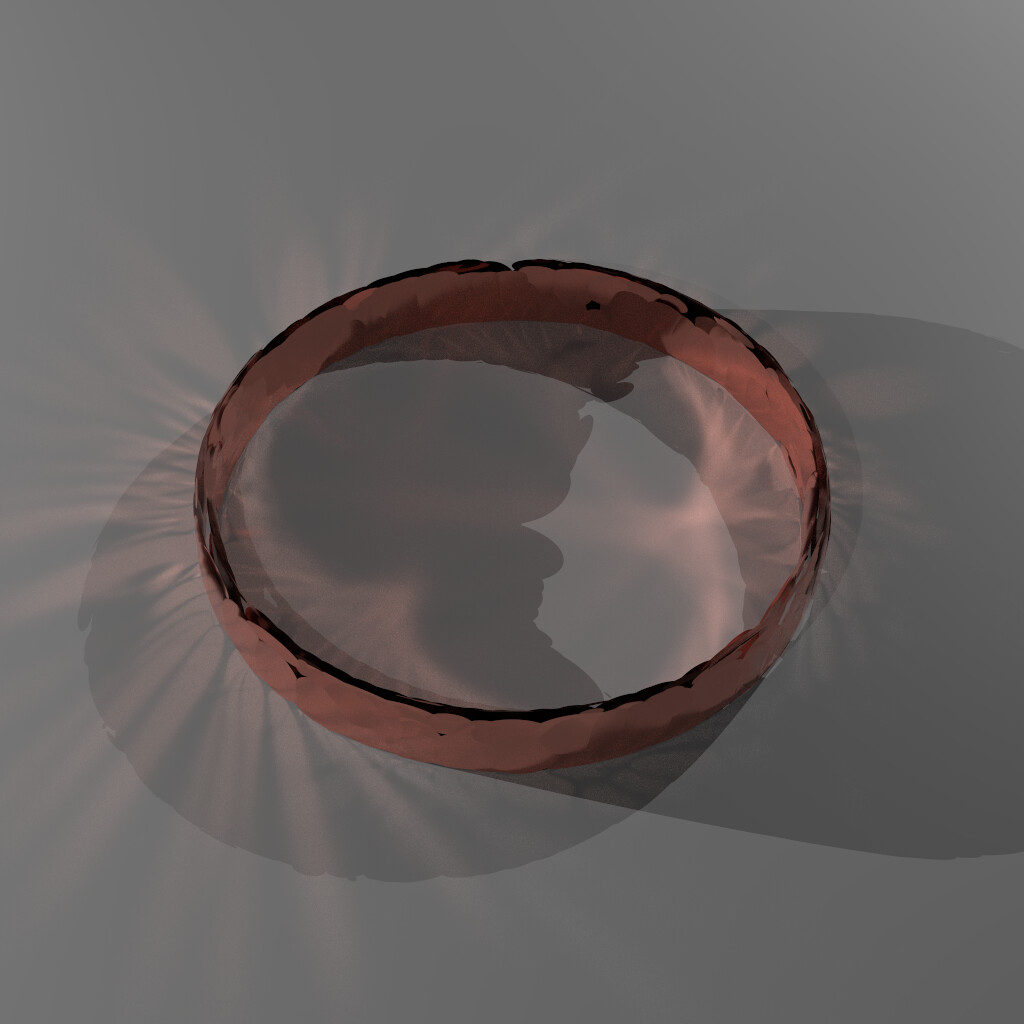}
        \\
		\multicolumn{1}{r}{\begin{overpic}[height=\heightMeshComp, frame,unit=1mm, frame]{resources/coolwarm_v.jpg}
			\put(-18, 91){\small 0.5}
			\put(-23, 1){\small -0.5}      
            \put(-80, 90){\footnotesize \emph{Caustic Ring}}                  
		\end{overpic}}
        &
        \includegraphics[height=\heightMeshComp, frame]{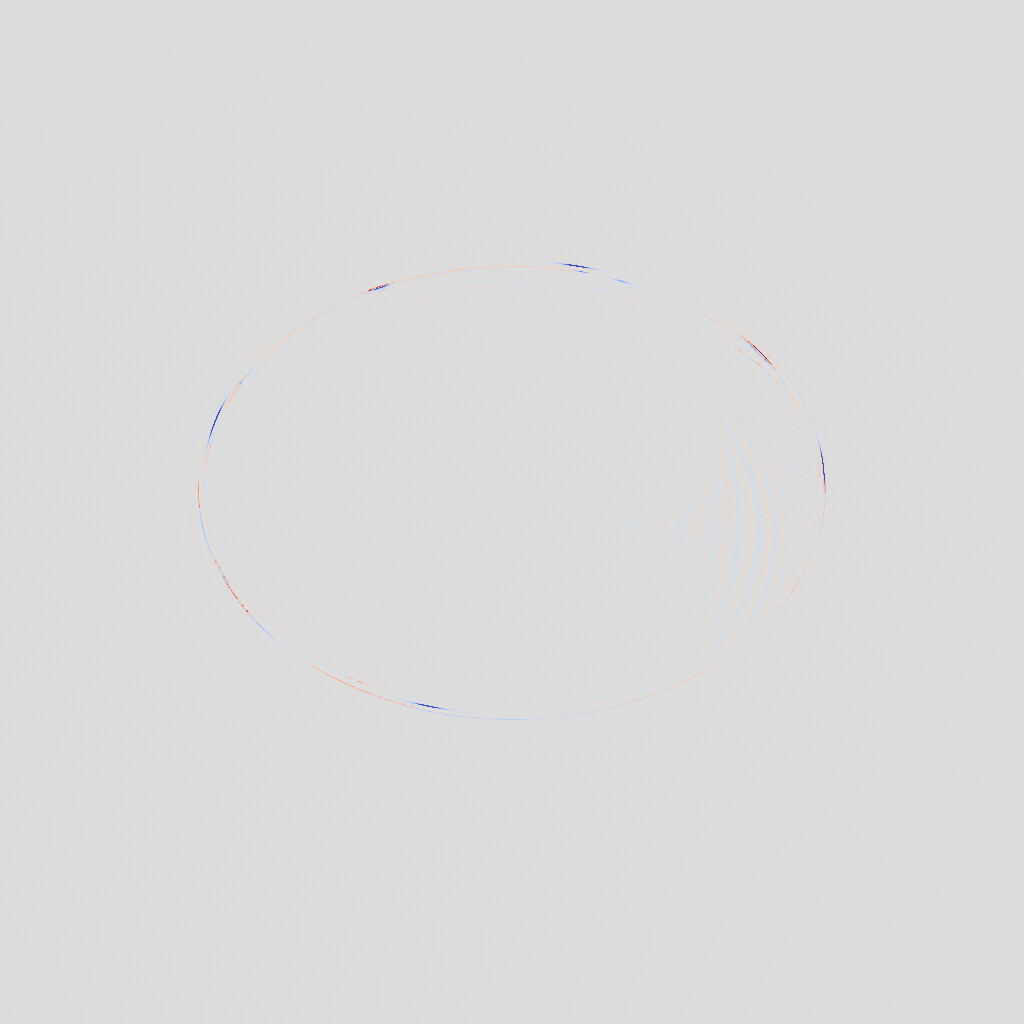}
        &
        \includegraphics[height=\heightMeshComp, frame]{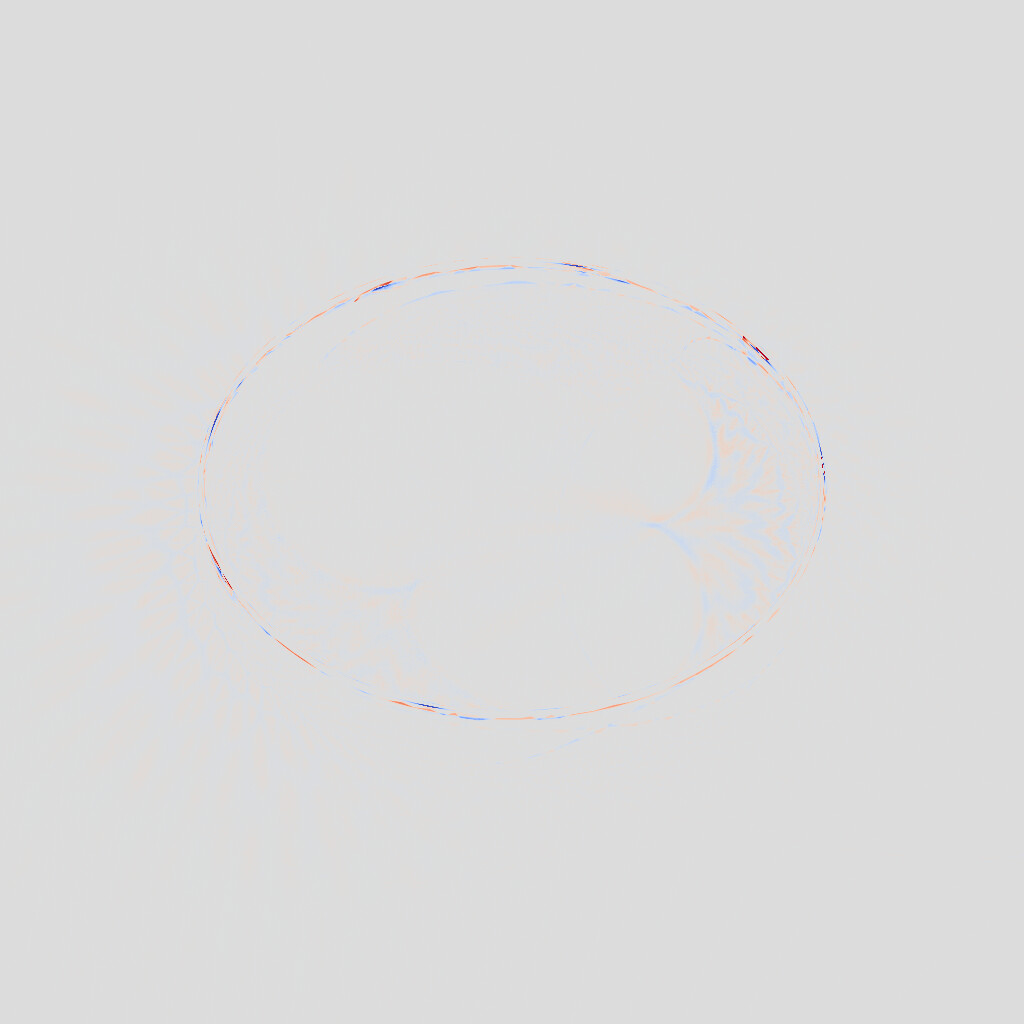}
        &
        \includegraphics[height=\heightMeshComp, frame]{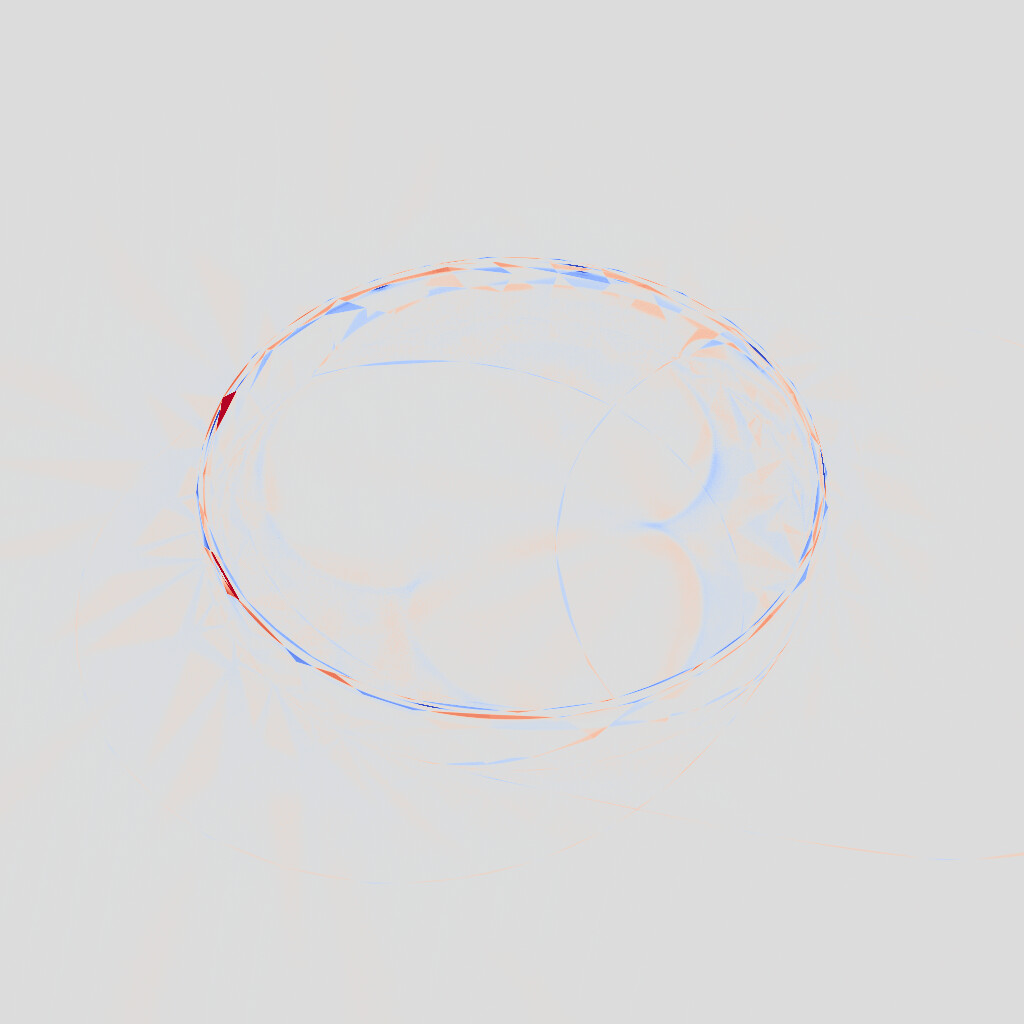}
        &
        \includegraphics[height=\heightMeshComp, frame]{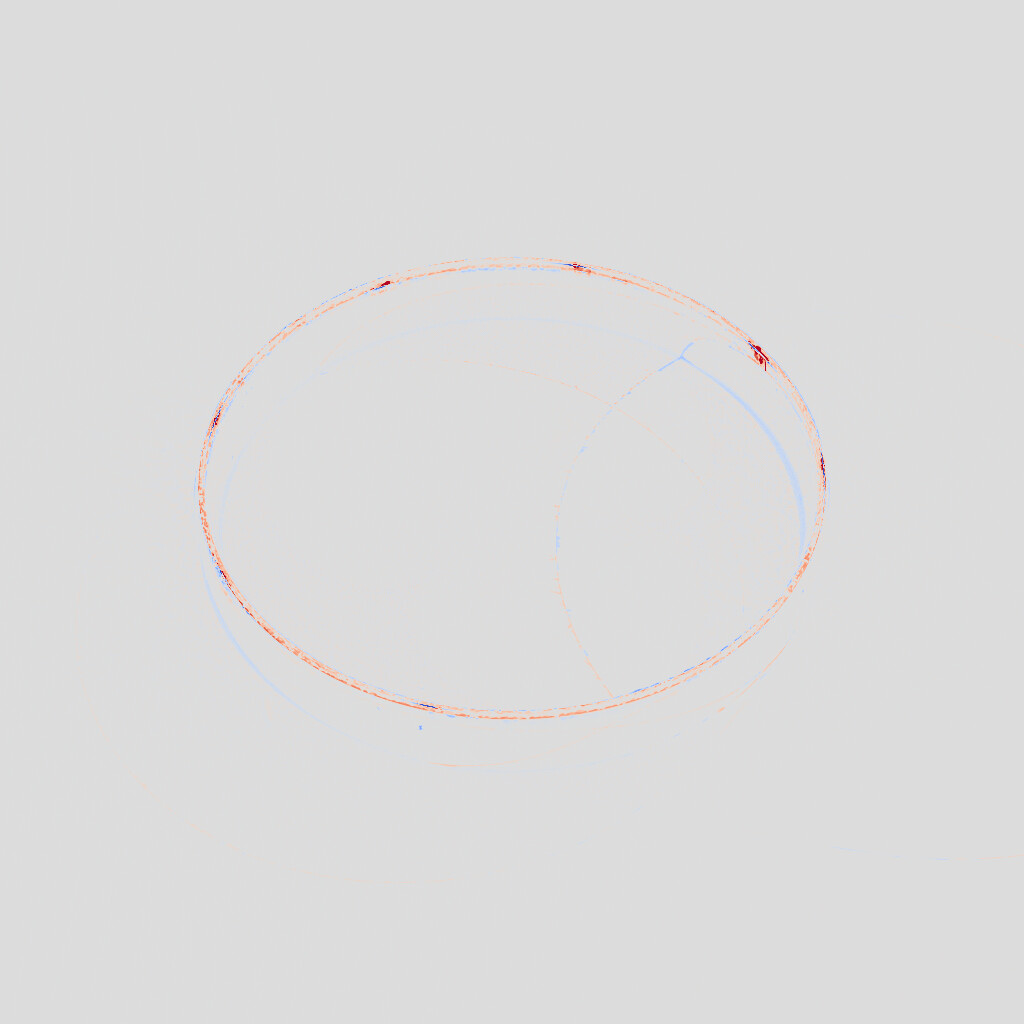}
        &
        \includegraphics[height=\heightMeshComp, frame]{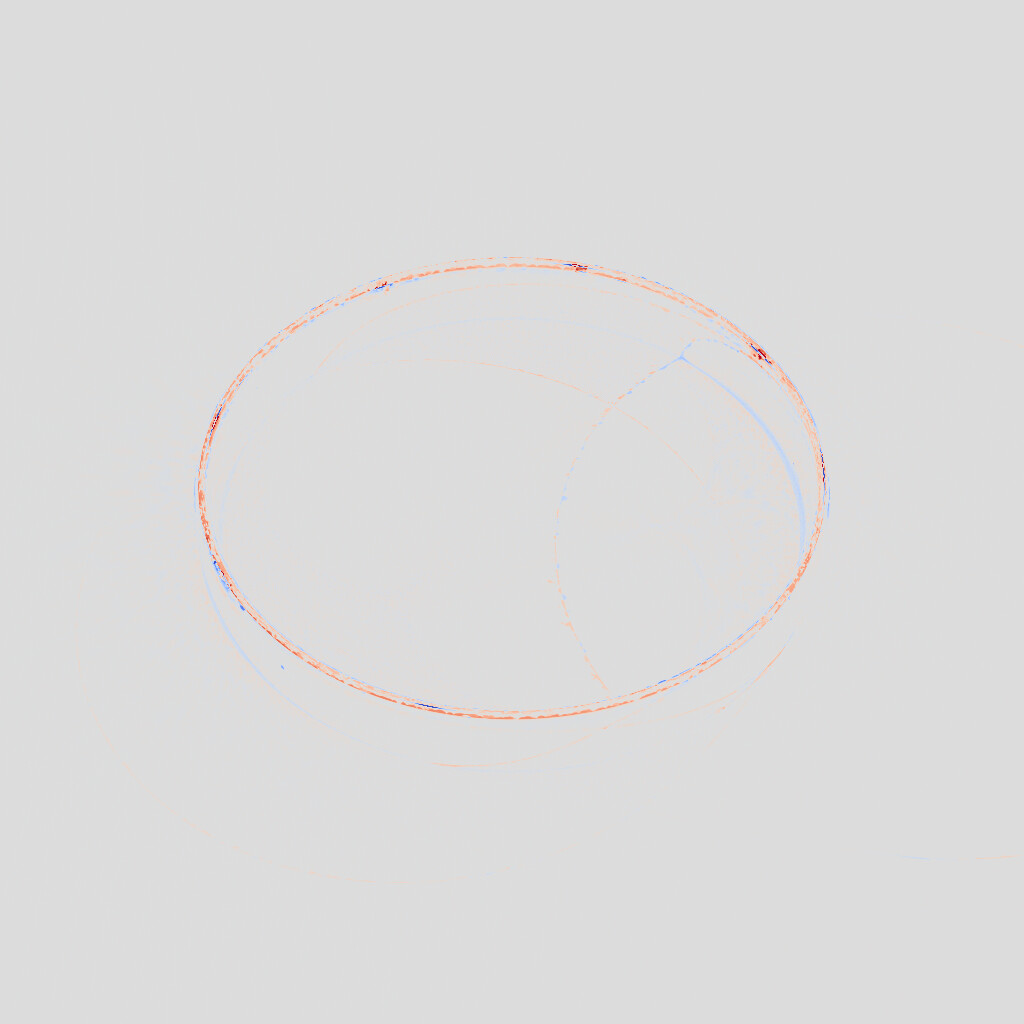}
        &
        \includegraphics[height=\heightMeshComp, frame]{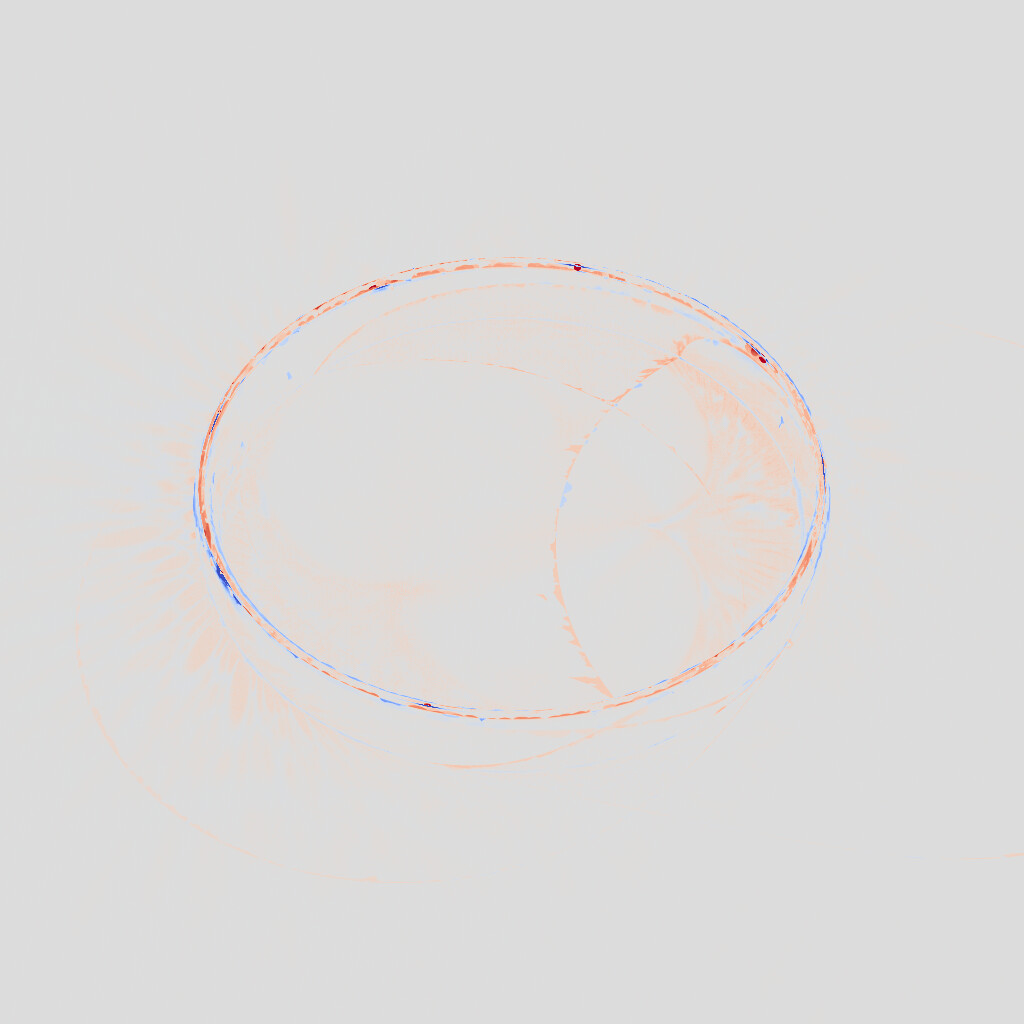}   
        &
        \includegraphics[height=\heightMeshComp, frame]{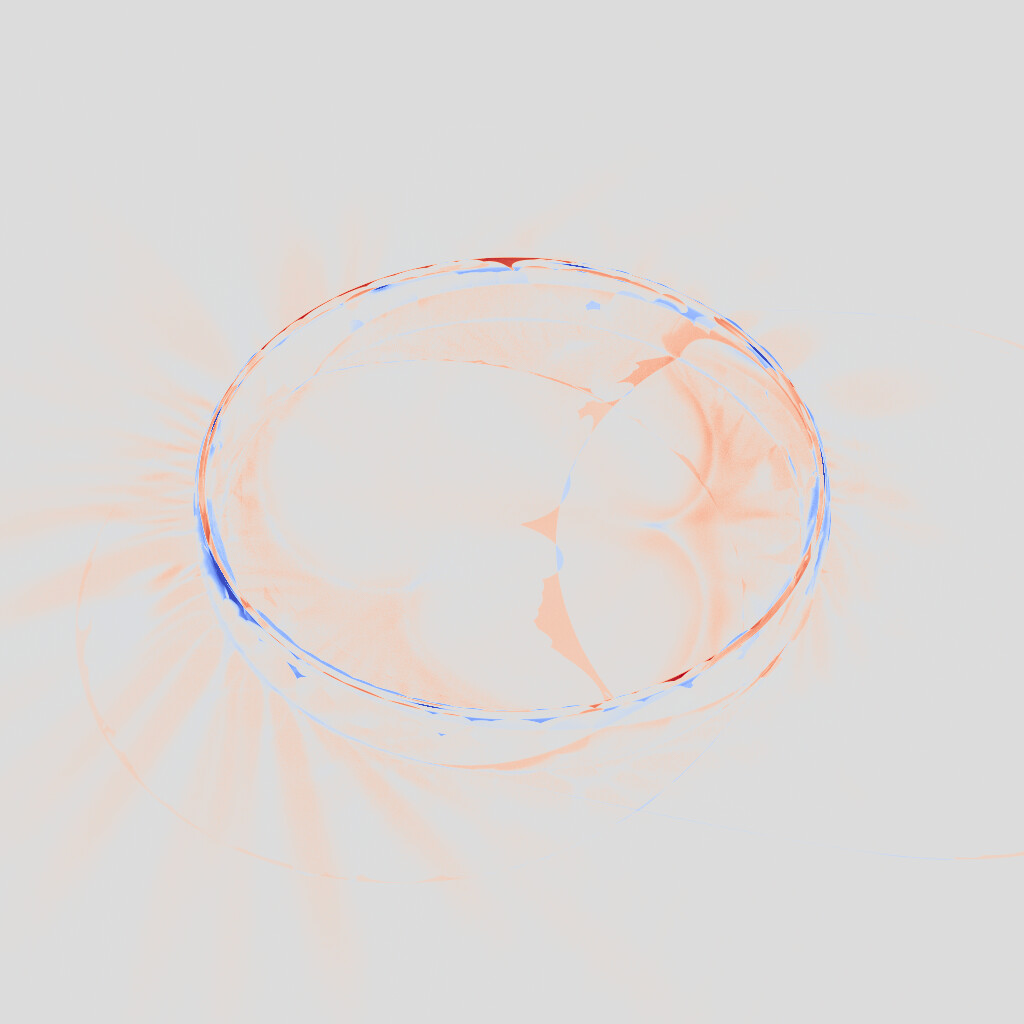}  
        \\
        \multicolumn{1}{l}{{\small{\textsf{PSNR$\uparrow$:}}}} & 
        {\small{\textsf{38.07}}} & 
        {\small{\textsf{36.13}}} & 
        {\small{\textsf{30.99}}} & 
        {\small{\textsf{34.27}}} & 
        {\small{\textsf{34.02}}} & 
        {\small{\textsf{32.84}}} & 
        {\small{\textsf{28.65}}} \\
        \multicolumn{1}{l}{{\small{\textsf{SSIM$\uparrow$:}}}} & 
        {\small{\textsf{0.970}}} & 
        {\small{\textsf{0.963}}} & 
        {\small{\textsf{0.946}}} & 
        {\small{\textsf{0.960}}} & 
        {\small{\textsf{0.956}}} & 
        {\small{\textsf{0.940}}} & 
        {\small{\textsf{0.927}}} \\
        \multicolumn{1}{l}{{\small{\textsf{LPIPS$\downarrow$:}}}} & 
        {\small{\textsf{0.022}}} & 
        {\small{\textsf{0.044}}} & 
        {\small{\textsf{0.066}}} & 
        {\small{\textsf{0.049}}} & 
        {\small{\textsf{0.054}}} & 
        {\small{\textsf{0.084}}} & 
        {\small{\textsf{0.117}}} \\
        \multicolumn{1}{l}{{\small{\textsf{\FLIP$\downarrow$:}}}} & 
        {\small{\textsf{0.016}}} & 
        {\small{\textsf{0.025}}} & 
        {\small{\textsf{0.041}}} & 
        {\small{\textsf{0.019}}} & 
        {\small{\textsf{0.022}}} & 
        {\small{\textsf{0.044}}} & 
        {\small{\textsf{0.080}}} \\        
        \multicolumn{1}{l}{{\small{\textsf{CPU Time$\downarrow$: 52.4}}}} & 
        {\small{\textsf{52.3}}} & 
        {\small{\textsf{52.3}}} & 
        {\small{\textsf{51.7}}} & 
        {\small{\textsf{67.1}}} & 
        {\small{\textsf{63.2}}} & 
        {\small{\textsf{60.8}}} & 
        {\small{\textsf{59.5}}} \\        
    \end{tabular}
    \caption{\label{fig:mesh_comp}
        Comparison between our Gaussian representation and meshes. Triangle and primitive counts are relative to the triangle count of the original 
        model (reference). We provide common quality metrics and {CPU time (ms/spp)} 
        measurements, and difference images between each render and the reference. PSNR values are calculated after clamping pixel intensities to $[0, 1]$.        
        }
\end{figure*}

Crucially, our framework supports full unbiased global illumination via Monte Carlo (volumetric) path tracing. 
\autoref{fig:teaser} presents two renders with drastically different lighting setups. 
In the ``daylight'' setup, the scene is illuminated by an area light and an environment light; 
in the ``night'' setup, the scene is lit by a logo, neon light bars, and candles, all modeled as emissive Gaussian primitives. 
The total number of emissive primitives exceeds 10K. 
Both renders showcase various global illumination effects, including soft shadows, color bleeding, and inter-reflections. 
Additional sequences with zoomed-in camera animations are provided in the supplemental video. 
Our ability to model full global illumination stands in contrast to rasterization approaches such as 3DGS and volumetric ray marching approaches such as NeRF. 
Both families of methods are limited to the first few dimensions of the path space and cannot solve the infinite-dimensional light transport integral. 
Even when compared with other recent particle-based representations designed with light transport in mind, our method offers unique versatility. 
While the method of \citet{tog/CondorSBBGDJ25} supports volumetric path tracing, its relatively simplistic phase function lacks the ability to model surface appearance. 
\citet{tog/JiangSLWLR25} achieve real-time rendering by simplifying both the geometric representation and the light transport formulation. 
However, their adoption of radiosity and radiance representation in the spherical harmonic coefficient space fundamentally limits their capability 
to model high-frequency reflectance and light transport effects. 
In contrast, we demonstrate challenging scenes with highly glossy appearance (\autoref{fig:teaser}) and caustics (\autoref{fig:mesh_comp}).

\begin{figure*}[tb]
	\newlength{\lenMaterialEdit}
	\setlength{\lenMaterialEdit}{0.23\linewidth}
    \addtolength{\tabcolsep}{-4pt}
    \renewcommand{\arraystretch}{0.5}
    \centering
    \begin{tabular}{ccccc}
        & \multicolumn{2}{c}{\emph{Dragon}} & \multicolumn{2}{c}{\emph{Blanket}} \\
        \raisebox{12pt}{\rotatebox{90}{\small{\textsf{Unmodified}}}}
        &
        \includegraphics[width=\lenMaterialEdit]{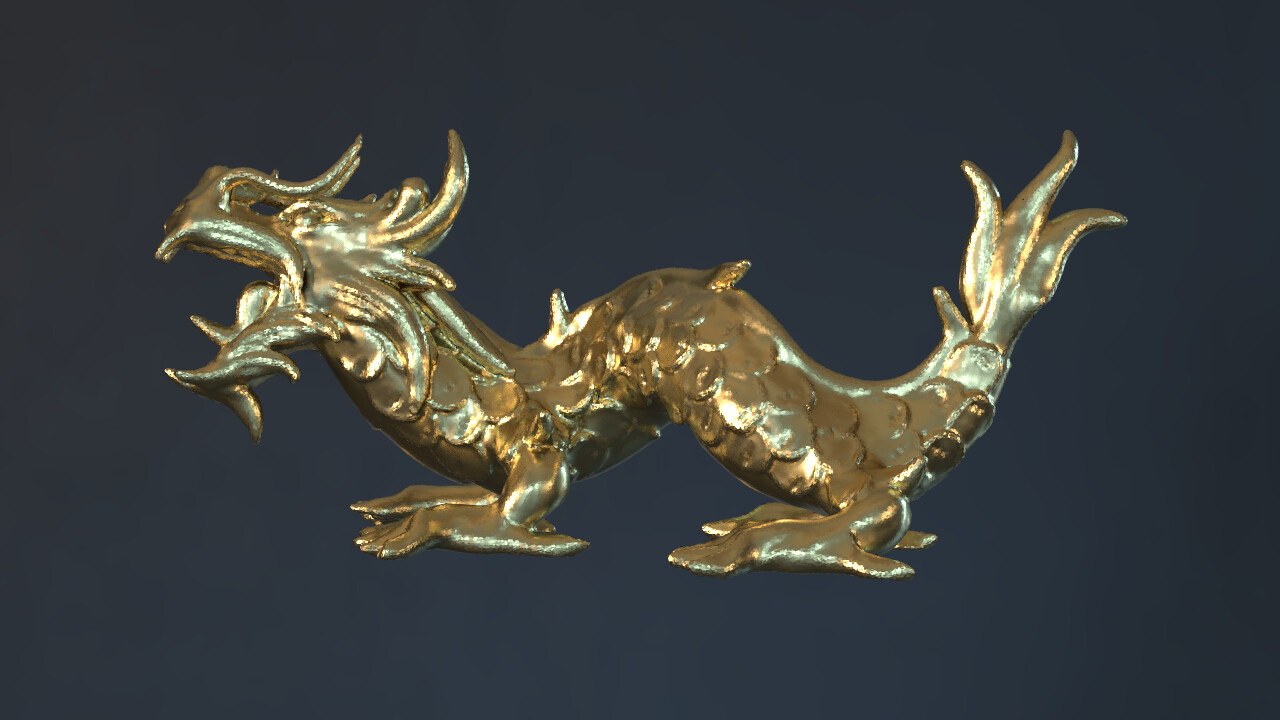}
        &
        \includegraphics[width=\lenMaterialEdit]{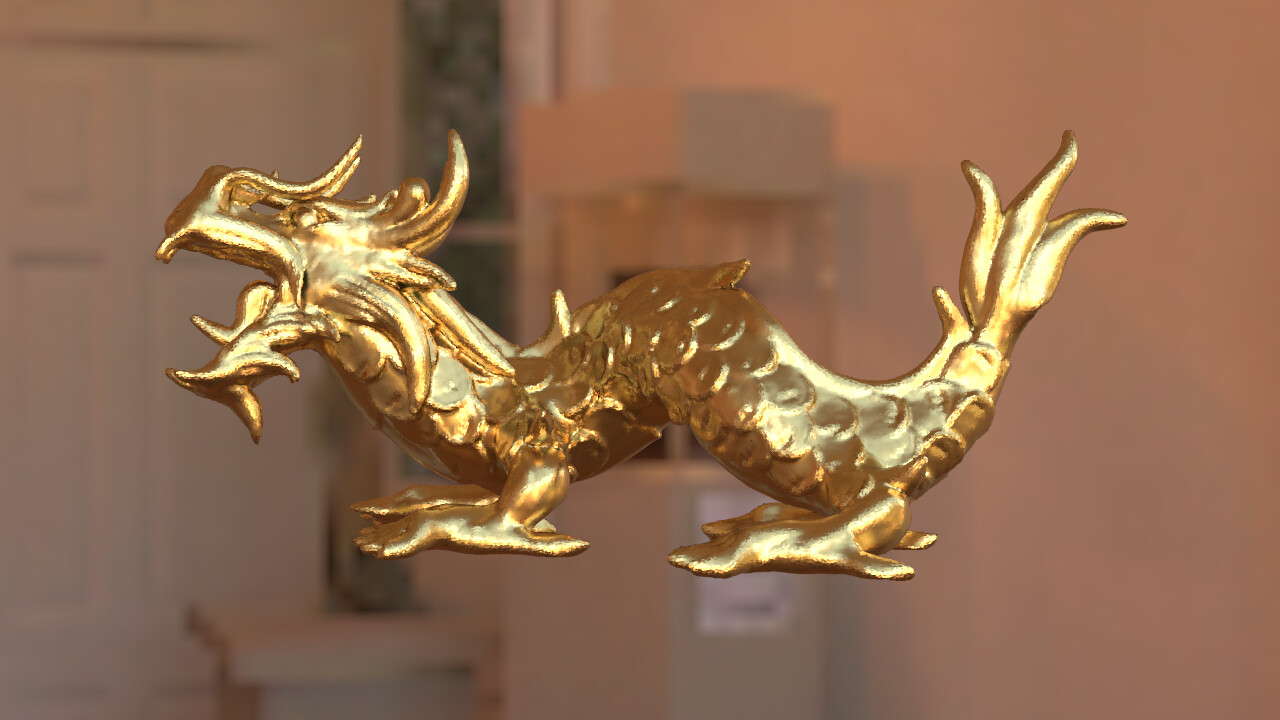}
        &
        \includegraphics[width=\lenMaterialEdit]{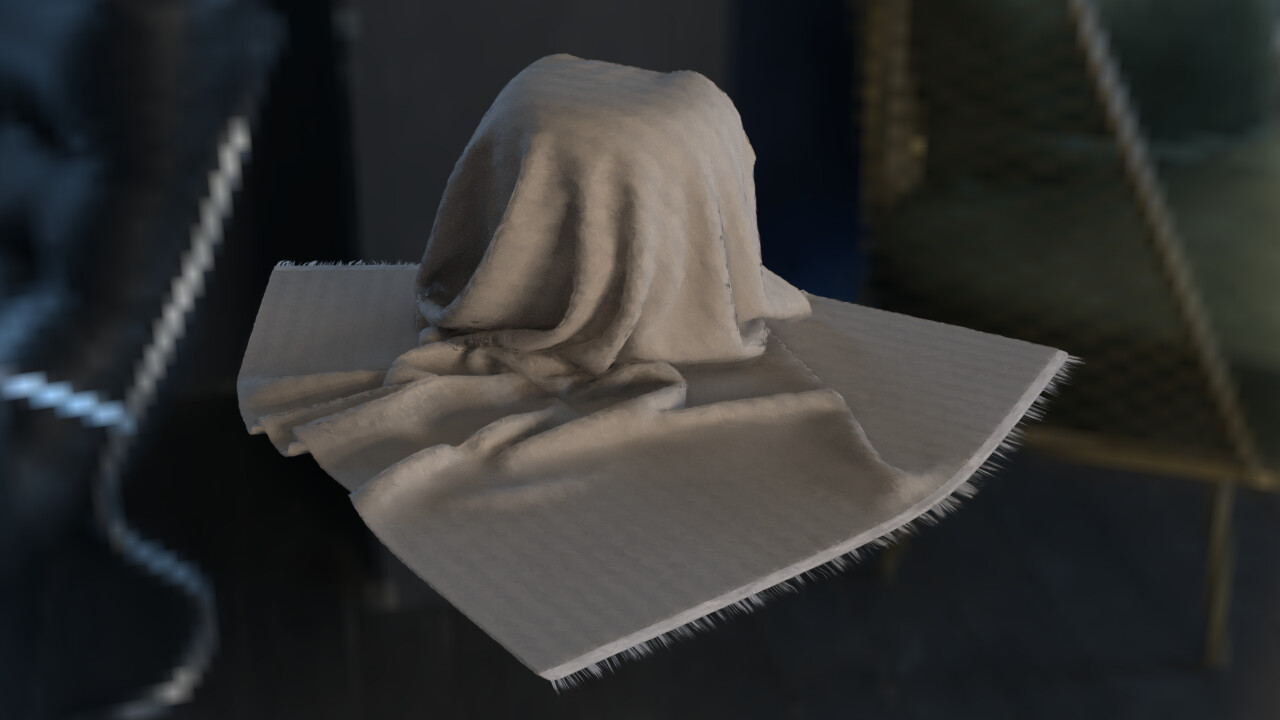}
        &
        \includegraphics[width=\lenMaterialEdit]{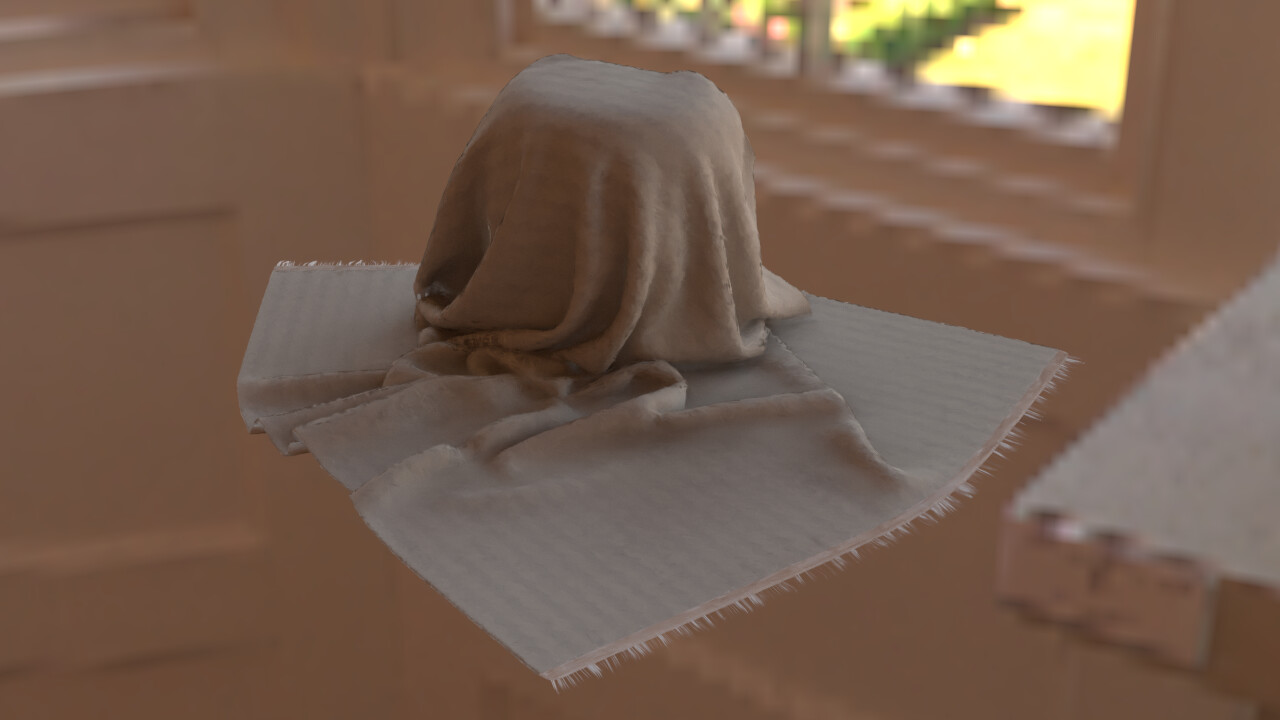}
        \\ 
        \raisebox{12pt}{\rotatebox{90}{\small{\textsf{Ext. triplanar}}}}
        &
        \includegraphics[width=\lenMaterialEdit]{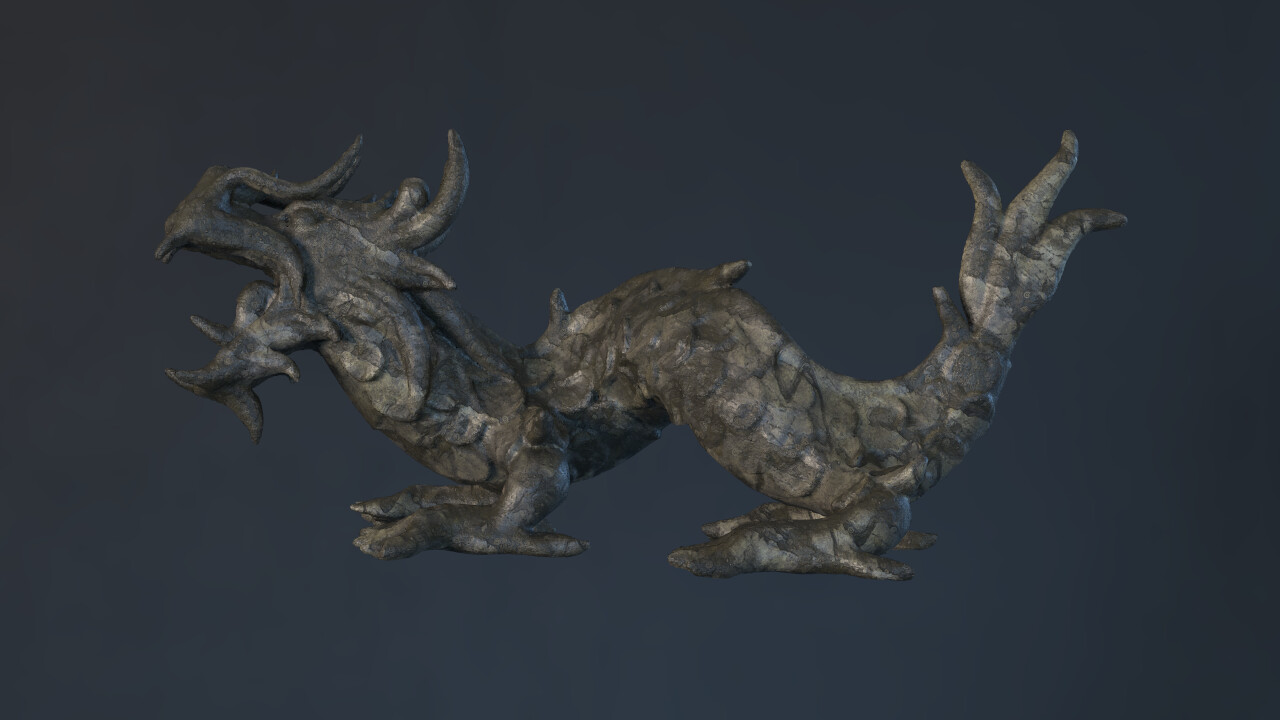}
        &
        \includegraphics[width=\lenMaterialEdit]{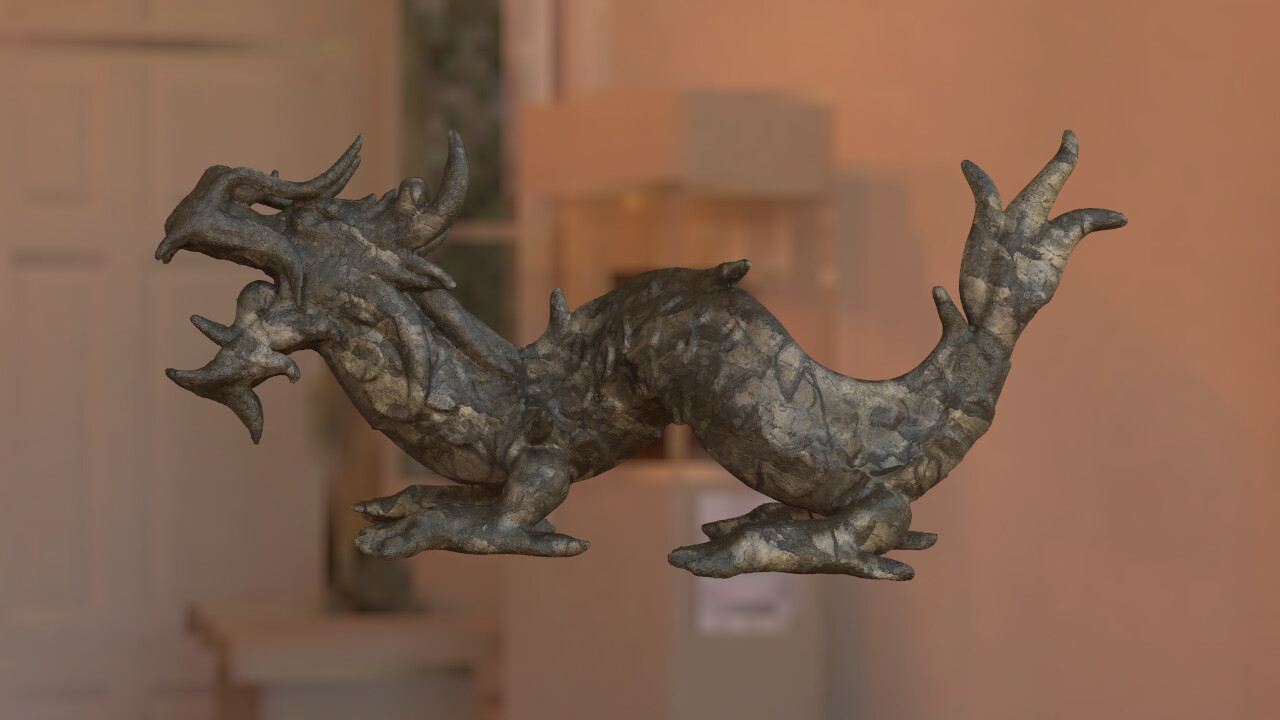}
        &
        \includegraphics[width=\lenMaterialEdit]{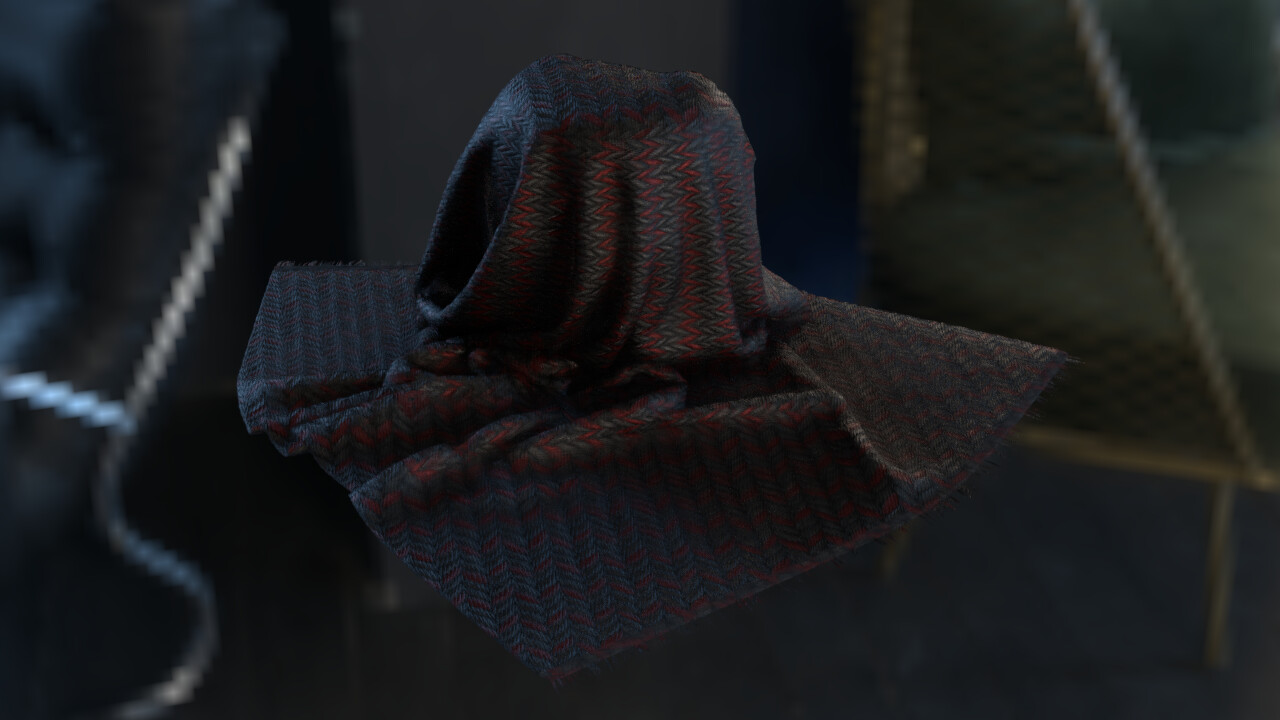}
        &
        \includegraphics[width=\lenMaterialEdit]{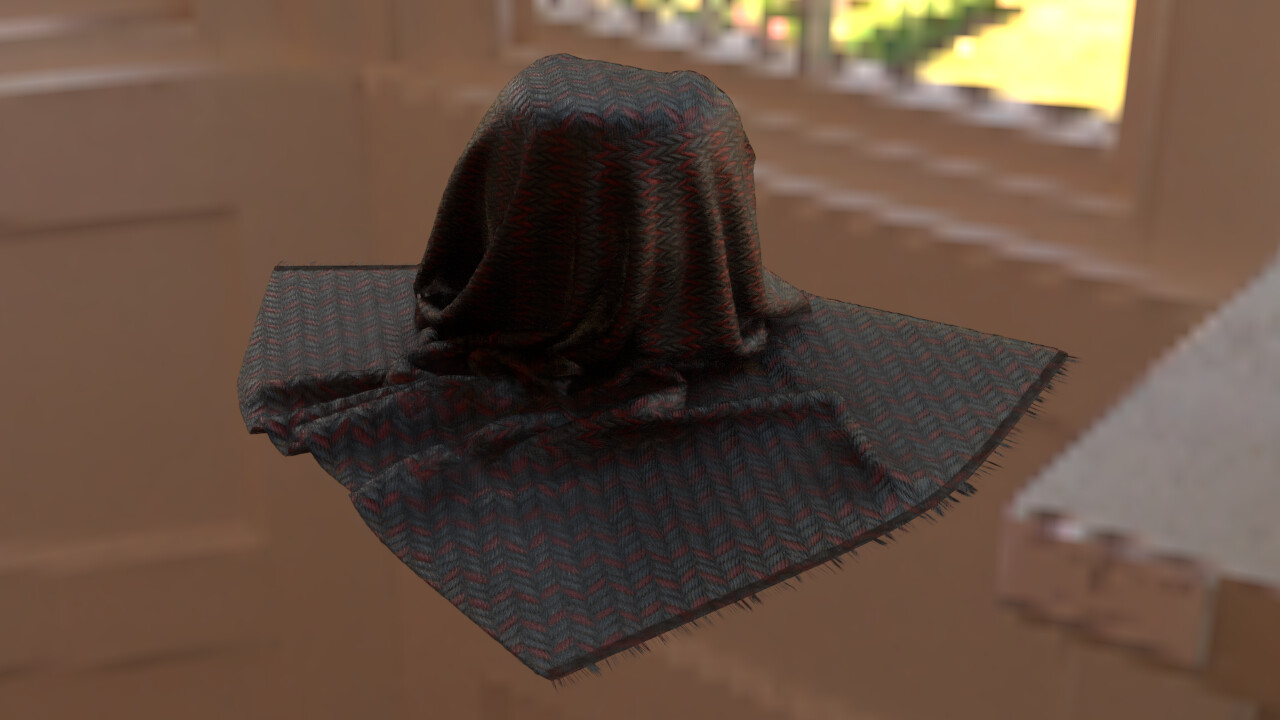}
        \\ 
        \raisebox{15pt}{\rotatebox{90}{\small{\textsf{Procedural}}}}
        &
        \includegraphics[width=\lenMaterialEdit]{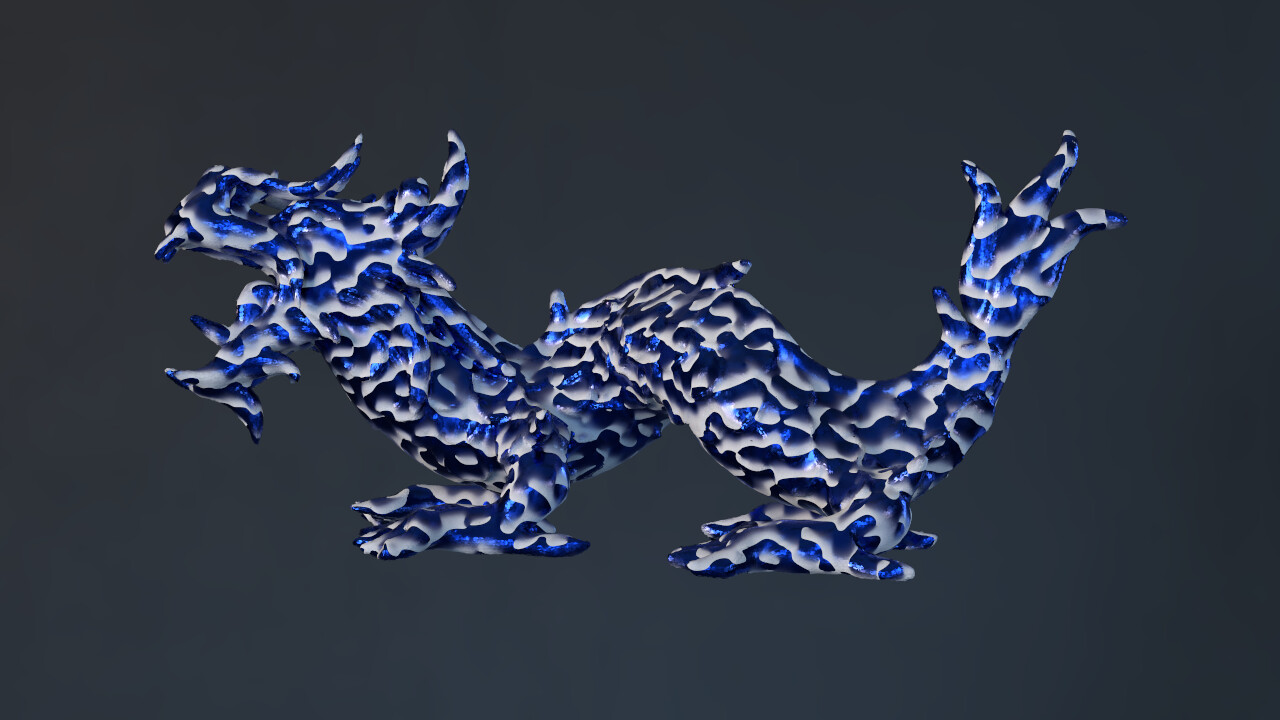}
        &
        \includegraphics[width=\lenMaterialEdit]{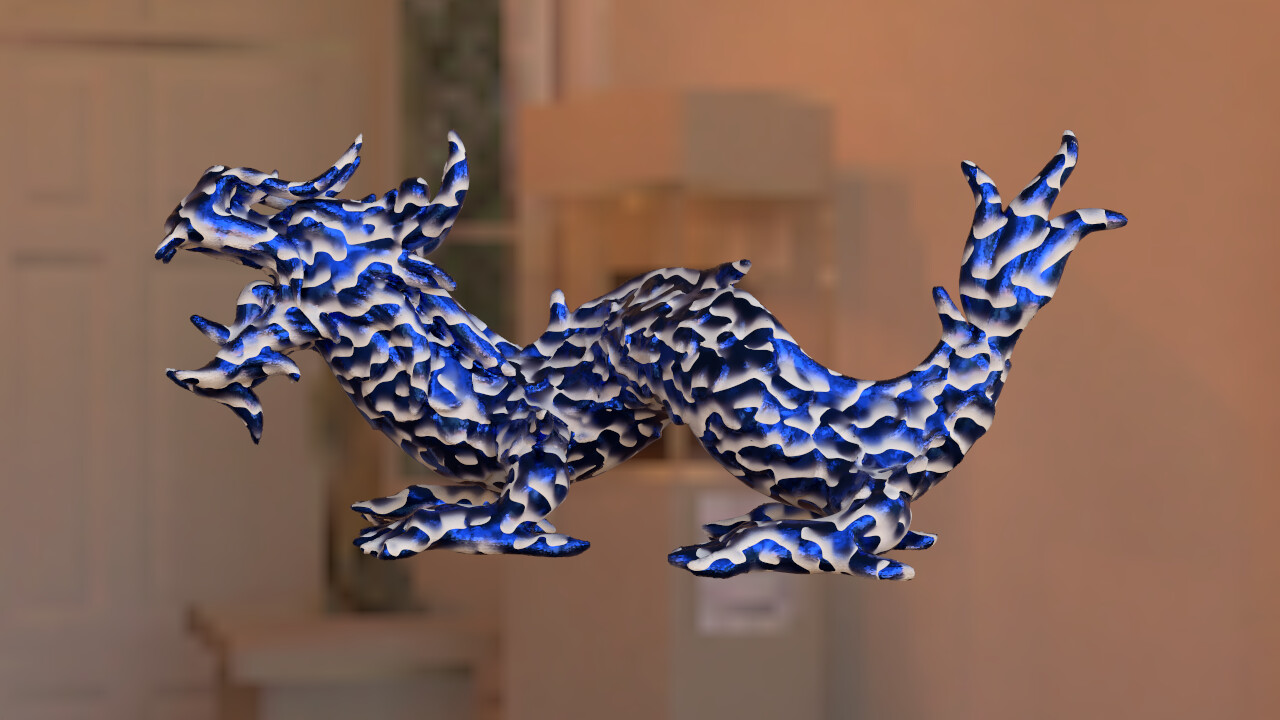}
        &
        \includegraphics[width=\lenMaterialEdit]{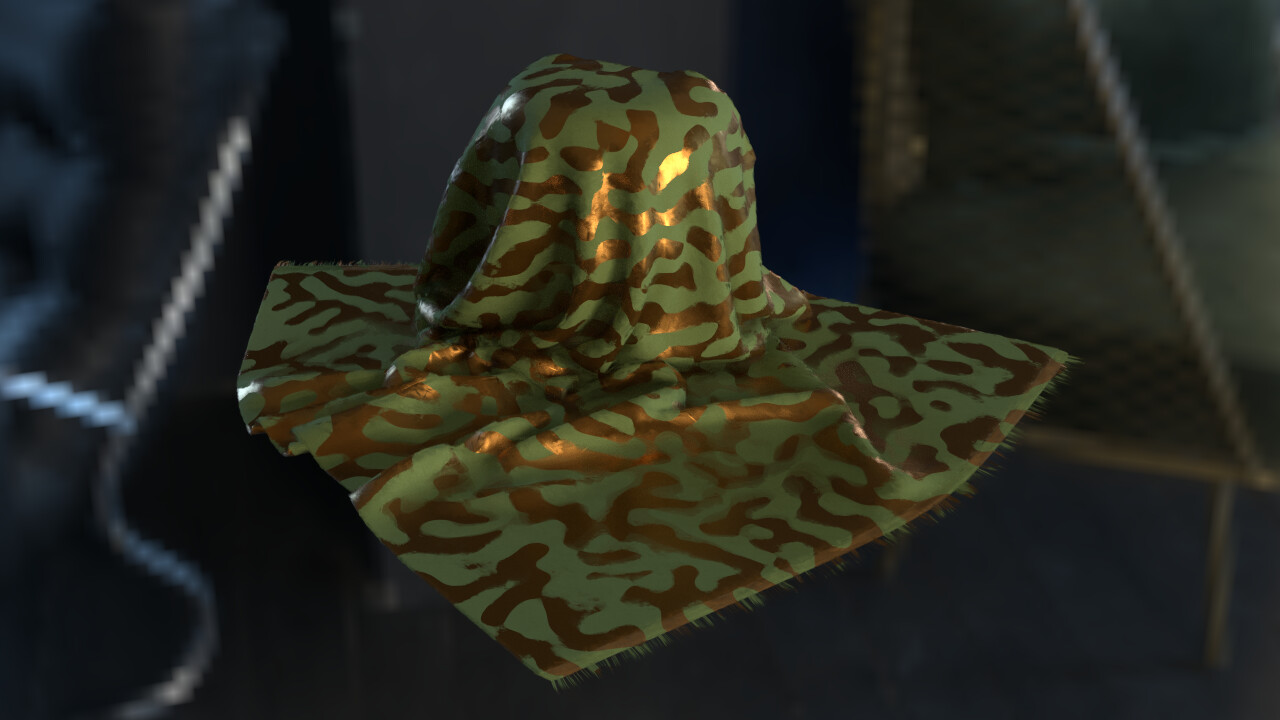}
        &
        \includegraphics[width=\lenMaterialEdit]{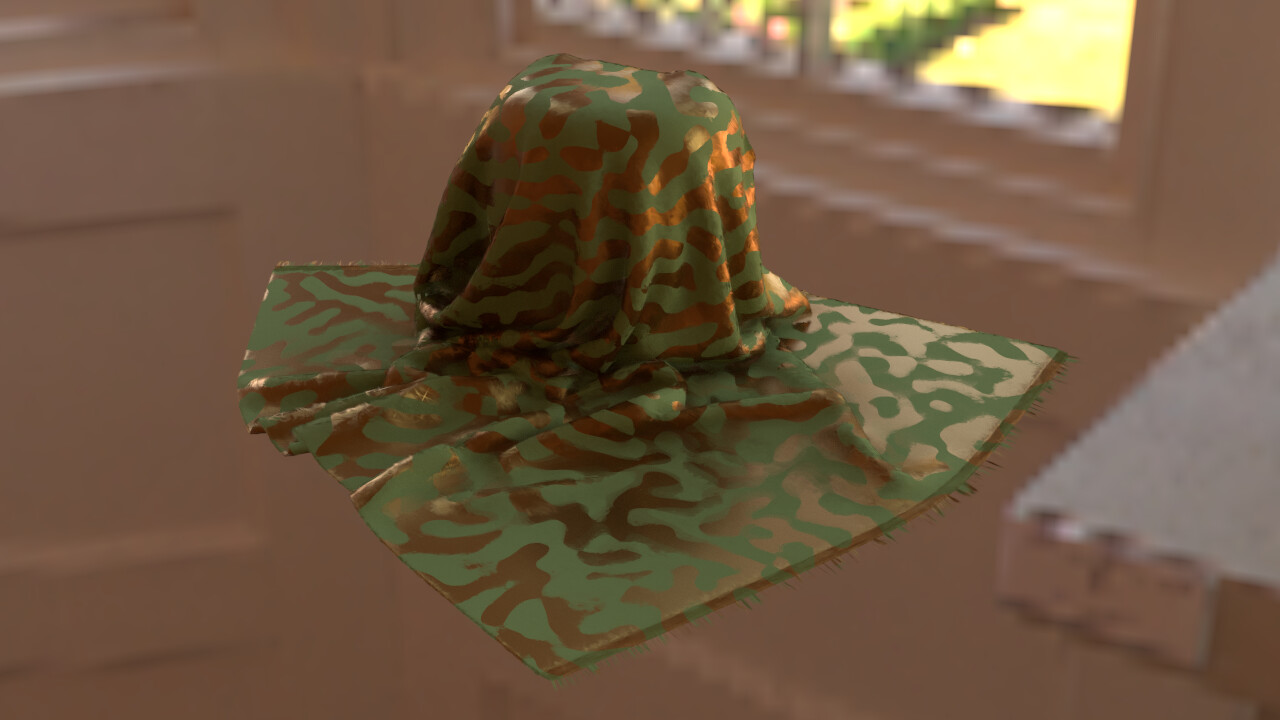}
        \\
        & \small{\textsf{Lighting 1}} & \small{\textsf{Lighting 2}} & \small{\textsf{Lighting 1}} & \small{\textsf{Lighting 2}}                             
    \end{tabular}
    \caption{\label{fig:material_edit}
        Editing the appearance of our models by different techniques. Both the base BSDF and the NDF (second row) can be altered to produce visually 
        interesting effects.
    }
\end{figure*}

\paragraph{Appearance Editing}
One of the goals of this work is to make the Gaussian primitive useful for general 3D content authoring. 
While meshes naturally support UV parameterization, there is generally no well-defined UV parameterization for volumetric representations.
To overcome this limitation and enable intuitive appearance editing, we adopt and extend UV-less texturing techniques.
\autoref{fig:material_edit} demonstrates two example techniques
on different models. The first technique, which we term \emph{extended triplanar mapping}, generalizes the well-known triplanar mapping for surfaces. The
traditional triplanar mapping projects a shading point to three axis-aligned planes, performs texture sampling on the planes, and blends the three samples based
on the surface normal. We can naturally generalize this for our representation by instead blending based on the projected area of the SGGX NDF
$\sigma(\omega) = \sqrt{\omega\trans S \omega}$, where $S$ is the SGGX matrix \citep{heitz2015sggx}. Moreover, we can edit the NDF itself by applying the extended triplanar
mapping with a normal map. This is achieved by defining the blended normal in the coordinate space formed by the SGGX eigenvectors, and rotating the dominant
SGGX eigenvector to it. The second row of \autoref{fig:material_edit} shows the texturing results using the extended triplanar mapping, including both base BSDF
and NDF editing to produce the bumpy effect.

Alternatively, we may apply 3D procedural noises to our representation just like other volumetric representations. In the third row of \autoref{fig:material_edit},
we use procedural phasor noise \citep{tricard2019procedural} to modulate the base color, roughness, and metallic parameters of our models to create patterns.

\begin{figure*}[tb]
	\newlength{\lenLinearVSExp}
	\setlength{\lenLinearVSExp}{0.118\linewidth}
    \addtolength{\tabcolsep}{-4pt}
    \renewcommand{\arraystretch}{0.5}
    \centering
    \begin{tabular}{cccccccc}
        \multicolumn{2}{c}{\small{\emph{Checkerboards} (5.9K Tri.)}} & 
        \multicolumn{2}{c}{\small{\emph{Office Chair} (297K Tri.)}} & 
        \multicolumn{2}{c}{\small{\emph{Plant} (1.99M Tri.)}} & 
        \multicolumn{2}{c}{\small{\emph{Color Tree} (151K Tri.)}}
        \\ [0.3ex]
        \multicolumn{8}{c}{\small{\textsf{Original}}}
        \\ [0.3ex]
        \multicolumn{2}{l}{\frame{\includegraphics[width=2.06\lenLinearVSExp]{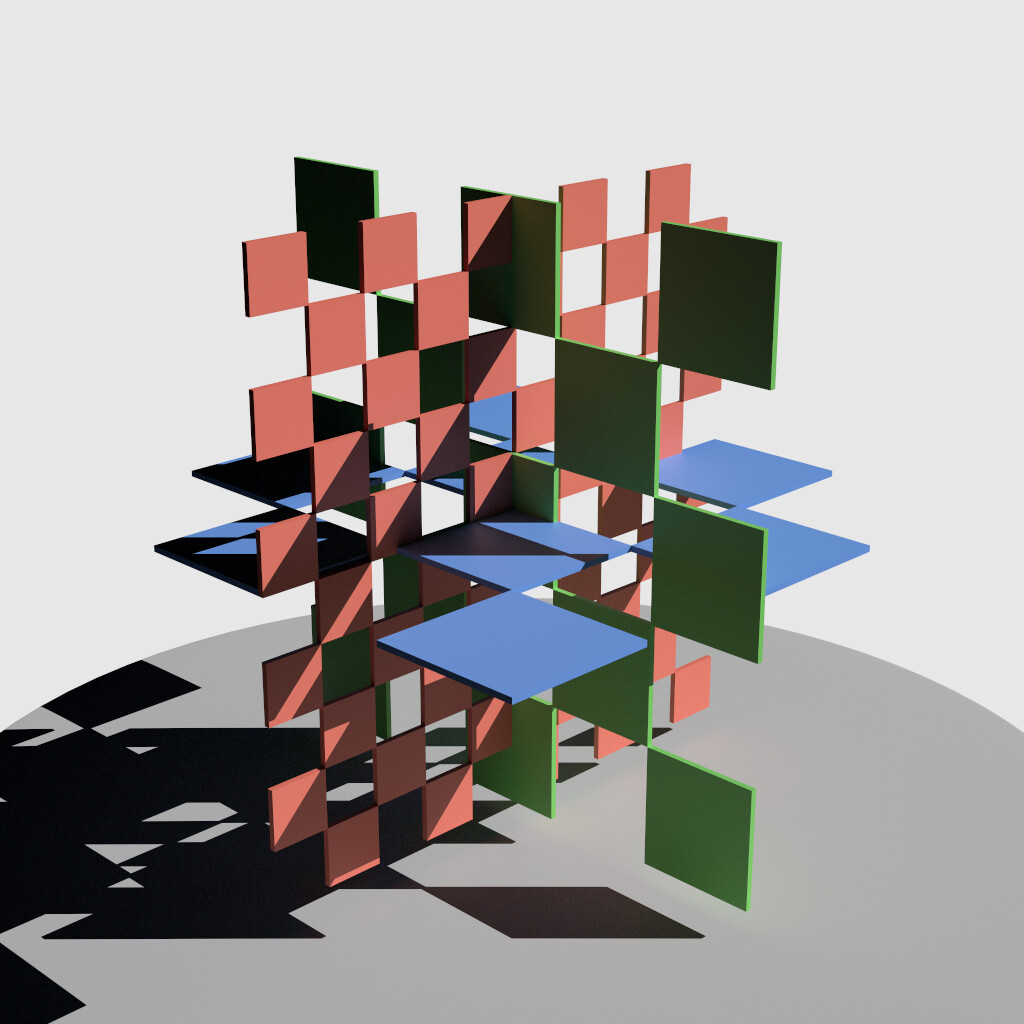}}}
        &
        \multicolumn{2}{l}{\frame{\includegraphics[width=2.06\lenLinearVSExp]{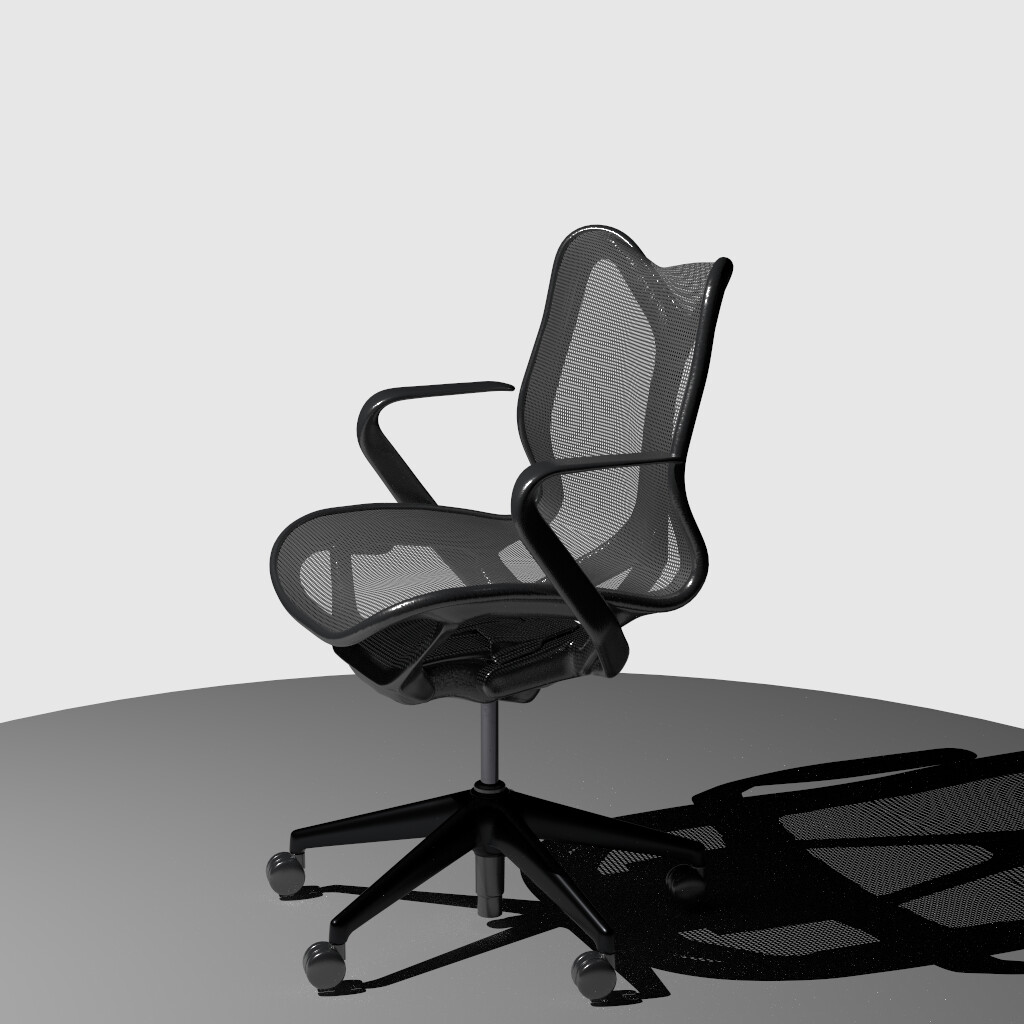}}}
        &
        \multicolumn{2}{l}{\frame{\includegraphics[width=2.06\lenLinearVSExp]{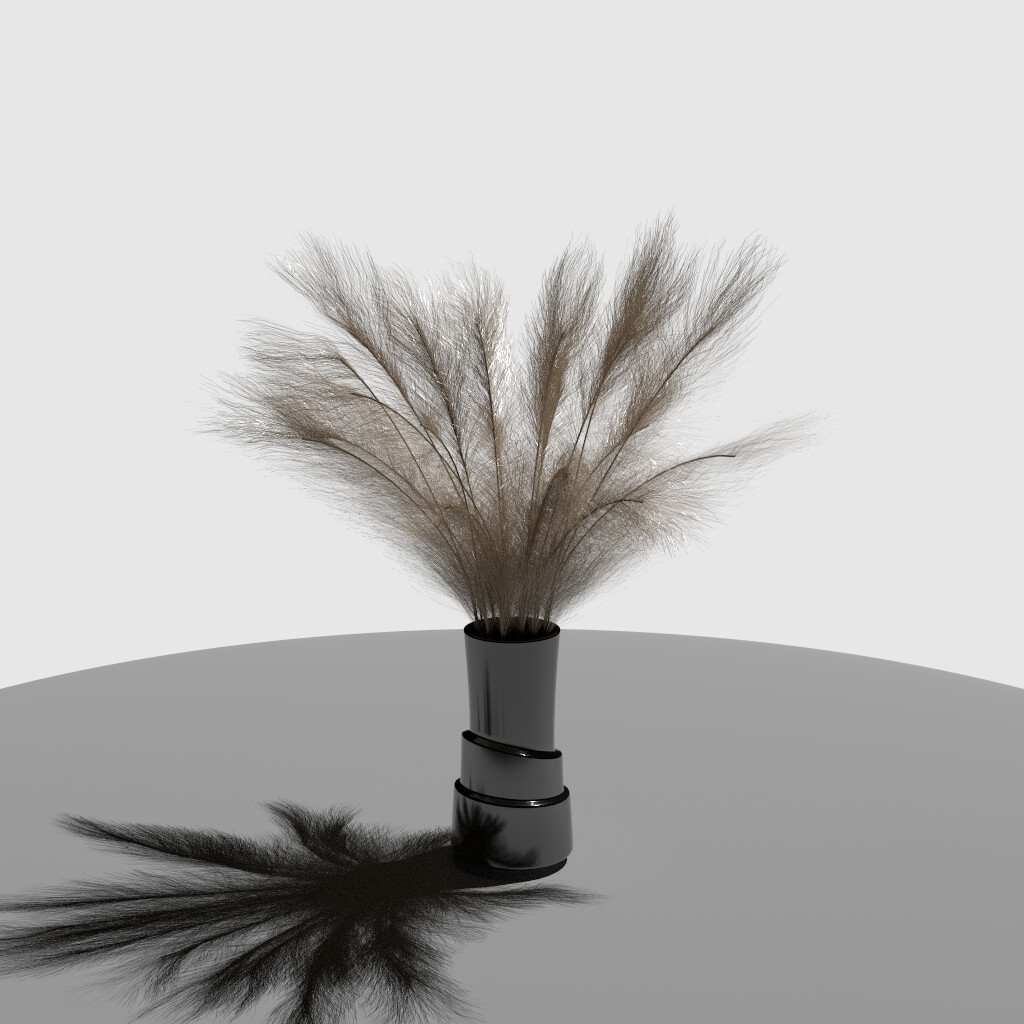}}}
        &
        \multicolumn{2}{l}{\frame{\includegraphics[width=2.06\lenLinearVSExp]{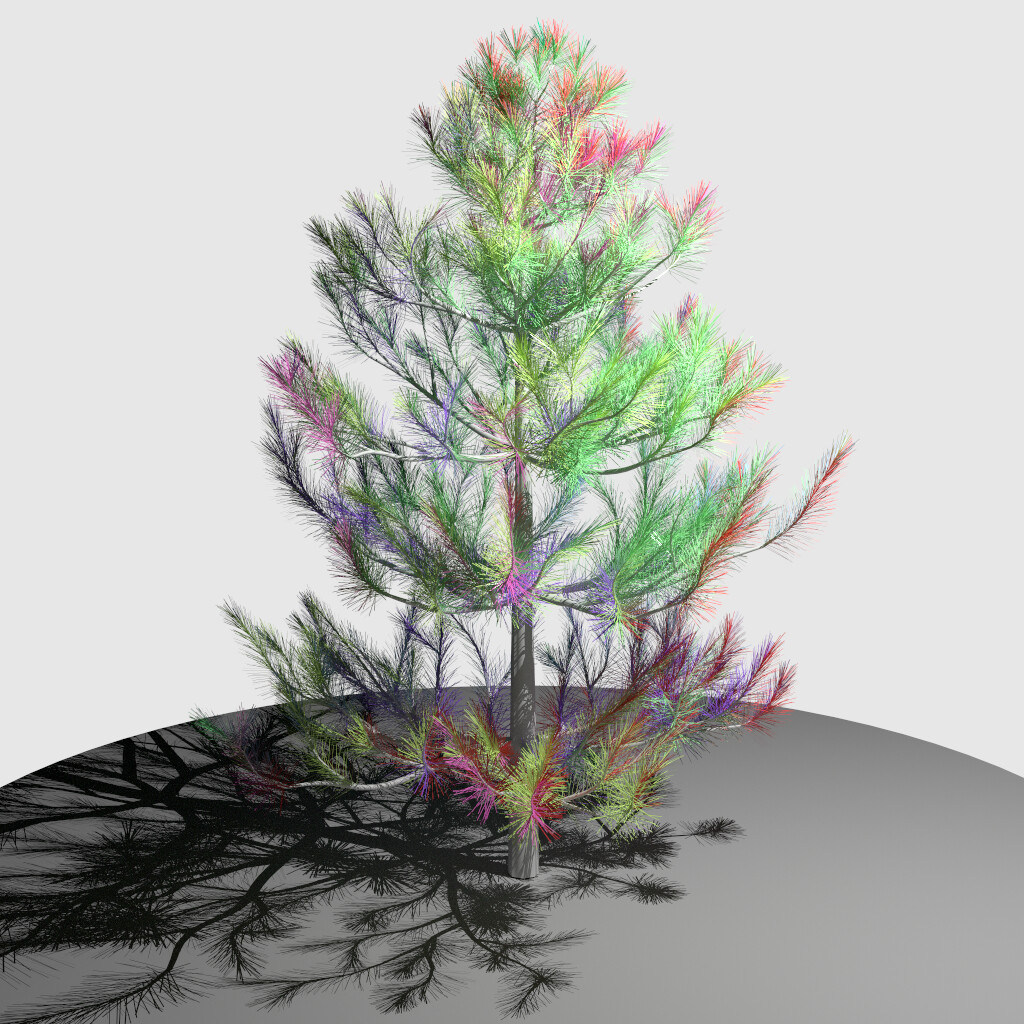}}}
        \\
        \small{\textsf{Linear 5K}} & \small{\textsf{PSNR: \textbf{28.31}}} &
        \small{\textsf{Linear 50K}} & \small{\textsf{PSNR: \textbf{30.05}}} &
        \small{\textsf{Linear 10K}} & \small{\textsf{PSNR: \textbf{26.80}}} &
        \small{\textsf{Linear 20K}} & \small{\textsf{PSNR: \textbf{21.28}}}
        \\
        \frame{\includegraphics[width=\lenLinearVSExp]{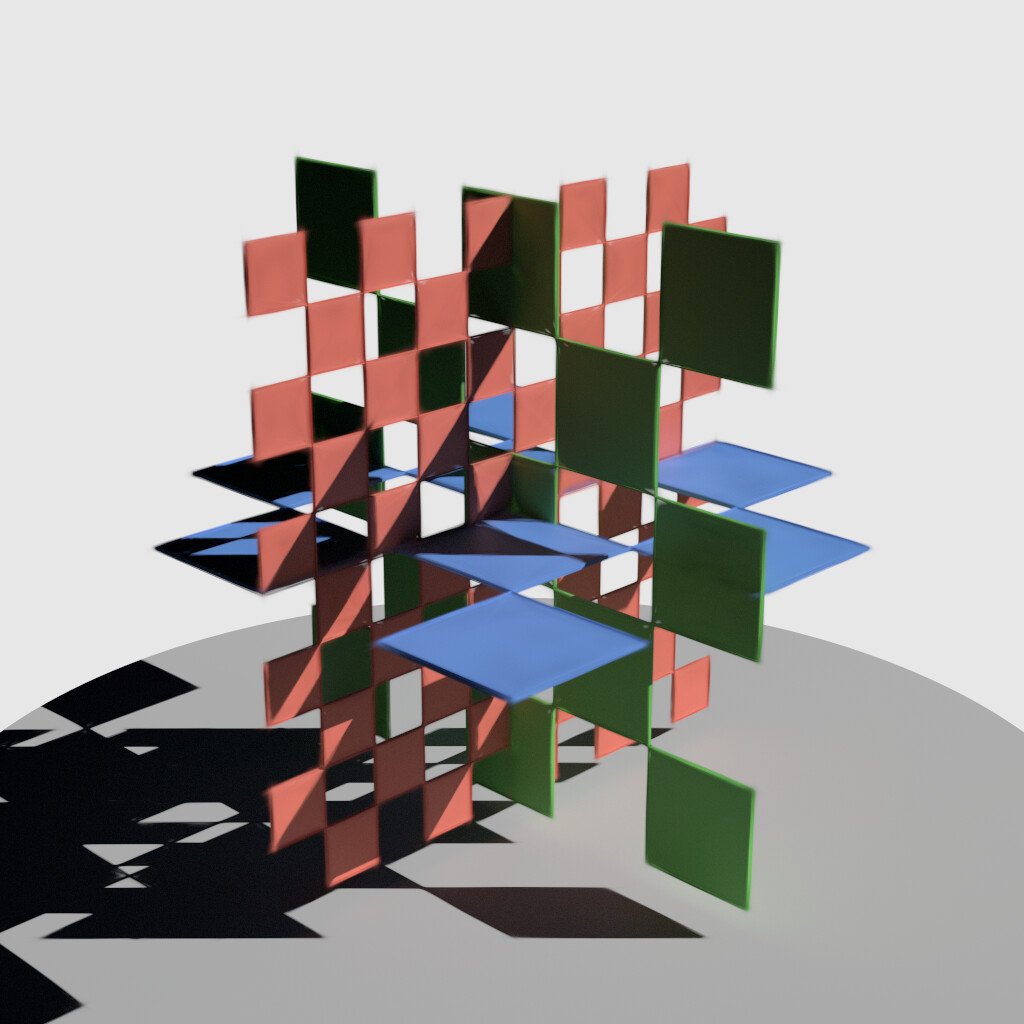}}
        &
        \frame{\includegraphics[width=\lenLinearVSExp]{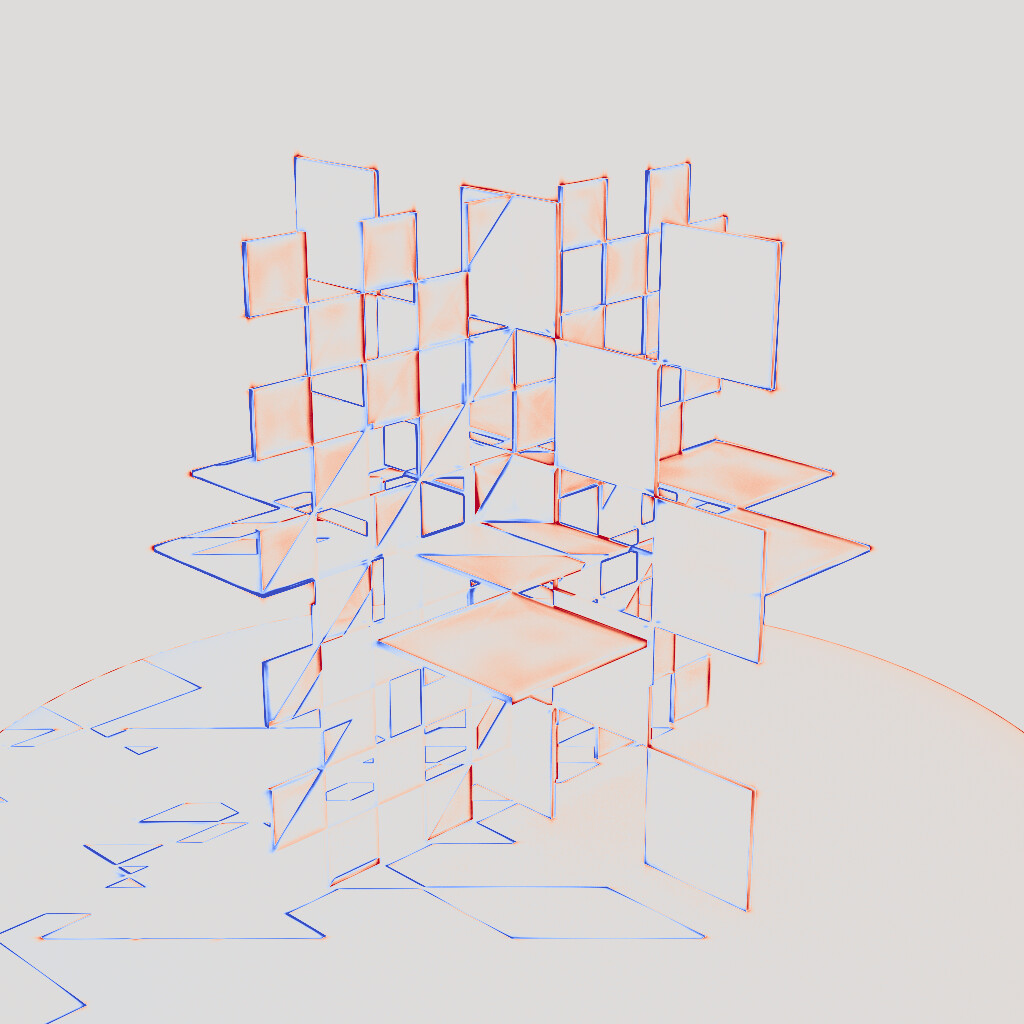}}
        &
        \frame{\includegraphics[width=\lenLinearVSExp]{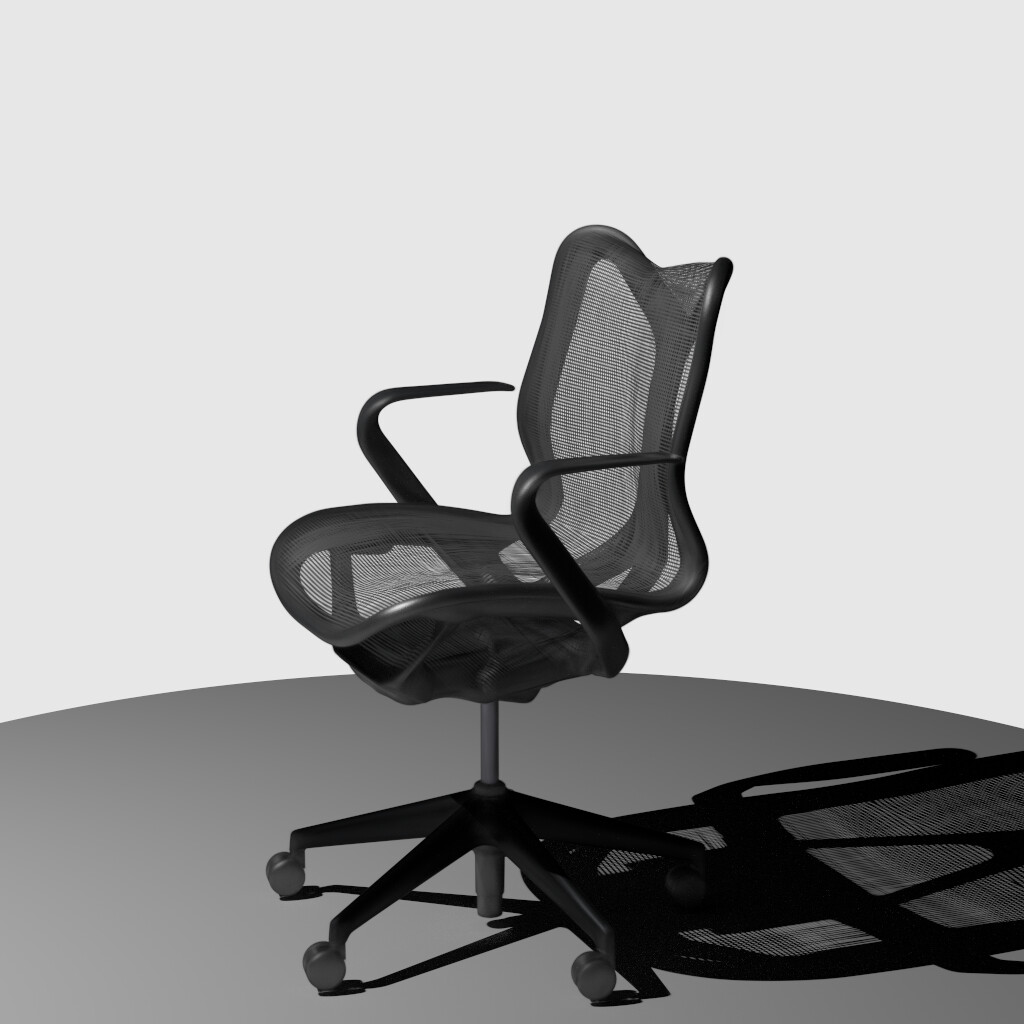}}
        &
        \frame{\includegraphics[width=\lenLinearVSExp]{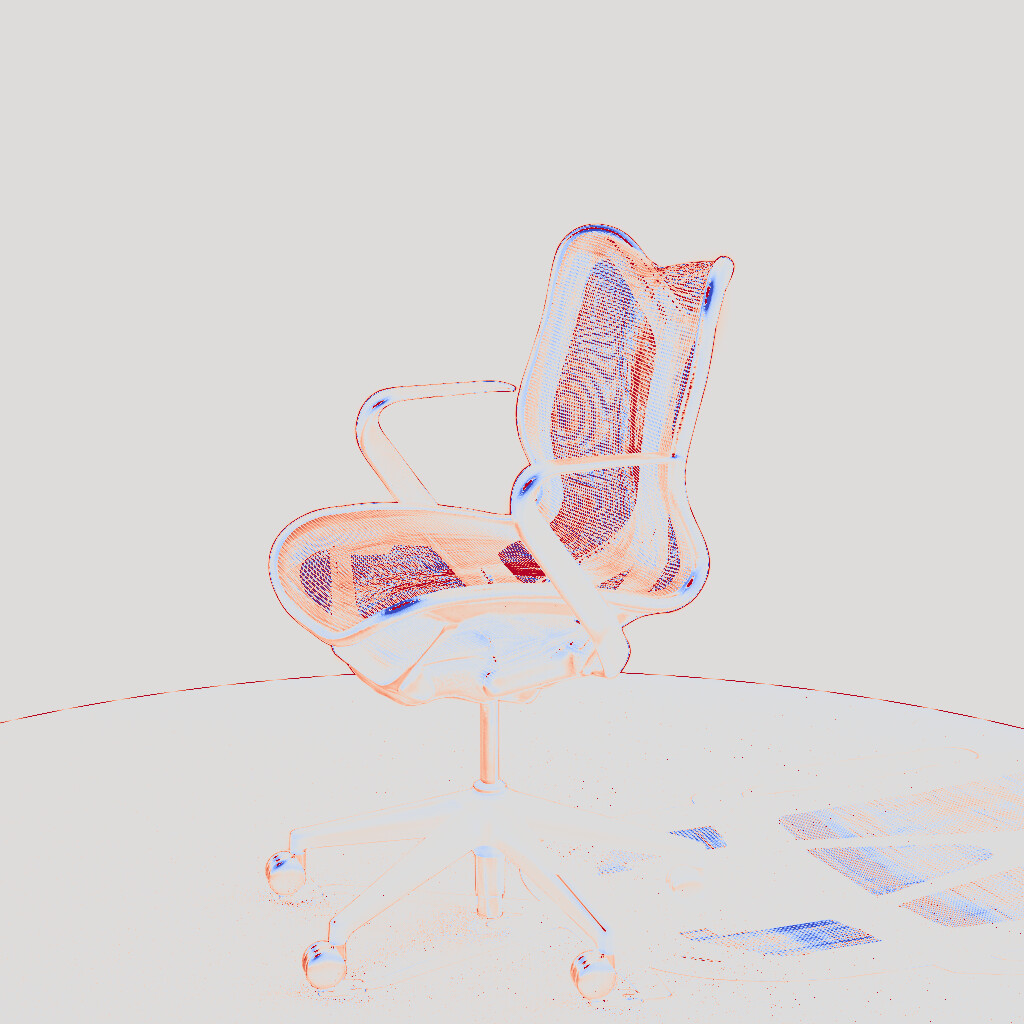}}
        &
        \frame{\includegraphics[width=\lenLinearVSExp]{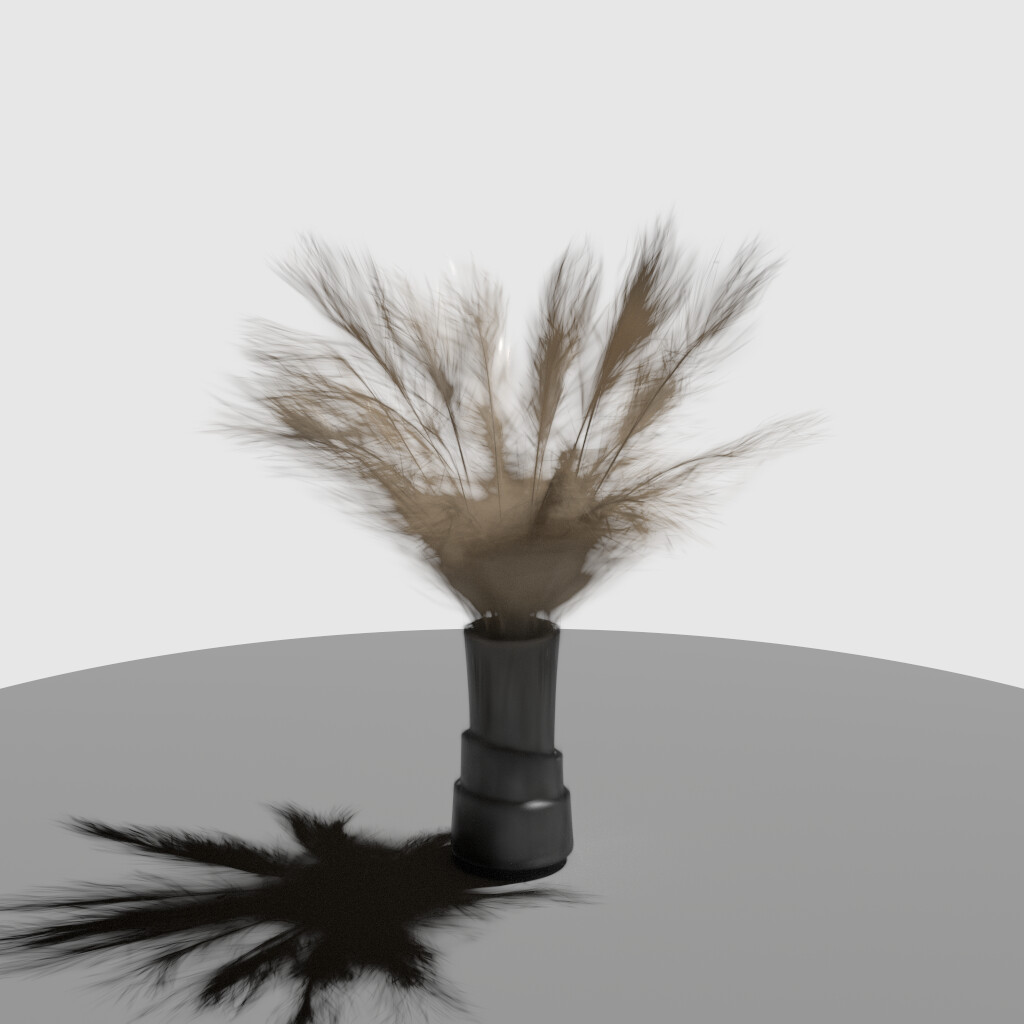}}
        &
        \frame{\includegraphics[width=\lenLinearVSExp]{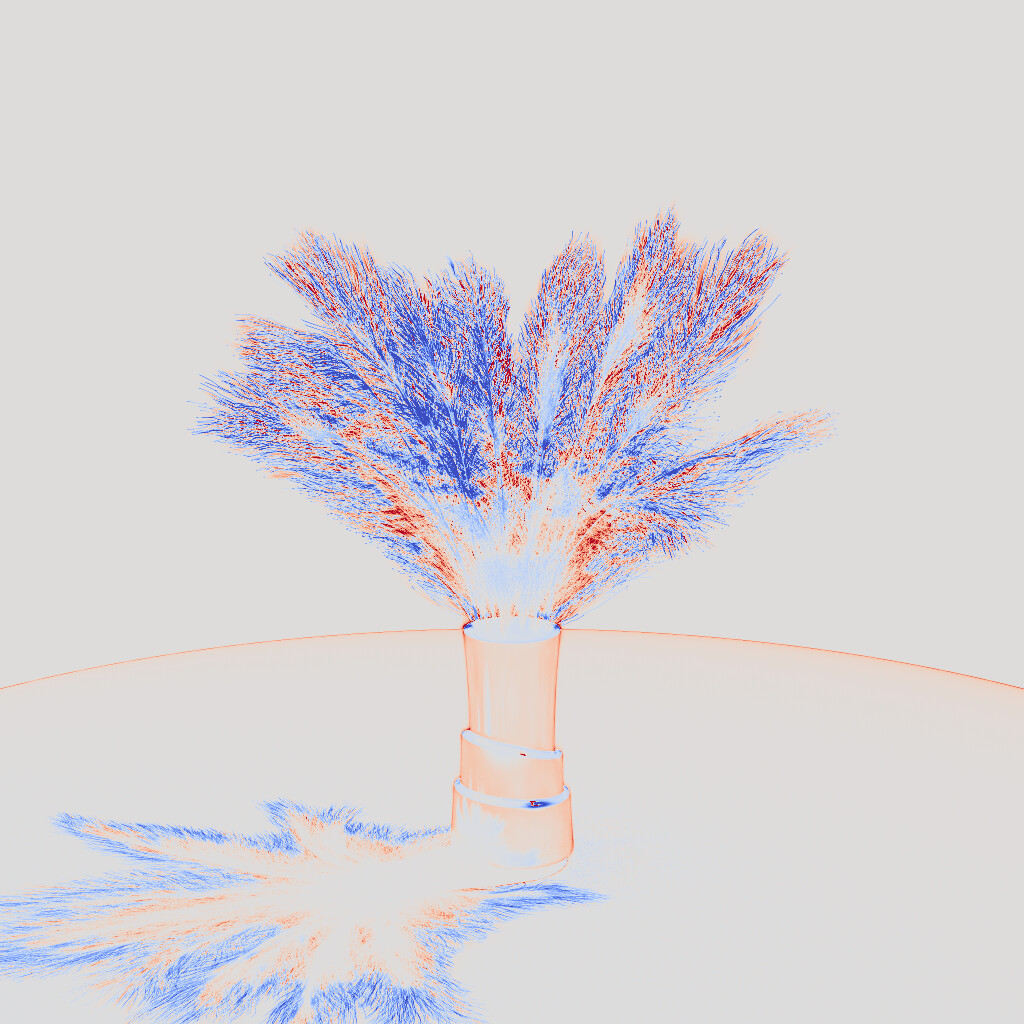}}
        &
        \frame{\includegraphics[width=\lenLinearVSExp]{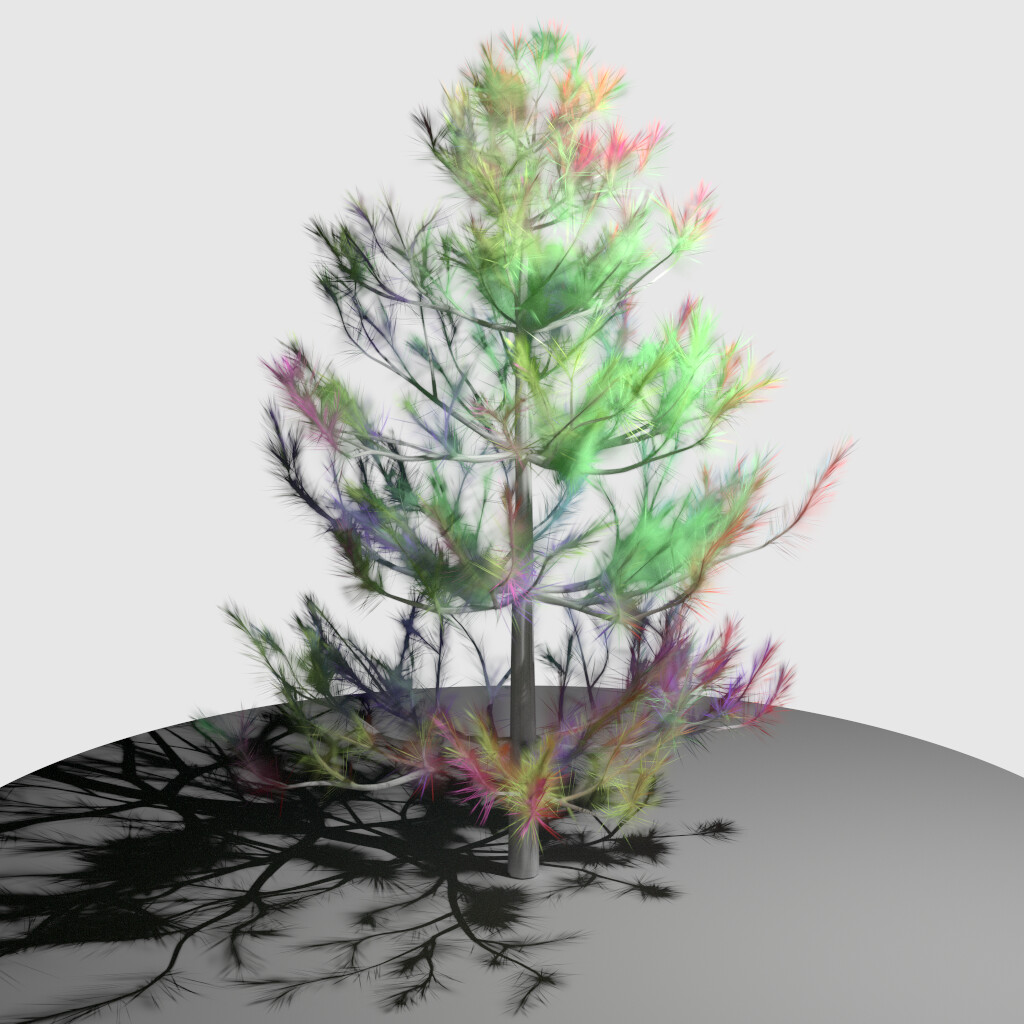}}
        &
        \frame{\includegraphics[width=\lenLinearVSExp]{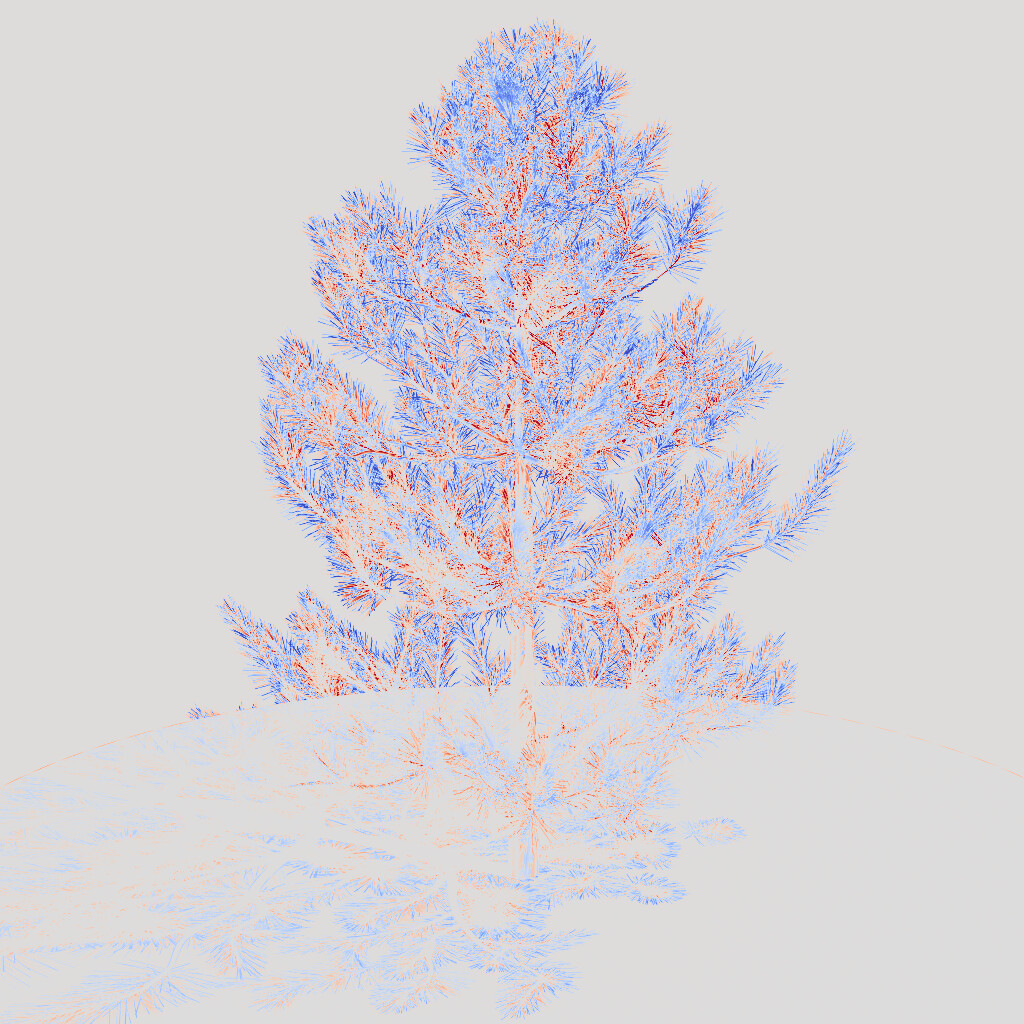}}
        \\
        \small{\textsf{Exponential 5K}} & \small{\textsf{PSNR: 25.10}} &
        \small{\textsf{Exponential 50K}} & \small{\textsf{PSNR: 28.03}} &
        \small{\textsf{Exponential 10K}} & \small{\textsf{PSNR: 26.59}} &
        \small{\textsf{Exponential 20K}} & \small{\textsf{PSNR: 21.10}}
        \\
        \frame{\includegraphics[width=\lenLinearVSExp]{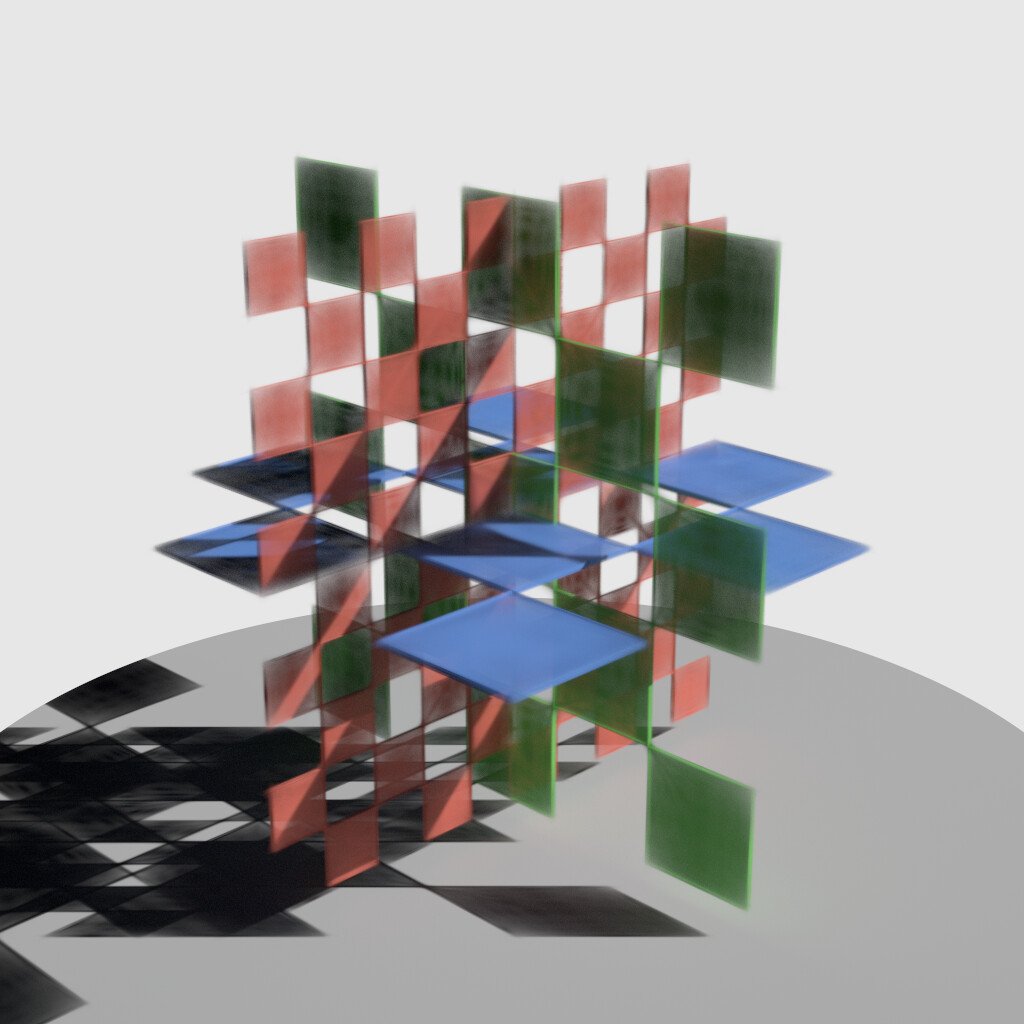}}
        &
        \frame{\includegraphics[width=\lenLinearVSExp]{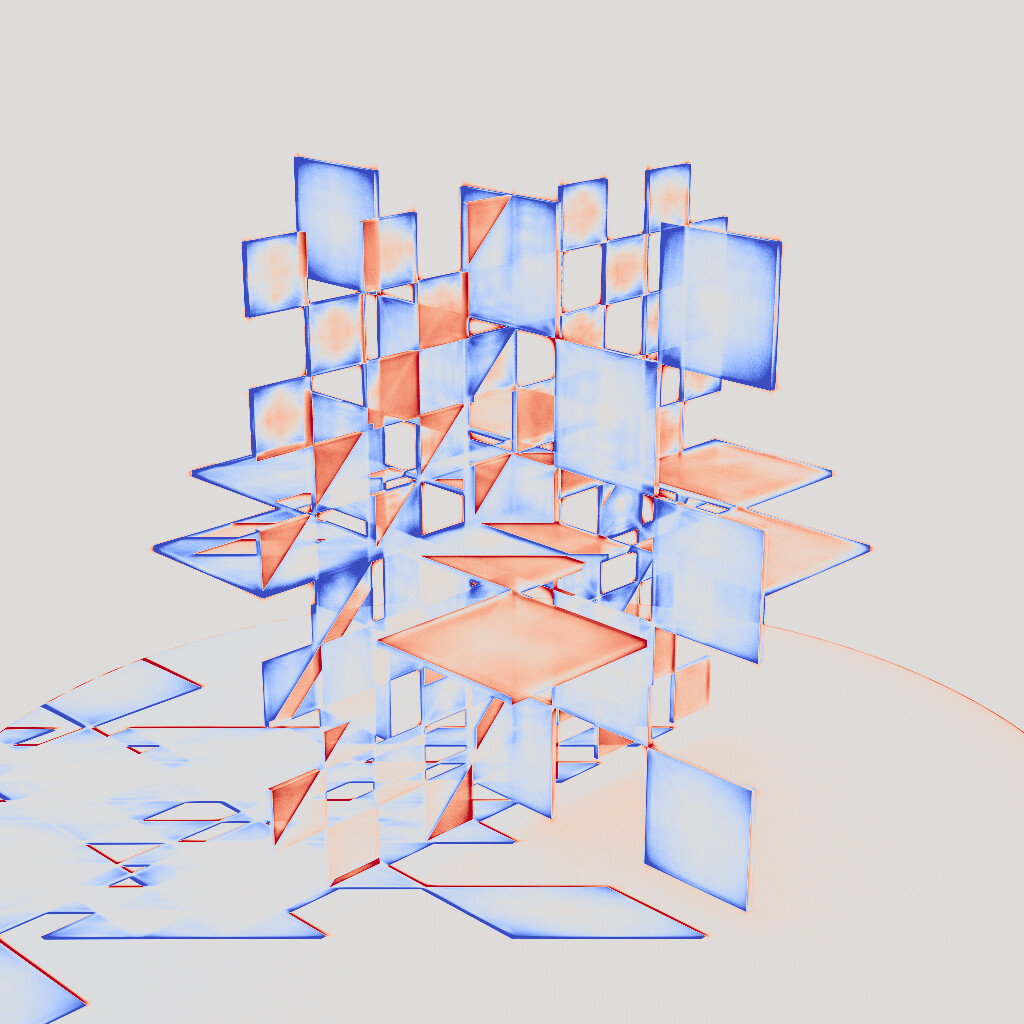}}
        &
        \frame{\includegraphics[width=\lenLinearVSExp]{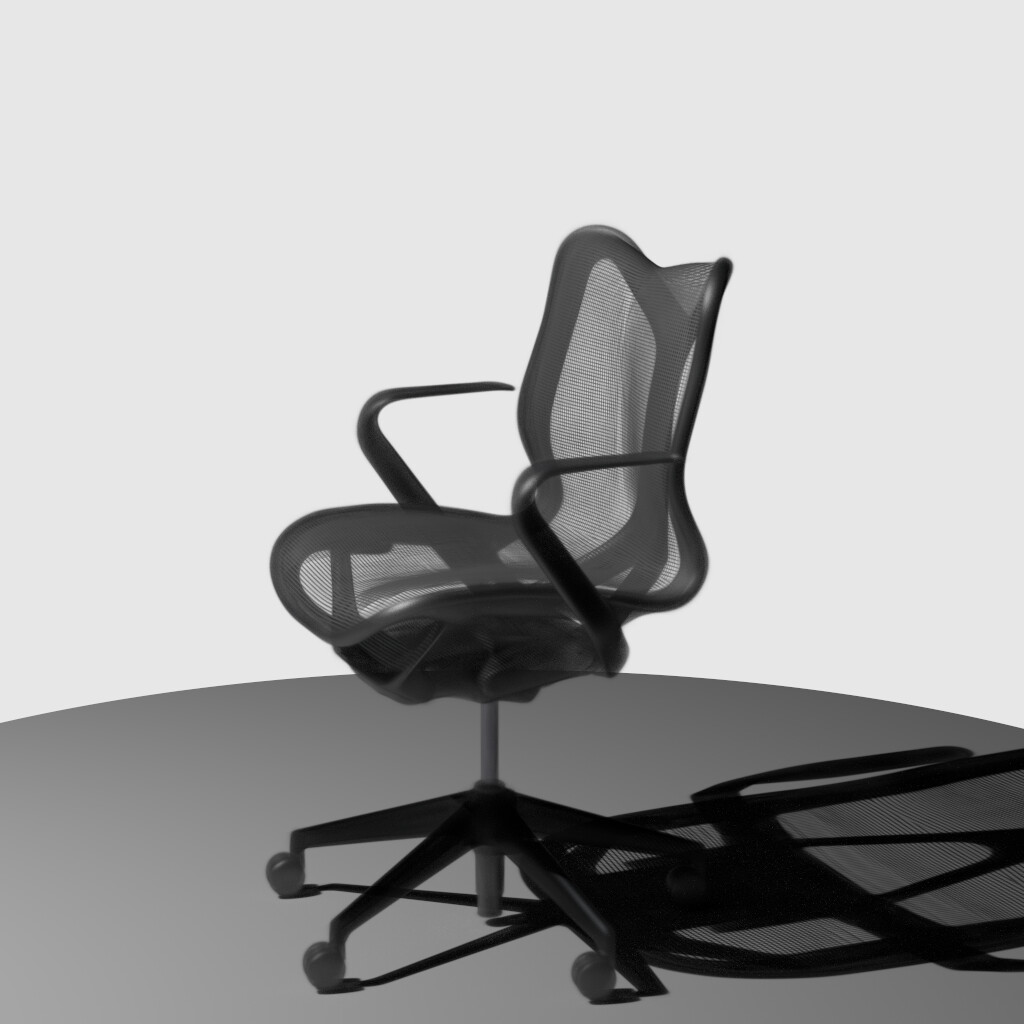}}
        &
        \frame{\includegraphics[width=\lenLinearVSExp]{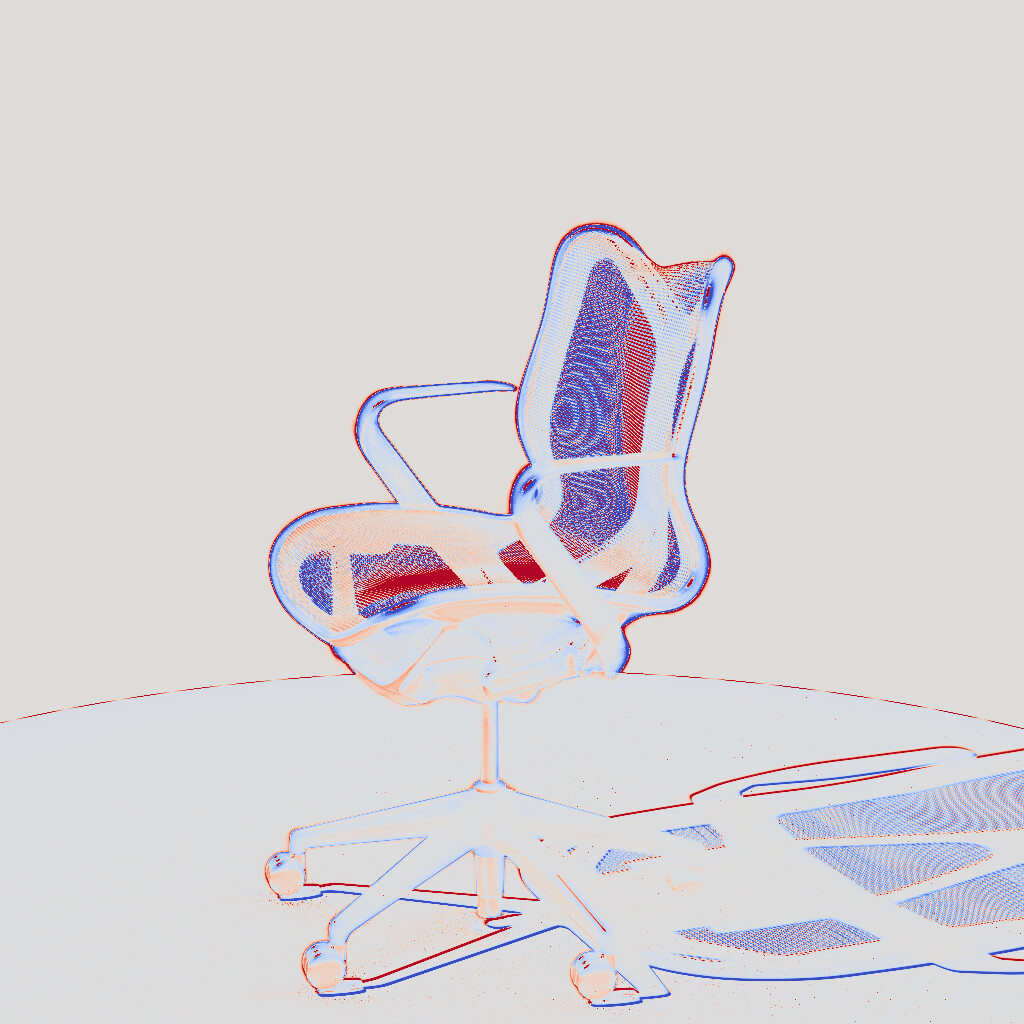}}
        &
        \frame{\includegraphics[width=\lenLinearVSExp]{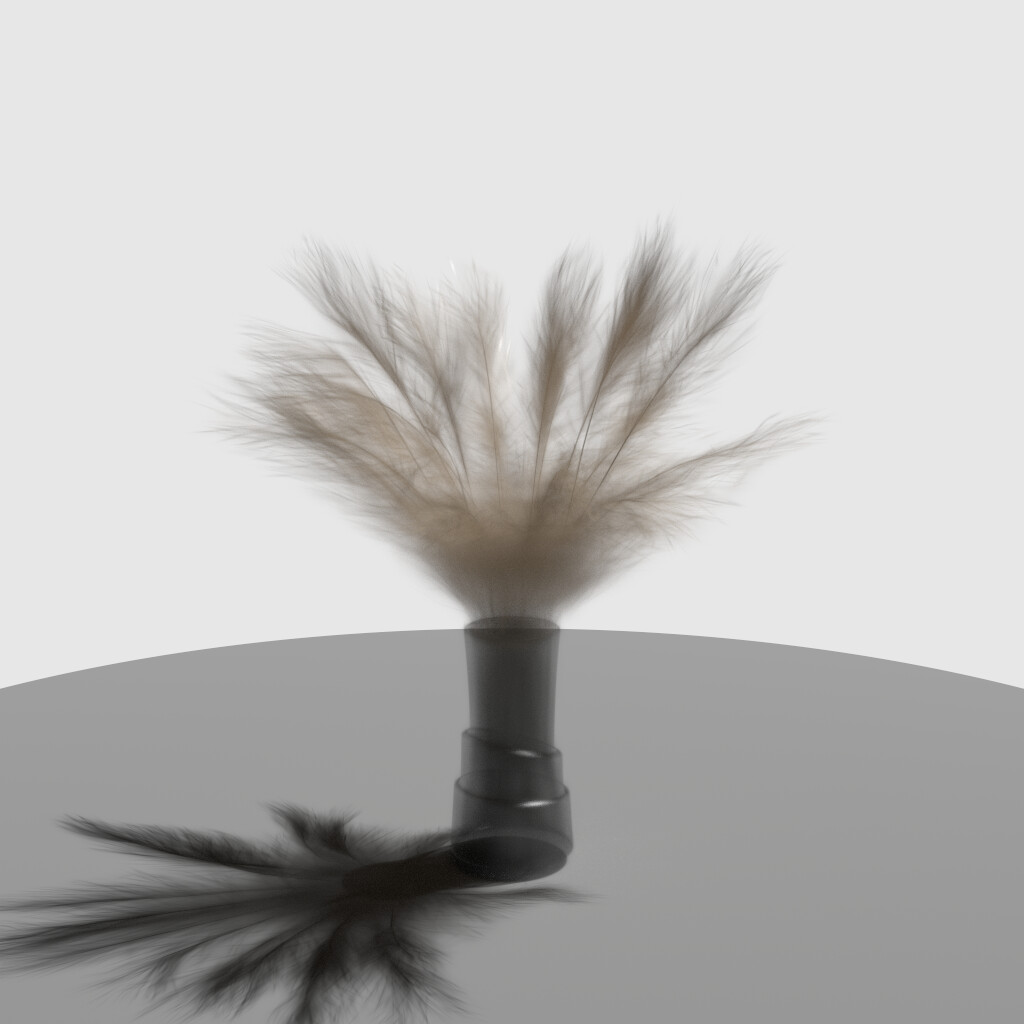}}
        &
        \frame{\includegraphics[width=\lenLinearVSExp]{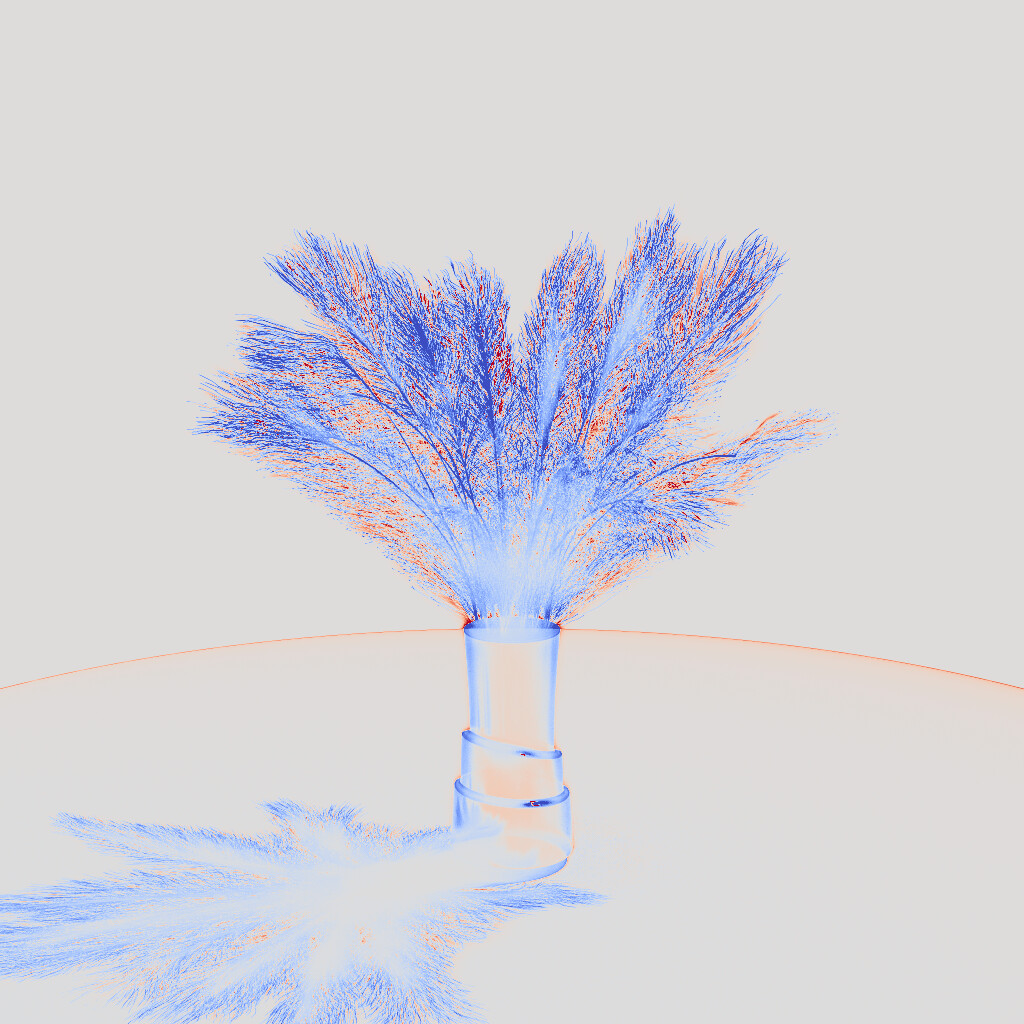}}
        &
        \frame{\includegraphics[width=\lenLinearVSExp]{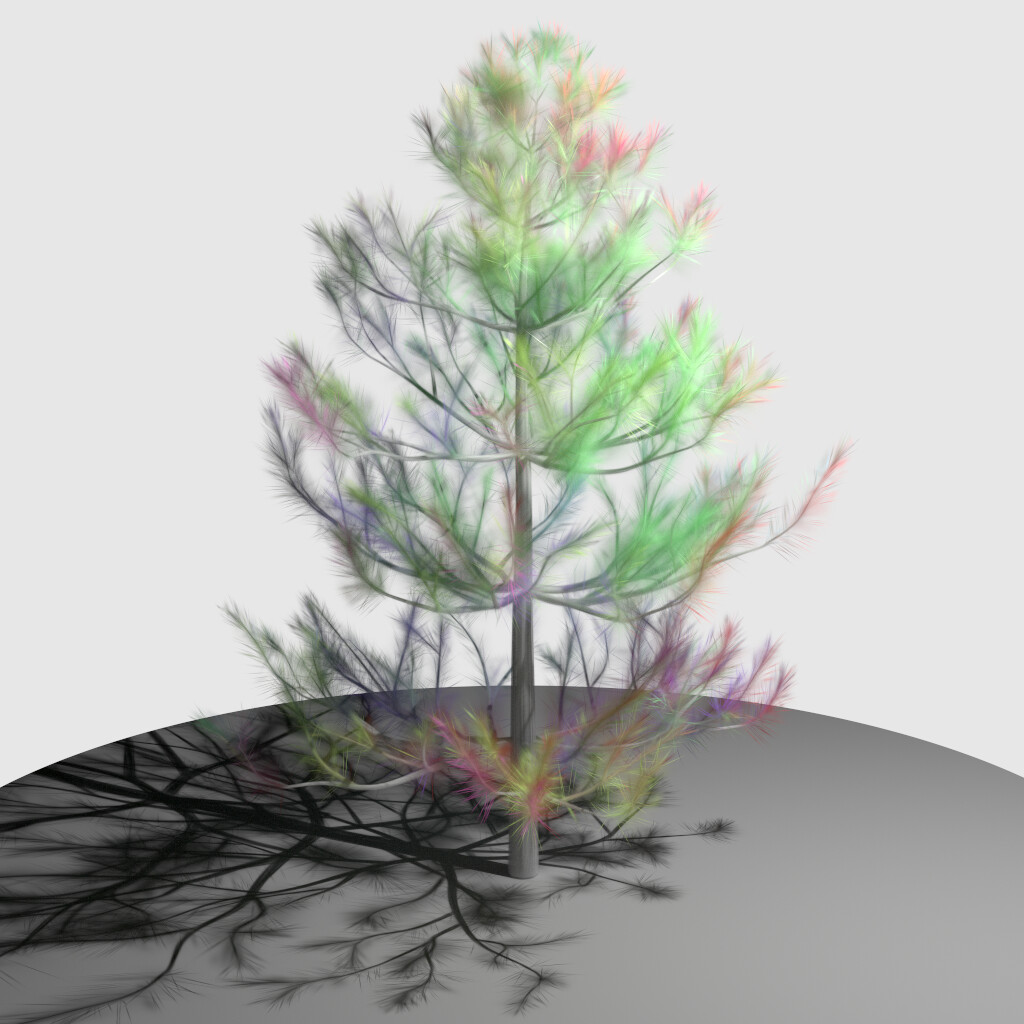}}
        &
        \frame{\includegraphics[width=\lenLinearVSExp]{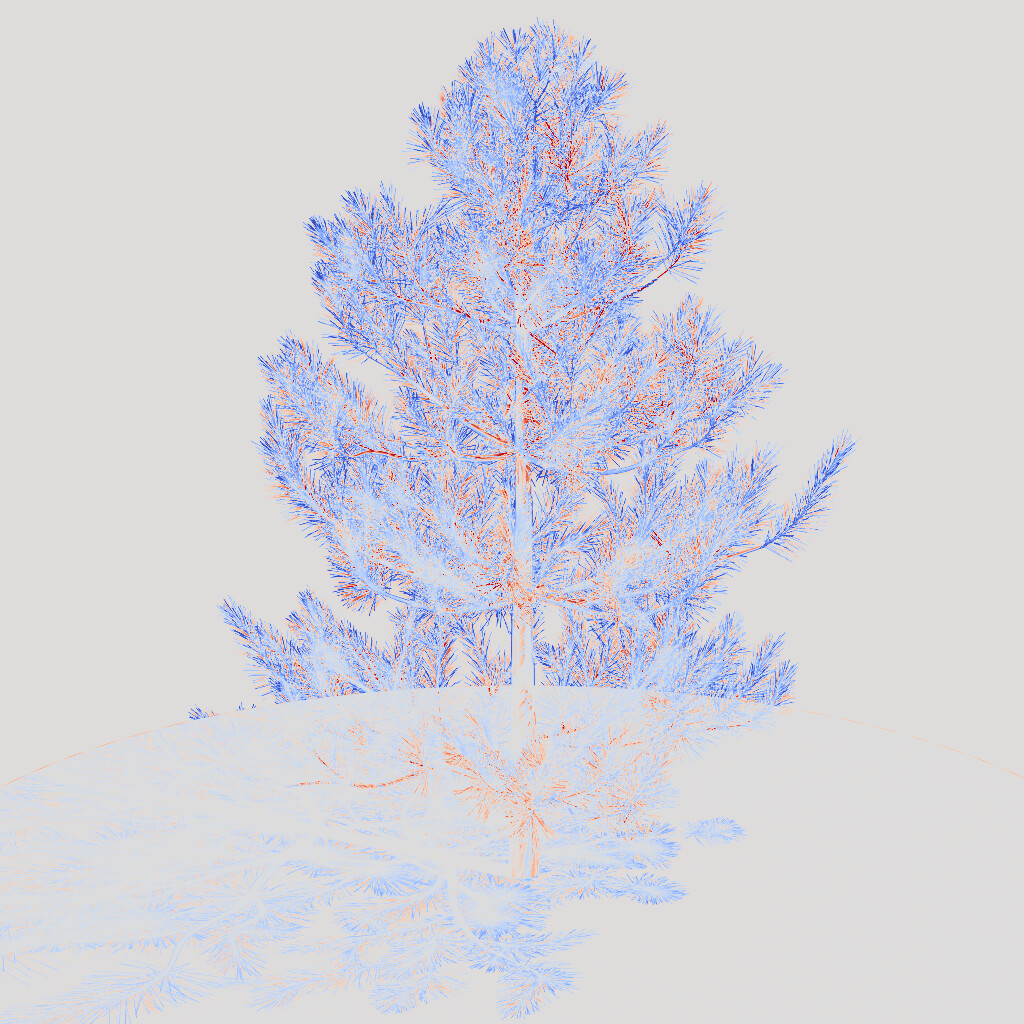}}
        \\
        \small{\textsf{Linear 25K}} & \small{\textsf{PSNR: \textbf{29.88}}} &
        \small{\textsf{Linear 250K}} & \small{\textsf{PSNR: \textbf{30.44}}} &
        \small{\textsf{Linear 50K}} & \small{\textsf{PSNR: 26.86}} &
        \small{\textsf{Linear 100K}} & \small{\textsf{PSNR: \textbf{22.95}}}
        \\
        \frame{\includegraphics[width=\lenLinearVSExp]{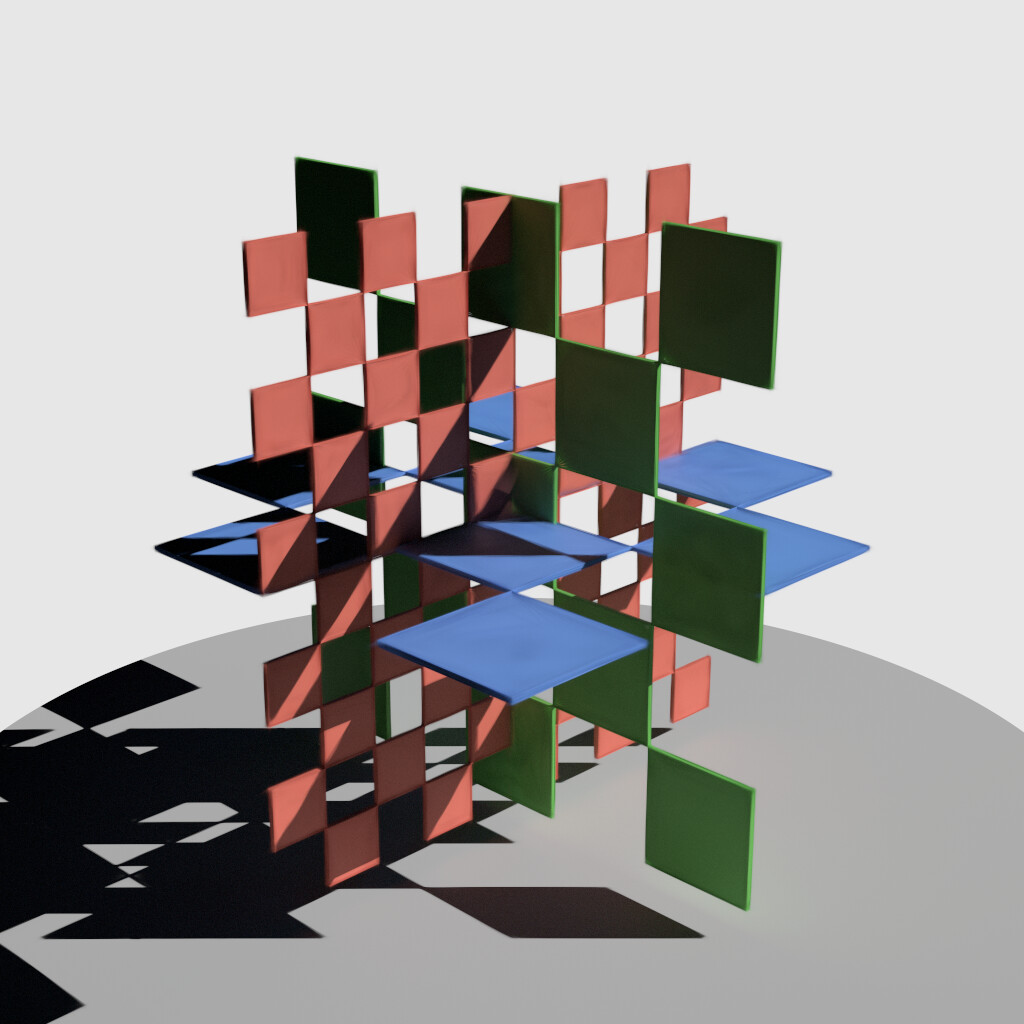}}
        &
        \frame{\includegraphics[width=\lenLinearVSExp]{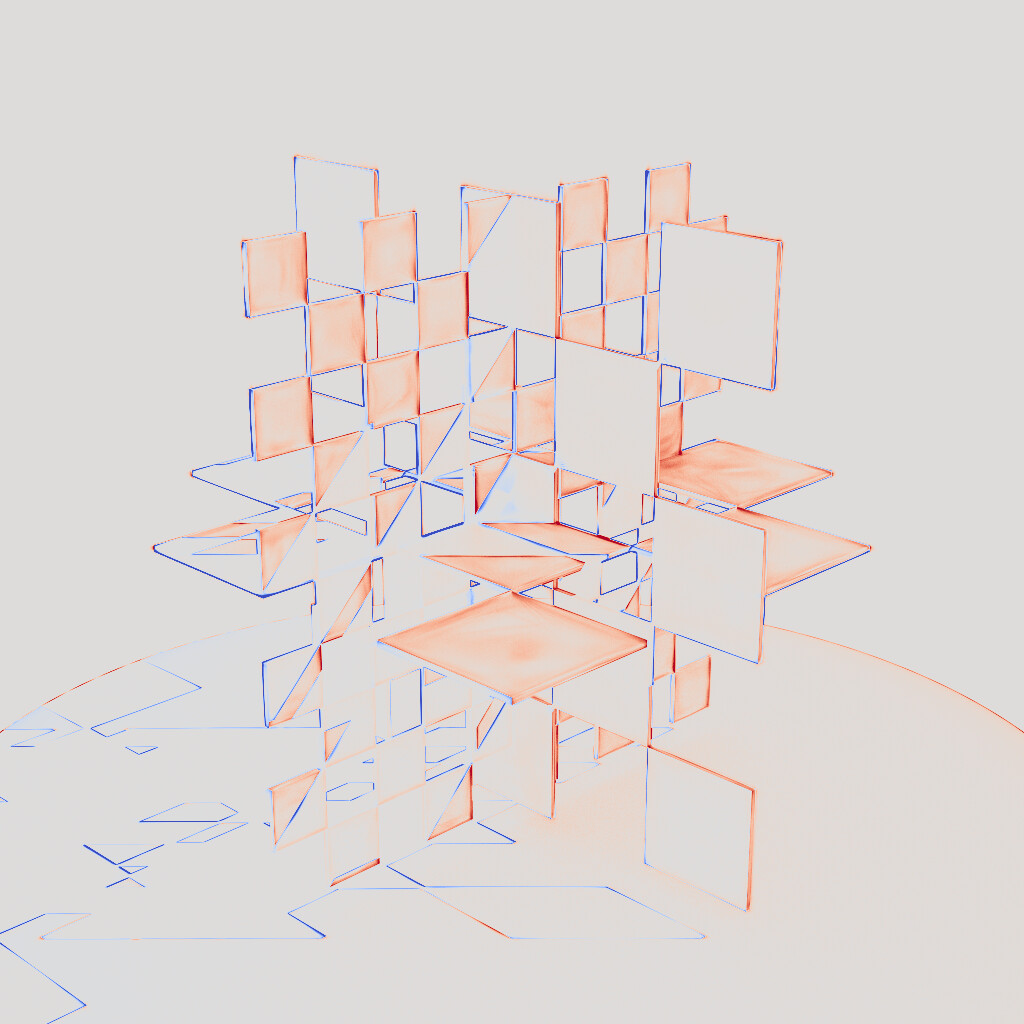}}
        &
        \frame{\includegraphics[width=\lenLinearVSExp]{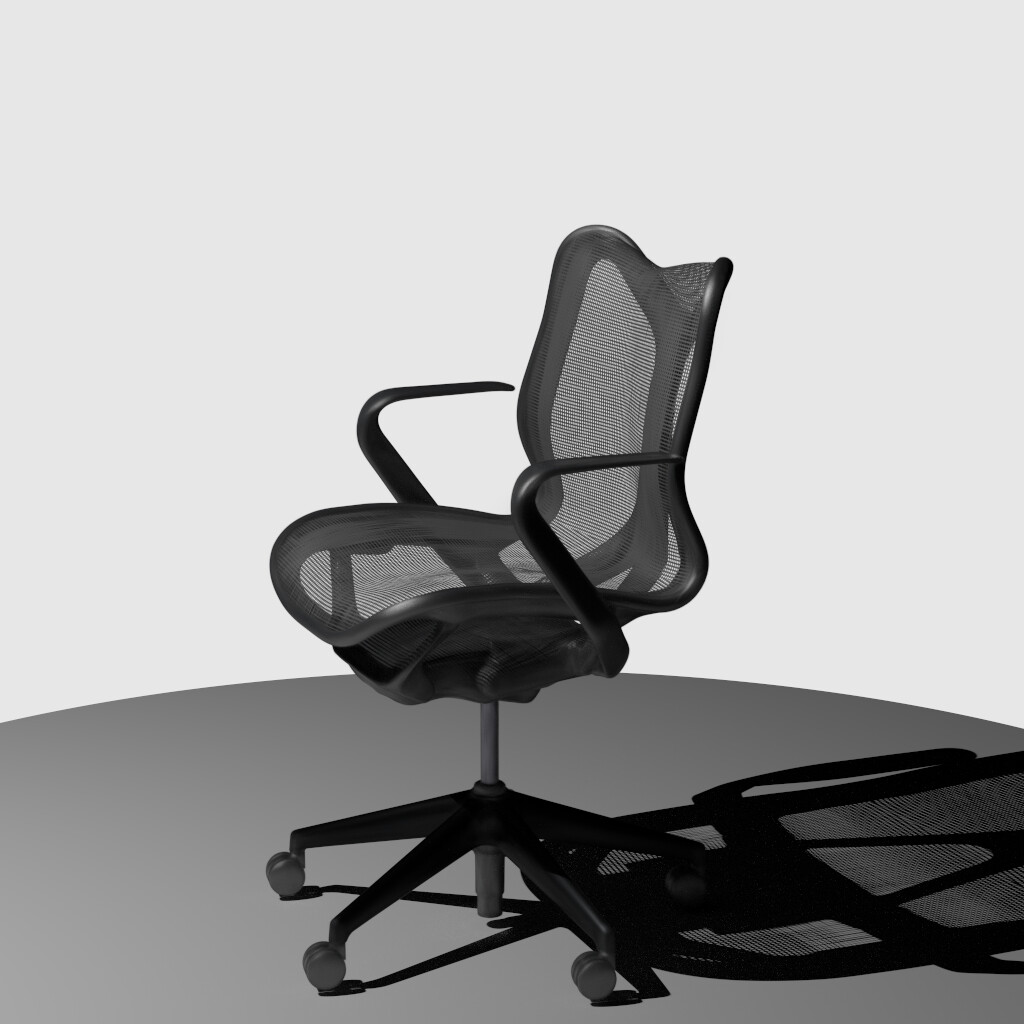}}
        &
        \frame{\includegraphics[width=\lenLinearVSExp]{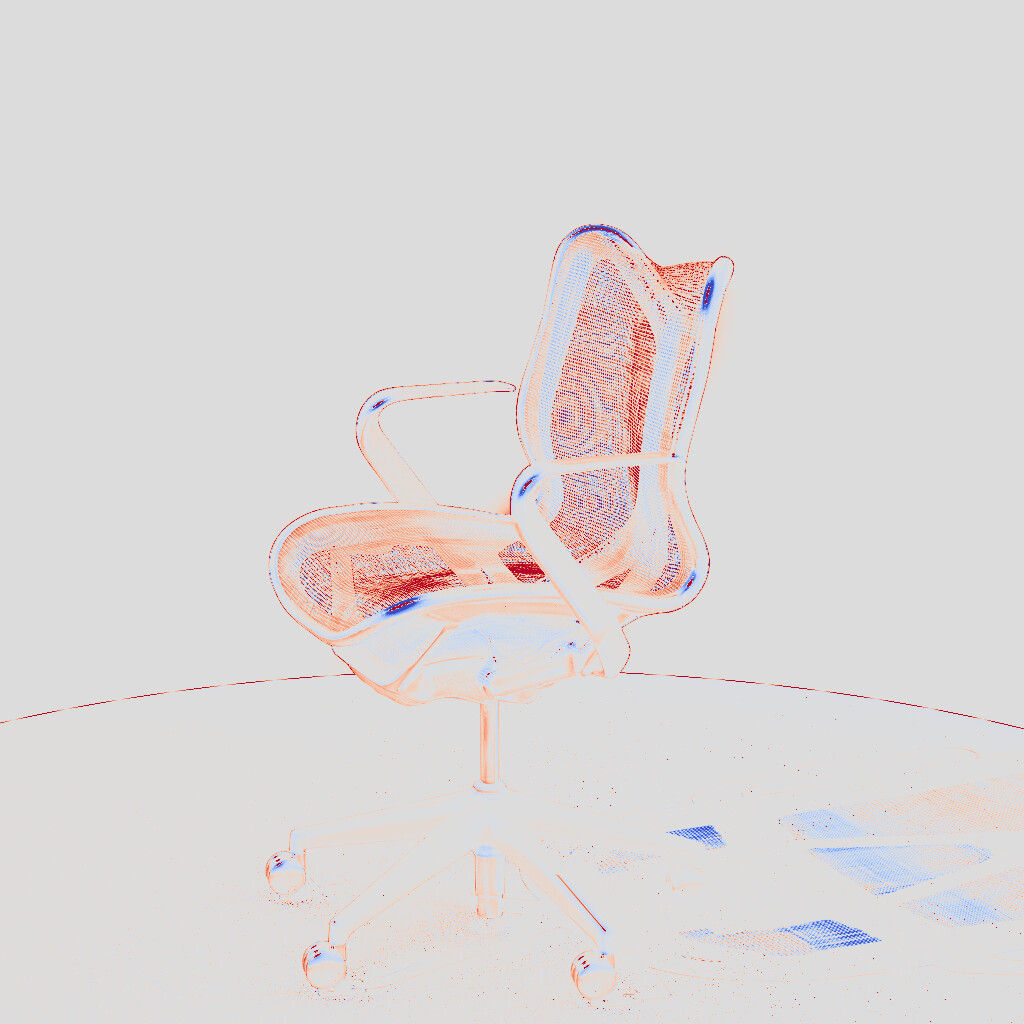}}
        &
        \frame{\includegraphics[width=\lenLinearVSExp]{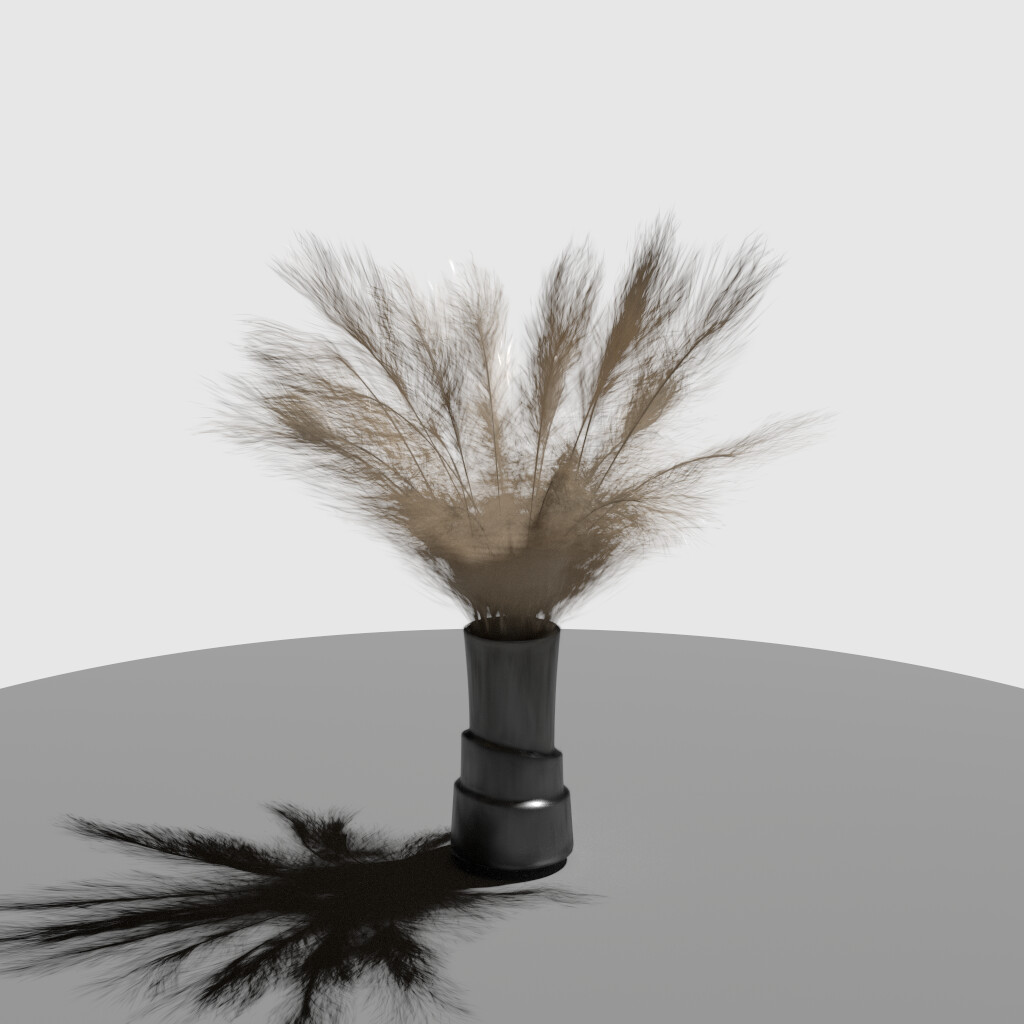}}
        &
        \frame{\includegraphics[width=\lenLinearVSExp]{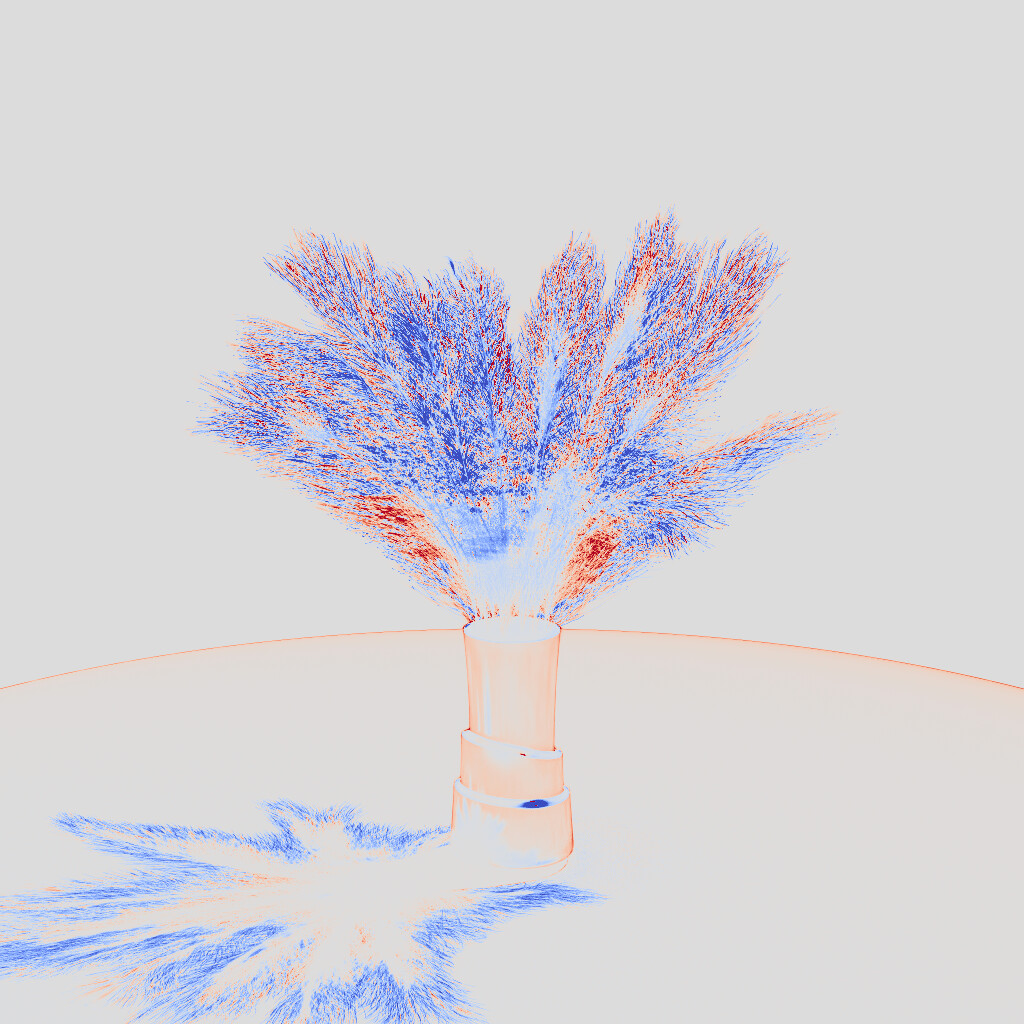}}
        &
        \frame{\includegraphics[width=\lenLinearVSExp]{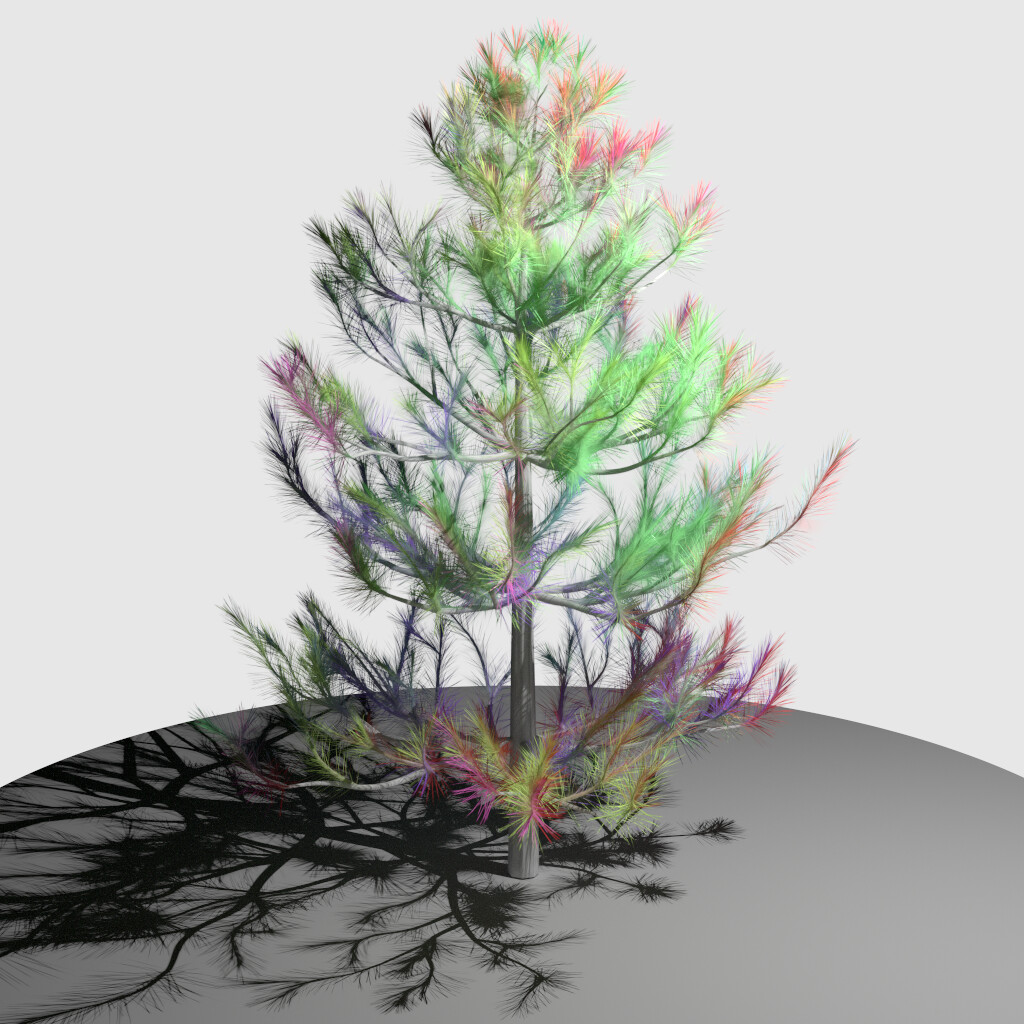}}
        &
        \frame{\includegraphics[width=\lenLinearVSExp]{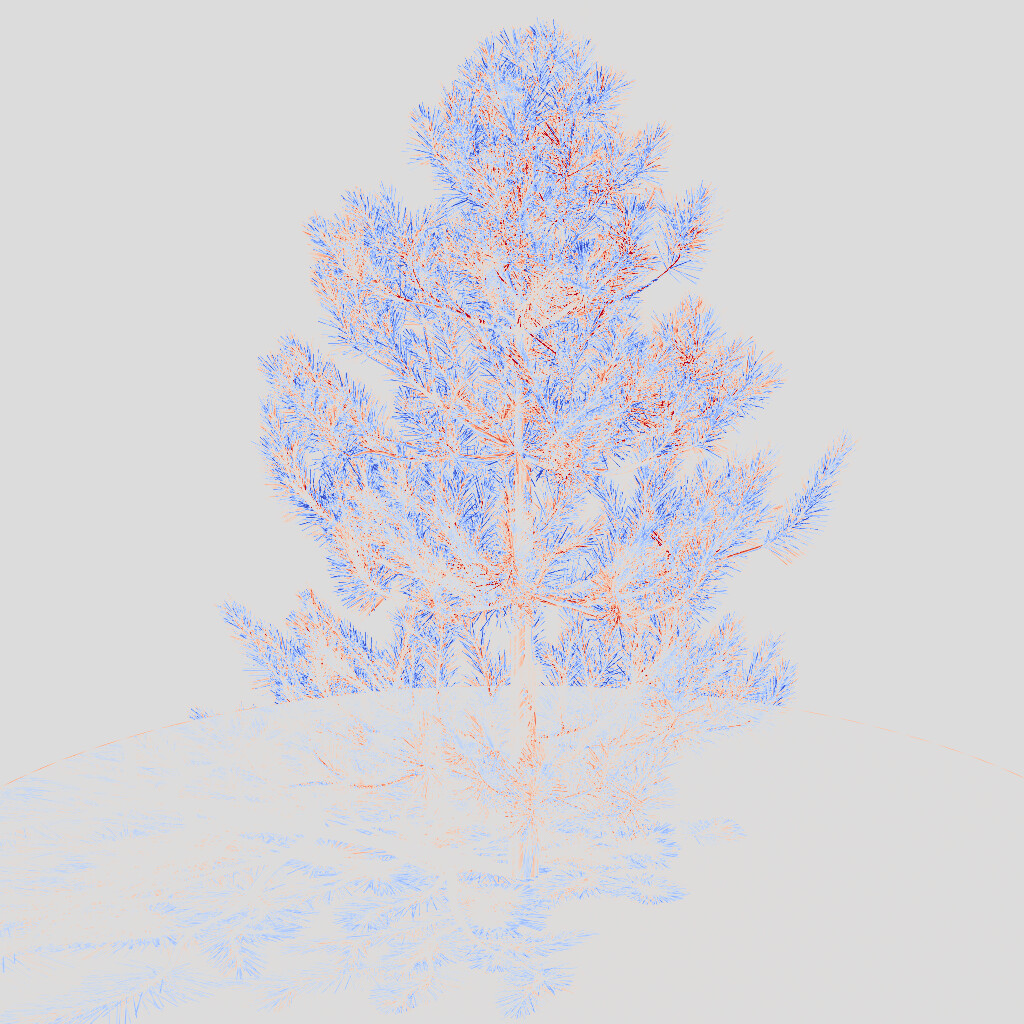}}
        \\
        \small{\textsf{Exponential 25K}} & \small{\textsf{PSNR: 26.43}} &
        \small{\textsf{Exponential 250K}} & \small{\textsf{PSNR: 28.41}} &
        \small{\textsf{Exponential 50K}} & \small{\textsf{PSNR: \textbf{27.44}}} &
        \small{\textsf{Exponential 100K}} & \small{\textsf{PSNR: 22.70}}
        \\
        \frame{\includegraphics[width=\lenLinearVSExp]{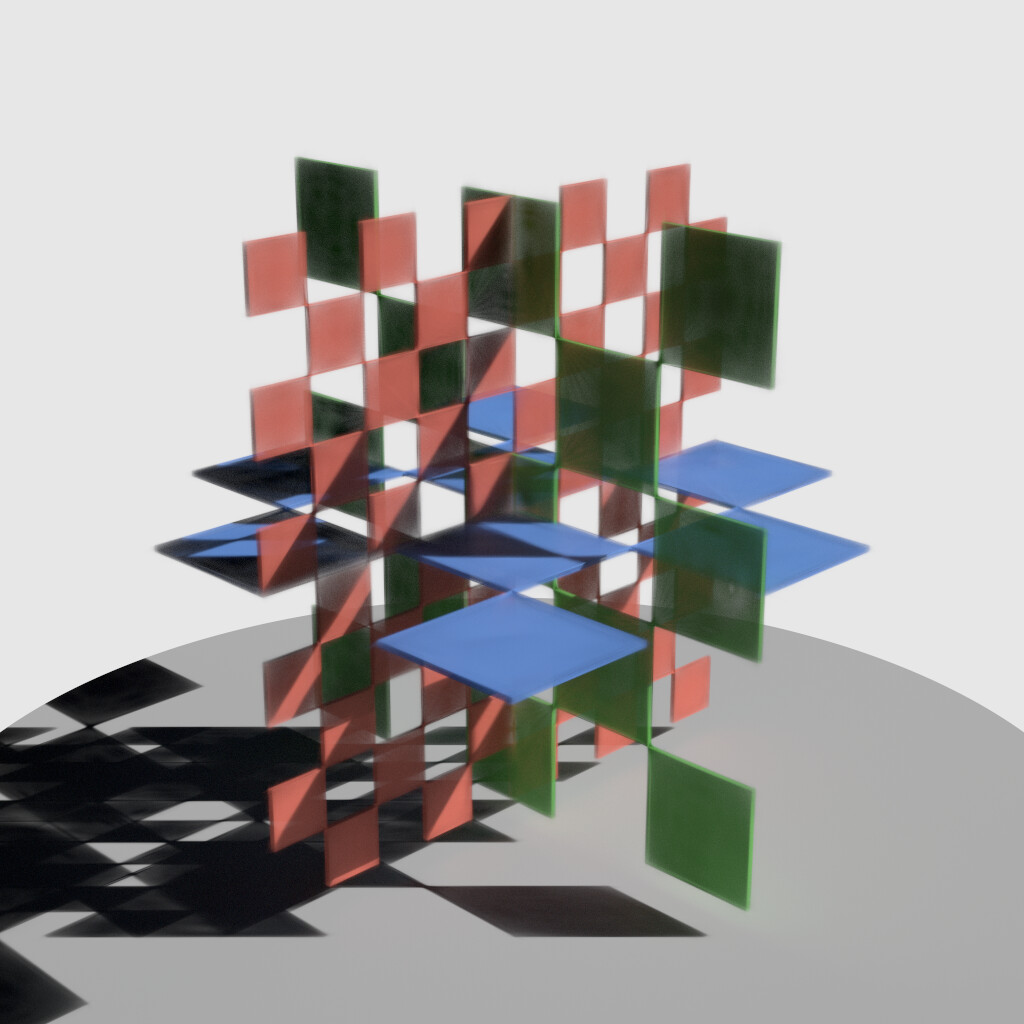}}
        &
        \frame{\includegraphics[width=\lenLinearVSExp]{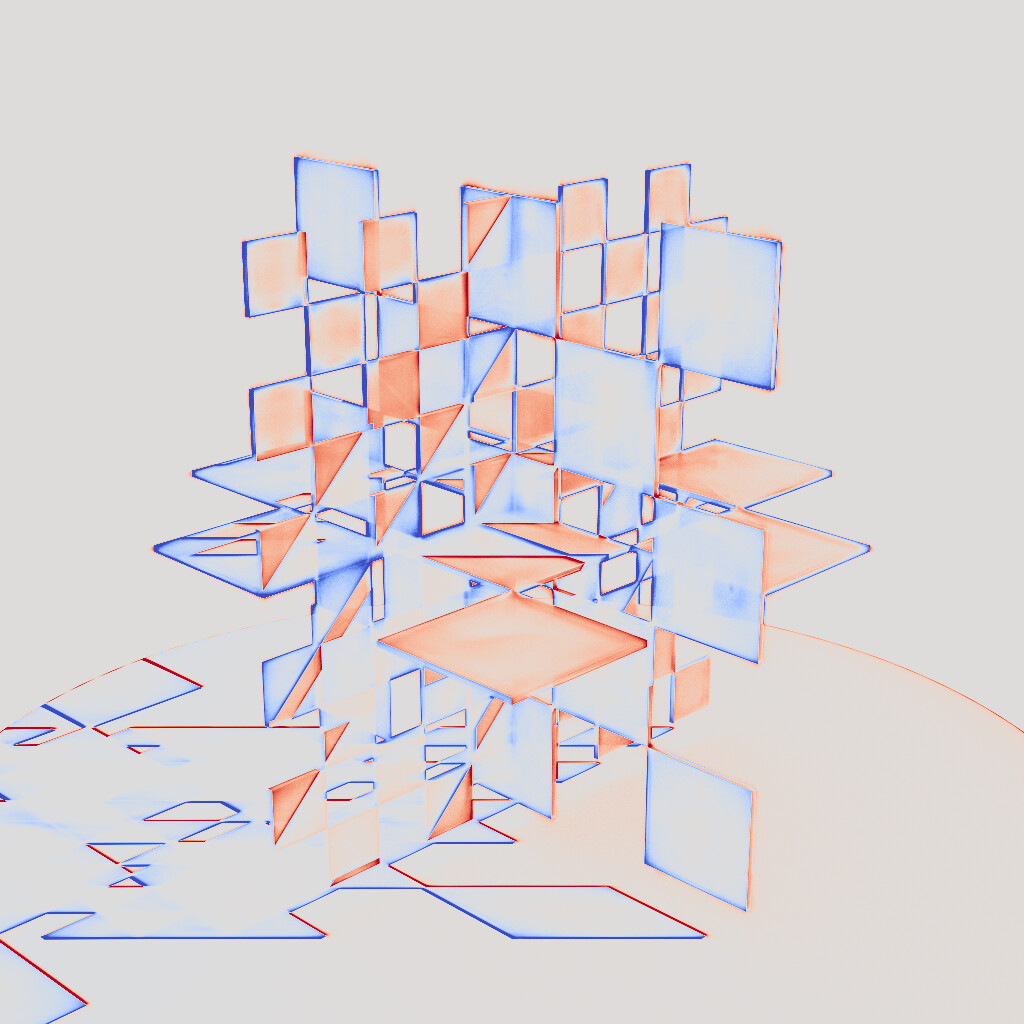}}
        &
        \frame{\includegraphics[width=\lenLinearVSExp]{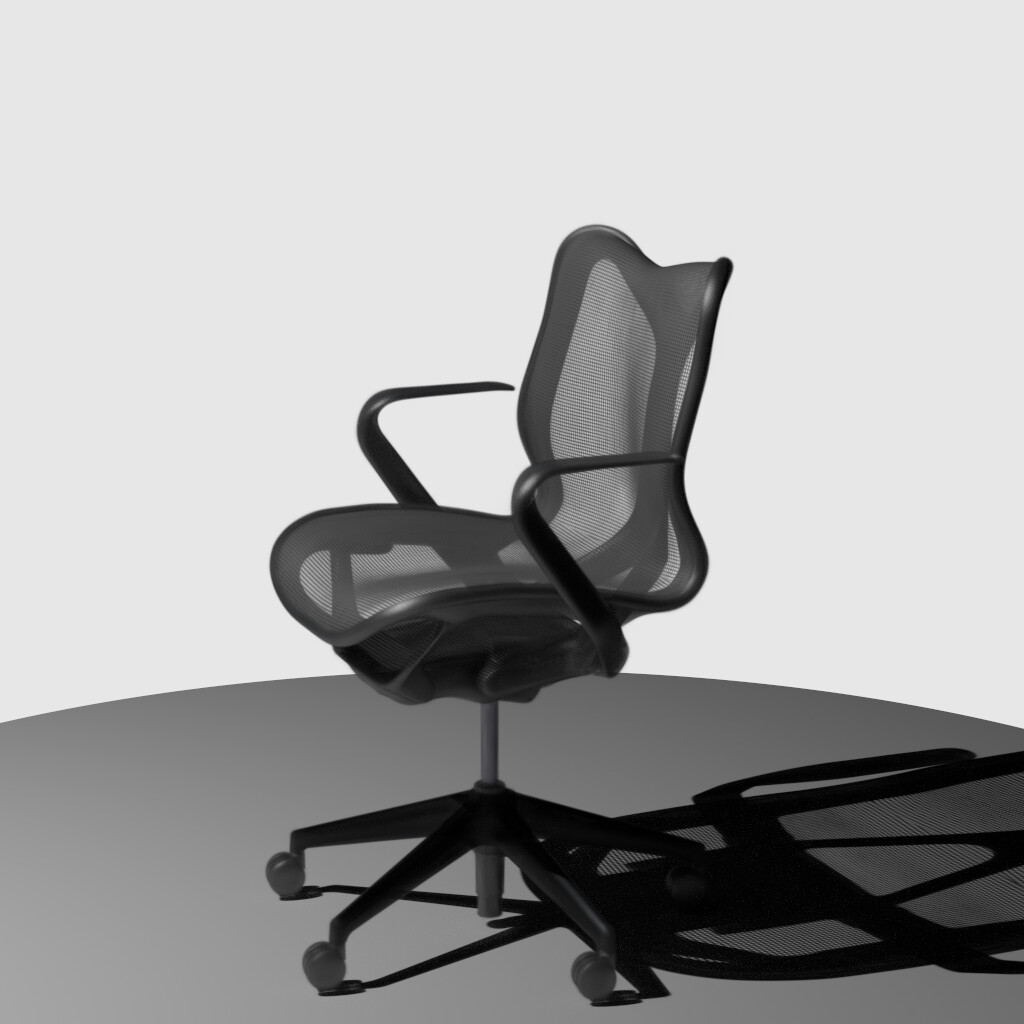}}
        &
        \frame{\includegraphics[width=\lenLinearVSExp]{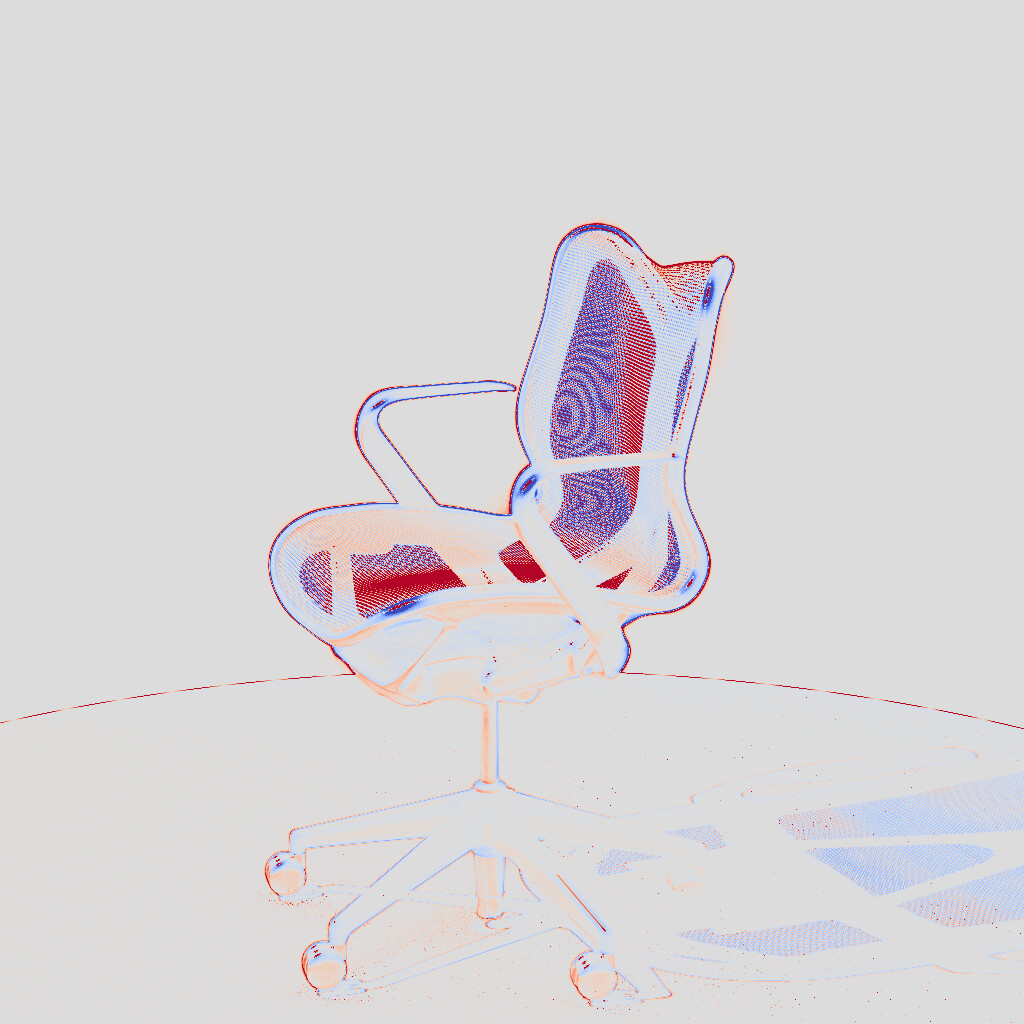}}
        &
        \frame{\includegraphics[width=\lenLinearVSExp]{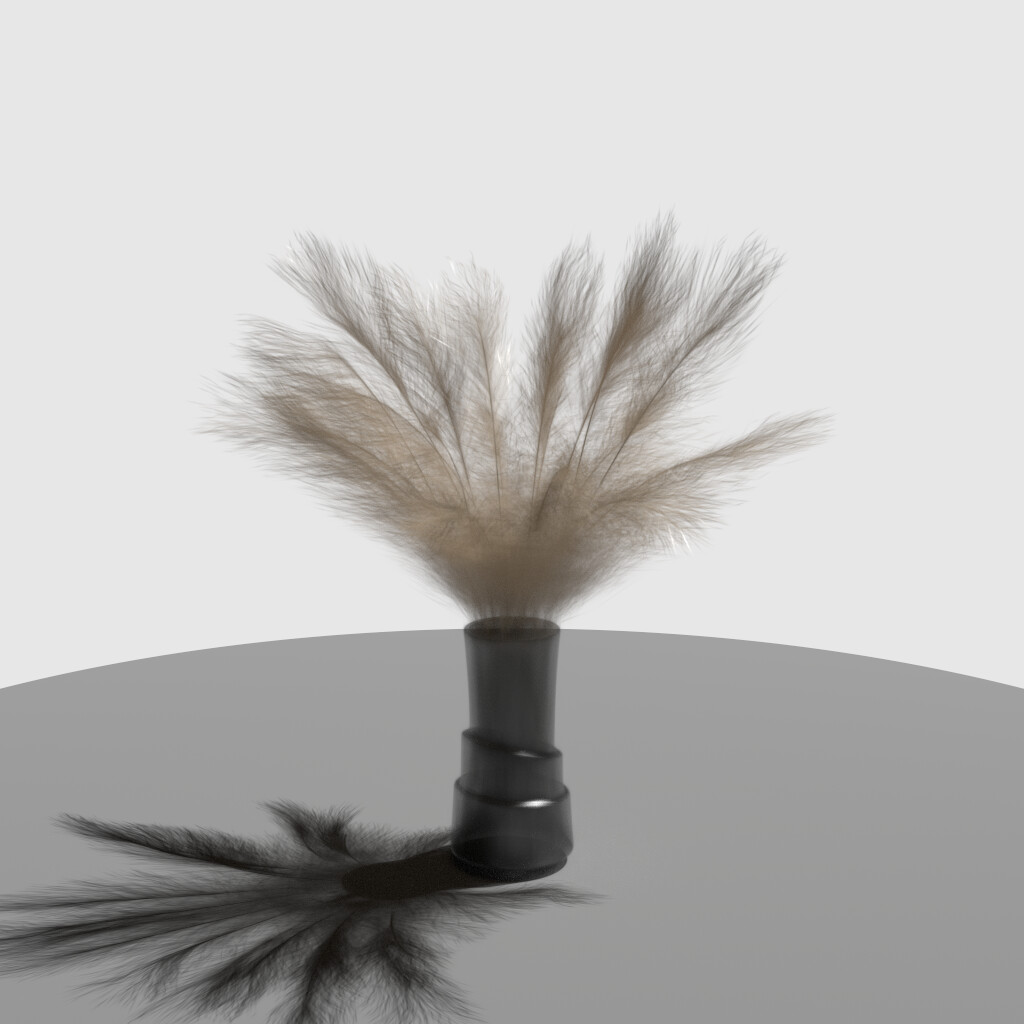}}
        &
        \frame{\includegraphics[width=\lenLinearVSExp]{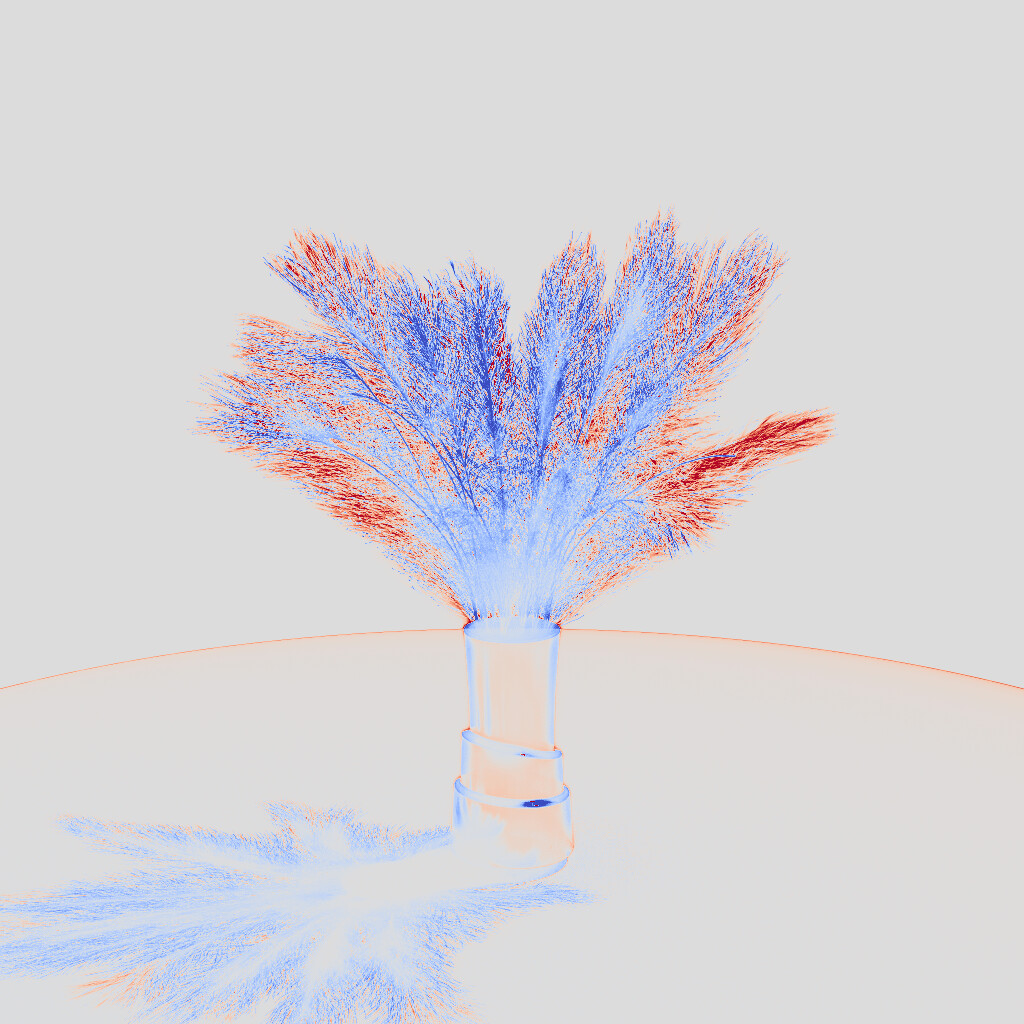}}
        &
        \frame{\includegraphics[width=\lenLinearVSExp]{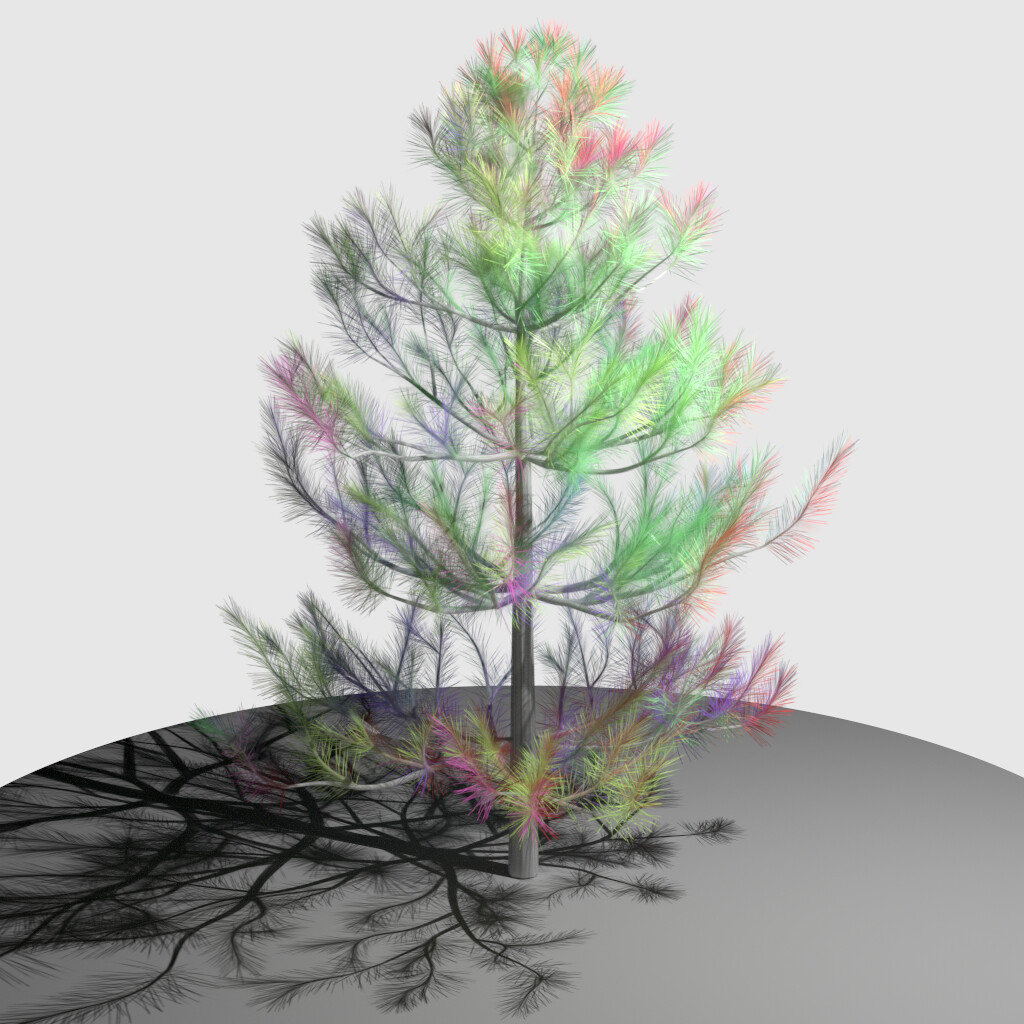}}
        &
        \frame{\includegraphics[width=\lenLinearVSExp]{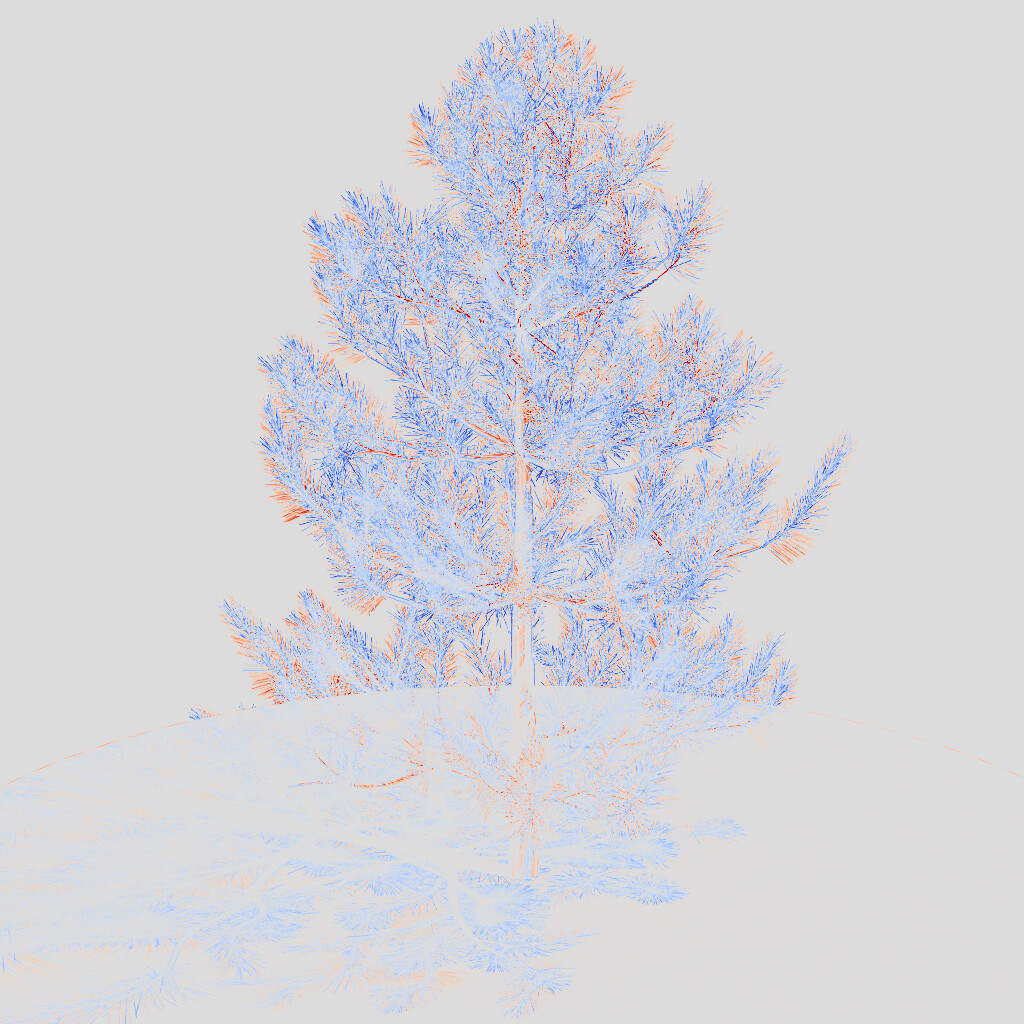}}
        \\
        &
        \begin{overpic}[width=\lenLinearVSExp, frame]{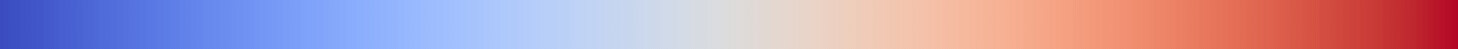}
			\put(0, -10){\small -0.2}
			\put(80, -10){\small 0.2}
		\end{overpic}
        & &
        \begin{overpic}[width=\lenLinearVSExp, frame]{resources/coolwarm_h.jpg}
			\put(0, -10){\small -0.1}
			\put(80, -10){\small 0.1}
		\end{overpic}
        & &
        \begin{overpic}[width=\lenLinearVSExp, frame]{resources/coolwarm_h.jpg}
			\put(0, -10){\small -0.2}
			\put(80, -10){\small 0.2}
		\end{overpic}
        & &
        \begin{overpic}[width=\lenLinearVSExp, frame]{resources/coolwarm_h.jpg}
			\put(0, -10){\small -0.4}
			\put(80, -10){\small 0.4}
		\end{overpic}
        \\
    \end{tabular}
    \caption{\label{fig:linear_vs_exp}
        Comparison between linear and exponential transmittance model with fixed number of Gaussians. Error maps and PSNRs with respect to renders of the
        original scenes are provided. The linear model helps avoid leaking and produce sharp silhouettes, making it favorable in most cases we focus on.
        Even for scenes with dense elements such as the \emph{Plant} and the \emph{Color Tree}, each Gaussian usually only covers one or a few opaque elements,
        making the exponential model less suitable at this granularity.
    }
\end{figure*}

\paragraph{Comparison between Transmittance Models}
In \autoref{fig:linear_vs_exp}, we compare the rendering results using both our linear transmittance and the traditional exponential transmittance, where each
Gaussian primitive defines a fraction of the extinction coefficient (\autoref{eq:exp_ff}) to be summed together. The exponential variants are obtained through a
conversion process similar to that described in \autoref{sec:data_acquisition}. The only difference is that we also switch the target transmittance 
model to exponential during the transmittance optimization (the optimization is still beneficial as 3DGS approximates overlapping primitives).
They are rendered by brute-force analog decomposition tracking~\citep{kutz2017spectral} (\autoref{sec:decomp_track}).
We also show the differences between those renders and renders using the
original scenes, while acknowledging that our method does not include a full inverse rendering pipeline to match the original renders exactly.

The selected scenes in \autoref{fig:linear_vs_exp} have different geometric characteristics. For the \emph{Checkerboards} and the \emph{Office Chair} that
consist of flat surfaces (the chair is textured by a high-resolution opacity mask), the linear model clearly excels at producing opaque appearance with
sharp silhouettes and shadows. The \emph{Plant} and the
\emph{Color Tree} contain dense, unstructured elements. This type of scenes can be abstracted as exponential volumes because \emph{at far field, with enough sample size,}
the average free-flight distribution through these elements approaches exponential. However, this is not necessarily true when we represent the scenes as
Gaussian primitives. To produce reasonable details, typically each primitive only covers one or a cluster of well-aligned opaque elements, which is the case
for those two scenes. At this granularity, the geometries inside a primitive still exhibit non-negligible spatial correlation. Therefore, the linear model is
roughly on par with the exponential model for those two scenes. The exponential model could be more suitable for true participating media content such as clouds and smoke.
Finally, the quality improves with more primitives regardless of the underlying transmittance model.

\paragraph{Comparison with Meshes}

In \autoref{fig:mesh_comp}, we evaluate the rendering quality and speed of our representation against meshes. Given a reference mesh-based model, 
we convert it to our representation using methods described in \autoref{sec:data_acquisition}. As it is possible to allocate varying number of 
primitives, we choose to compare both representations at matching primitive (triangle) count, which serves as a reasonable indicator of model quality.
Meshes are simplified based on the standard quadric error metric~\citep{siggraph/GarlandH97}. For meshes with textures, textures are kept at original 
resolutions. For this comparison, we select two scenes that represent two distinctive categories of objects.

The first example, the \emph{Color Tree} (originally 151K triangles), represents objects made of aggregated elements. This is where our Gaussian primitives excel as they 
can effectively represent thin geometries with high-frequency details. At reduced primitive count, a single primitive instead covers a cluster of small elements. 
The overall quality degrades gracefully and the general appearance is maintained. In contrast, meshes fail catastrophically as triangle count decreases. 
This is not only because mesh simplification cannot perform meaningful edge collapses on disconnected topology, but also because meshes are fundamentally 
inefficient at representing partial coverage from dense elements. 
We further show how this superiority translates to practical performance advantages later in \autoref{fig:lod_fan}.

The second example, the \emph{Caustic Ring} (originally 64K triangles), represents objects made of smooth manifold surfaces. Non-reference 
meshes are rendered with geometric normals for a fair comparison. This example is the opposite case where meshes can easily achieve watertight surfaces, but point-based representations 
require either dense packing or hole filling~\citep{pfister2000surfels}. 
Nevertheless, our representation still produces reasonable results after transmittance optimization (\autoref{subsec:transmittance_optim}). 
We would also like to focus on the challenging caustic pattern formed from global illumination. As the primitive (triangle) count decreases, 
the pattern is gradually distorted due to discretization.
Notably, our representation can preserve it reasonably well until the primitive count drops to 1$\%$, at which point the corresponding mesh 
also fails to do so.
We observe the highest error around high-curvature regions with limited primitive count, as expected.
An interesting future direction would be to develop non-convex particle kernels that better adapt to curved shapes.

In summary, the choice between meshes and our Gaussian primitives is ultimately governed by the geometric character of the content. 
Meshes remain preferable for connected manifold hard surfaces. 
Our primitives provide a single representation that unifies hard surfaces and aggregates under full light transport. 
In particular, they are advantageous for scenes dominated by dense, thin, or disconnected aggregate geometry, 
where mesh simplification fails catastrophically.

\begin{figure*}[tb]
	\newlength{\lenVoxelCompare}
	\setlength{\lenVoxelCompare}{0.16\linewidth}
    \addtolength{\tabcolsep}{-4pt}
    \renewcommand{\arraystretch}{0.5}
    \centering
    \begin{tabular}{cccccc}
         & \emph{Color Tree} &  &  & \emph{Plant} &  \\
        \includegraphics[width=\lenVoxelCompare, frame]{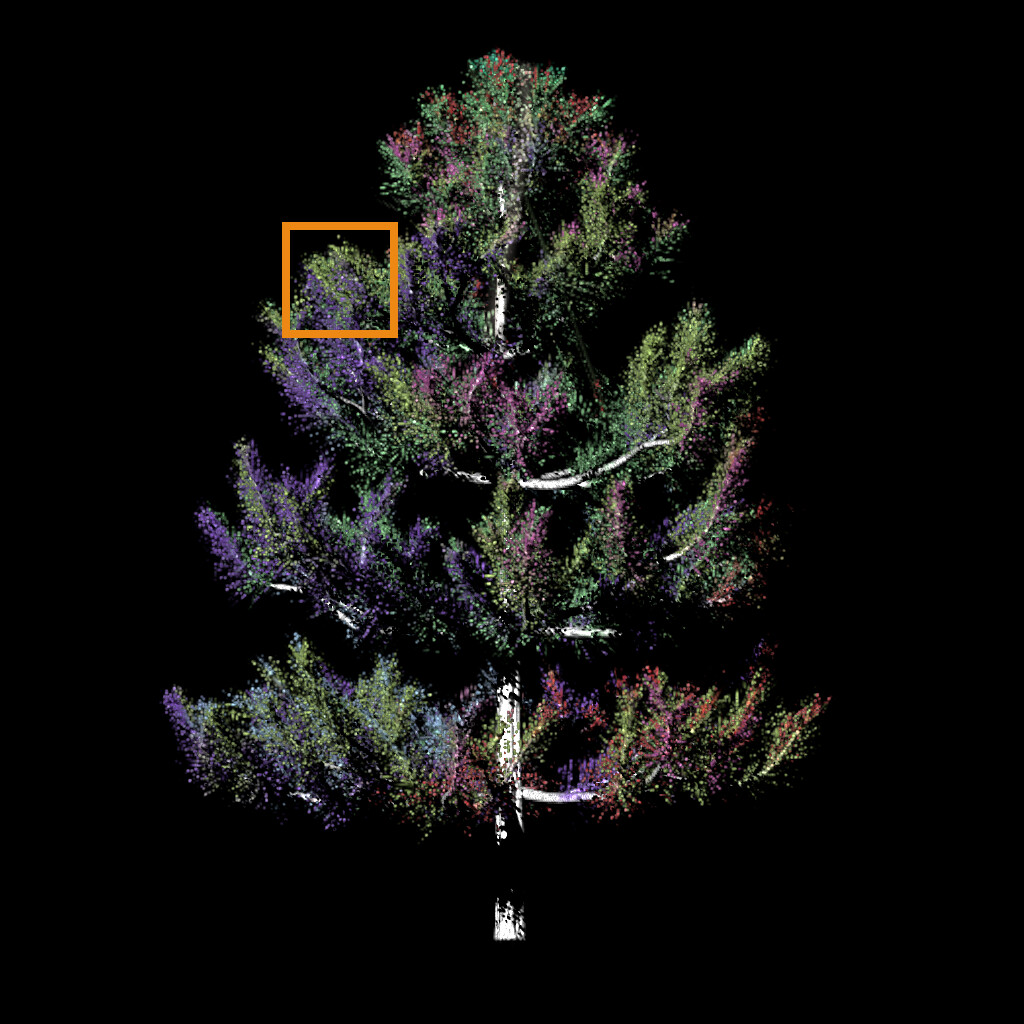}
        &
        \includegraphics[width=\lenVoxelCompare, frame]{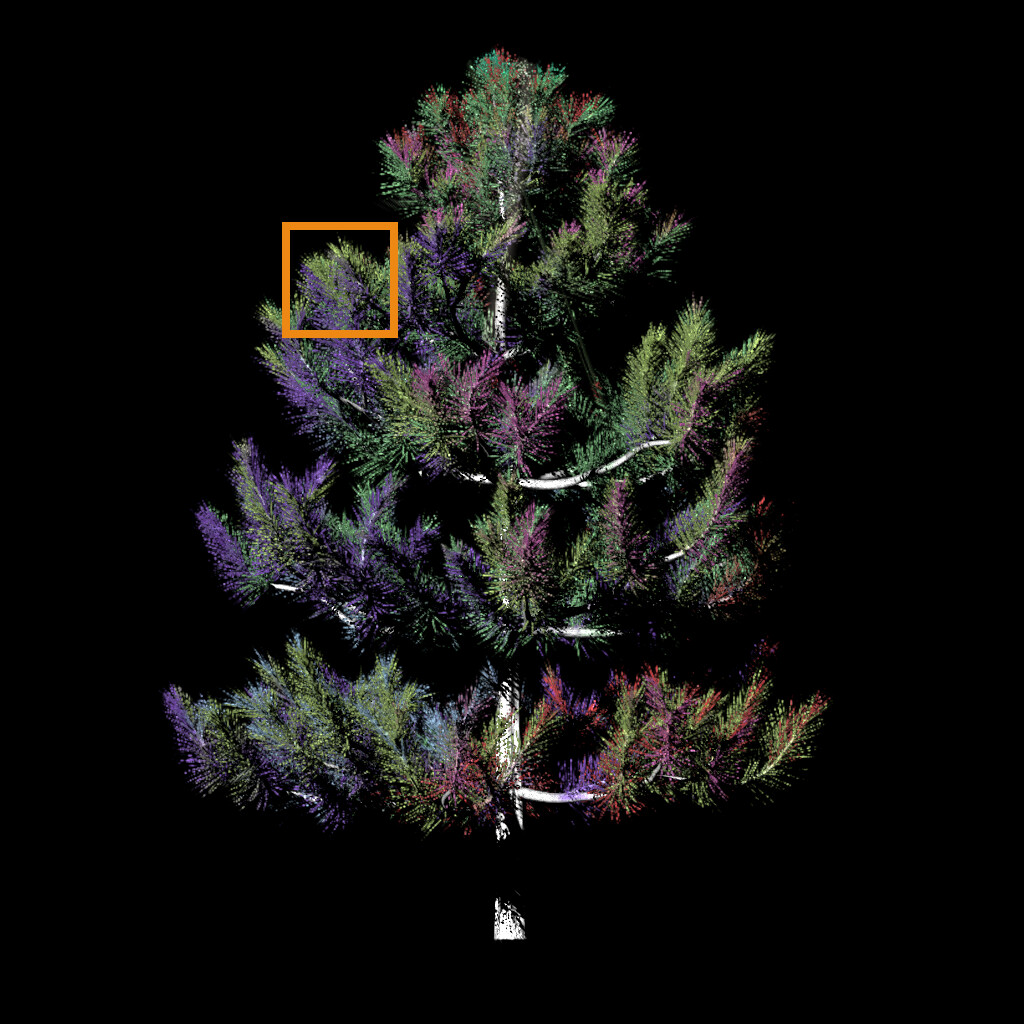}
        &
        \includegraphics[width=\lenVoxelCompare, frame]{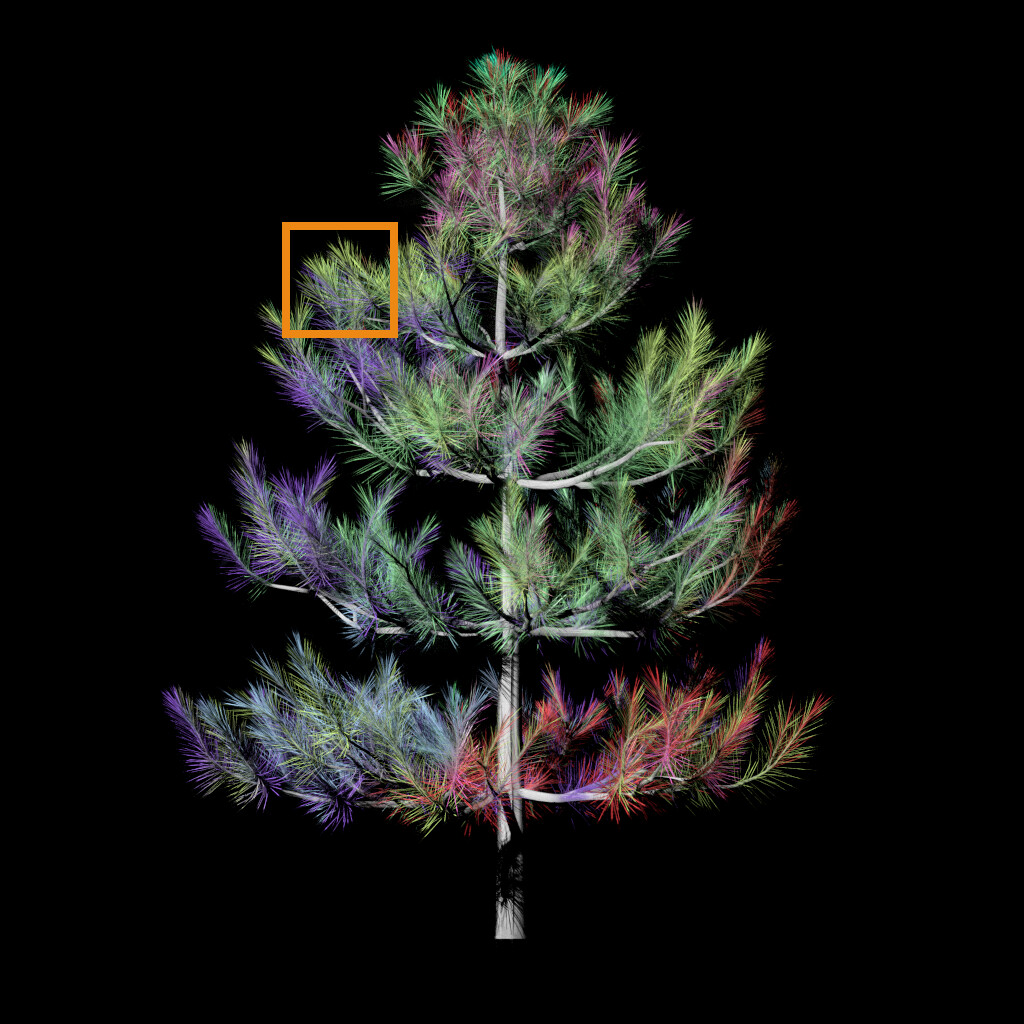}
        &
        \includegraphics[width=\lenVoxelCompare, frame]{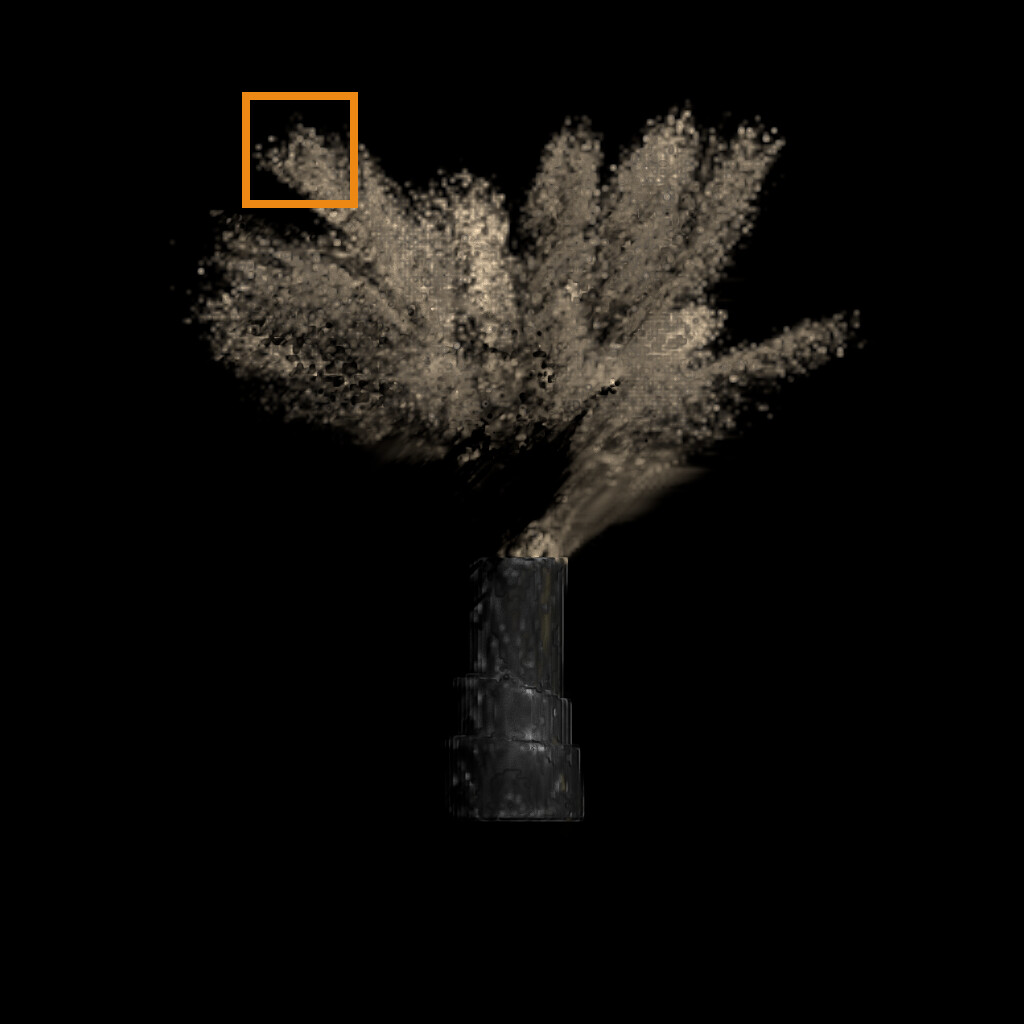}
        &
        \includegraphics[width=\lenVoxelCompare, frame]{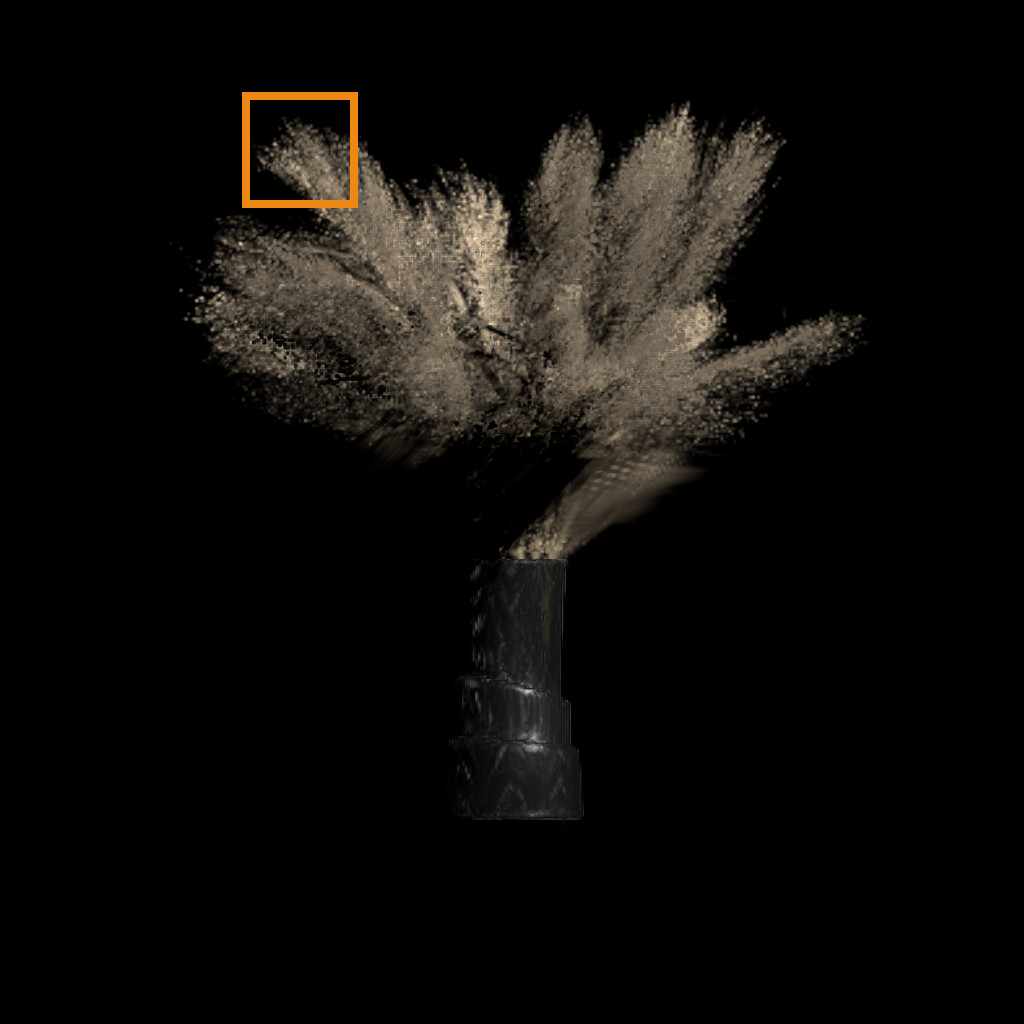}
        &
        \includegraphics[width=\lenVoxelCompare, frame]{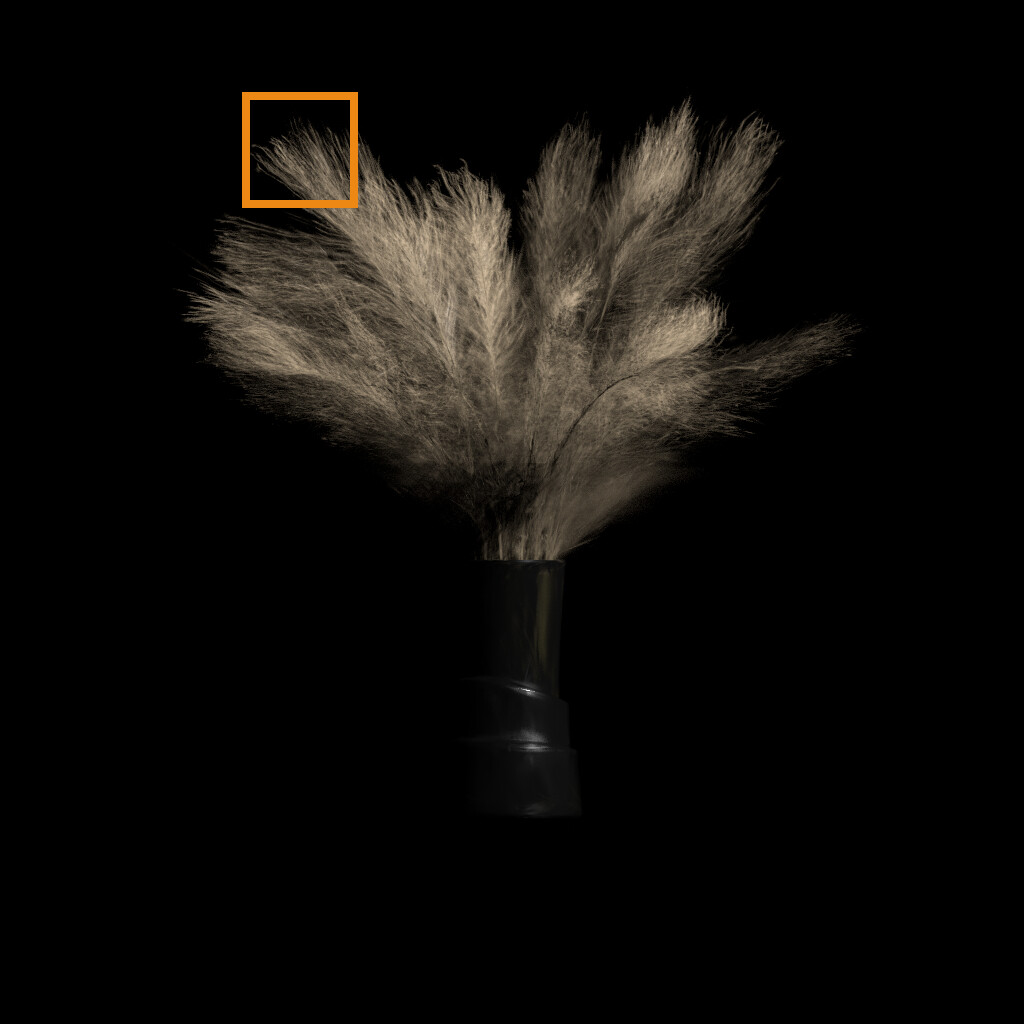}
        \\
        \includegraphics[width=\lenVoxelCompare, frame]{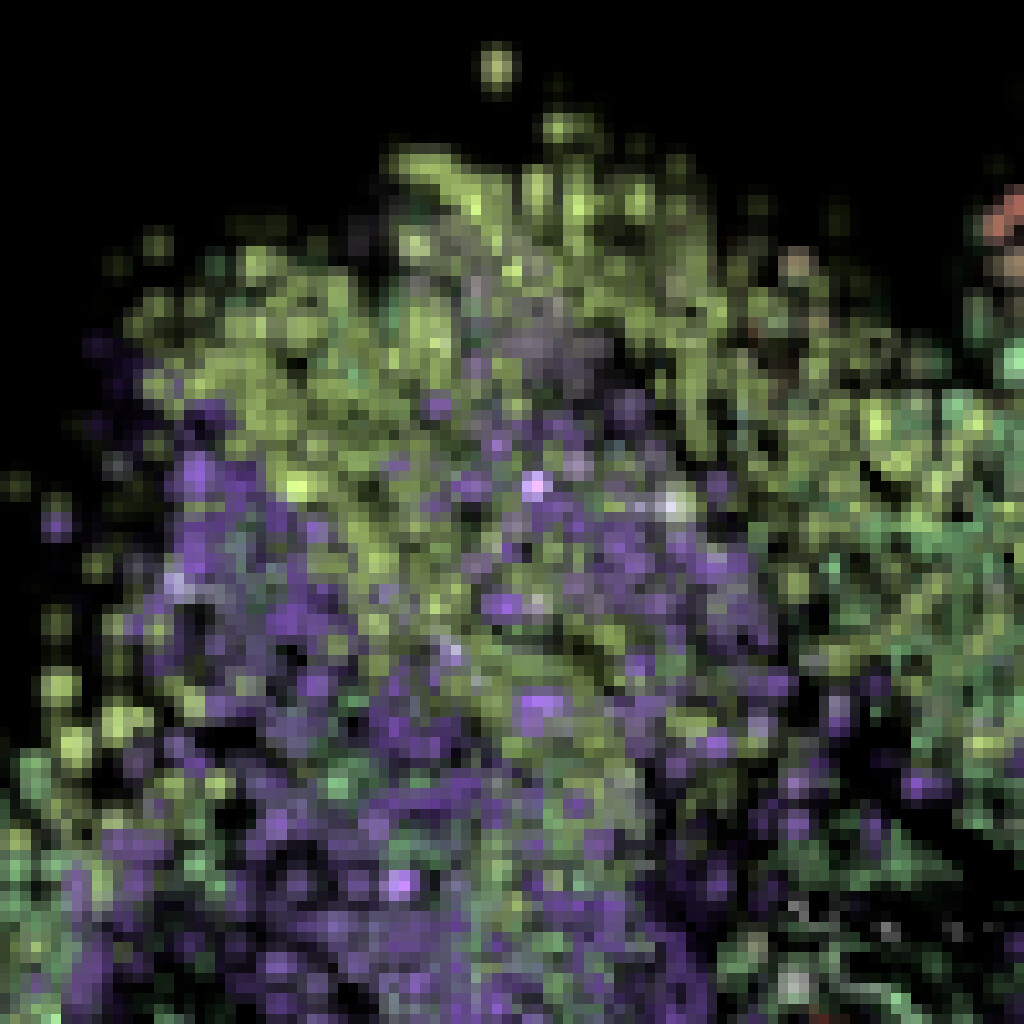}
        &
        \includegraphics[width=\lenVoxelCompare, frame]{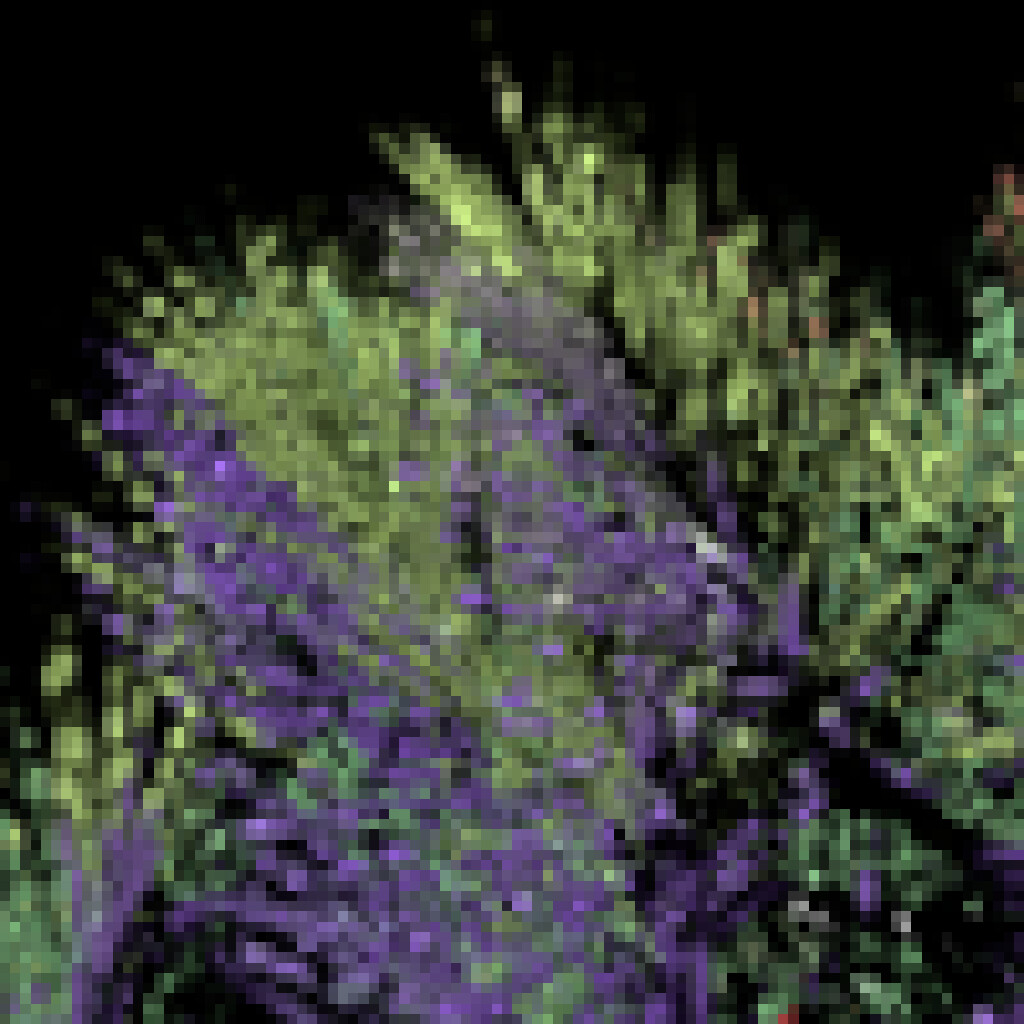}
        &
        \includegraphics[width=\lenVoxelCompare, frame]{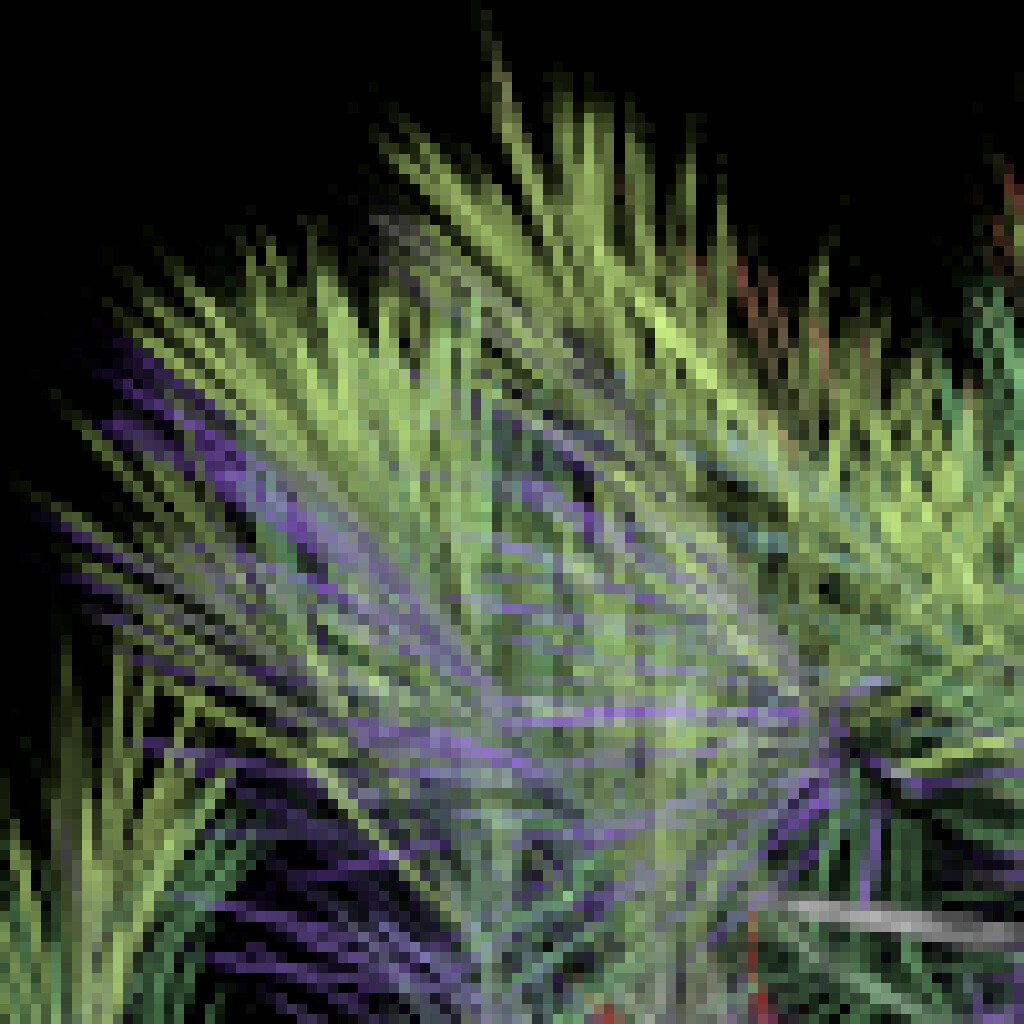}
        &
        \includegraphics[width=\lenVoxelCompare, frame]{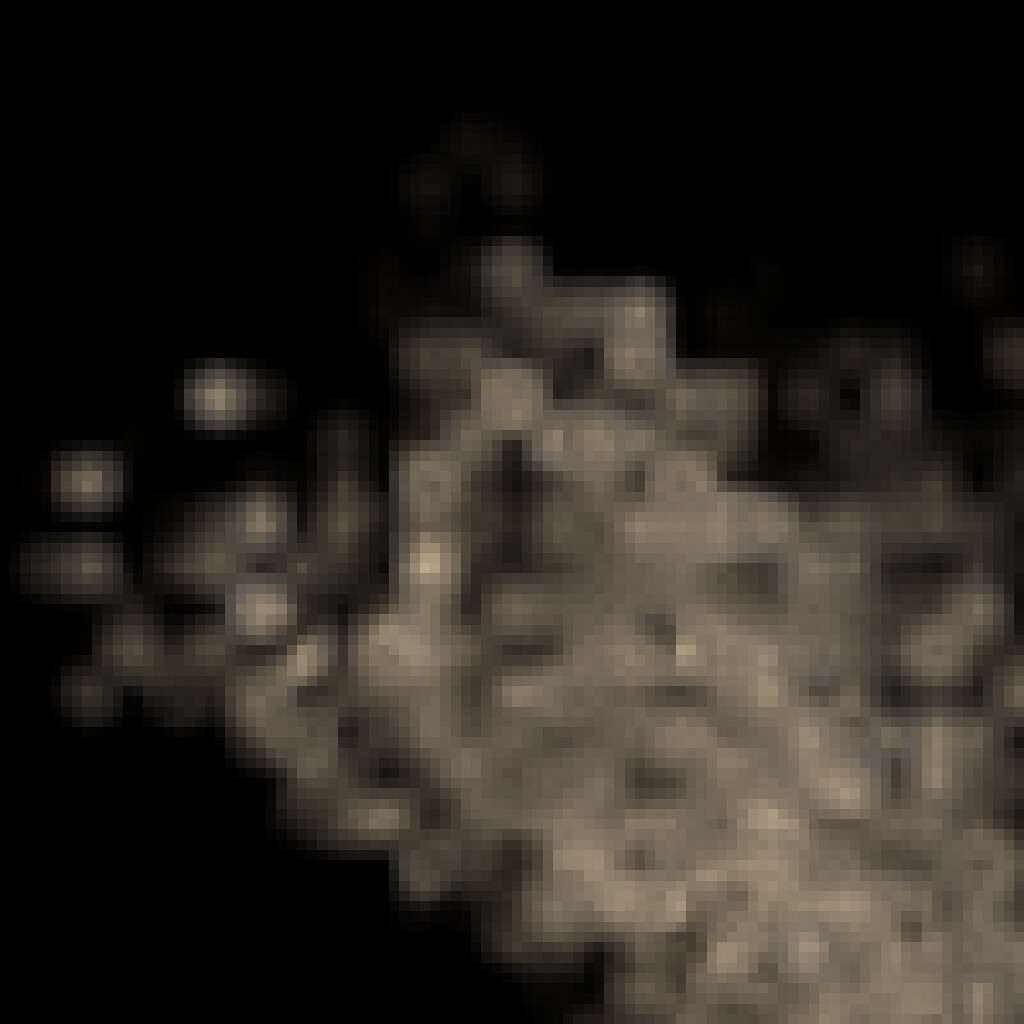}
        &
        \includegraphics[width=\lenVoxelCompare, frame]{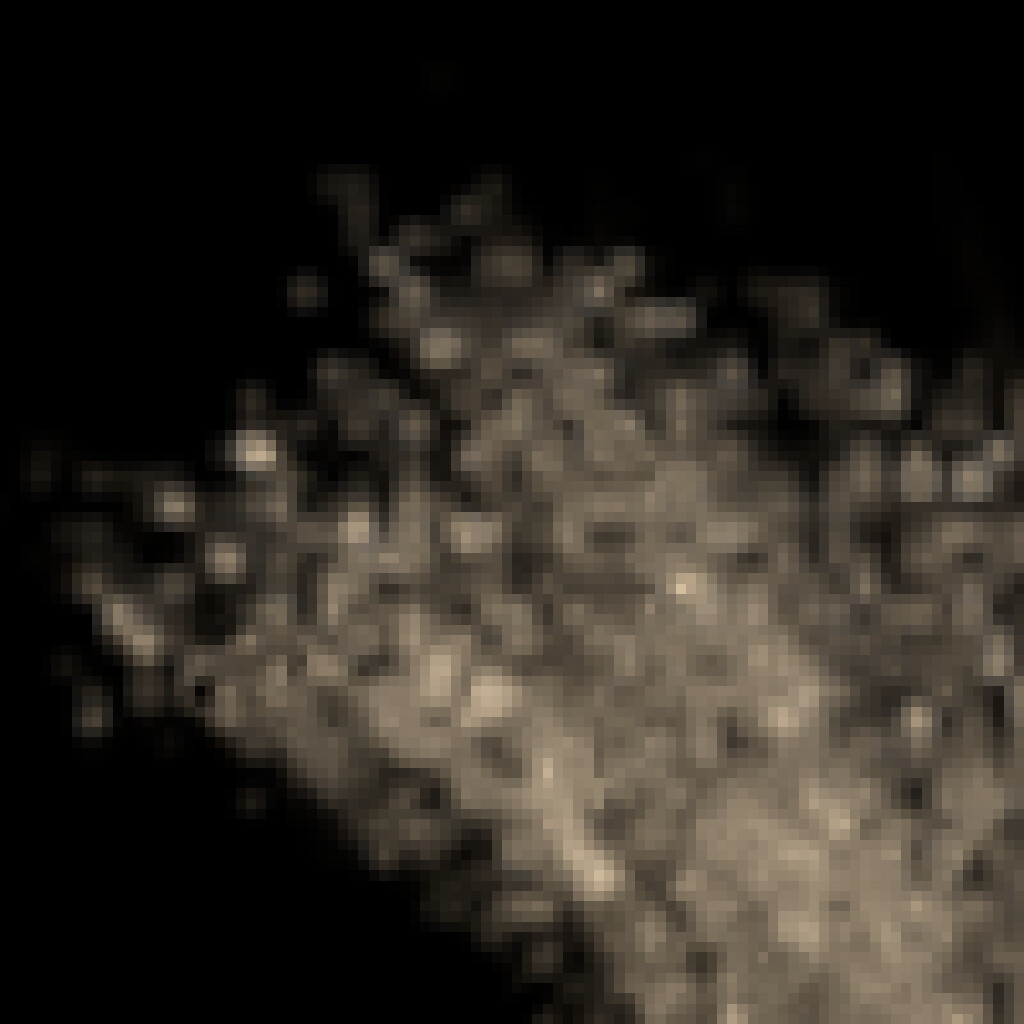}
        &
        \includegraphics[width=\lenVoxelCompare, frame]{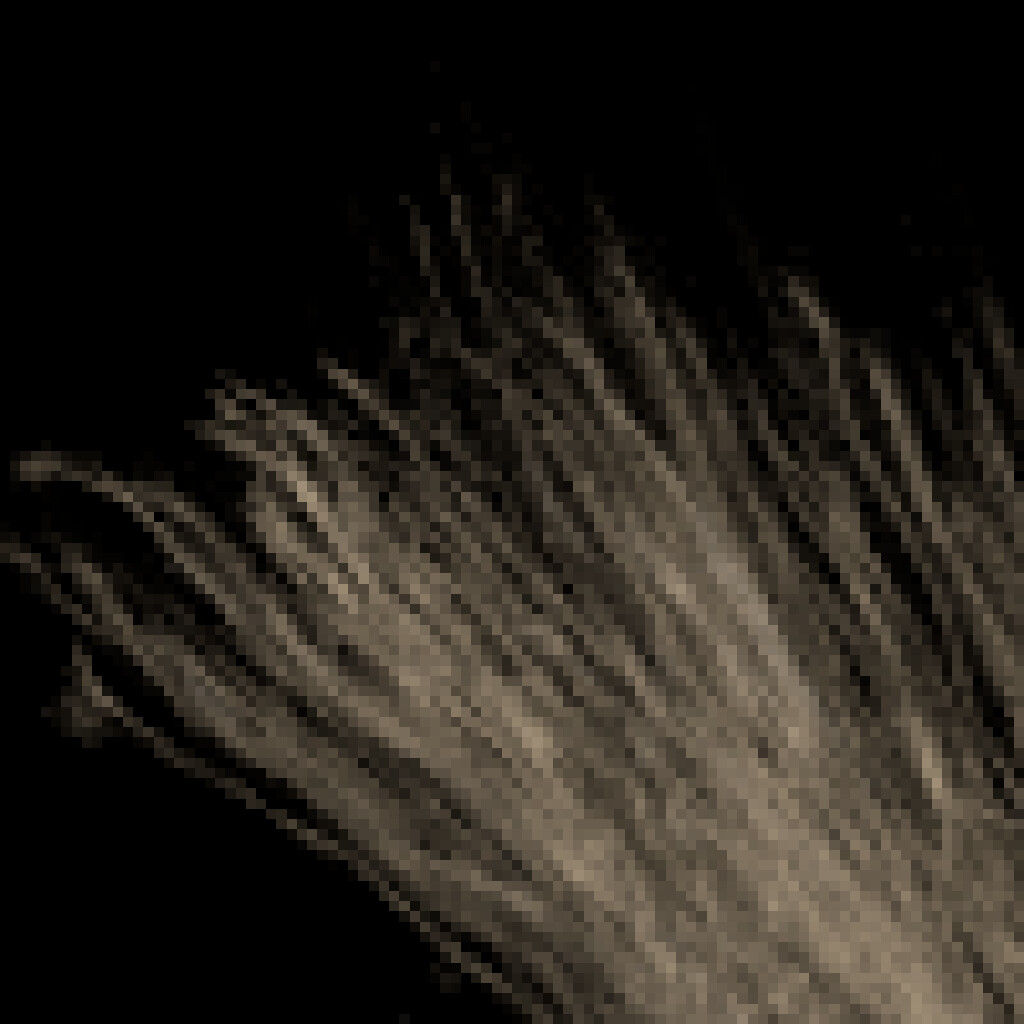}
        \\
        \small{\textsf{Eq. num voxels}} & \small{\textsf{8$\times$ num voxels}} & \small{\textsf{Gaussians (\(\sim\)750K)}} &
        \small{\textsf{Eq. num voxels}} & \small{\textsf{8$\times$ num voxels}} & \small{\textsf{Gaussians (\(\sim\)220K)}}
    \end{tabular}
    \caption{\label{fig:voxel_compare}
        Comparison between our Gaussian models and voxelized models at different resolutions. Only \textbf{non-empty} voxels are counted. Gaussian primitives are much more effective at representing sharp,
        detailed structure than voxels. Note that our current simple voxelization causes self-shadowing that leads to slight darkening. Direct illumination only.
    }
\end{figure*}

\paragraph{Comparison with Voxel-based Representations}
In \autoref{fig:voxel_compare}, we compare our representation to the traditional volume representation consisting of regular voxels. We perform a simple
voxelization of our models by resampling the free-flight PDF. For each voxel, we evaluate all overlapping Gaussians at the voxel center. The phase function
parameters are similarly resampled. Only non-empty voxels are stored in a sparse voxel grid. The voxel grid is then linearly interpolated and rendered by ray marching. We evaluate the reconstruction quality using
different voxel resolutions. Even using 8$\times$ more non-empty voxels than the number of Gaussians, the reconstruction quality is still significantly inferior.
This is expected because unlike Gaussian mixtures that can approximate signals at arbitrary frequency, regular grid sampling is limited by the Nyquist-Shannon
sampling theorem and the resolution must be at least twice the signal bandwidth to avoid aliasing. It would require an impractical amount of storage to properly
represent the thin structures common in vegetation. Conversely, Gaussian primitives are much more effective at capturing fine geometry
details.

\paragraph{Re-rendering under Different Lighting Conditions}
Our forward renderer supports full light transport, naturally allowing the same scene to react to different lighting 
conditions. This appears to be similar to other relightable representations~\citep{cvpr/LiangZFSJ24, tog/JiangSLWLR25} 
that partially recover unknown scene and material properties from multi-view observations for subsequent rendering. However, 
our method is fundamentally different in that we assume a ground truth scene is known. 

\begin{figure*}[tb]
	\newlength{\heightRerender}
	\setlength{\heightRerender}{0.13\linewidth}
    \addtolength{\tabcolsep}{-4pt}
    \renewcommand{\arraystretch}{0.5}
    \centering
    \begin{tabular}{cccccc}
        \small{\textsf{Reference}} &
        \small{\textsf{Ours}} & 
        \small{\textsf{GS-IR}} & 
        \multicolumn{3}{c}{\small{\textsf{Composition}}} 
        \\
        \includegraphics[draft=false, height=\heightRerender, frame]{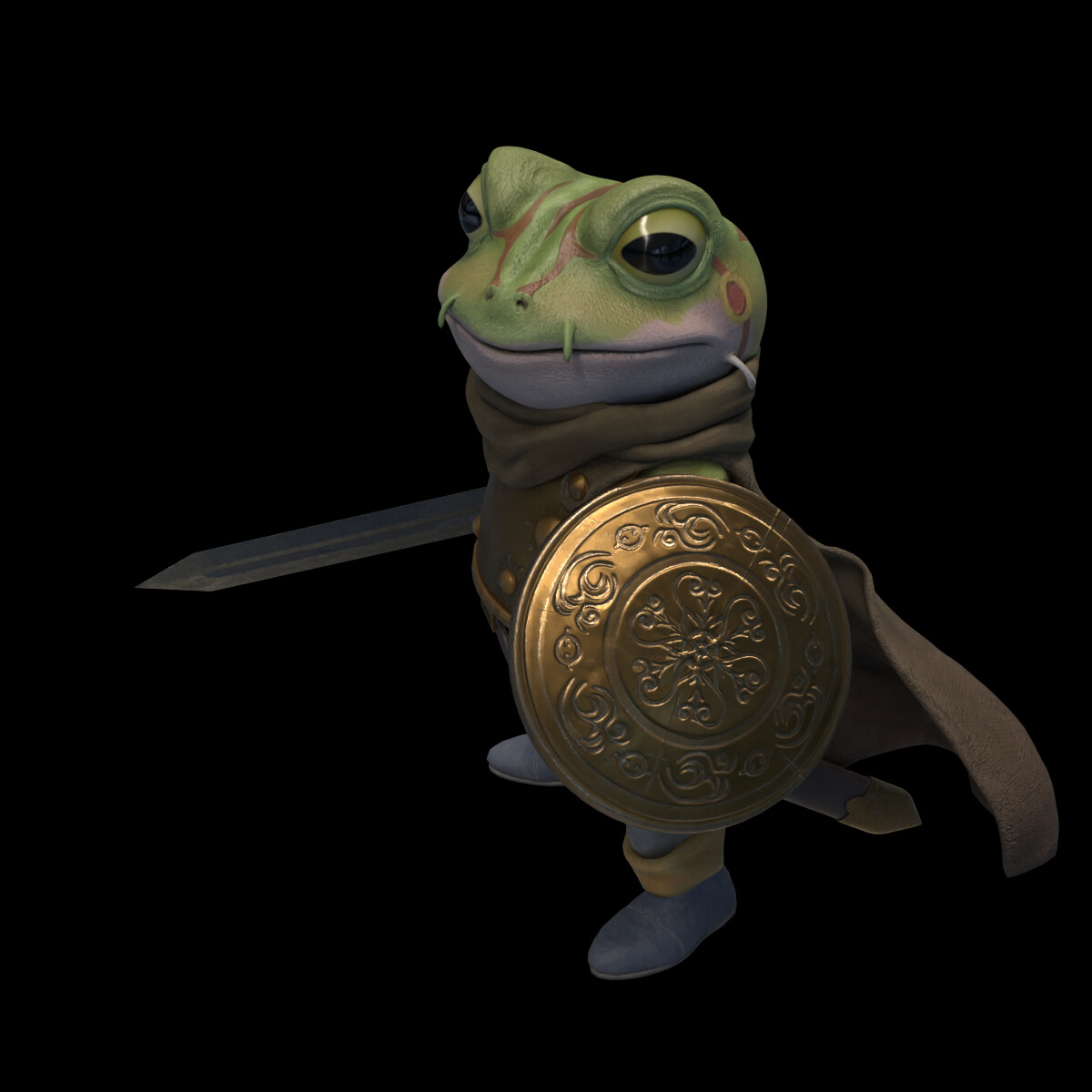}
        &
        \includegraphics[draft=false, height=\heightRerender, frame]{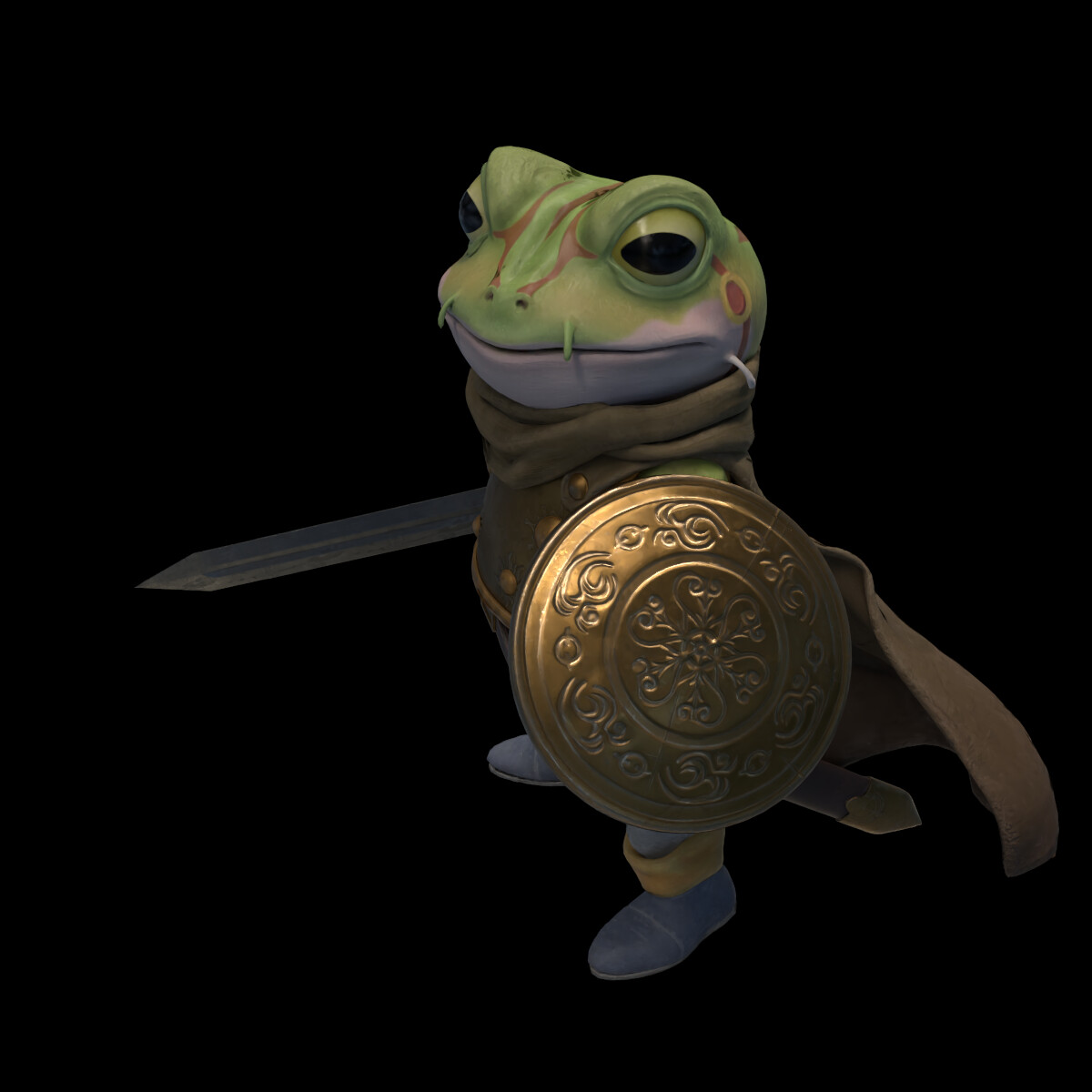}
        &
        \includegraphics[draft=false, height=\heightRerender, frame]{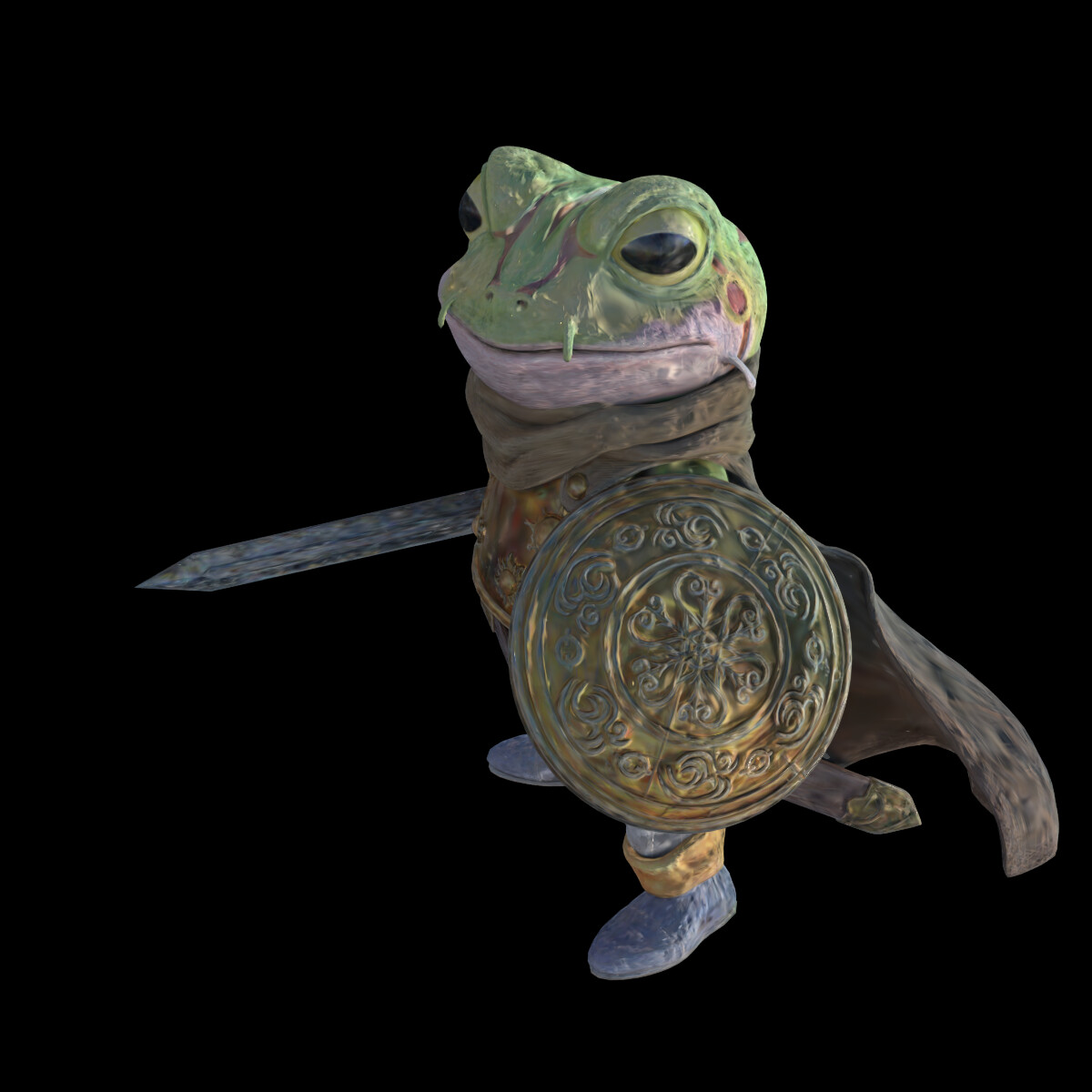}
        &
        \multicolumn{3}{c}{\multirow{2}{*}[0.945\heightRerender]{\includegraphics[draft=false, height=2.04\heightRerender, frame]{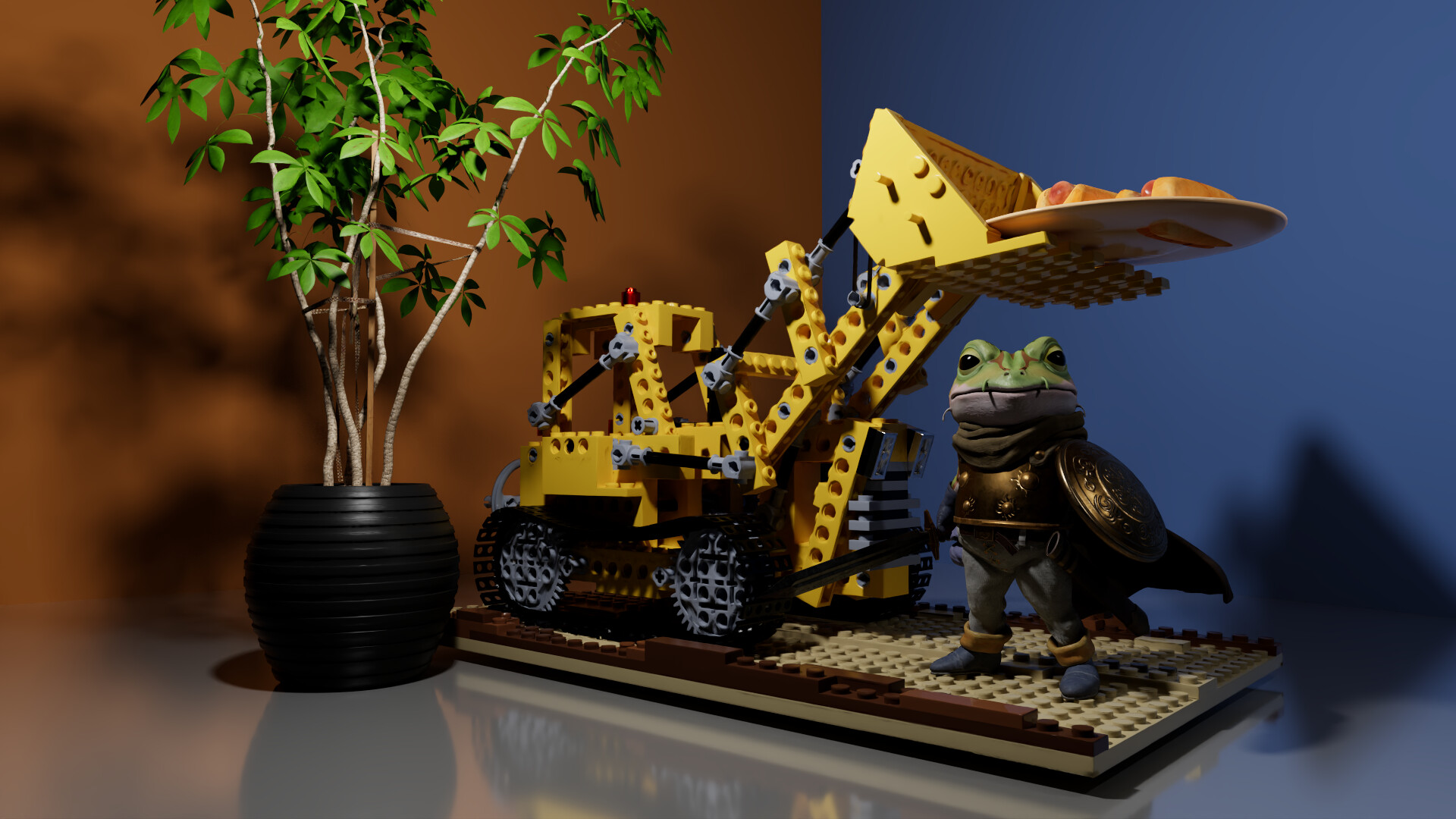}}}       
        \\
        \includegraphics[draft=false, height=\heightRerender, frame]{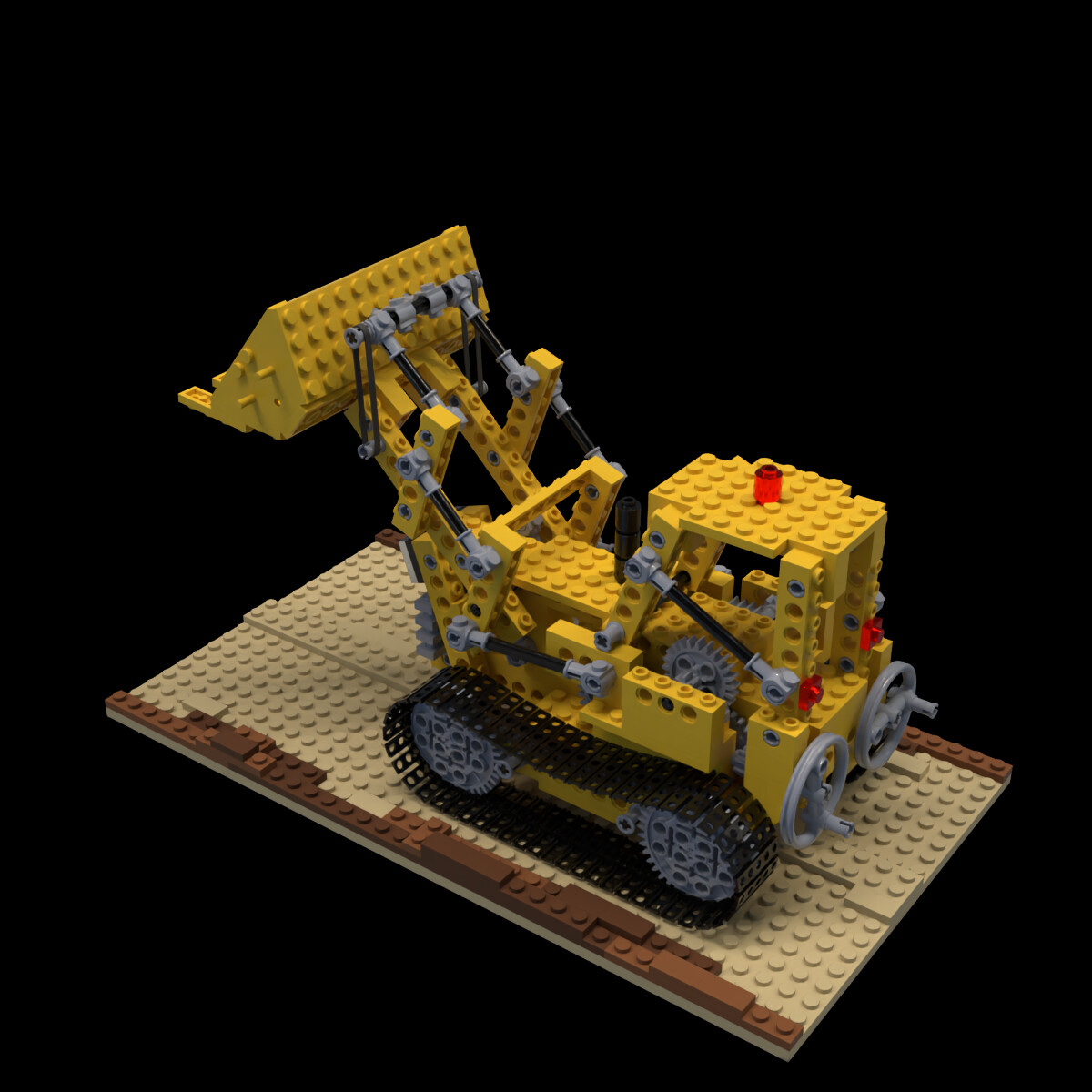}
        &
        \includegraphics[draft=false, height=\heightRerender, frame]{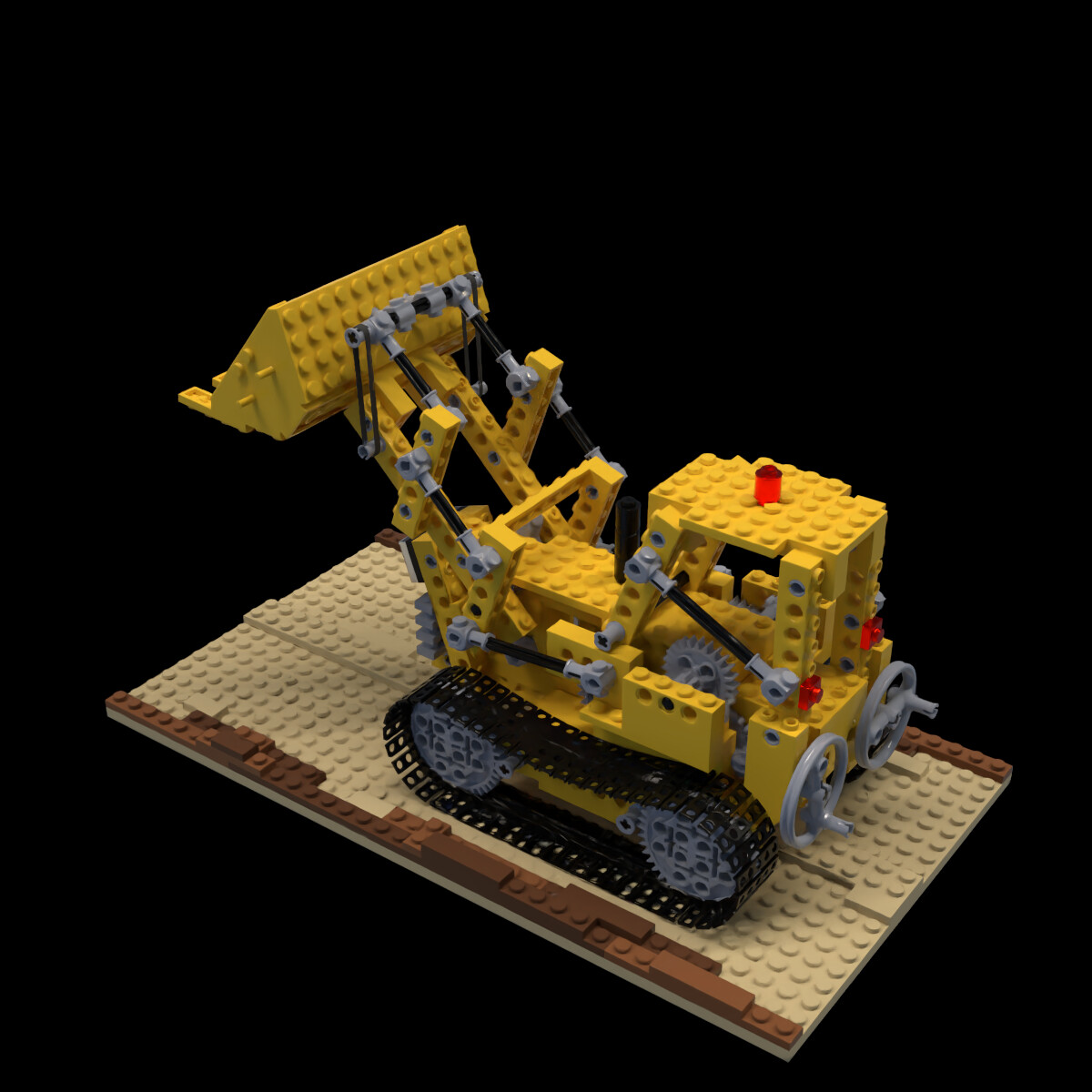}
        &
        \includegraphics[draft=false, height=\heightRerender, frame]{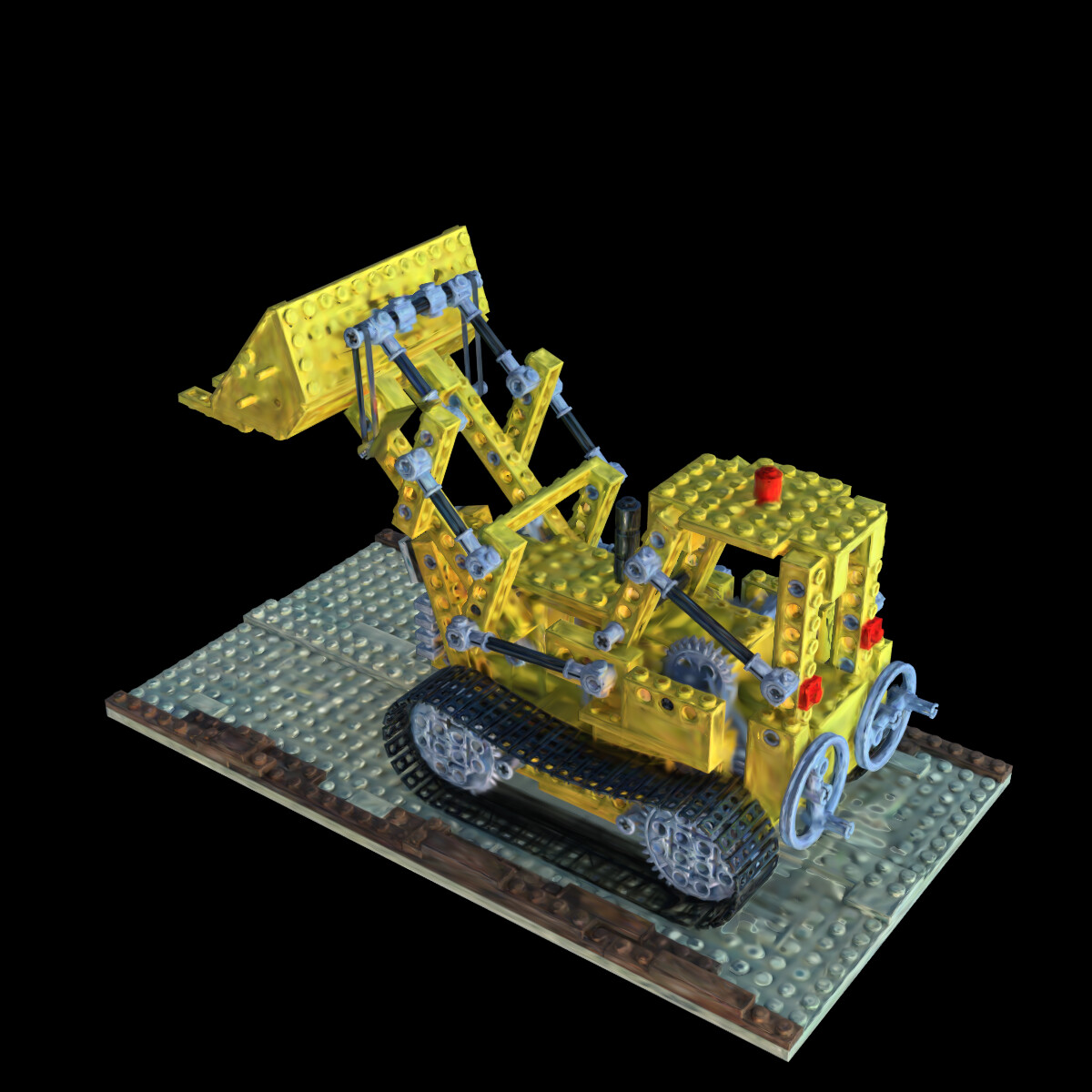}
        &
        \multicolumn{3}{c}{}
        \\
        \includegraphics[draft=false, height=\heightRerender, frame]{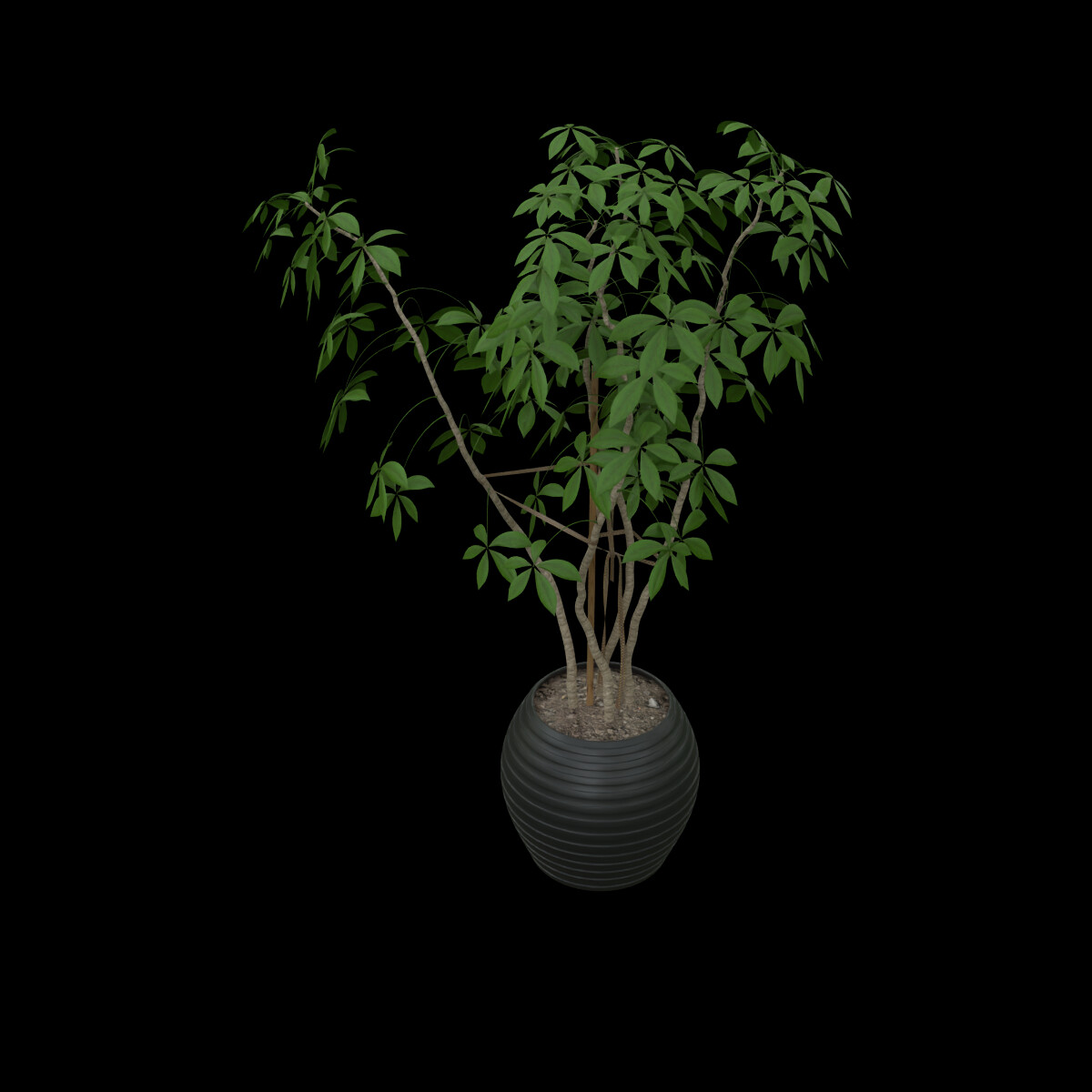}
        &
        \includegraphics[draft=false, height=\heightRerender, frame]{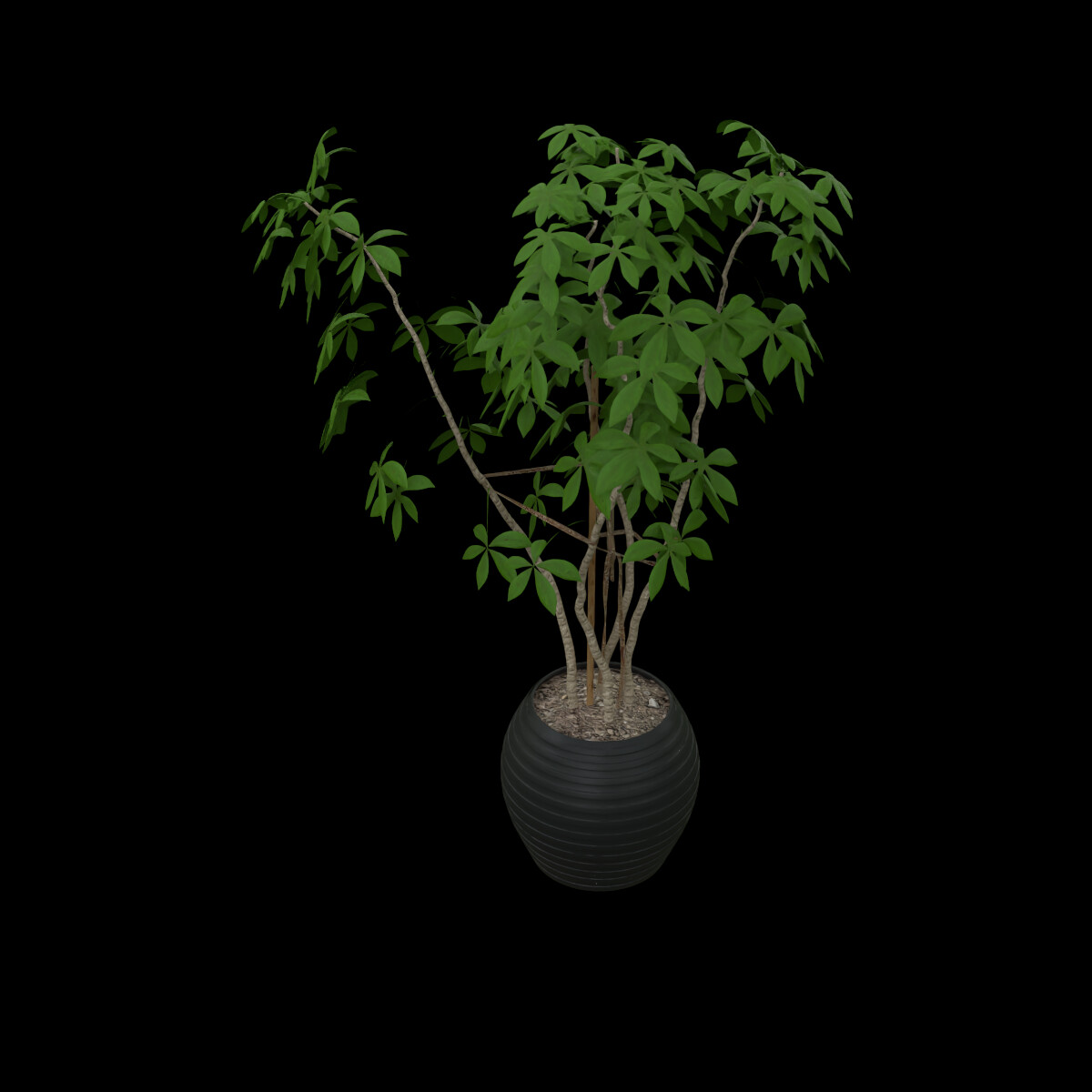}
        &
        \includegraphics[draft=false, height=\heightRerender, frame]{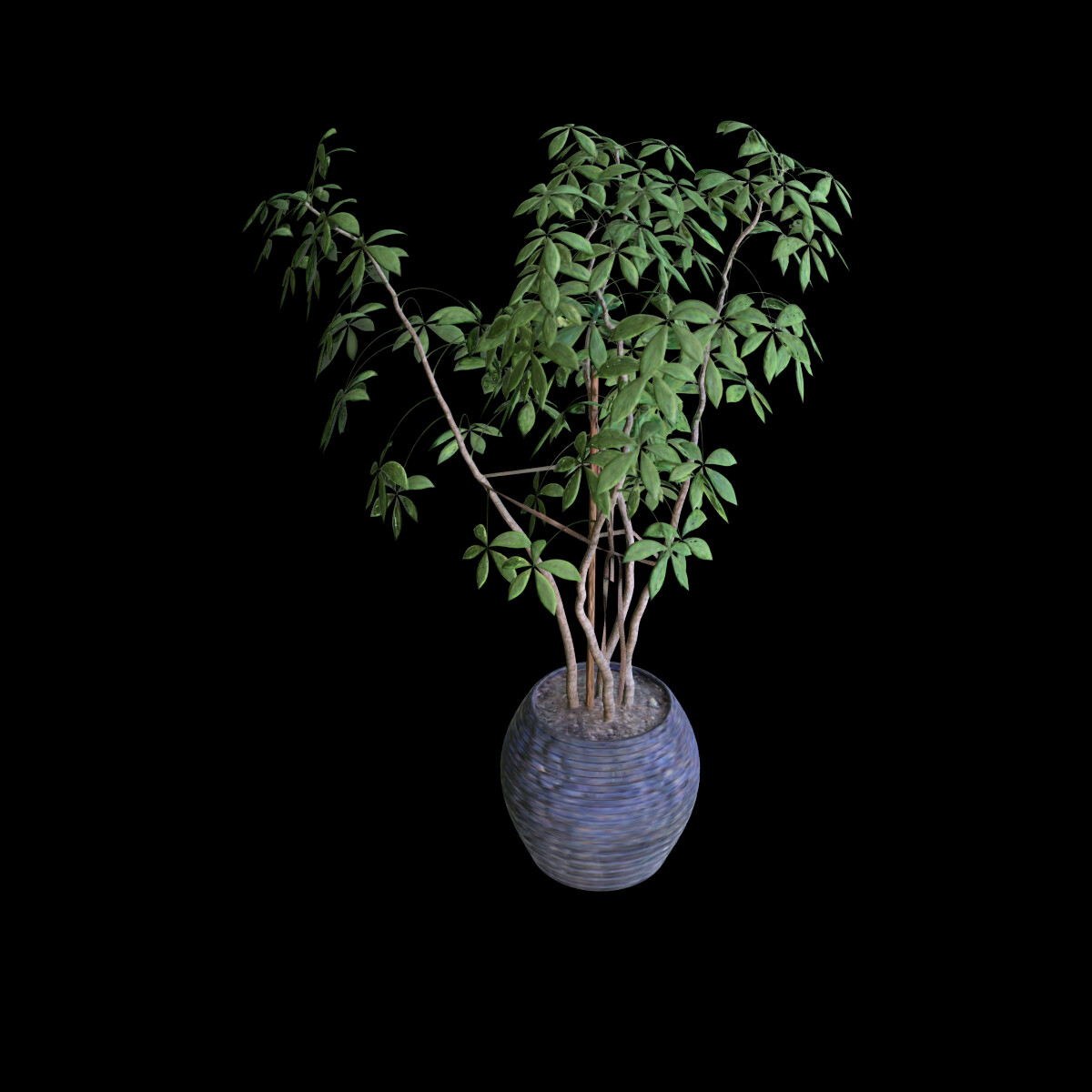}
        &
        \multicolumn{3}{c}{\multirow{2}{*}[0.945\heightRerender]{\includegraphics[draft=false, height=2.04\heightRerender, frame]{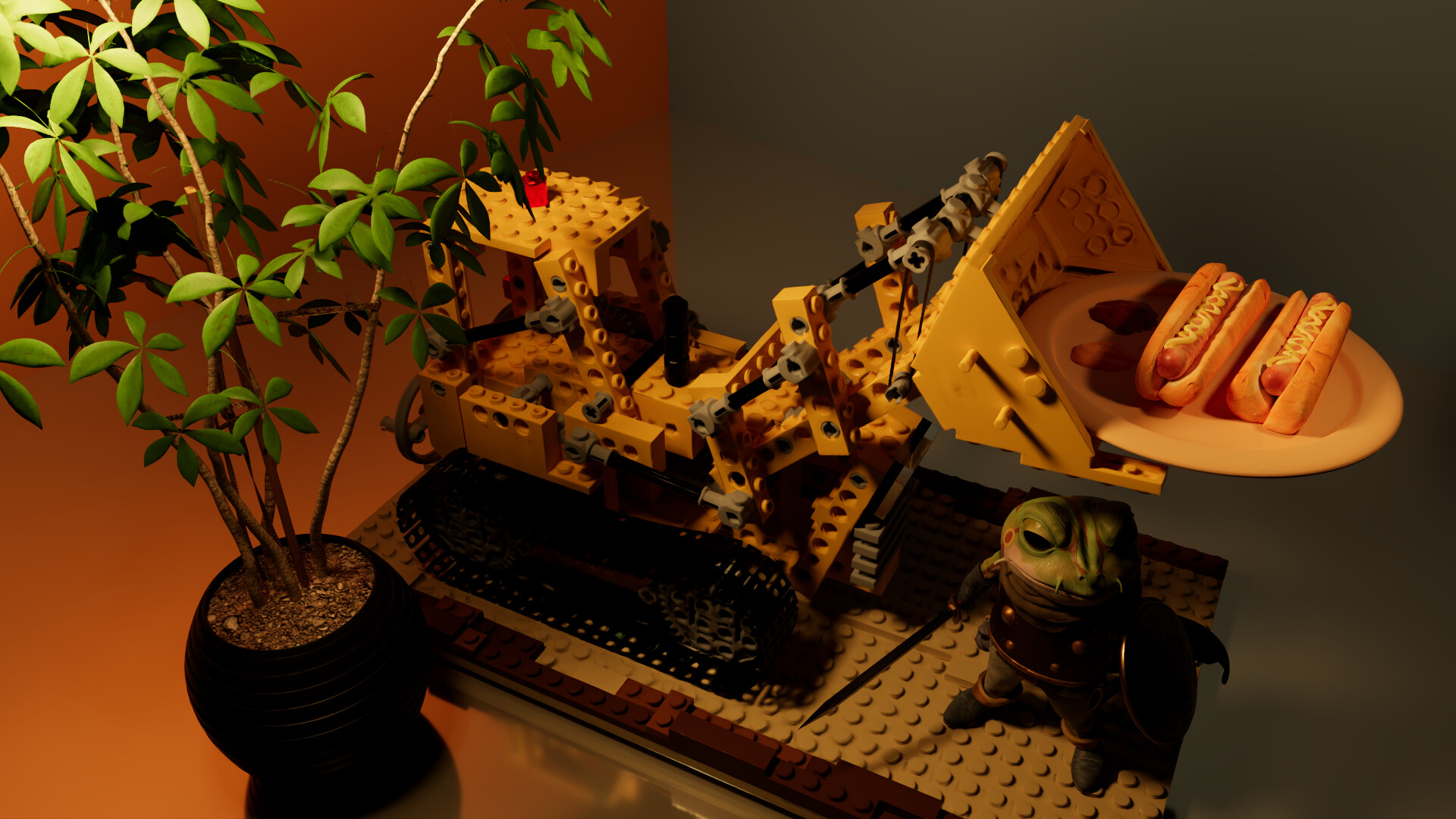}}}       
        \\
        \includegraphics[draft=false, height=\heightRerender, frame]{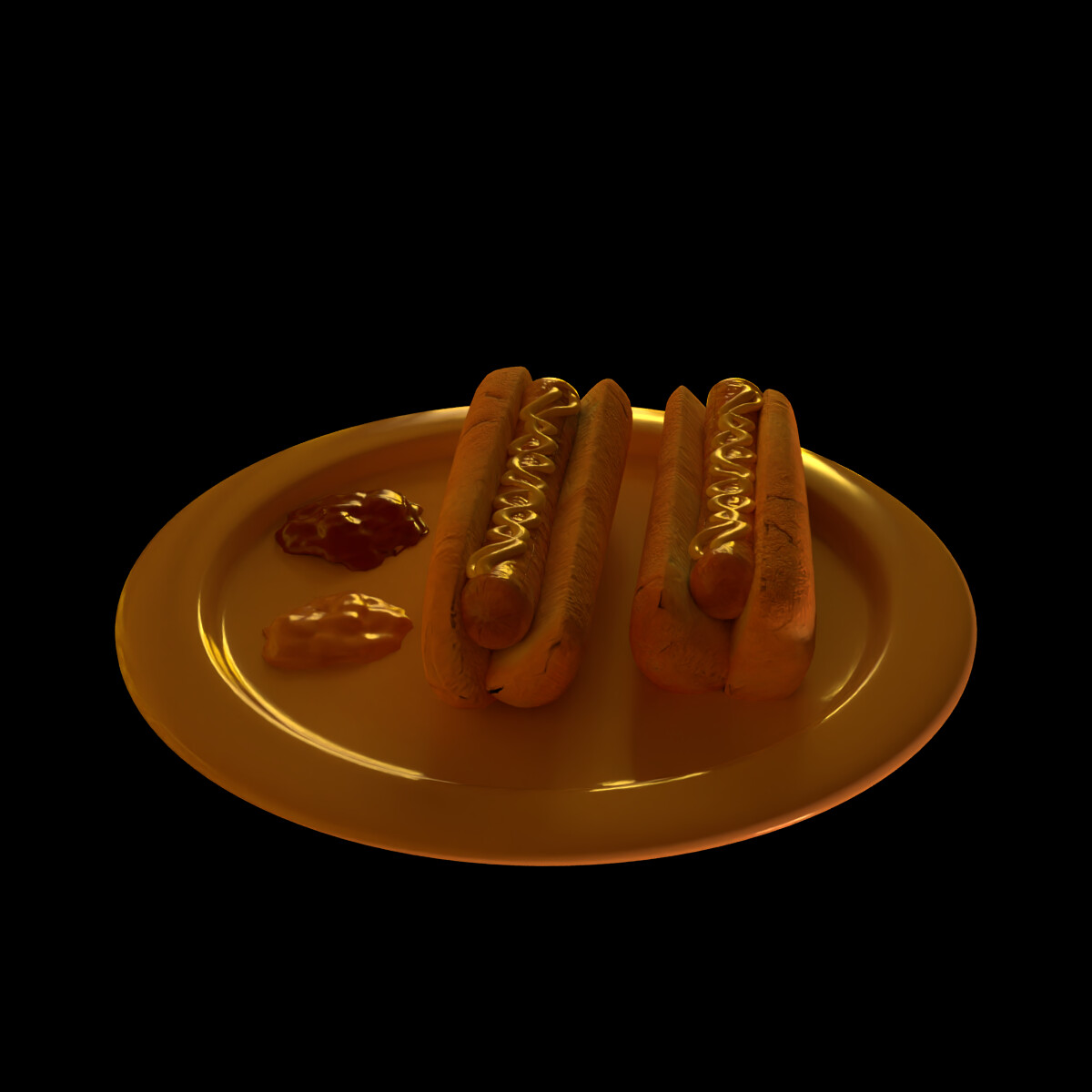}
        &
        \includegraphics[draft=false, height=\heightRerender, frame]{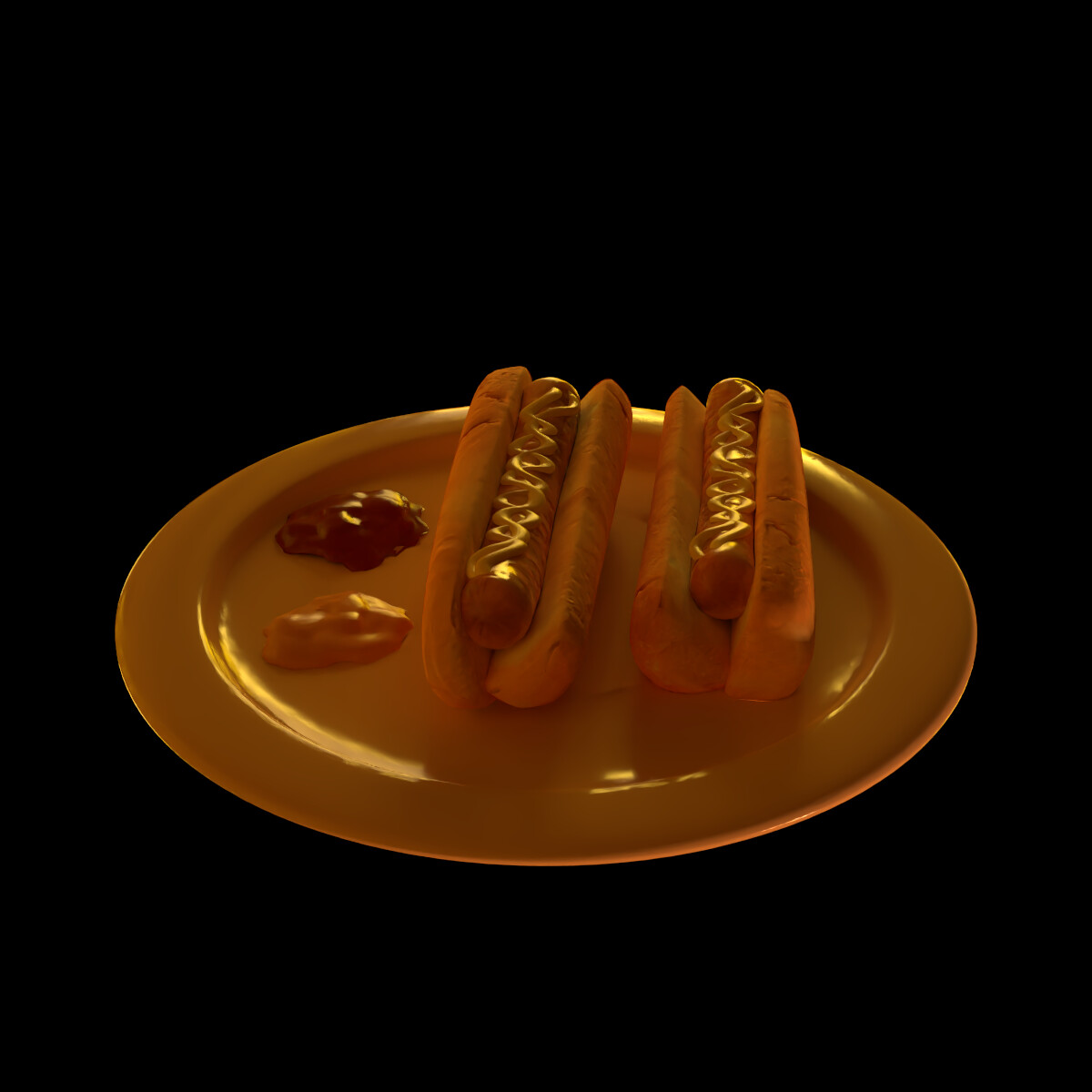}
        &
        \includegraphics[draft=false, height=\heightRerender, frame]{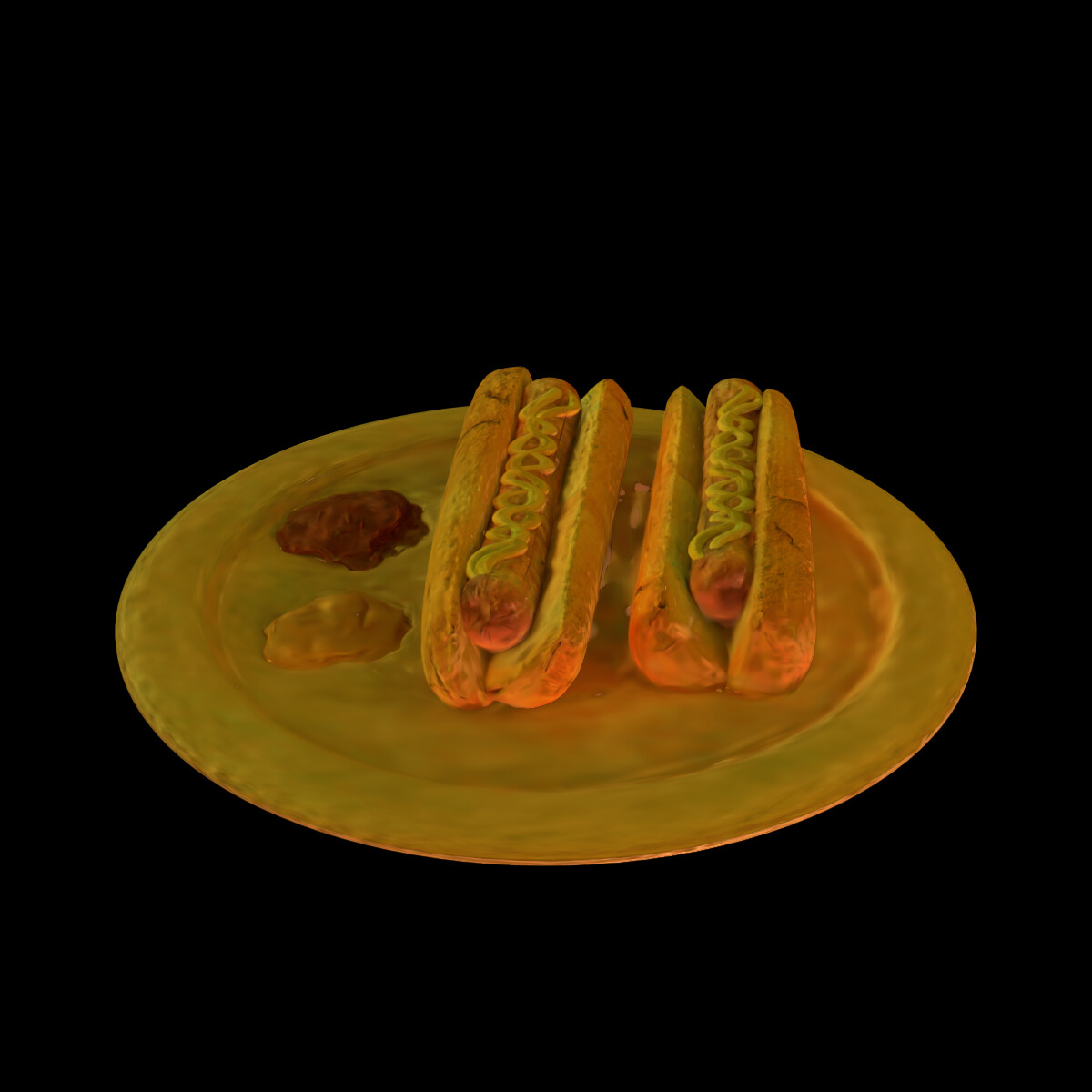}
        &
        \multicolumn{3}{c}{}                    
    \end{tabular}
    \caption{\label{fig:re-render}
        Re-rendering comparison between our method and GS-IR~\citep{cvpr/LiangZFSJ24}. Our results closely match the 
        references, whereas GS-IR fails to remove baked-in lighting and to reproduce important shading effects. We further 
        demonstrate accurate scene-level global illumination (``Composition''). See \autoref{table:re-render_accuracy} for
        quantitative metrics.        
    }
\end{figure*}

We demonstrate that current inverse rendering-based relightable representations are still far from reaching the acceptable 
quality bar for production rendering. On the other hand, our representation acquired through conversion 
(\autoref{sec:data_acquisition}) can readily reach high visual fidelity on par with the original. In \autoref{fig:re-render} and 
\autoref{table:re-render_accuracy}, we conduct comparisons between our method and GS-IR~\citep{cvpr/LiangZFSJ24}, a 
representative Gaussian-based relighting method, on the TensoIR synthetic dataset~\citep{cvpr/JinLXZHBZX023} augmented 
with one more model (\emph{Sir Frog}) that features greater texture and material complexity. Following the original 
protocol, GS-IR is trained with multi-view observations lit by the same environment light, and the metrics of both 
methods are calculated over 5 other environment lights. As shown in the figure, GS-IR fails to remove the baked-in 
training-time lighting, resulting in poor overall shading quality. In contrast, our results faithfully 
reproduce all-frequency effects, including glossy reflections and soft shadows. 
Beyond per-object renders, we further demonstrate high-fidelity, scene-level global 
illumination: the right panel of \autoref{fig:re-render} shows a composed scene 
rendered with multiple sets of area lights. This scene exhibits non-local effects 
such as inter-reflections and shadows cast from one object onto another -- all 
made possible only by accurate light transport simulation and beyond the capability 
of GS-IR.

\begin{figure*}[tb]
	\newlength{\lenGSLit}
	\setlength{\lenGSLit}{0.32\linewidth}
    \addtolength{\tabcolsep}{-4.5pt}
    \renewcommand{\arraystretch}{0.5}
    \centering
    \begin{tabular}{cccc}
        \raisebox{35pt}{\rotatebox{90}{\small{\emph{Garden}}}}
        &
        \includegraphics[width=\lenGSLit]{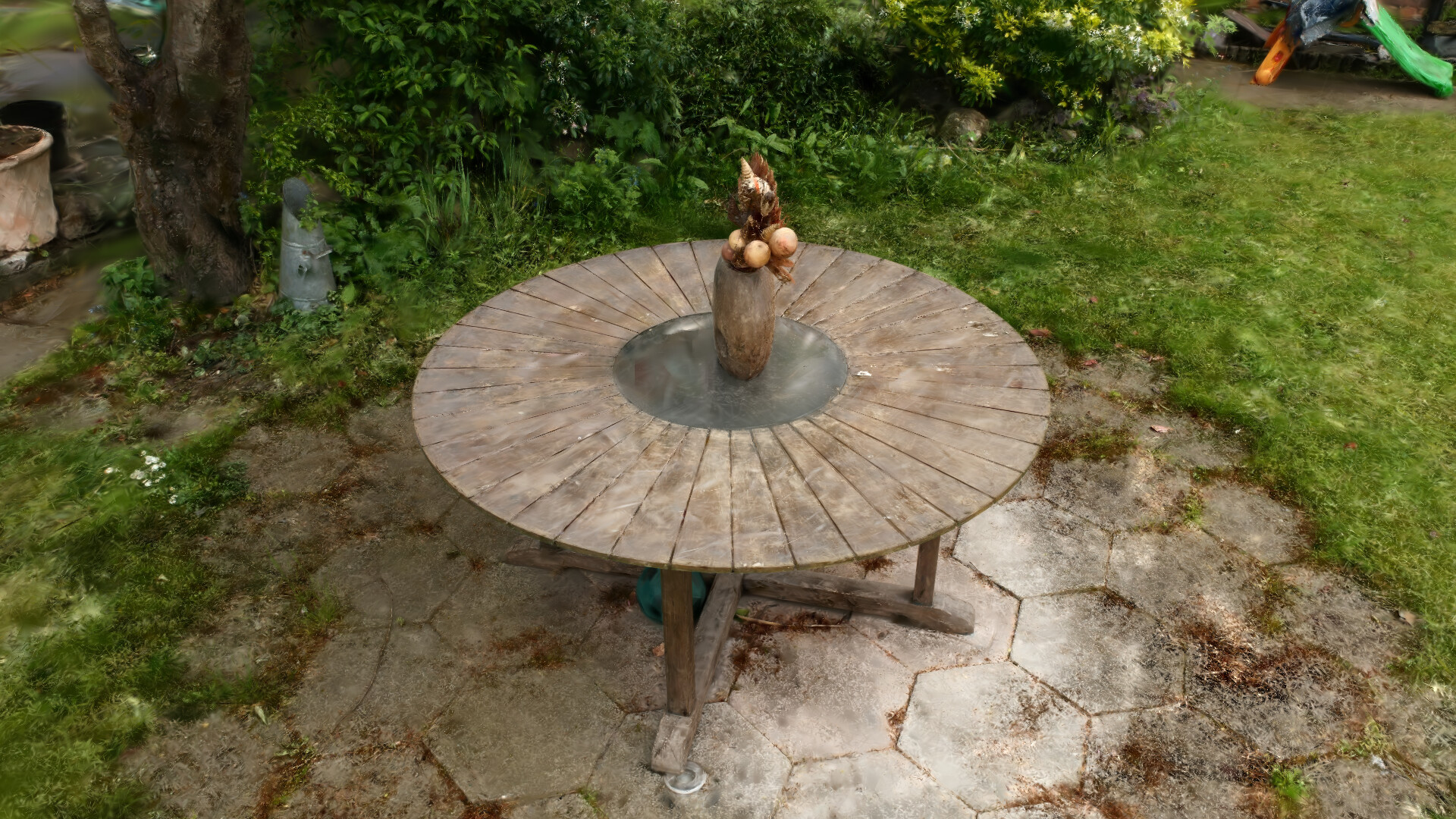}
        &
        \includegraphics[width=\lenGSLit]{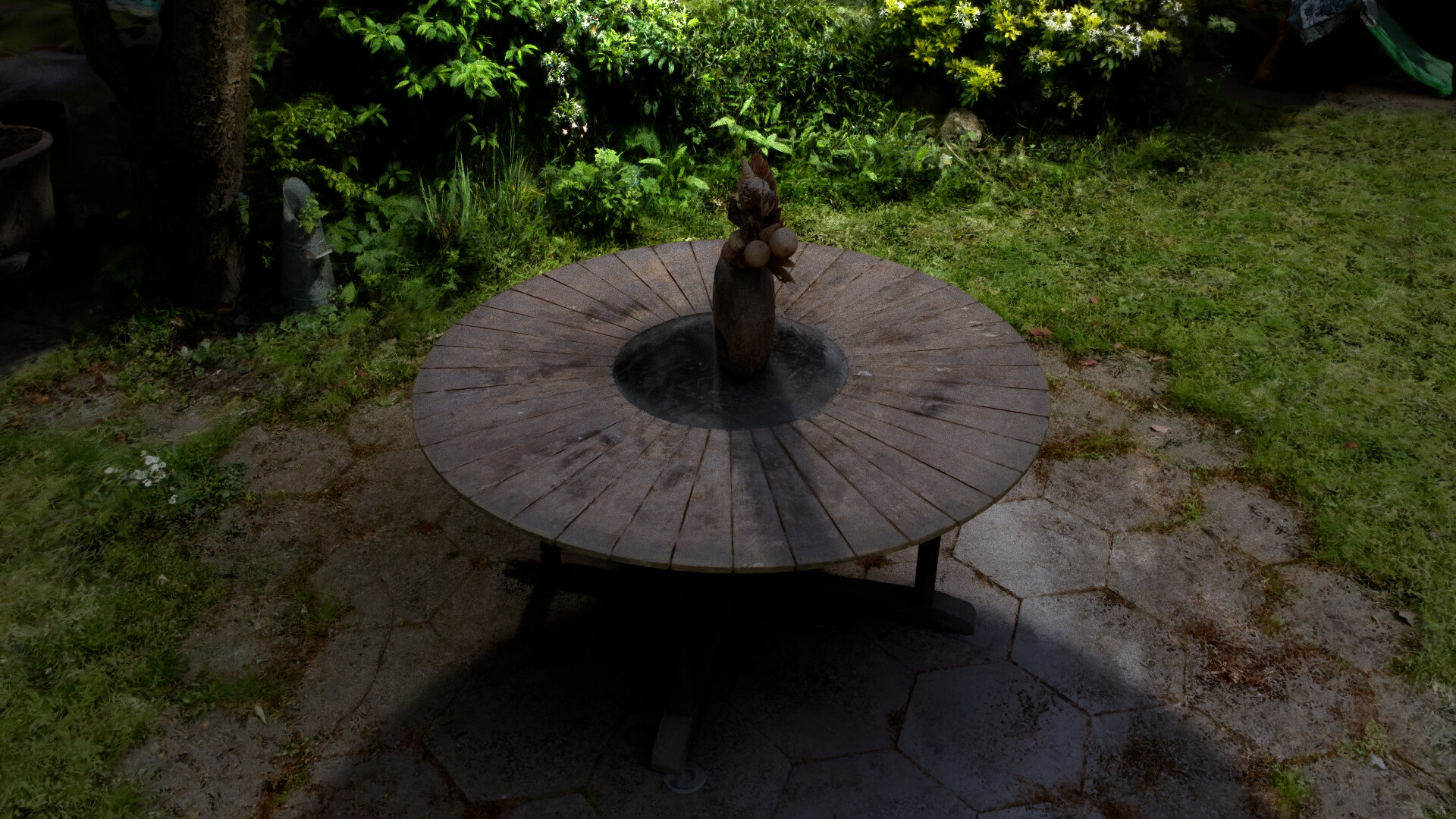}
        &
        \includegraphics[width=\lenGSLit]{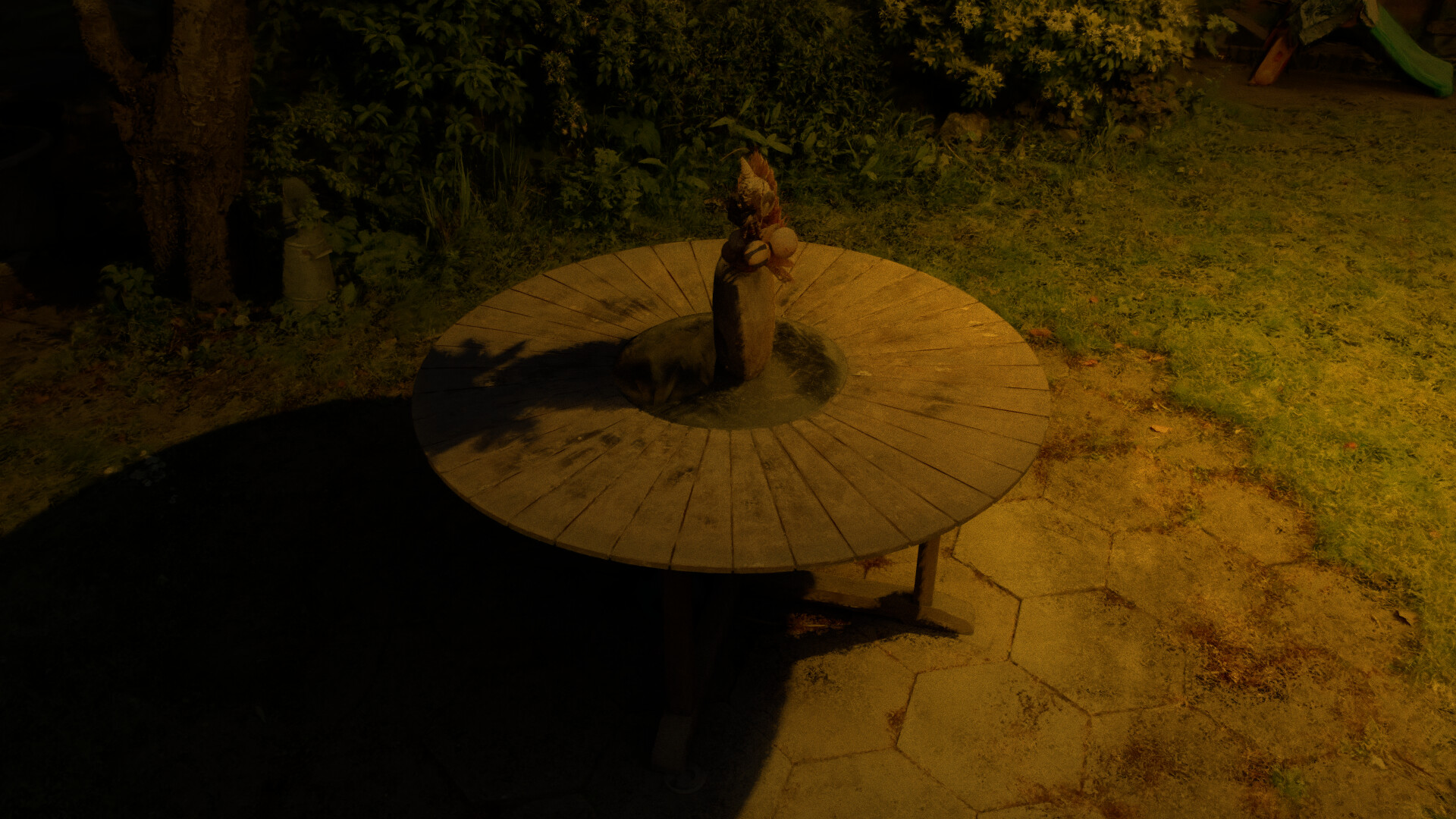}
        \\        
        \raisebox{35pt}{\rotatebox{90}{\small{\emph{Stump}}}}
        &
        \includegraphics[width=\lenGSLit]{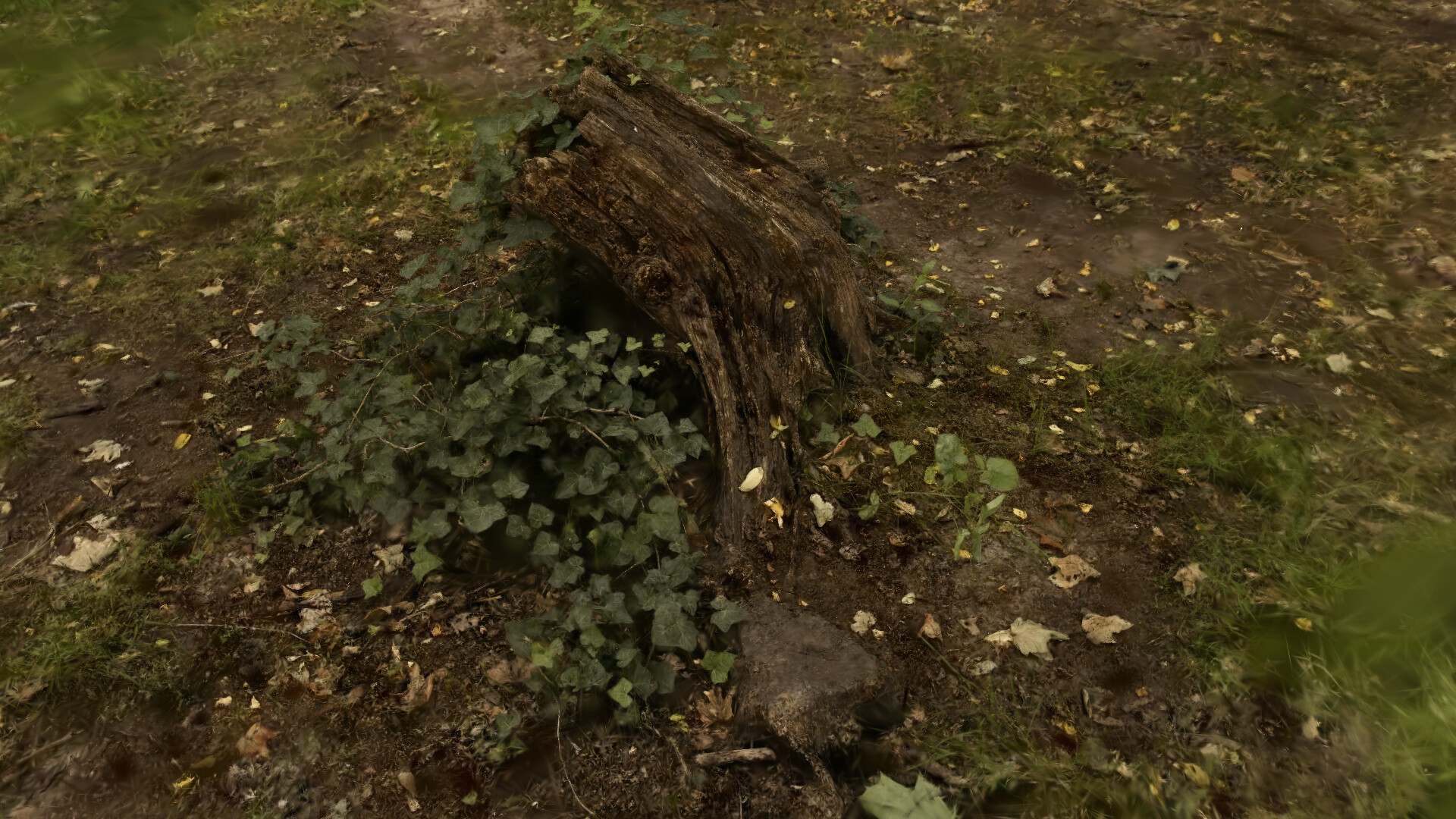}
        &
        \includegraphics[width=\lenGSLit]{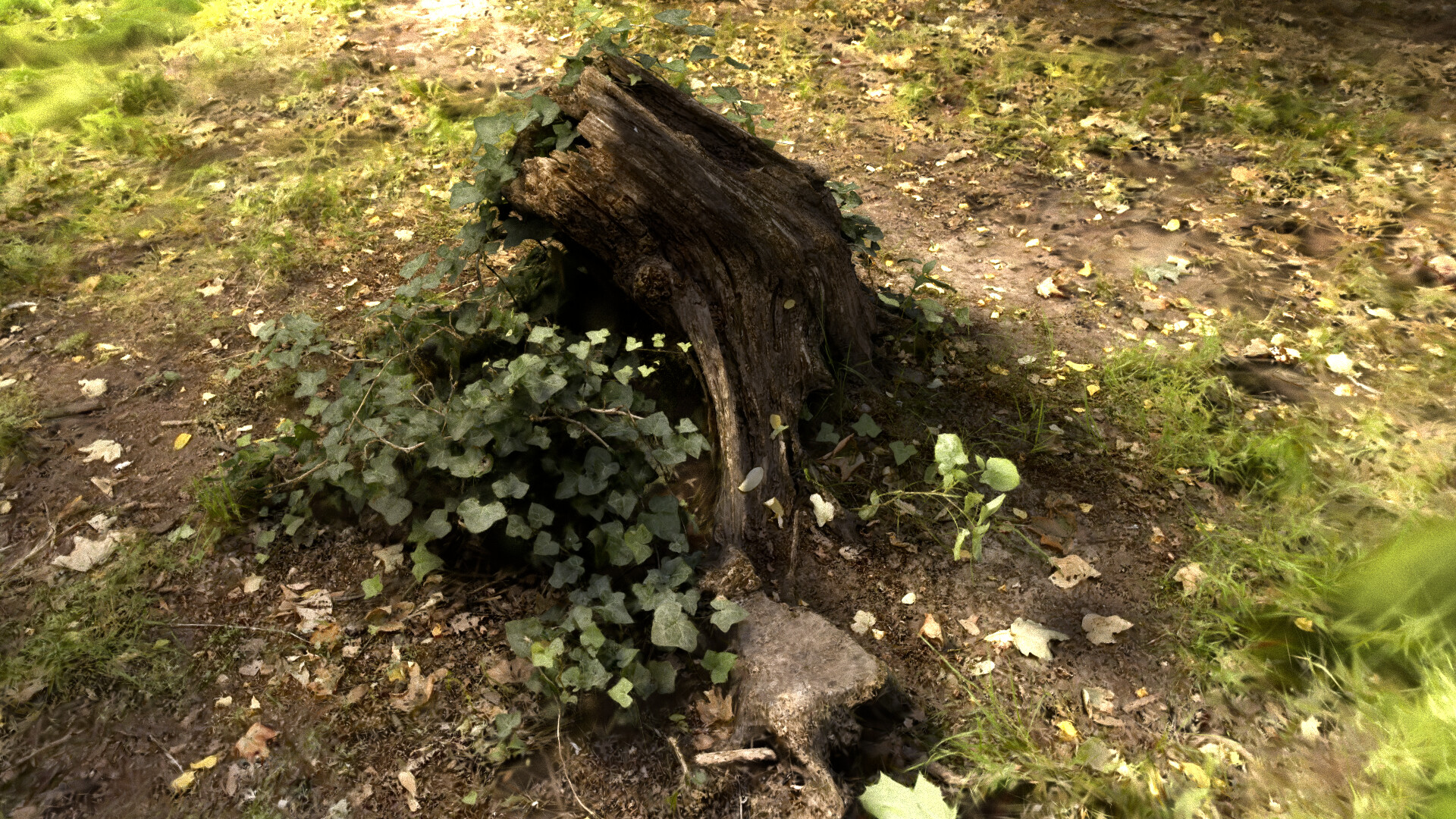}
        &
        \includegraphics[width=\lenGSLit]{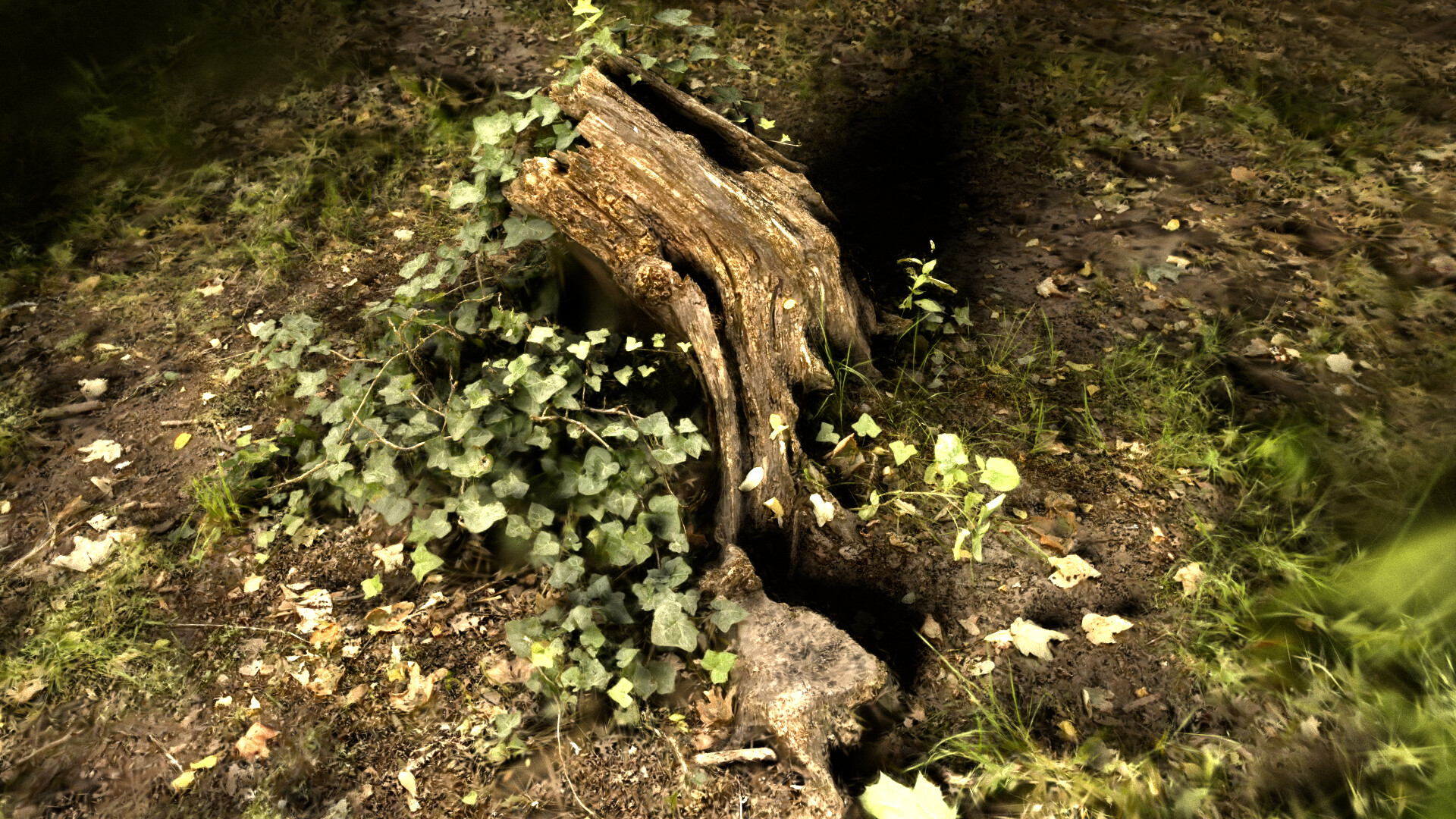}
        \\
        &
        \small{\textsf{(a) Original}}
        &
        \small{\textsf{(b) Lighting 1}}
        &
        \small{\textsf{(c) Lighting 2}}
    \end{tabular}
    \caption{\label{fig:gs_lit}
        3DGS scenes trained from the Mip-NeRF 360 dataset ((a)) empirically converted into our representation and rendered with different lighting conditions ((b) and (c)).
        Note the soft shadows introduced by the new light sources that are not present in the original radiance fields.
    }
\end{figure*}

For real-world scenes, our method cannot simulate light transport without ground truth 3D information. 
Nonetheless, we demonstrate compatibility by performing an empirical conversion as described in \autoref{sec:data_acquisition}. 
The converted scenes can then be rendered under arbitrary lighting conditions, as demonstrated in \autoref{fig:gs_lit}.
Both the \emph{Garden} and the \emph{Stump} scenes created from the Mip-NeRF 360 dataset are rendered with new area lights and environment lights, producing
plausible lighting and soft shadows. Additional rendering sequences are provided in the supplemental video.

\begin{table}
	\centering
	\caption{
        Re-rendering accuracy on the augmented TensoIR dataset. We report common metrics averaged over the testing 
        set of all scenes and all environment lights.
    }
	\begin{tabular}{lllll}
		\Xhline{1pt}
		\textbf{Method} & \textbf{PSNR}$\uparrow$ & \textbf{SSIM}$\uparrow$ & \textbf{LPIPS}$\downarrow$ & \textbf{\FLIP}$\downarrow$\\
        \hline
        Ours & 35.31 & 0.970 & 0.045 & 0.033\\        
        GS-IR & 22.37 & 0.890 & 0.100 & 0.137\\
		\Xhline{1pt}
	\end{tabular}
	\label{table:re-render_accuracy} 
\end{table}


\begin{figure}[h]
	\newlength{\lenLeaf}
	\setlength{\lenLeaf}{0.48\linewidth}
    \addtolength{\tabcolsep}{-4pt}
    \renewcommand{\arraystretch}{0.5}
    \centering
    \begin{tabular}{cc}
      \includegraphics[width=\lenLeaf]{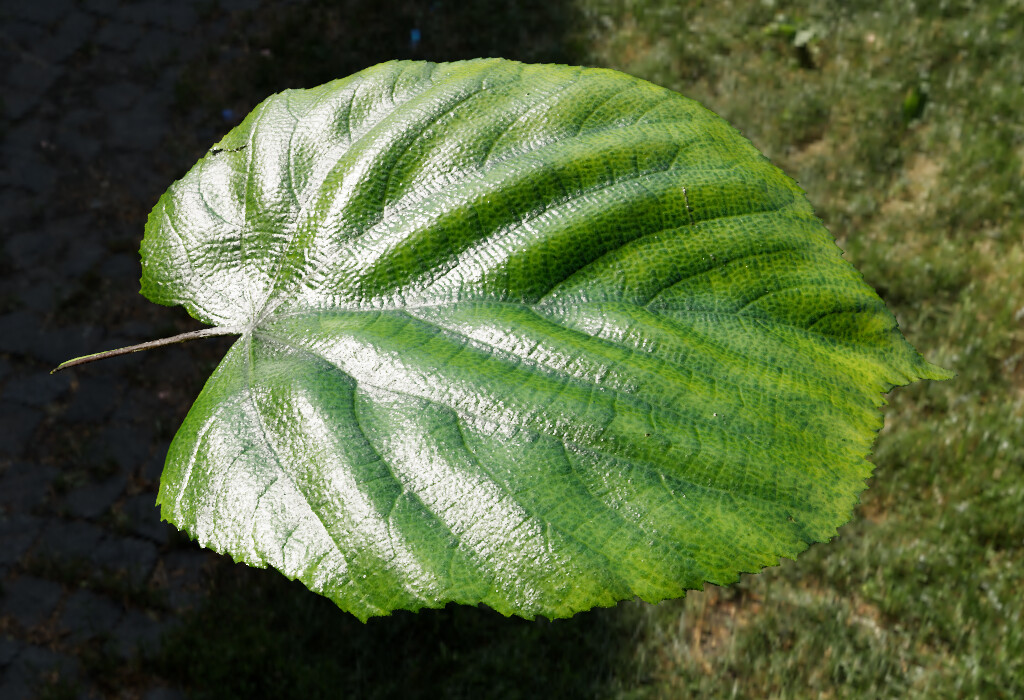}
      &
      \includegraphics[width=\lenLeaf]{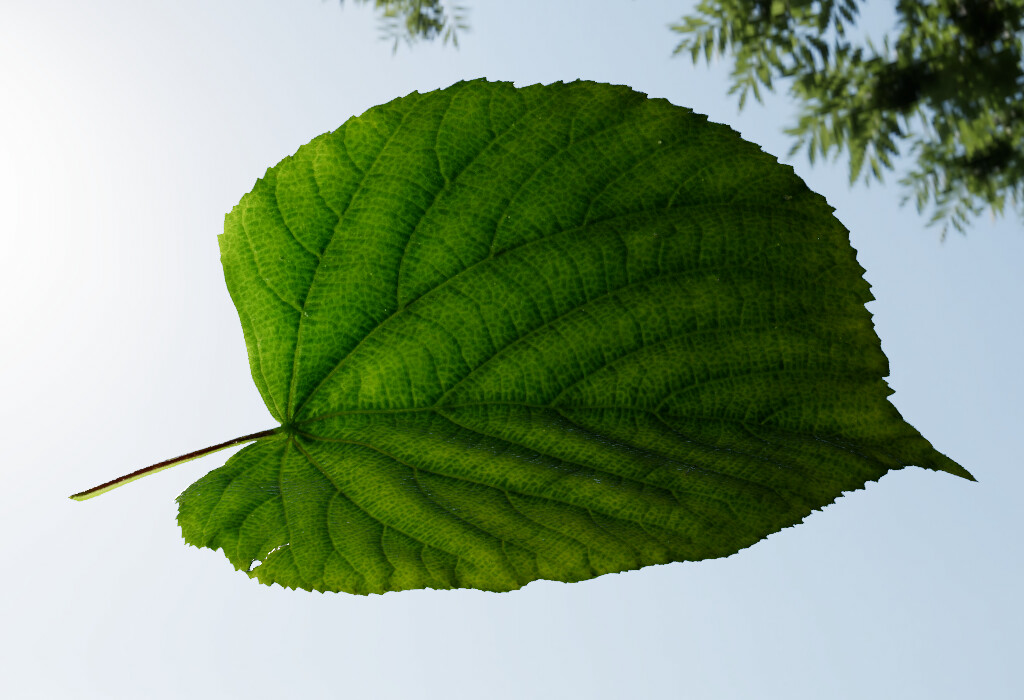}
      \\
      \small{\textsf{Front view}}
      &
      \small{\textsf{Back view}}
    \end{tabular}
    \caption{\label{fig:leaf}
        Our phase function is compatible with different base BSDFs.
        Here we show a leaf represented by our primitives and a translucent thin-surface base BSDF.
    }
\end{figure}

\paragraph{Radiance Fields}
We provide more implementation details of our radiance field reconstruction.
Other than losses and initialization (\autoref{sec:radiance_field}), our optimization scheme is overall 
similar to that of 3DGS. Following EDGS~\citep{kotovenko2025edgs}, we no longer perform densification during optimization, but only pruning. 
We employ the Adam optimizer~\citep{kingma2014adam} and train on one randomized view per iteration. 
Learning rates follow those of 3DGS training, with the exception of the magnitude parameter, for which we use a smaller value of $0.005$. 
The blending $\beta$ term (\autoref{eq:rf_full_loss}) is linearly annealed from $0$ to $1$ over the first 2000 iterations. 
We note that hyperparameters were kept constant across all scenes without per-scene tuning, which may offer opportunities for further quality improvements.

\begin{figure*}[tb]
	\newlength{\lenRFCompare}
	\setlength{\lenRFCompare}{0.19\linewidth}
    \addtolength{\tabcolsep}{-5pt}
    \renewcommand{\arraystretch}{0.5}

    \centering
    \begin{tabular}{cccccc}
        & \emph{Chair} & \emph{Drums} & \emph{Ficus} & \emph{Lego} & \emph{Ship} \\
        \raisebox{35pt}{\rotatebox{90}{\footnotesize{\textsf{Reference}}}}
        &
        \begin{overpic}[width=\lenRFCompare]{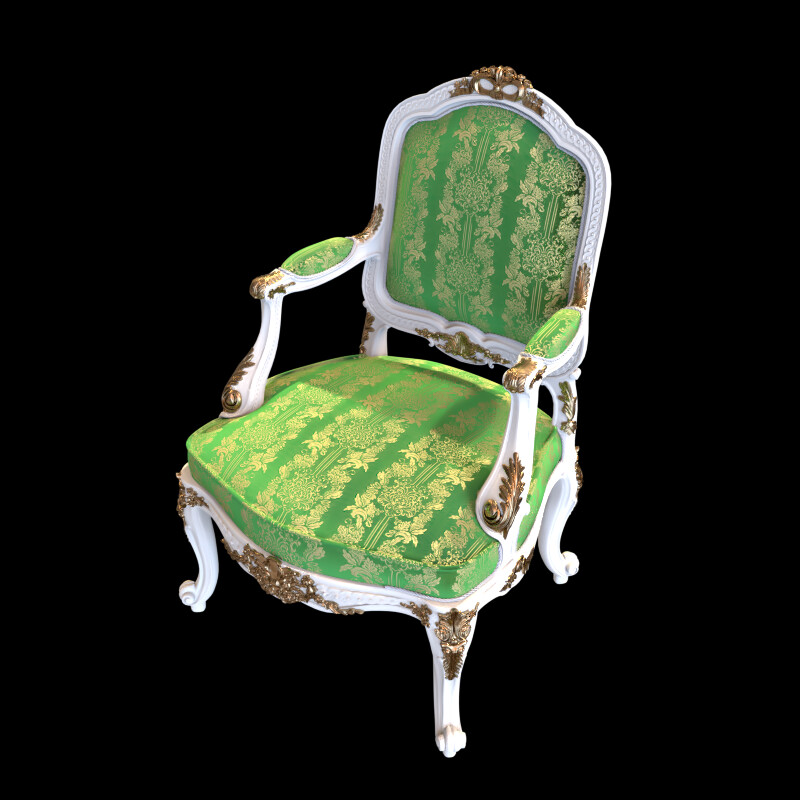}\end{overpic}
        &
        \begin{overpic}[width=\lenRFCompare]{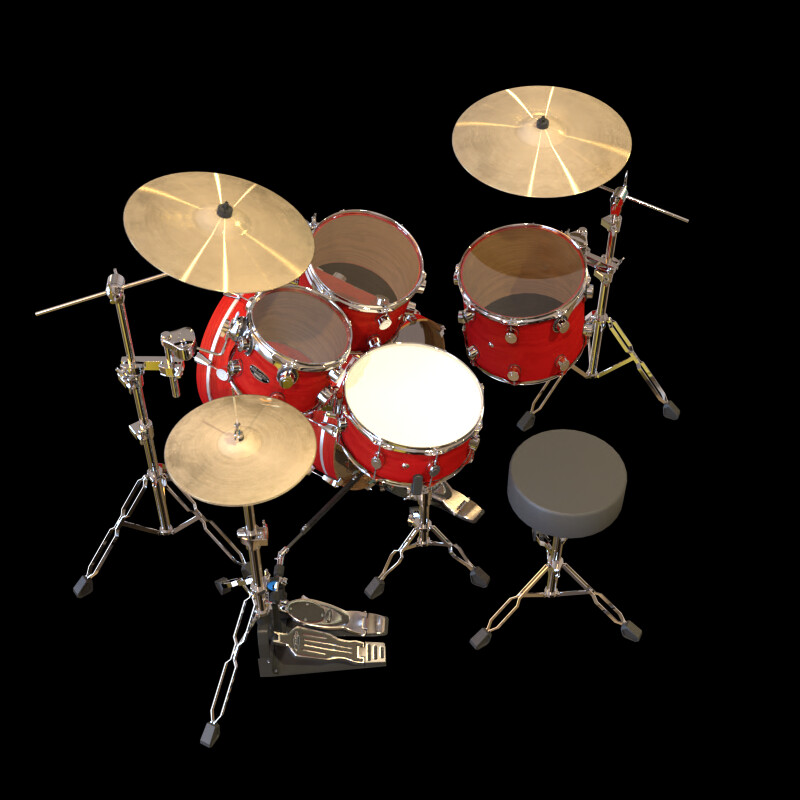}\end{overpic}
        &
        \begin{overpic}[width=\lenRFCompare]{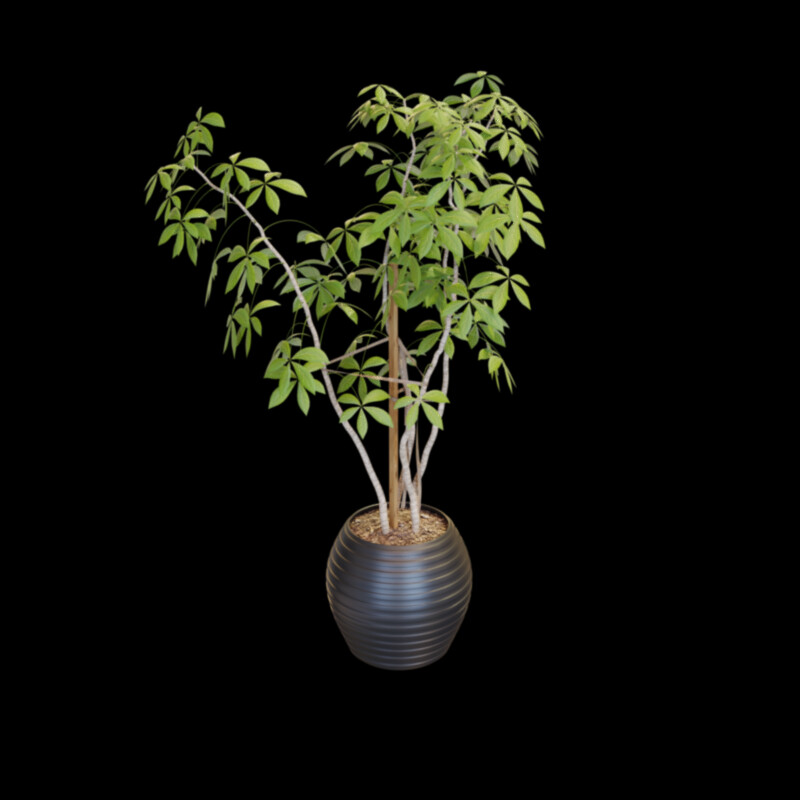}\end{overpic}
        &
        \begin{overpic}[width=\lenRFCompare]{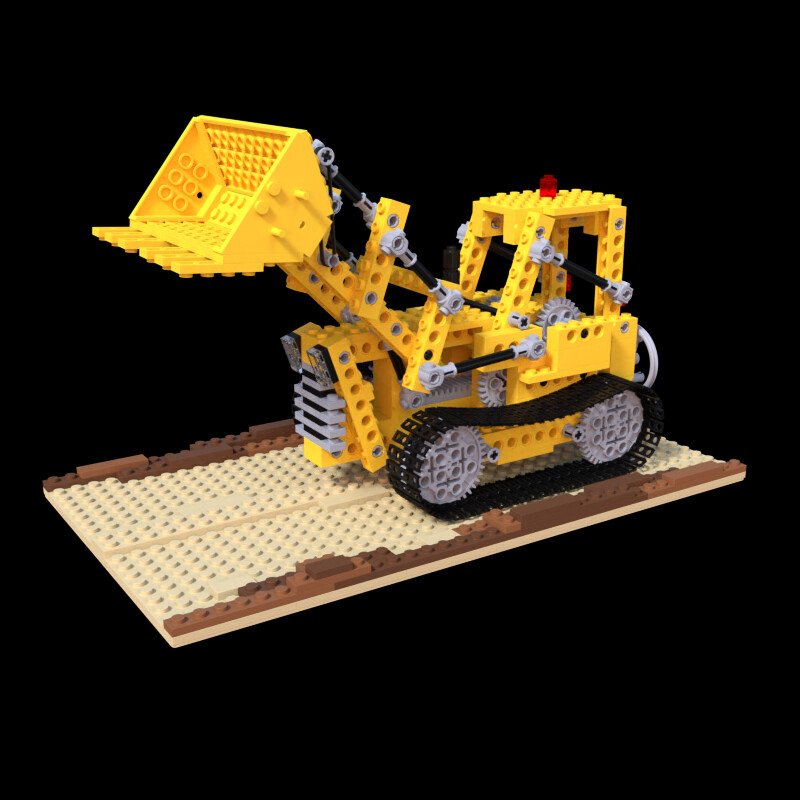}\end{overpic}
        &
        \begin{overpic}[width=\lenRFCompare]{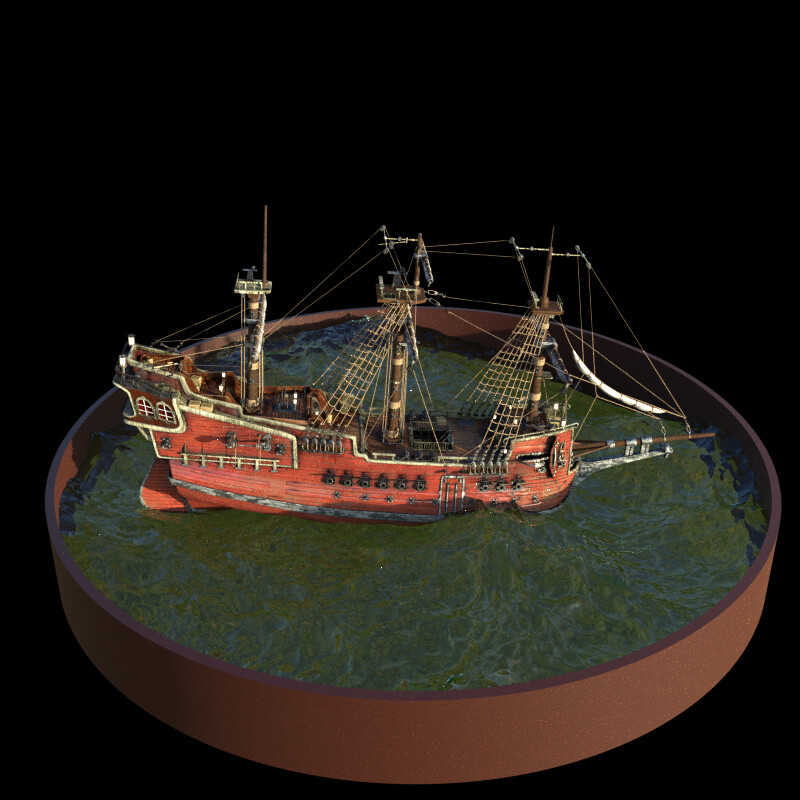}\end{overpic}
        \\
        \raisebox{40pt}{\rotatebox{90}{\footnotesize{\textsf{Ours}}}}
        &        
        \begin{overpic}[width=\lenRFCompare]{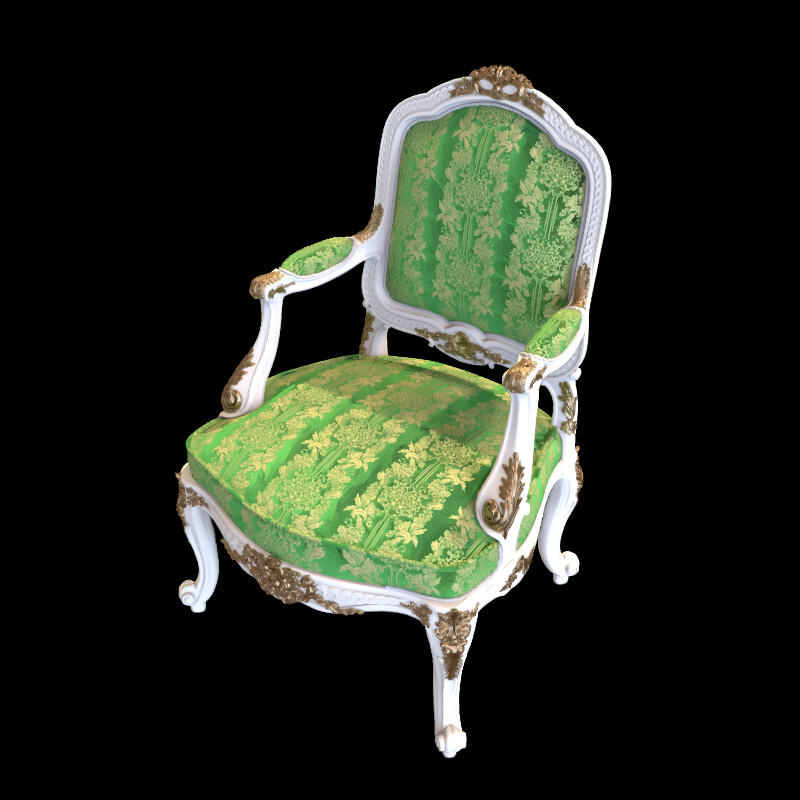}
            \put(68, 0.5){%
            \setlength{\fboxrule}{0.5pt}  
            \setlength{\fboxsep}{0pt}   
            \fcolorbox{white}{white}{\includegraphics[width=0.05\textwidth]{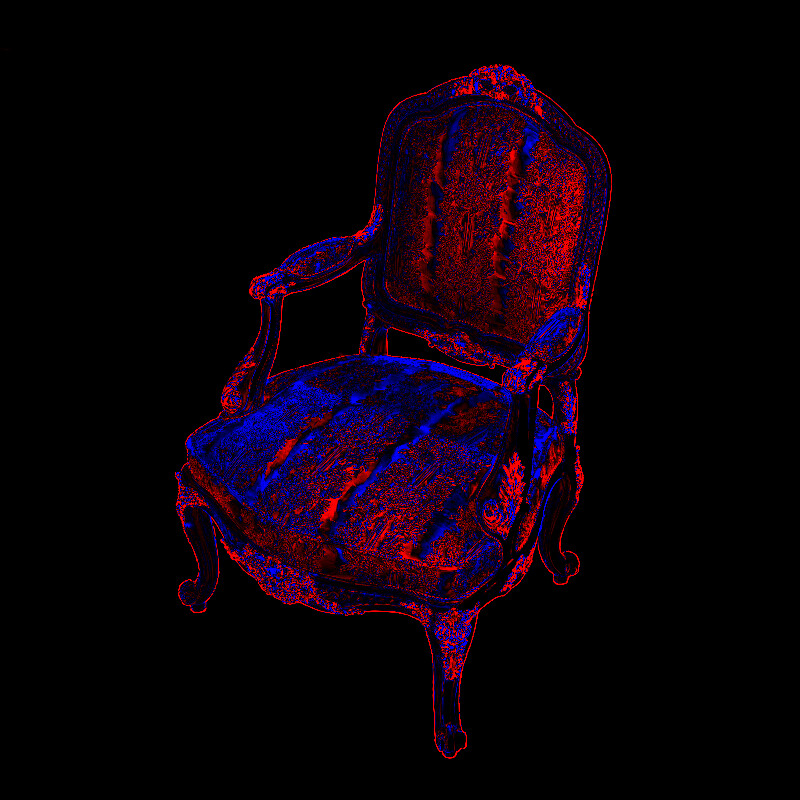}}
            }            
        \end{overpic}
        &
        \begin{overpic}[width=\lenRFCompare]{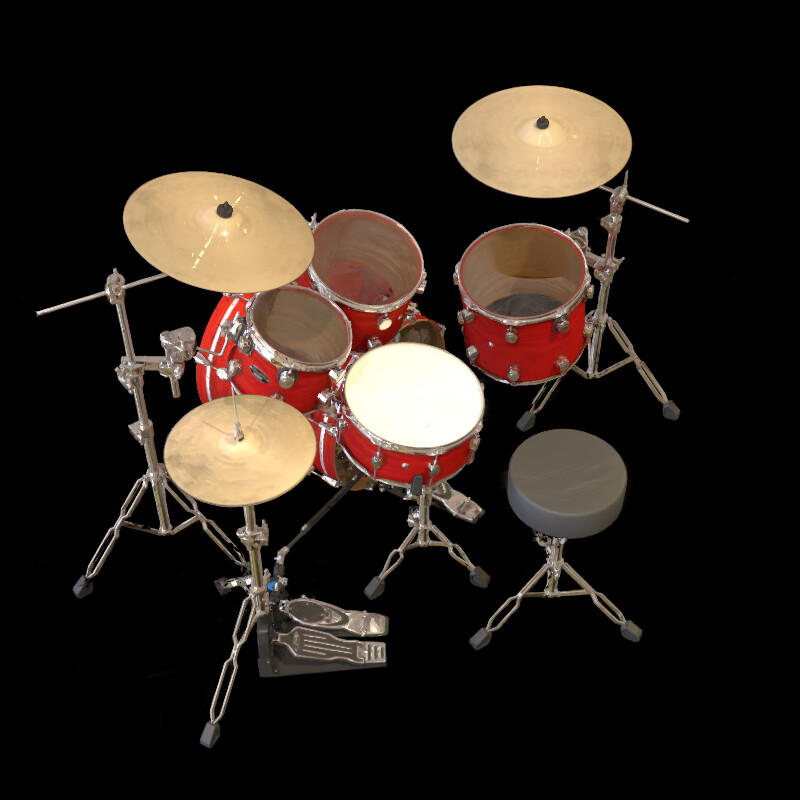}\put(68, 0.5){%
            \setlength{\fboxrule}{0.5pt}  
            \setlength{\fboxsep}{0pt}   
            \fcolorbox{white}{white}{\includegraphics[width=0.05\textwidth]{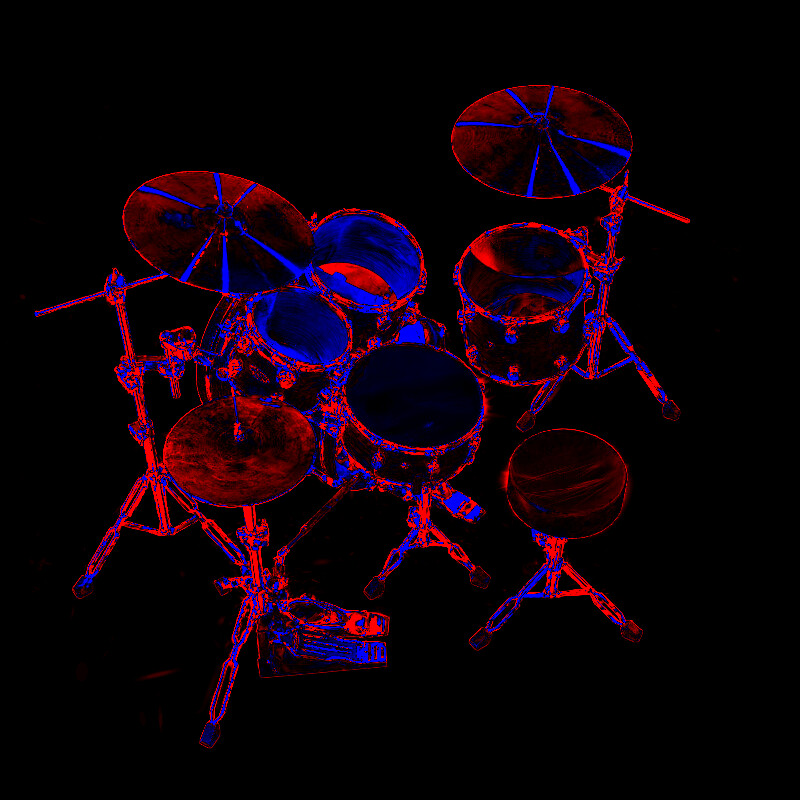}}
            }            
        \end{overpic}
        &
        \begin{overpic}[width=\lenRFCompare]{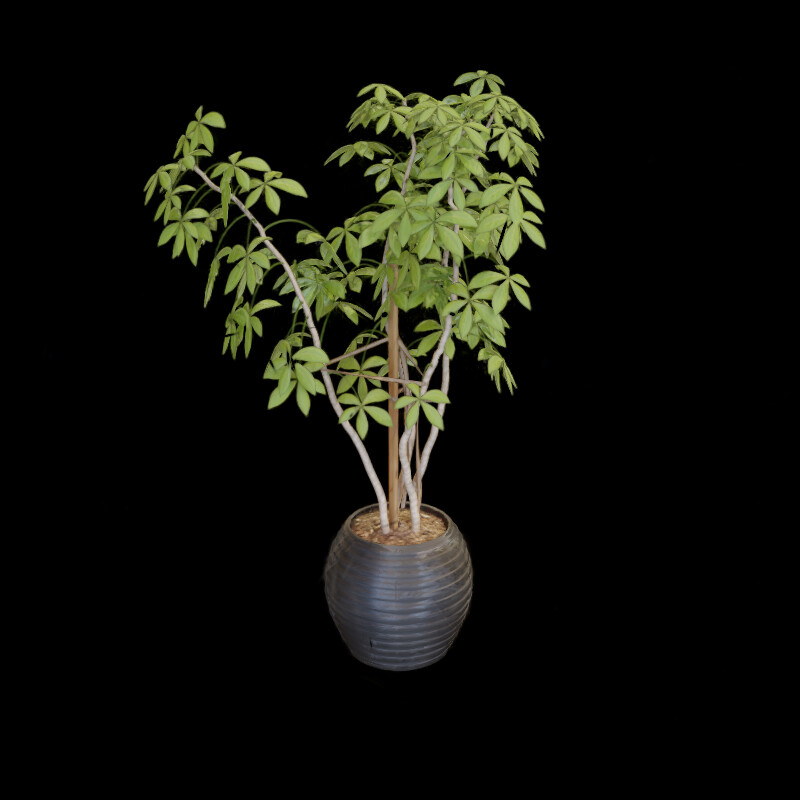}\put(68, 0.5){%
            \setlength{\fboxrule}{0.5pt}  
            \setlength{\fboxsep}{0pt}   
            \fcolorbox{white}{white}{\includegraphics[width=0.05\textwidth]{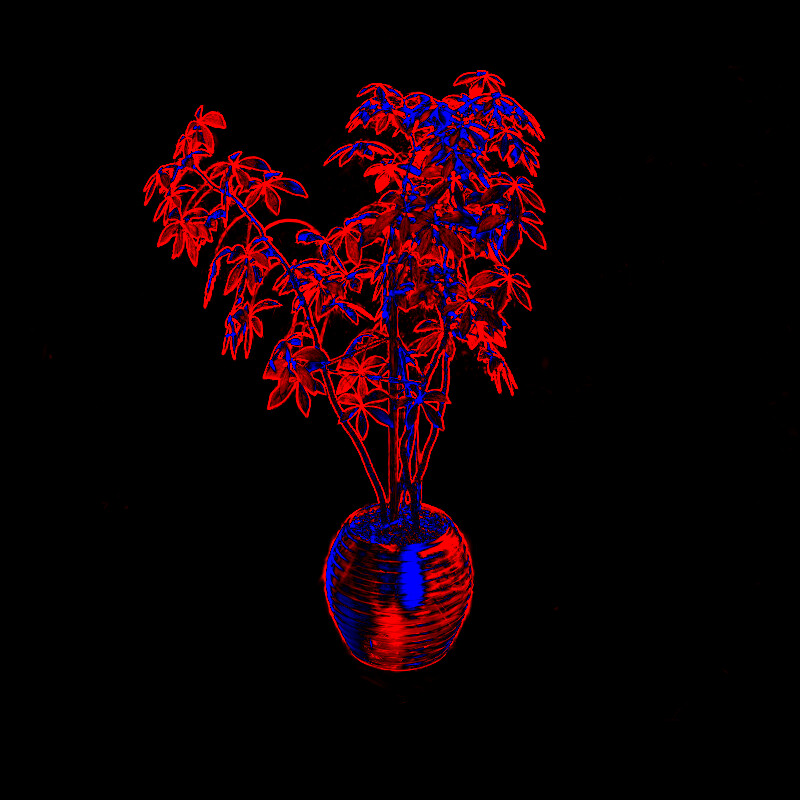}}
            }            
        \end{overpic}
        &
        \begin{overpic}[width=\lenRFCompare]{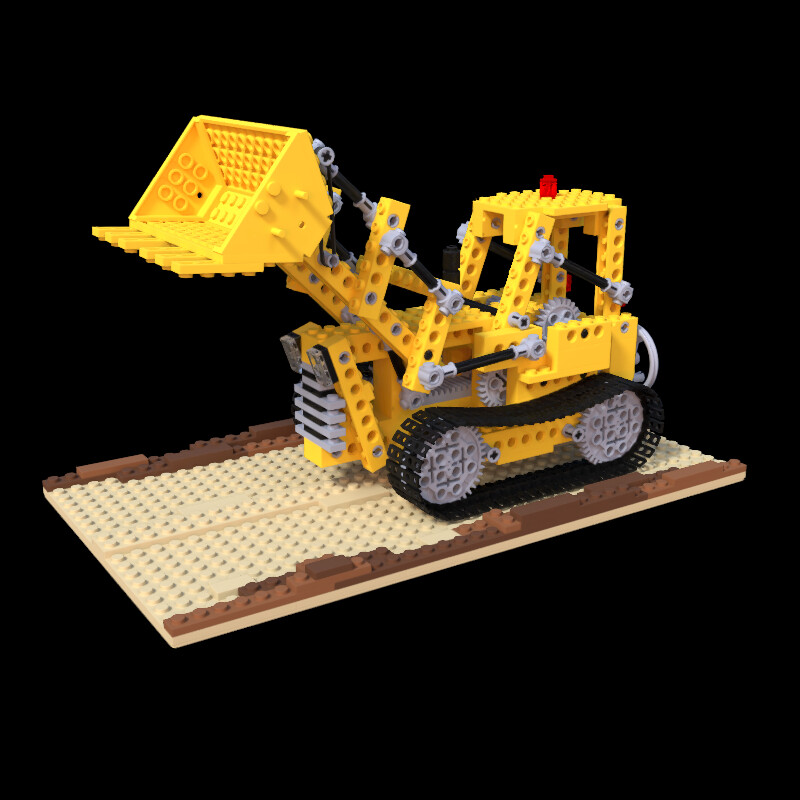}\put(68, 0.5){%
            \setlength{\fboxrule}{0.5pt}  
            \setlength{\fboxsep}{0pt}   
            \fcolorbox{white}{white}{\includegraphics[width=0.05\textwidth]{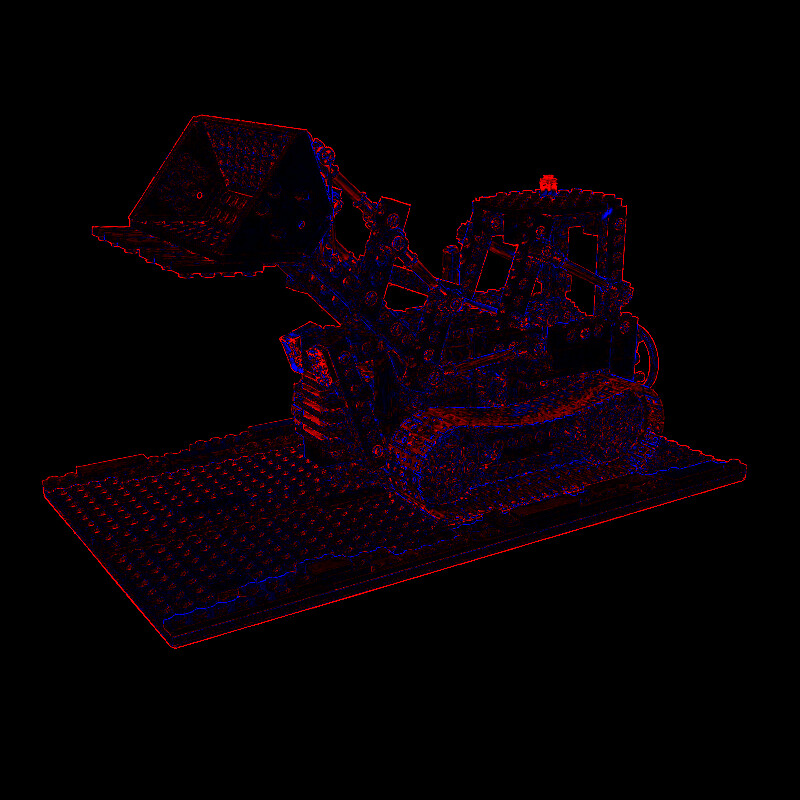}}
            }            
        \end{overpic}
        &
        \begin{overpic}[width=\lenRFCompare]{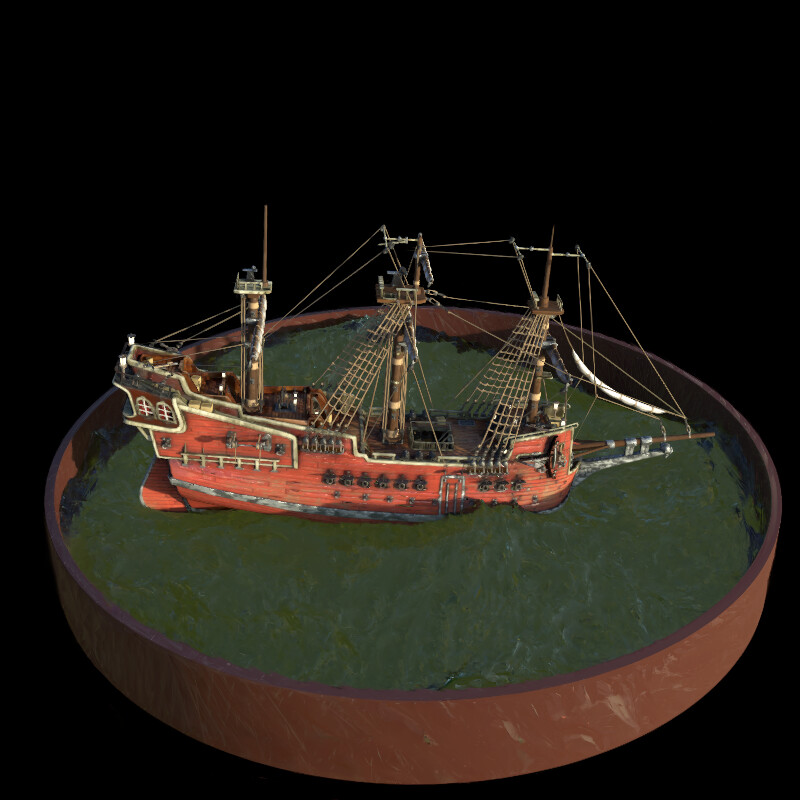}\put(68, 73.2){%
            \setlength{\fboxrule}{0.5pt}  
            \setlength{\fboxsep}{0pt}   
            \fcolorbox{white}{white}{\includegraphics[width=0.05\textwidth]{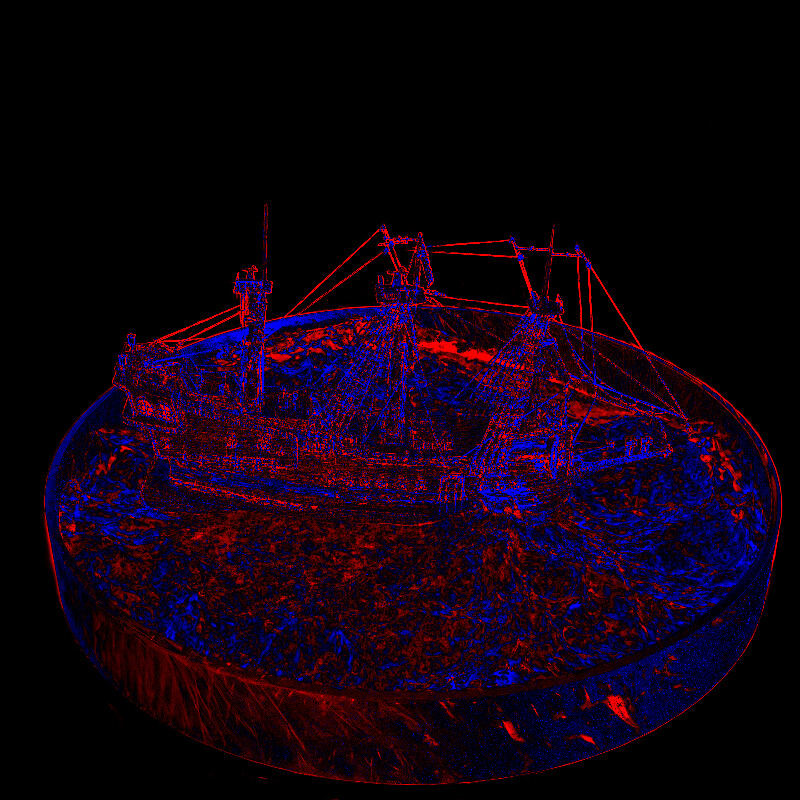}}
            }            
        \end{overpic}
        \\  
        \raisebox{30pt}{\rotatebox{90}{\footnotesize{\textsf{Instant-NGP}}}}
        &                      
        \begin{overpic}[width=\lenRFCompare]{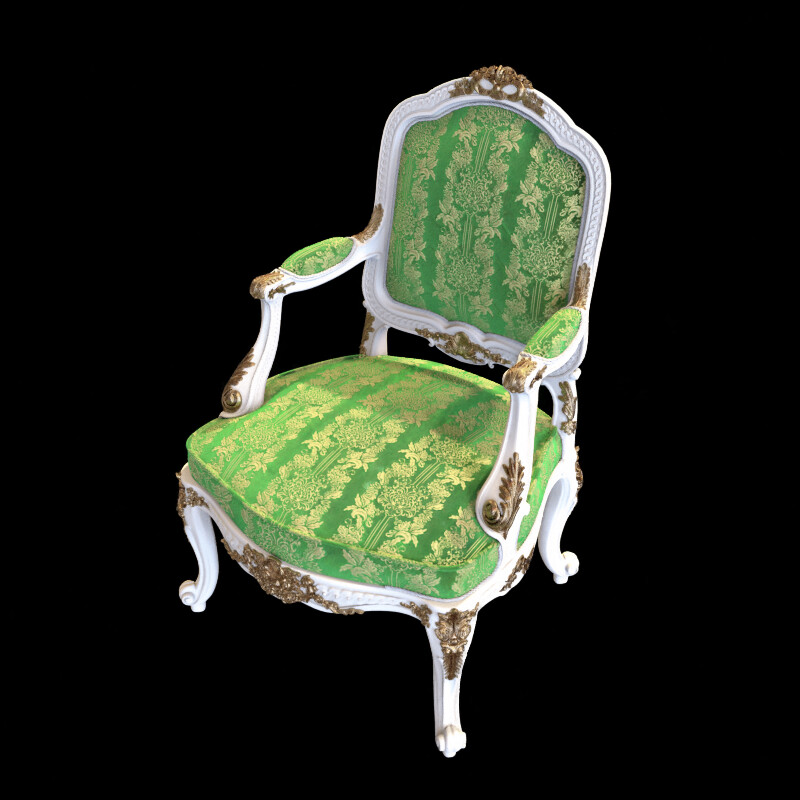}\put(68, 0.5){%
            \setlength{\fboxrule}{0.5pt}  
            \setlength{\fboxsep}{0pt}   
            \fcolorbox{white}{white}{\includegraphics[width=0.05\textwidth]{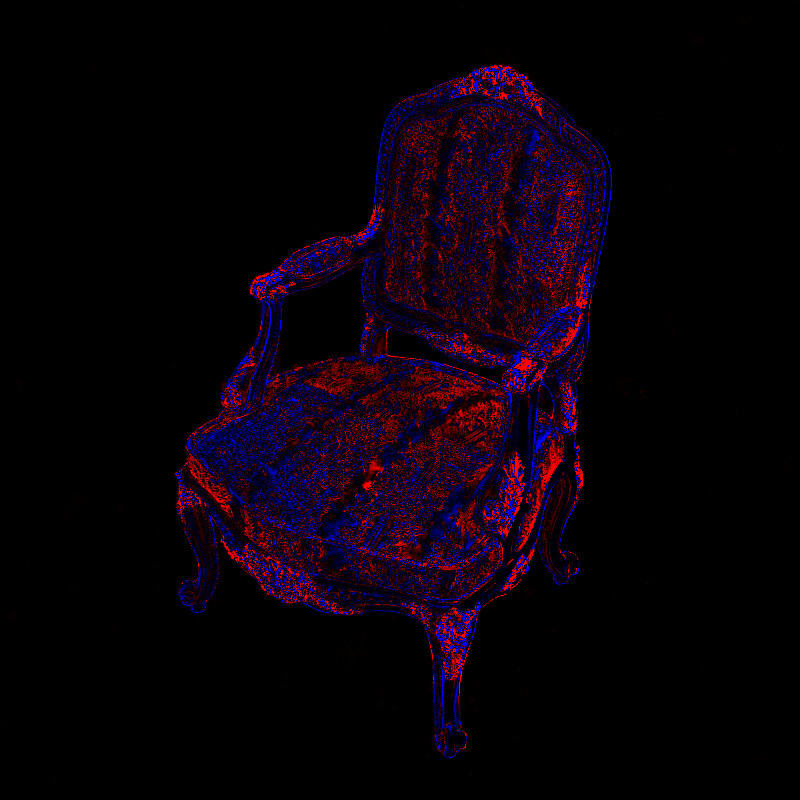}}
            }            
        \end{overpic}
        &
        \begin{overpic}[width=\lenRFCompare]{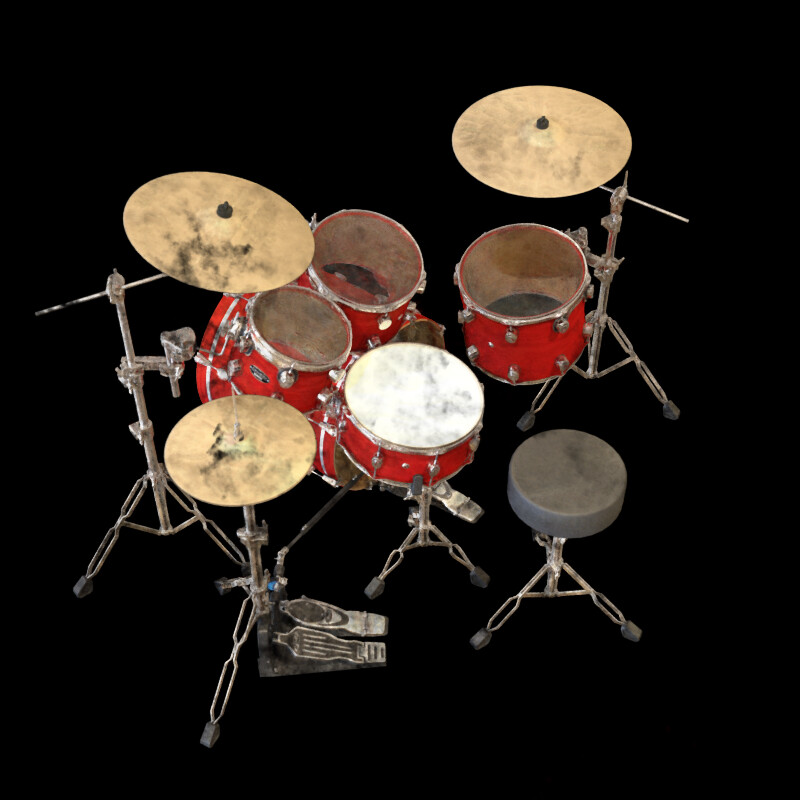}\put(68, 0.5){%
            \setlength{\fboxrule}{0.5pt}  
            \setlength{\fboxsep}{0pt}   
            \fcolorbox{white}{white}{\includegraphics[width=0.05\textwidth]{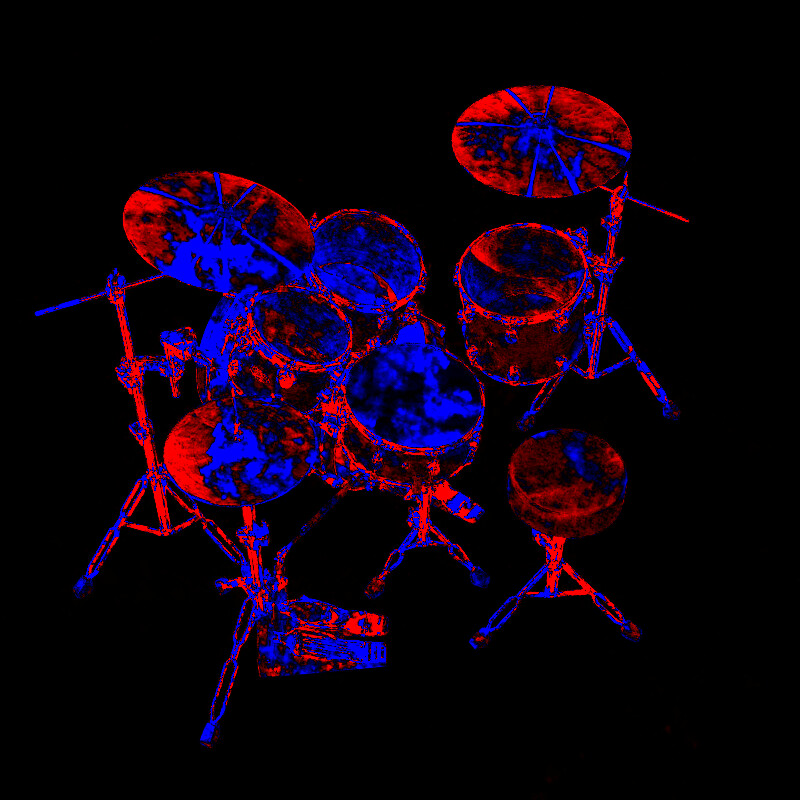}}
            }            
        \end{overpic}
        &
        \begin{overpic}[width=\lenRFCompare]{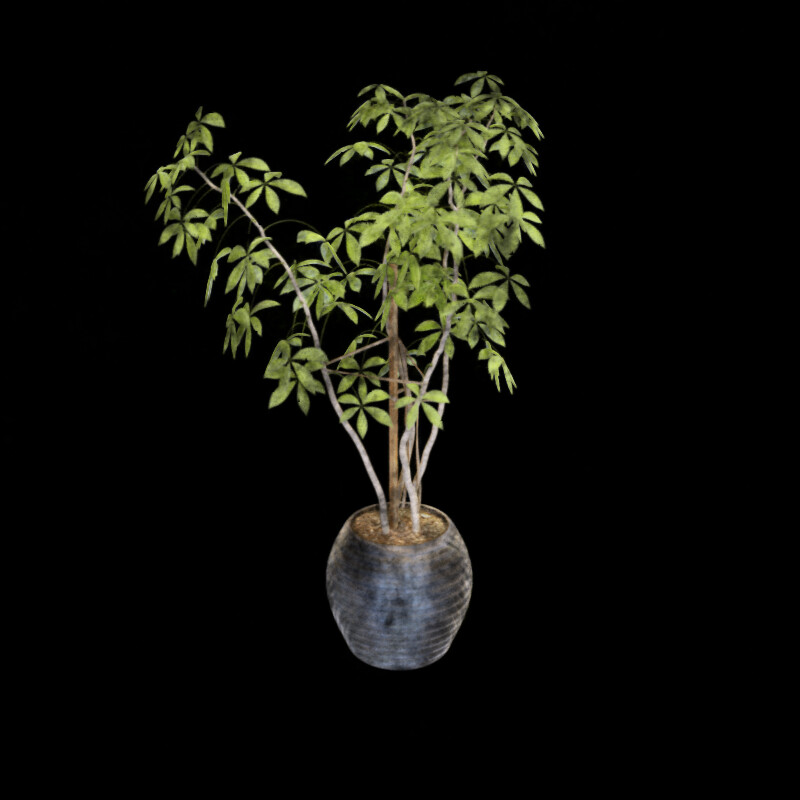}\put(68, 0.5){%
            \setlength{\fboxrule}{0.5pt}  
            \setlength{\fboxsep}{0pt}   
            \fcolorbox{white}{white}{\includegraphics[width=0.05\textwidth]{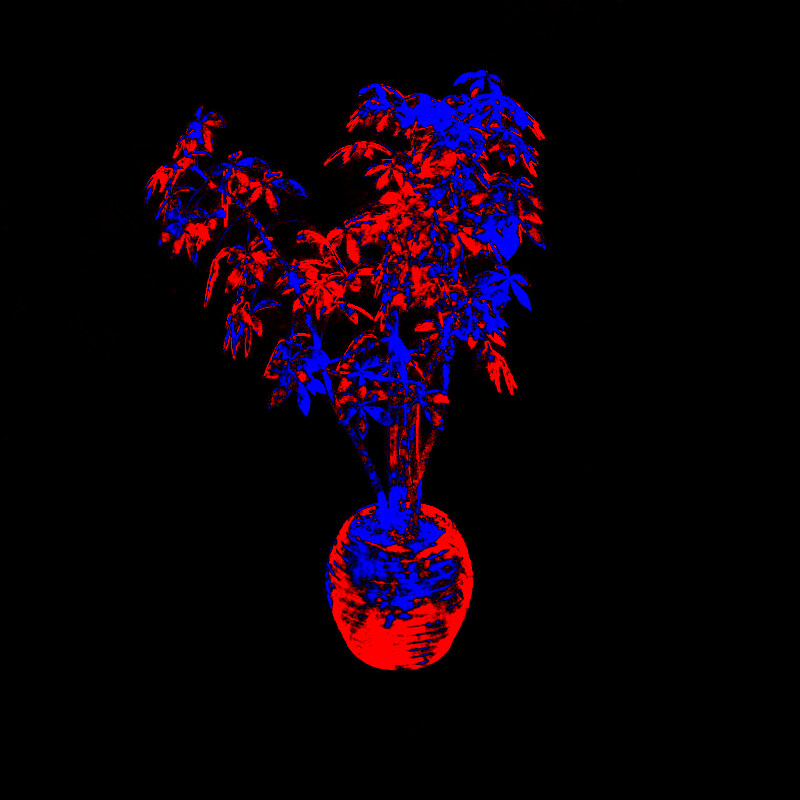}}
            }            
        \end{overpic}
        &
        \begin{overpic}[width=\lenRFCompare]{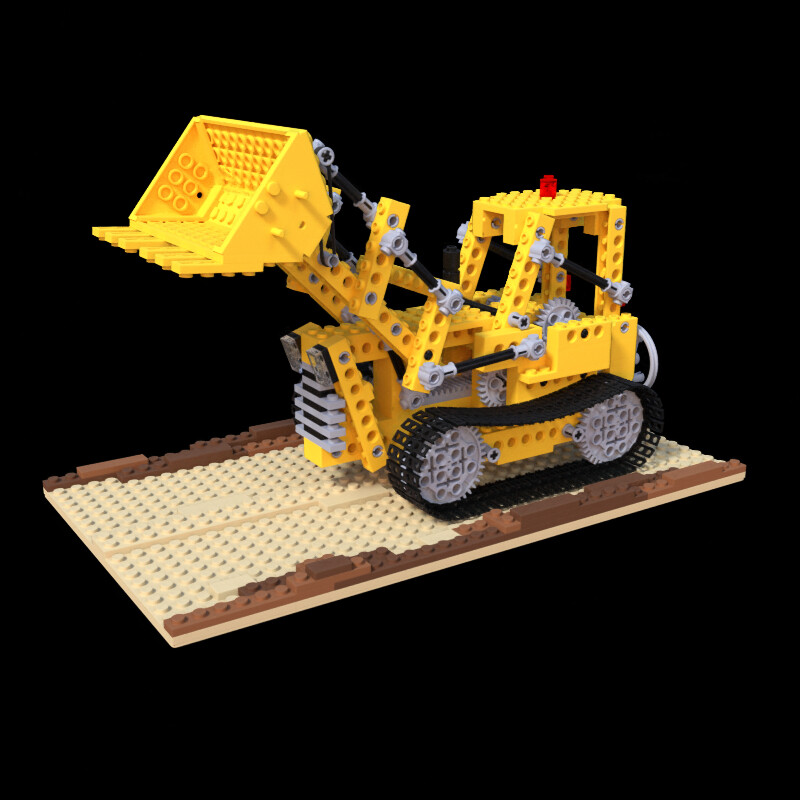}\put(68, 0.5){%
            \setlength{\fboxrule}{0.5pt}  
            \setlength{\fboxsep}{0pt}   
            \fcolorbox{white}{white}{\includegraphics[width=0.05\textwidth]{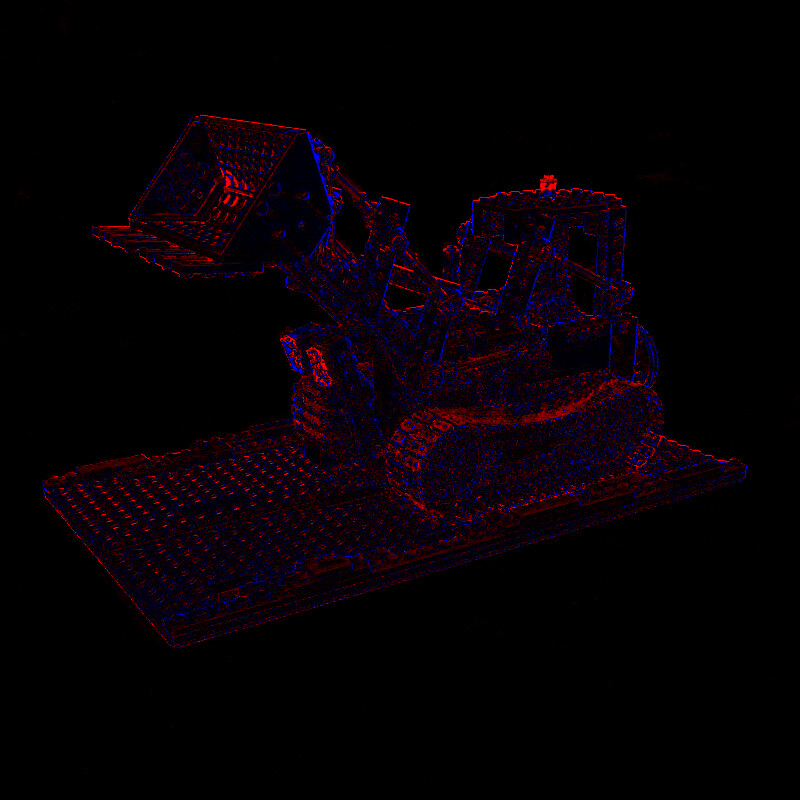}}
            }            
        \end{overpic}
        &
        \begin{overpic}[width=\lenRFCompare]{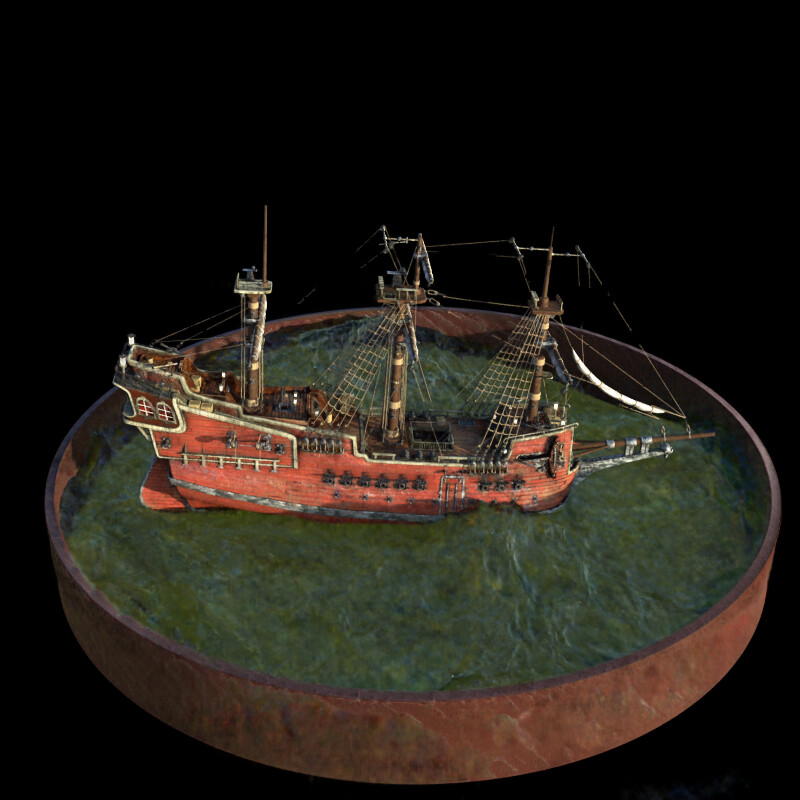}\put(68, 73.2){%
            \setlength{\fboxrule}{0.5pt}  
            \setlength{\fboxsep}{0pt}   
            \fcolorbox{white}{white}{\includegraphics[width=0.05\textwidth]{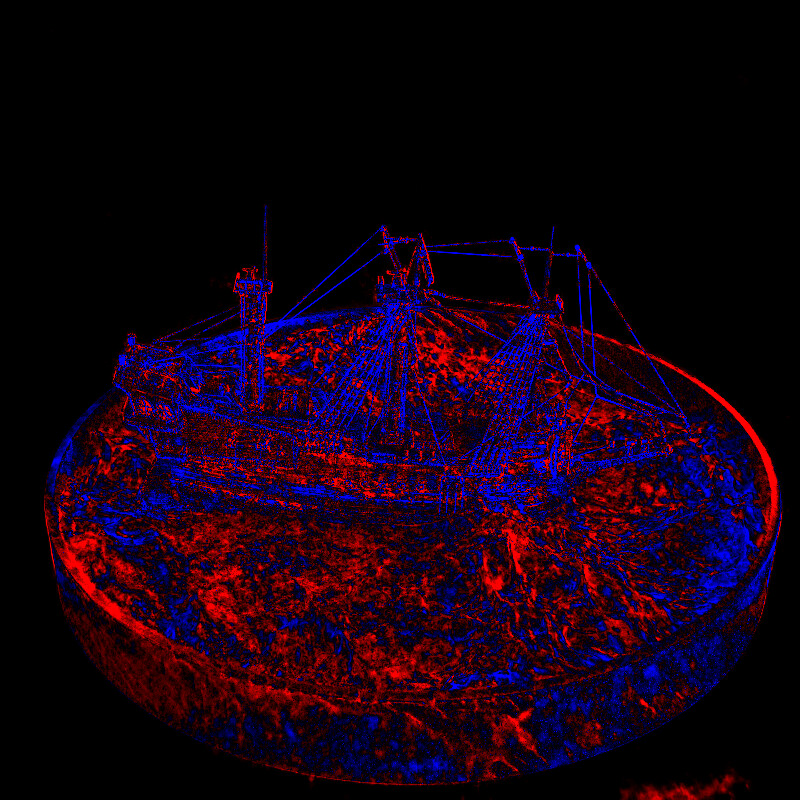}}
            }            
        \end{overpic}        
        \\
        \raisebox{40pt}{\rotatebox{90}{\footnotesize{\textsf{3DGS}}}}
        &        
        \begin{overpic}[width=\lenRFCompare]{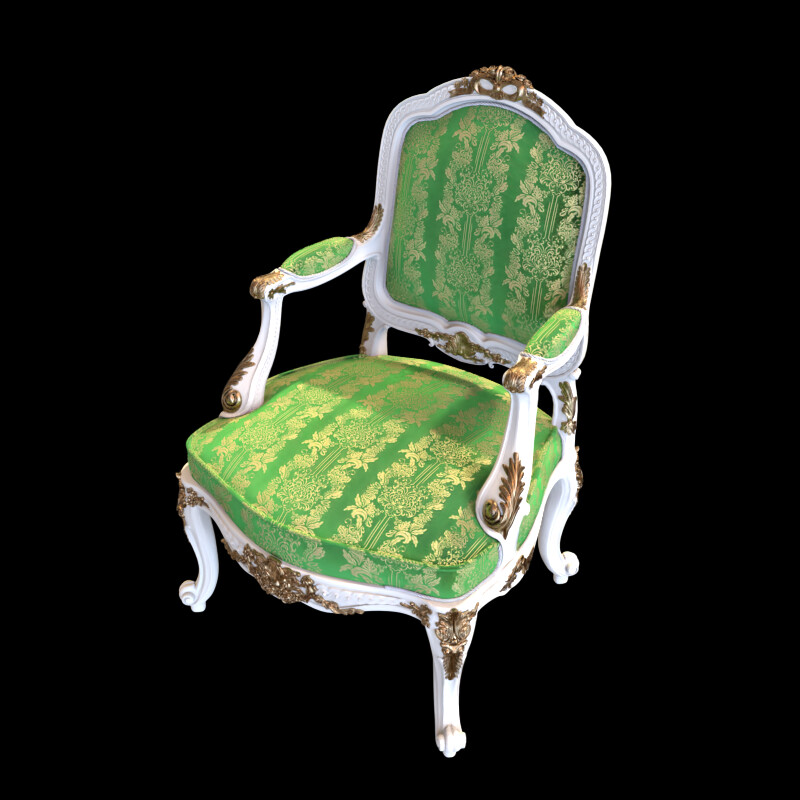}\put(68, 0.5){%
            \setlength{\fboxrule}{0.5pt}  
            \setlength{\fboxsep}{0pt}   
            \fcolorbox{white}{white}{\includegraphics[width=0.05\textwidth]{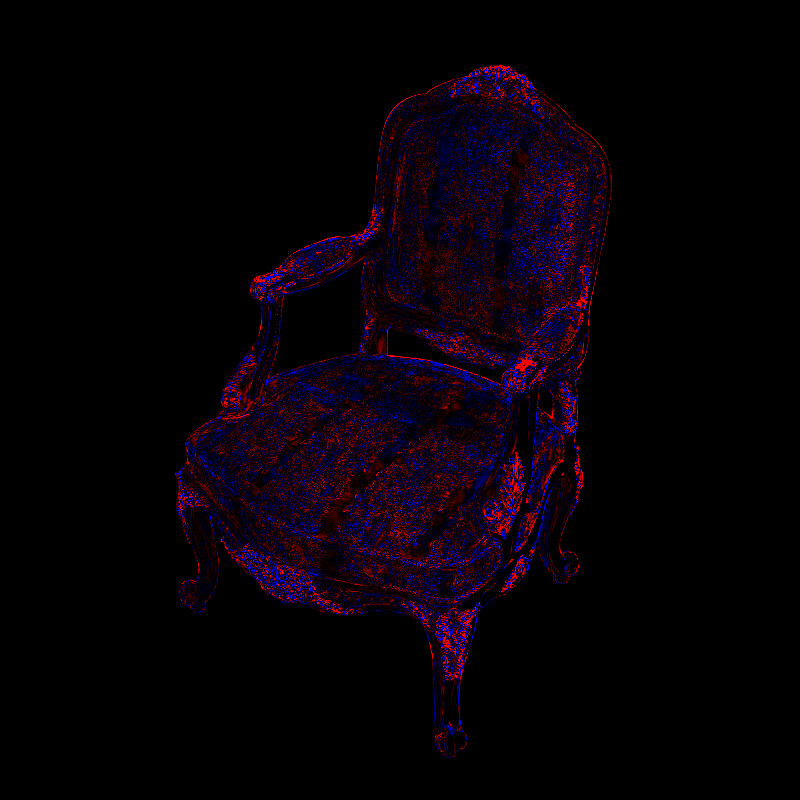}}
            }            
        \end{overpic}
        &
        \begin{overpic}[width=\lenRFCompare]{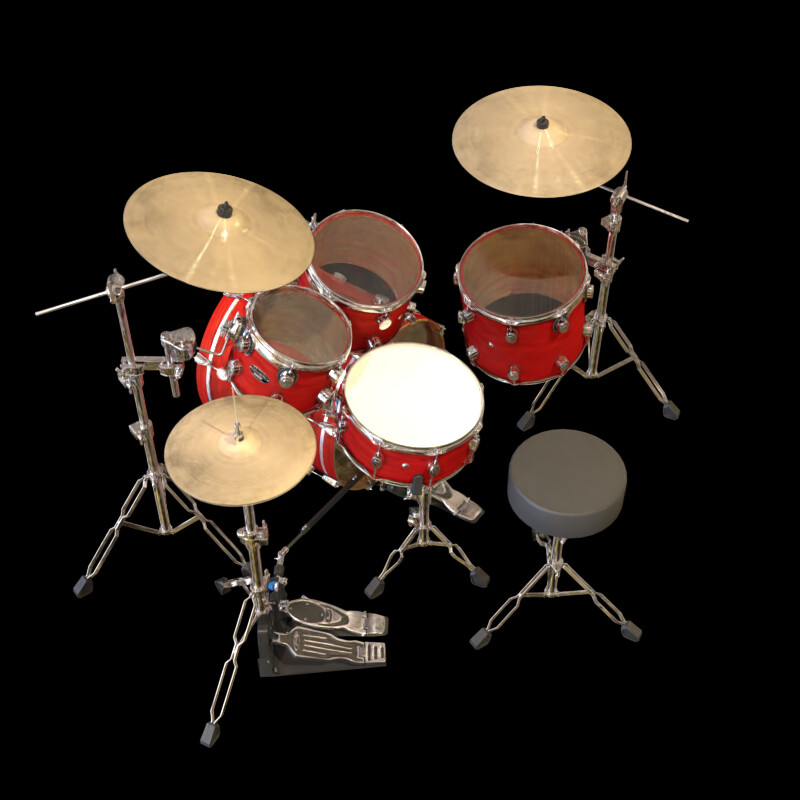}\put(68, 0.5){%
            \setlength{\fboxrule}{0.5pt}  
            \setlength{\fboxsep}{0pt}   
            \fcolorbox{white}{white}{\includegraphics[width=0.05\textwidth]{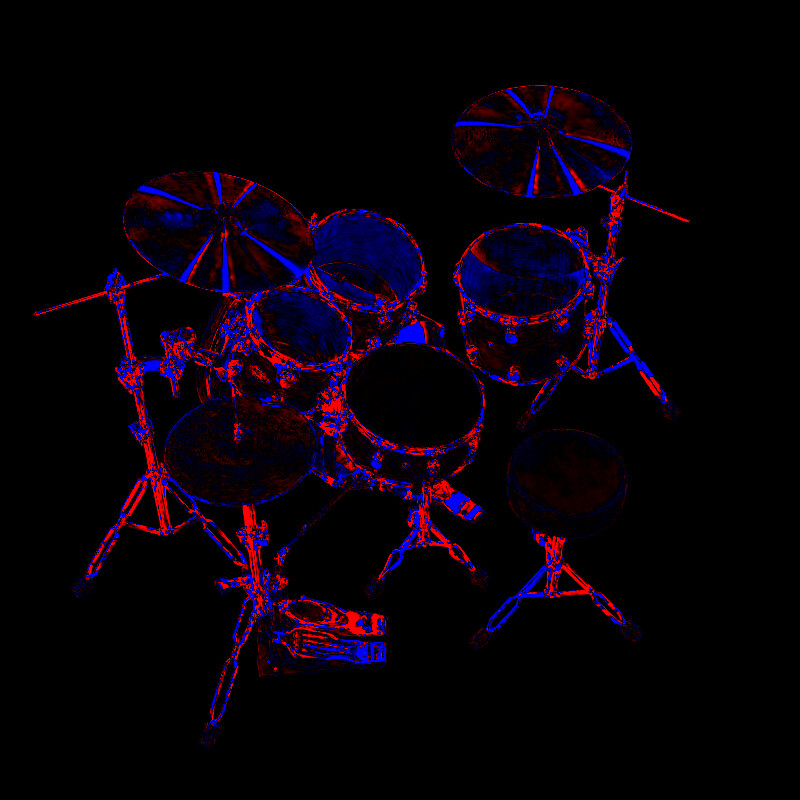}}
            }            
        \end{overpic}
        &
        \begin{overpic}[width=\lenRFCompare]{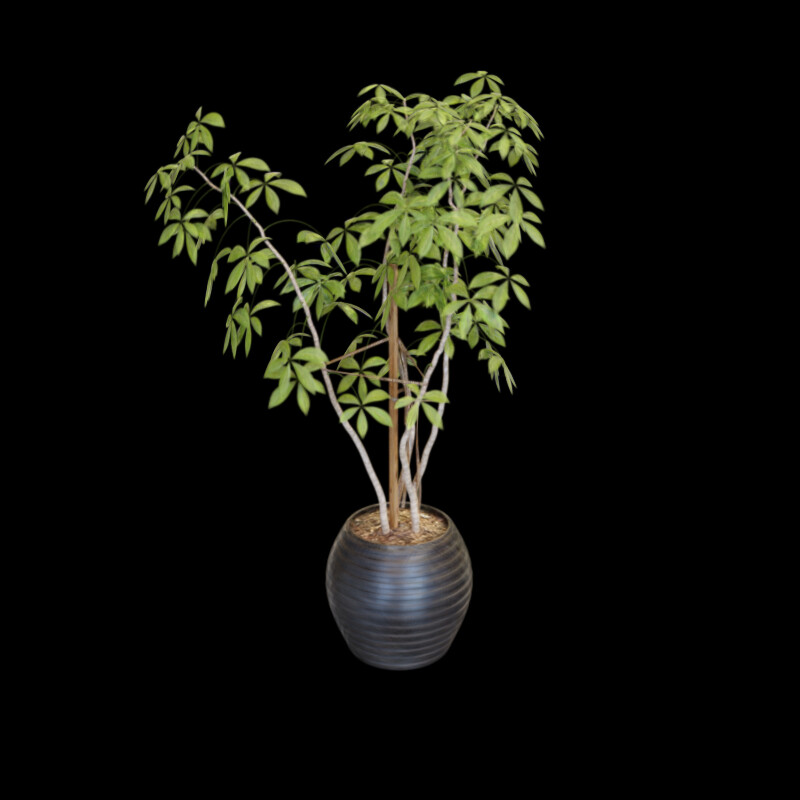}\put(68, 0.5){%
            \setlength{\fboxrule}{0.5pt}  
            \setlength{\fboxsep}{0pt}   
            \fcolorbox{white}{white}{\includegraphics[width=0.05\textwidth]{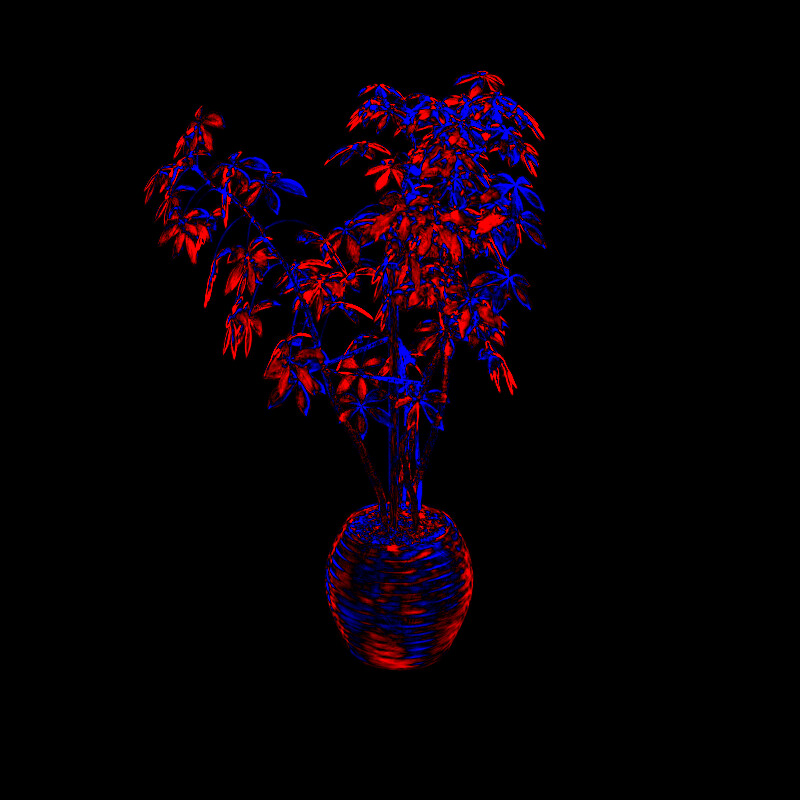}}
            }            
        \end{overpic}
        &
        \begin{overpic}[width=\lenRFCompare]{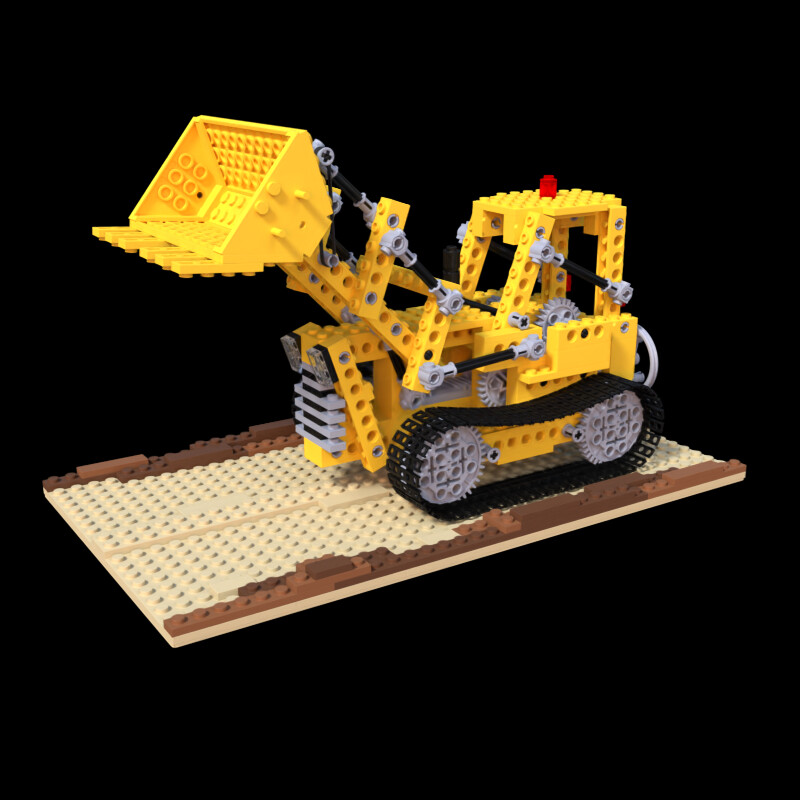}\put(68, 0.5){%
            \setlength{\fboxrule}{0.5pt}  
            \setlength{\fboxsep}{0pt}   
            \fcolorbox{white}{white}{\includegraphics[width=0.05\textwidth]{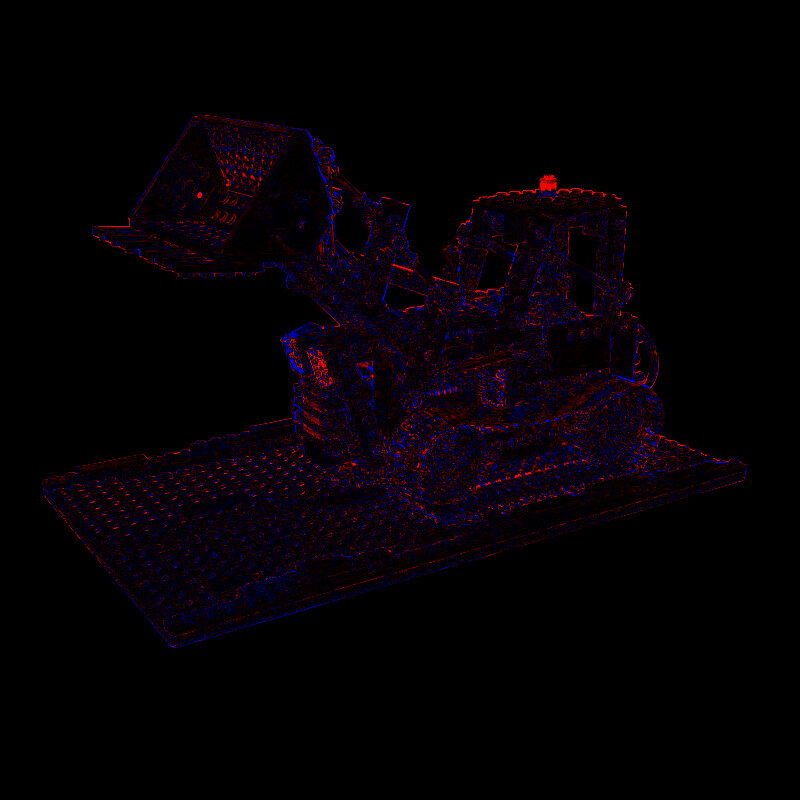}}
            }            
        \end{overpic}
        &
        \begin{overpic}[width=\lenRFCompare]{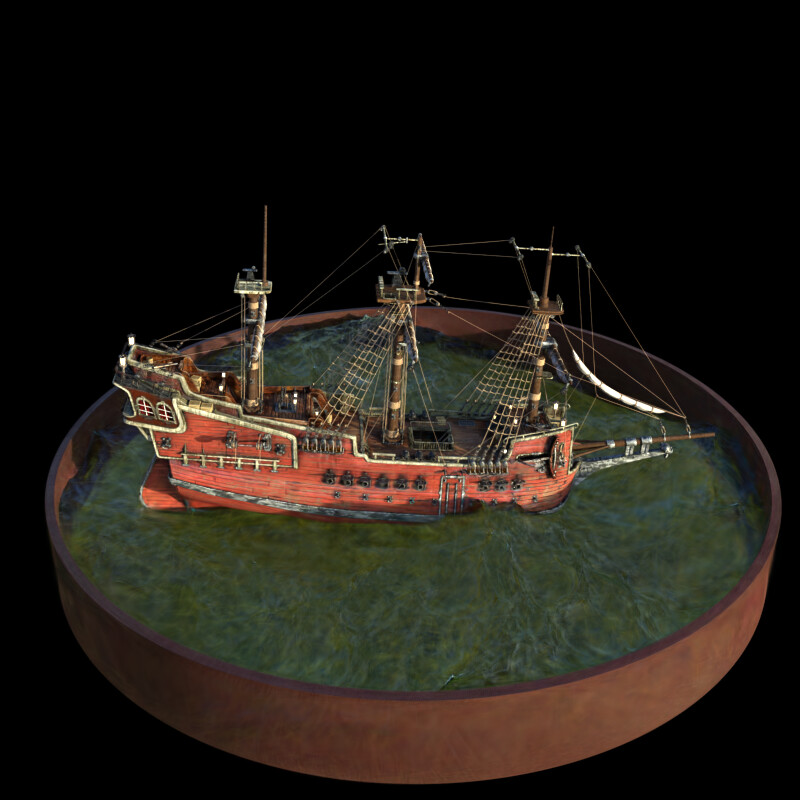}\put(68, 73.2){%
            \setlength{\fboxrule}{0.5pt}  
            \setlength{\fboxsep}{0pt}   
            \fcolorbox{white}{white}{\includegraphics[width=0.05\textwidth]{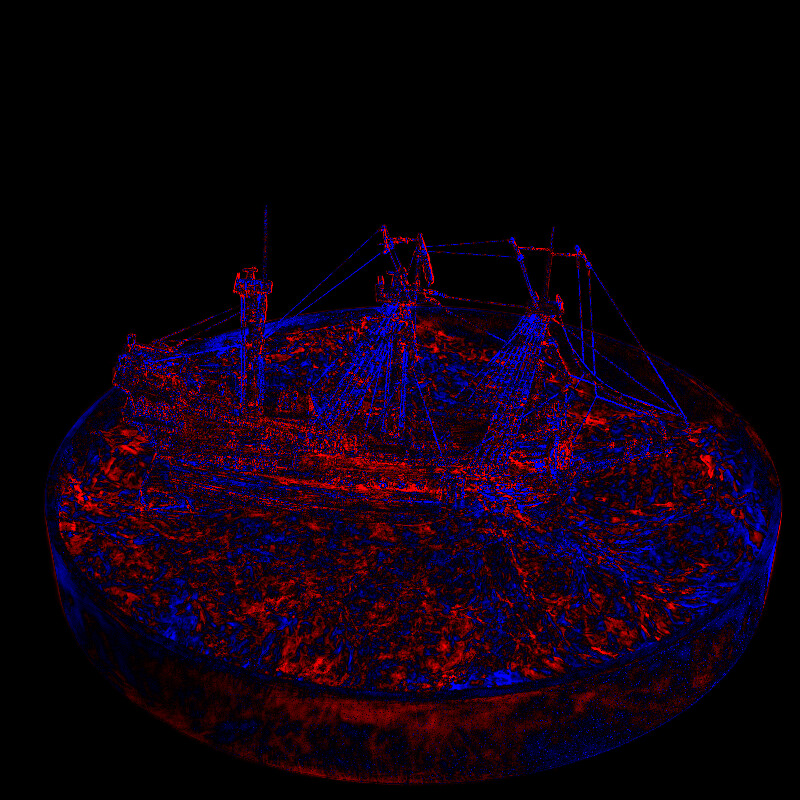}}
            }            
        \end{overpic}        
    \end{tabular}
    \caption{\label{fig:rf_compare}
        Visual comparison of radiance field reconstruction quality on the NeRF Blender dataset~\citep{mildenhall2020nerf}. 
        Our method is compared against Instant-NGP~\citep{muller2022instant} and 3DGS~\citep{kerbl20233d} on unseen test views.
        Insets show difference images ($10\times$ signed difference; blue indicates negative values, red indicates positive values).         
        See \autoref{table:rf_quality} for quantitative metrics.
    }
\end{figure*}

In \autoref{fig:rf_compare} and \autoref{table:rf_quality}, we show our results of radiance field reconstruction and compare our approach 
with representative previous works~\citep{muller2022instant, kerbl20233d}. Overall, our reconstruction quality edges out Instant-NGP, but trails 
3DGS by approximately 0.5-3 dB in PSNR. While the quality is slightly behind the state of the art, our method still brings advantages as a ray tracing framework even when 
specialized to radiance field rendering, including accurate ordering and support for arbitrary camera models. These benefits have been extensively discussed 
in prior works~\citep{moenne20243d,tog/CondorSBBGDJ25}.

Note that to match our straightforward implementation, we have disabled the view-dependent emission capability (spherical harmonics coefficients or directional 
encodings, respectively) of previous works. This explains why our measurements of previous works are worse than what was reported originally. 
This is orthogonal to our discussion because view-dependent emission is shown to benefit different radiance field methods in the same way 
(e.g. improving highlights and glossy reflections). Our conclusion on quality comparison remains the same.

\citet{tog/CondorSBBGDJ25} reported a PSNR of 32.11 dB and an SSIM score of 0.957 on the NeRF Blender dataset \emph{with} spherical harmonics. 
Unfortunately, we were not able to adapt their code release to a view-independent configuration for direct comparison.
Nevertheless, considering the extra quality boost from spherical 
harmonics, we believe it is reasonable to conclude that our method and theirs achieve similar quality under equivalent assumptions.
We also note that their optimization requires a warm start from existing 3DGS training, which makes their optimization overall an easier problem.

\begin{table*}[tb]
    \setlength{\tabcolsep}{3pt}%
	\centering
	\caption{
        Quantitative comparison of radiance field reconstruction quality with \emph{view-independent} emission on the NeRF Blender dataset~\citep{mildenhall2020nerf}.
        We report common metrics including \textcolor{solarized_orange}{PSNR$\uparrow$}, \textcolor{solarized_yellow}{SSIM$\uparrow$}, 
        \textcolor{solarized_cyan}{LPIPS$\downarrow$}, and \textcolor{solarized_blue}{\FLIP$\downarrow$}~\citep{pacmcgit/AnderssonNAOAF20} 
        averaged over the testing set of each scene.
        We achieve comparable quality even though our method is not natively designed for this task.
    }        
	\begin{tabular}{l|l|l|l|l|l|l|l|l|l}
		\Xhline{1pt}
		\textbf{Method} & \emph{Chair} & \emph{Drums} & \emph{Ficus} & \emph{Hotdog} & \emph{Lego} & \emph{Materials} & \emph{Mic} & \emph{Ship} & Avg. \\
        \hline
        Ours        &         
        {\setlength{\tabcolsep}{3pt}\begin{tabular}{@{}cc@{}} \textcolor{solarized_orange}{31.67} & \textcolor{solarized_yellow}{0.976} \\ \textcolor{solarized_cyan}{0.017} & \textcolor{solarized_blue}{0.037} \end{tabular}}
        & 
        {\setlength{\tabcolsep}{3pt}\begin{tabular}{@{}cc@{}} \textcolor{solarized_orange}{23.89} & \textcolor{solarized_yellow}{0.925} \\ \textcolor{solarized_cyan}{0.062} & \textcolor{solarized_blue}{0.070} \end{tabular}}
        & 
        {\setlength{\tabcolsep}{3pt}\begin{tabular}{@{}cc@{}} \textcolor{solarized_orange}{28.05} & \textcolor{solarized_yellow}{0.950} \\ \textcolor{solarized_cyan}{0.040} & \textcolor{solarized_blue}{0.052} \end{tabular}}
        & 
        {\setlength{\tabcolsep}{3pt}\begin{tabular}{@{}cc@{}} \textcolor{solarized_orange}{35.09} & \textcolor{solarized_yellow}{0.974} \\ \textcolor{solarized_cyan}{0.020} & \textcolor{solarized_blue}{0.038} \end{tabular}}
        &
        {\setlength{\tabcolsep}{3pt}\begin{tabular}{@{}cc@{}} \textcolor{solarized_orange}{34.46} & \textcolor{solarized_yellow}{0.981} \\ \textcolor{solarized_cyan}{0.011} & \textcolor{solarized_blue}{0.028} \end{tabular}}
        &
        {\setlength{\tabcolsep}{3pt}\begin{tabular}{@{}cc@{}} \textcolor{solarized_orange}{25.27} & \textcolor{solarized_yellow}{0.902} \\ \textcolor{solarized_cyan}{0.082} & \textcolor{solarized_blue}{0.067} \end{tabular}}
        &
        {\setlength{\tabcolsep}{3pt}\begin{tabular}{@{}cc@{}} \textcolor{solarized_orange}{30.90} & \textcolor{solarized_yellow}{0.967} \\ \textcolor{solarized_cyan}{0.042} & \textcolor{solarized_blue}{0.028} \end{tabular}}
        &
        {\setlength{\tabcolsep}{3pt}\begin{tabular}{@{}cc@{}} \textcolor{solarized_orange}{29.17} & \textcolor{solarized_yellow}{0.896} \\ \textcolor{solarized_cyan}{0.110} & \textcolor{solarized_blue}{0.072} \end{tabular}}       
        &
        {\setlength{\tabcolsep}{3pt}\begin{tabular}{@{}cc@{}} \textcolor{solarized_orange}{29.81} & \textcolor{solarized_yellow}{0.946} \\ \textcolor{solarized_cyan}{0.048} & \textcolor{solarized_blue}{0.049} \end{tabular}}               
        \\
        \hline
        Instant-NGP & 
        {\setlength{\tabcolsep}{3pt}\begin{tabular}{@{}cc@{}} \textcolor{solarized_orange}{32.29} & \textcolor{solarized_yellow}{0.974} \\ \textcolor{solarized_cyan}{0.018} & \textcolor{solarized_blue}{0.037} \end{tabular}}
        & 
        {\setlength{\tabcolsep}{3pt}\begin{tabular}{@{}cc@{}} \textcolor{solarized_orange}{23.55} & \textcolor{solarized_yellow}{0.901} \\ \textcolor{solarized_cyan}{0.079} & \textcolor{solarized_blue}{0.087} \end{tabular}}
        & 
        {\setlength{\tabcolsep}{3pt}\begin{tabular}{@{}cc@{}} \textcolor{solarized_orange}{27.88} & \textcolor{solarized_yellow}{0.956} \\ \textcolor{solarized_cyan}{0.035} & \textcolor{solarized_blue}{0.058} \end{tabular}}
        & 
        {\setlength{\tabcolsep}{3pt}\begin{tabular}{@{}cc@{}} \textcolor{solarized_orange}{34.40} & \textcolor{solarized_yellow}{0.966} \\ \textcolor{solarized_cyan}{0.035} & \textcolor{solarized_blue}{0.038} \end{tabular}}
        &
        {\setlength{\tabcolsep}{3pt}\begin{tabular}{@{}cc@{}} \textcolor{solarized_orange}{34.11} & \textcolor{solarized_yellow}{0.975} \\ \textcolor{solarized_cyan}{0.013} & \textcolor{solarized_blue}{0.033} \end{tabular}}
        &
        {\setlength{\tabcolsep}{3pt}\begin{tabular}{@{}cc@{}} \textcolor{solarized_orange}{22.65} & \textcolor{solarized_yellow}{0.870} \\ \textcolor{solarized_cyan}{0.109} & \textcolor{solarized_blue}{0.100} \end{tabular}}
        &
        {\setlength{\tabcolsep}{3pt}\begin{tabular}{@{}cc@{}} \textcolor{solarized_orange}{30.25} & \textcolor{solarized_yellow}{0.962} \\ \textcolor{solarized_cyan}{0.042} & \textcolor{solarized_blue}{0.033} \end{tabular}}
        &
        {\setlength{\tabcolsep}{3pt}\begin{tabular}{@{}cc@{}} \textcolor{solarized_orange}{28.09} & \textcolor{solarized_yellow}{0.849} \\ \textcolor{solarized_cyan}{0.133} & \textcolor{solarized_blue}{0.081} \end{tabular}}
        &
        {\setlength{\tabcolsep}{3pt}\begin{tabular}{@{}cc@{}} \textcolor{solarized_orange}{29.15} & \textcolor{solarized_yellow}{0.932} \\ \textcolor{solarized_cyan}{0.058} & \textcolor{solarized_blue}{0.058} \end{tabular}}                
        \\
        \hline
        3DGS        &
        {\setlength{\tabcolsep}{3pt}\begin{tabular}{@{}cc@{}} \textcolor{solarized_orange}{34.23} & \textcolor{solarized_yellow}{0.984} \\ \textcolor{solarized_cyan}{0.010} & \textcolor{solarized_blue}{0.028} \end{tabular}}
        & 
        {\setlength{\tabcolsep}{3pt}\begin{tabular}{@{}cc@{}} \textcolor{solarized_orange}{25.53} & \textcolor{solarized_yellow}{0.948} \\ \textcolor{solarized_cyan}{0.046} & \textcolor{solarized_blue}{0.062} \end{tabular}}
        & 
        {\setlength{\tabcolsep}{3pt}\begin{tabular}{@{}cc@{}} \textcolor{solarized_orange}{32.37} & \textcolor{solarized_yellow}{0.978} \\ \textcolor{solarized_cyan}{0.016} & \textcolor{solarized_blue}{0.036} \end{tabular}}
        & 
        {\setlength{\tabcolsep}{3pt}\begin{tabular}{@{}cc@{}} \textcolor{solarized_orange}{36.67} & \textcolor{solarized_yellow}{0.982} \\ \textcolor{solarized_cyan}{0.016} & \textcolor{solarized_blue}{0.028} \end{tabular}}
        &
        {\setlength{\tabcolsep}{3pt}\begin{tabular}{@{}cc@{}} \textcolor{solarized_orange}{34.99} & \textcolor{solarized_yellow}{0.980} \\ \textcolor{solarized_cyan}{0.013} & \textcolor{solarized_blue}{0.029} \end{tabular}}
        &
        {\setlength{\tabcolsep}{3pt}\begin{tabular}{@{}cc@{}} \textcolor{solarized_orange}{28.50} & \textcolor{solarized_yellow}{0.946} \\ \textcolor{solarized_cyan}{0.039} & \textcolor{solarized_blue}{0.049} \end{tabular}}
        &
        {\setlength{\tabcolsep}{3pt}\begin{tabular}{@{}cc@{}} \textcolor{solarized_orange}{34.58} & \textcolor{solarized_yellow}{0.988} \\ \textcolor{solarized_cyan}{0.013} & \textcolor{solarized_blue}{0.017} \end{tabular}}
        &
        {\setlength{\tabcolsep}{3pt}\begin{tabular}{@{}cc@{}} \textcolor{solarized_orange}{29.98} & \textcolor{solarized_yellow}{0.884} \\ \textcolor{solarized_cyan}{0.108} & \textcolor{solarized_blue}{0.067} \end{tabular}}
        &
        {\setlength{\tabcolsep}{3pt}\begin{tabular}{@{}cc@{}} \textcolor{solarized_orange}{32.11} & \textcolor{solarized_yellow}{0.961} \\ \textcolor{solarized_cyan}{0.033} & \textcolor{solarized_blue}{0.040} \end{tabular}} 
        \\
		\Xhline{1pt}
	\end{tabular}
	\label{table:rf_quality} 
\end{table*}

\begin{figure}[tb]
	\newlength{\lenRFOverdraw}
	\setlength{\lenRFOverdraw}{0.3\linewidth}
    \addtolength{\tabcolsep}{-5pt}
    \renewcommand{\arraystretch}{0.5}

    \centering
    \begin{tabular}{ccccc}
        & \small{\textsf{Setup}} & \small{\textsf{Ours}} & \small{\textsf{3DGS}} &
        \vspace{0.5pt}
        \\
        \raisebox{30pt}{\rotatebox{90}{\footnotesize{\emph{Chair}}}}
        &
        \begin{overpic}[height=\lenRFOverdraw]{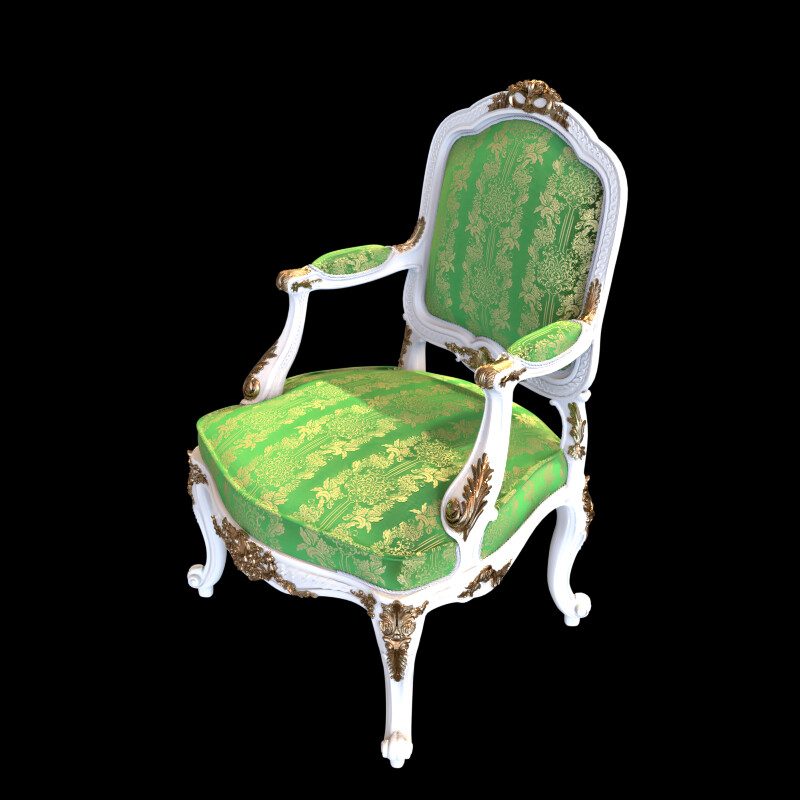}\end{overpic}
        &
        \begin{overpic}[height=\lenRFOverdraw]{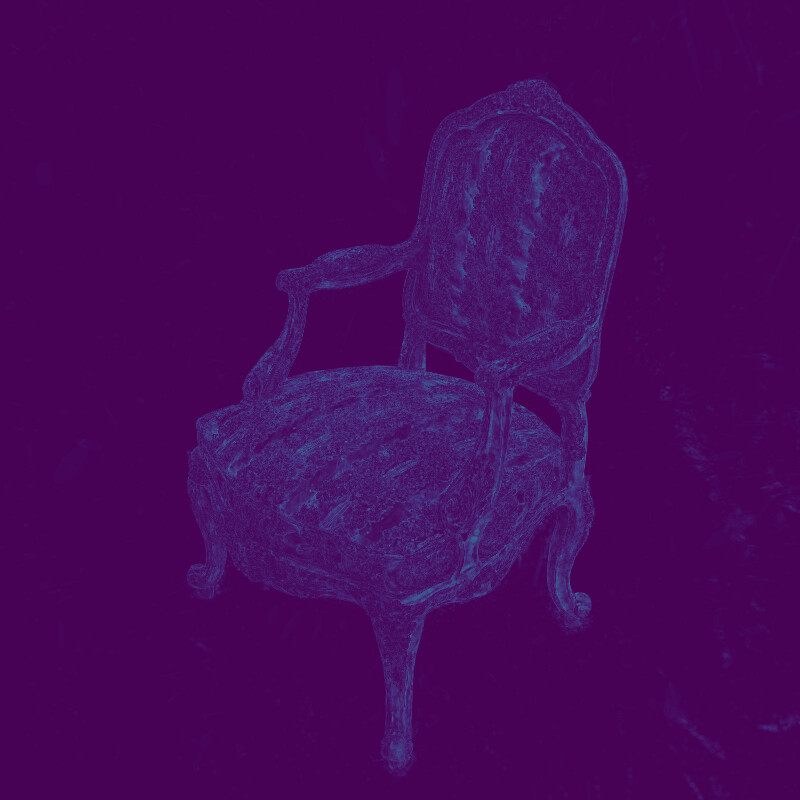}\end{overpic}
        &
        \begin{overpic}[height=\lenRFOverdraw]{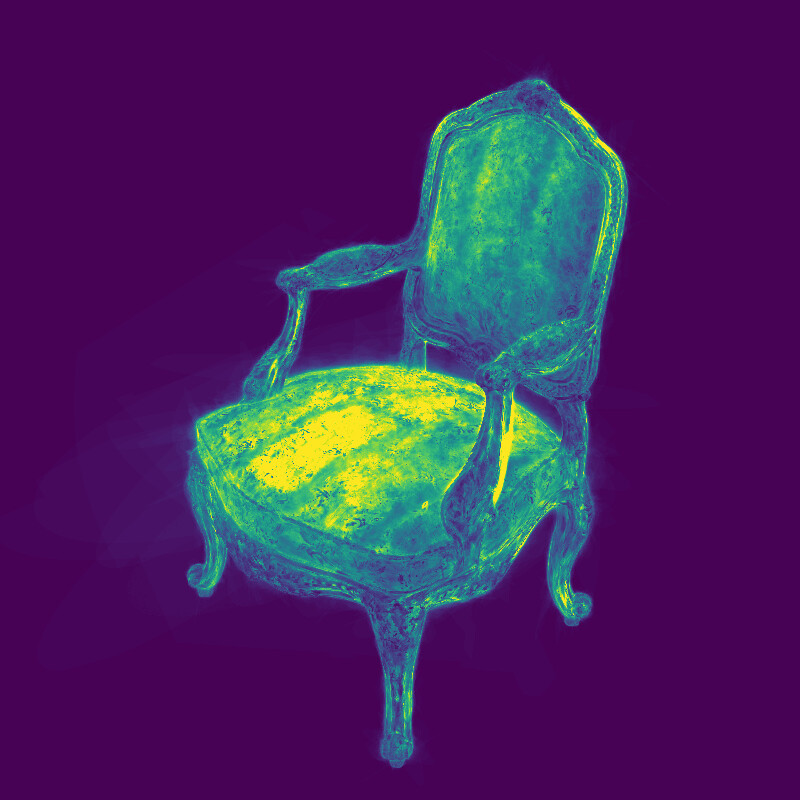}\end{overpic}
        &
        \begin{overpic}[height=0.99\lenRFOverdraw, frame]{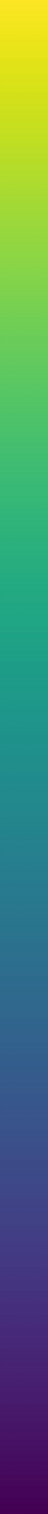}
            \put(5, 0){\footnotesize{0}}
            \put(5, 92){\footnotesize{100}}
        \end{overpic}
        \\
        \raisebox{30pt}{\rotatebox{90}{\footnotesize{\emph{Lego}}}}
        &        
        \begin{overpic}[height=\lenRFOverdraw]{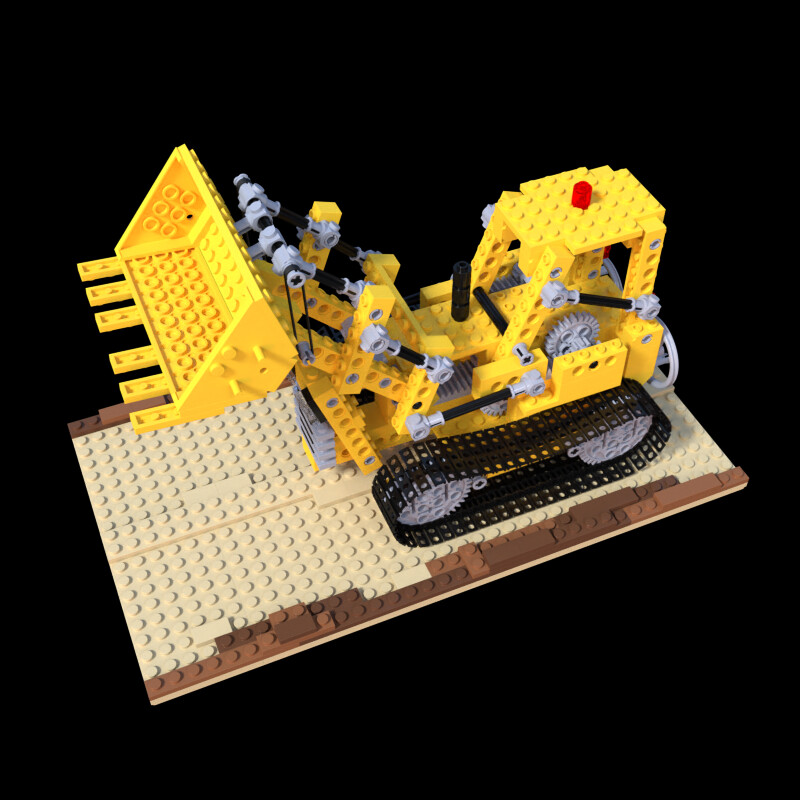}\end{overpic}
        &
        \begin{overpic}[height=\lenRFOverdraw]{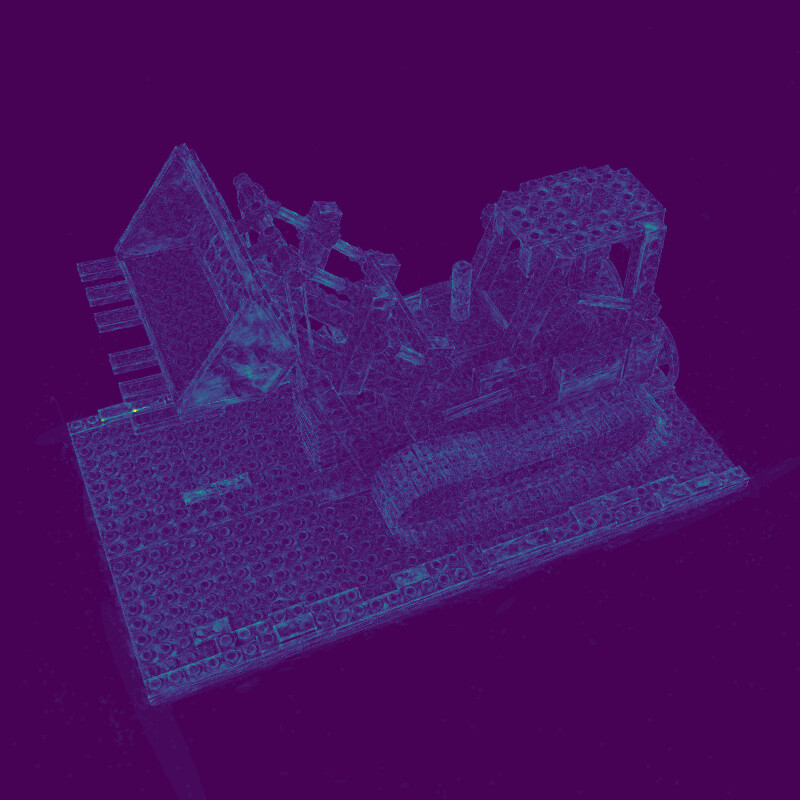}\end{overpic}
        &
        \begin{overpic}[height=\lenRFOverdraw]{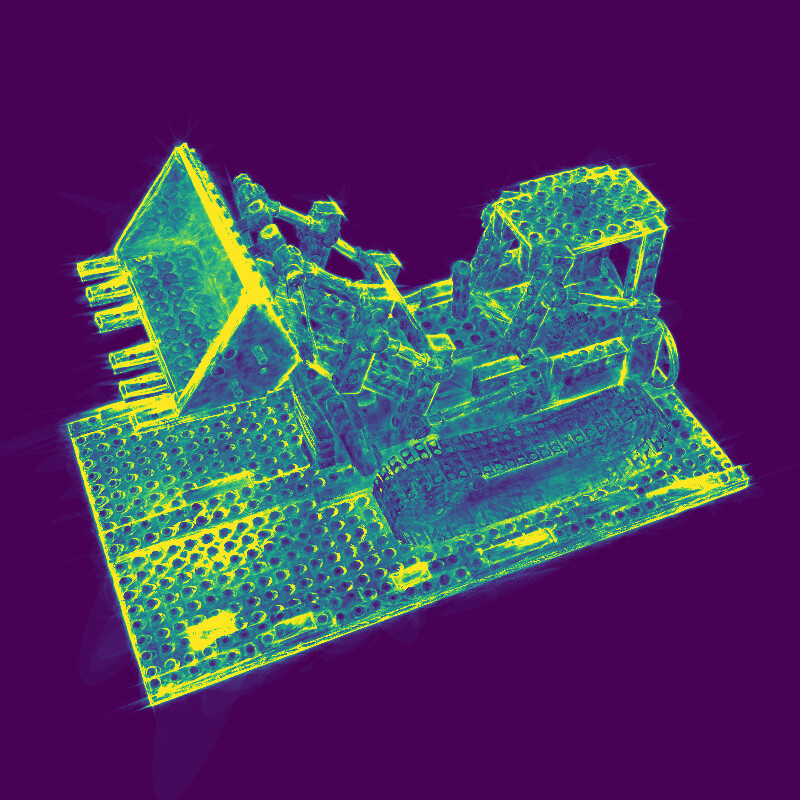}\end{overpic}  
        &
        \begin{overpic}[height=0.99\lenRFOverdraw, frame]{resources/viridis_h.jpg}
            \put(5, 0){\footnotesize{0}}
            \put(5, 92){\footnotesize{50}}
        \end{overpic}  
    \end{tabular}
    \caption{\label{fig:rf_overdraw}
        Primitive overdraw visualization in radiance field rendering. 3DGS optimization generates excessive overdraw even for 
        opaque-looking assets. With similar primitive counts, our optimization substantially alleviates this issue.
    }
\end{figure}

\paragraph{Performance}

We first discuss the performance of our forward path tracer. To demonstrate the full capability of our representation, 
we choose to implement it in a ``reference'' fashion. We refrain from practical techniques such as path reuse, temporal accumulation, 
or denoising to avoid artifacts such as correlated patterns, ghosting, bias, and overblurring. The only exception is the 
supplemental video, where we provide denoised sequences using the NVIDIA OptiX denoiser~\citep{nvidia_optix_denoiser}.

\autoref{table:fwd_timings} provides timings of representative forward renders.
We have used our method to render various scenes with primitive count ranging from 100K to 5.8M, using up to 4K samples per pixel (spp).
Rendering a converged frame requires minutes to hours, as is typical for a high-quality Monte Carlo renderer, while per-spp frame rates achieve interactive 
performance on small-scale scenes.
Our GPU renderer generally outperforms the CPU counterpart with the amount of speedup depending on actual scenes.
Typically, the majority (\(\sim\)85\% on CPU) of rendering
time is spent on kd-tree traversal, bounding ellipsoid intersection test, and ray integral evaluation. The disambiguation step in \autoref{list:sample_free_flight}
takes less than 5\% of time. This distribution is even more pronounced on GPU as GPUs are more prone to divergence. It is possible to consider alternative 
wavefront-based architectures to further improve GPU performance. 

At comparable scene complexity, our current implementation does not yet achieve the rendering performance of mesh-based representations. 
\autoref{fig:mesh_comp} presents CPU timing comparisons between our method and mesh rendering. In the \emph{Color Tree} scene, 
where scene traversal and intersection tests dominate the computation, our method is 2--4$\times$ slower.
However, our method does achieve higher or similar quality using a much smaller primitive count, thus it is faster in that sense.
The difference is less pronounced in the \emph{Caustic Ring} scene, where other operations such as shading and material sampling dominate.
This disparity is expected: our primitive operations, such as ray sampling and integration, are more expensive than the mathematically simpler 
ray-triangle intersection, and mesh renderers have benefited from decades of engineering effort.
Our CPU implementation is not as thoroughly optimized as the mesh ray tracer, which is built upon Embree and benefits 
from its highly optimized SIMD traversal and intersection routines~\citep{tog/WaldWBJE14}. Similarly, our GPU performance is limited due to the absence of 
dedicated hardware acceleration for kd-tree on mainstream GPUs (unlike bounding volume hierarchies). 
We anticipate specialized hardware architectures could substantially alleviate this limitation.

Similar to many primitive-based representations, the memory requirement of our representation is linear with respect to the number of primitives.
The attributes of each primitive can be directly accessed. This is in contrast to the more implicit neural fields, which are 
typically much more compact but come at the cost of a much heavier decoding process.
Specifically, for each primitive, we store its shape (scale, rotation, translation), magnitude, SGGX NDF (6 real numbers)~\citep{heitz2015sggx} and the base material parameters (12 real numbers in our implementation of the Disney BSDF), 
resulting in an uncompressed size of 116 bytes with 32-bit floats.
The memory consumption of each scene can thus be calculated as shown in \autoref{table:fwd_timings}. Note that this is already more compact than 
3DGS which typically stores 4 bands of SH coefficients for RGB channels (48 numbers).
The per-primitive attribute storage is analogous to per-vertex attribute storage for meshes. While this may introduce redundancy compared to textured meshes, 
many compression methods designed for 3DGS~\citep{cgf/BagdasarianKLBHEM25} are compatible with our representation due to the very similar nature 
of their storage and memory layouts.
Alternatively, it is possible to enhance our representation by using either compact neural fields~\citep{baatz2021nerftex}
or additionally learned textures~\citep{Chao_2025_CVPR}, such that each primitive can model spatially varying appearance. 
We expect this to remove much of the per-primitive redundancy, given proper balancing between primitive count and network 
capacity or texture resolution. However, these techniques would introduce the additional cost of network inference or 
texture lookups.



\begin{figure}[tb]
	\newlength{\lenLoDFan}
	\setlength{\lenLoDFan}{\linewidth}
    \addtolength{\tabcolsep}{-4pt}
    \renewcommand{\arraystretch}{0.5}
    \centering
    \begin{tabular}{c}             
         \includegraphics[draft=false, width=\lenLoDFan]{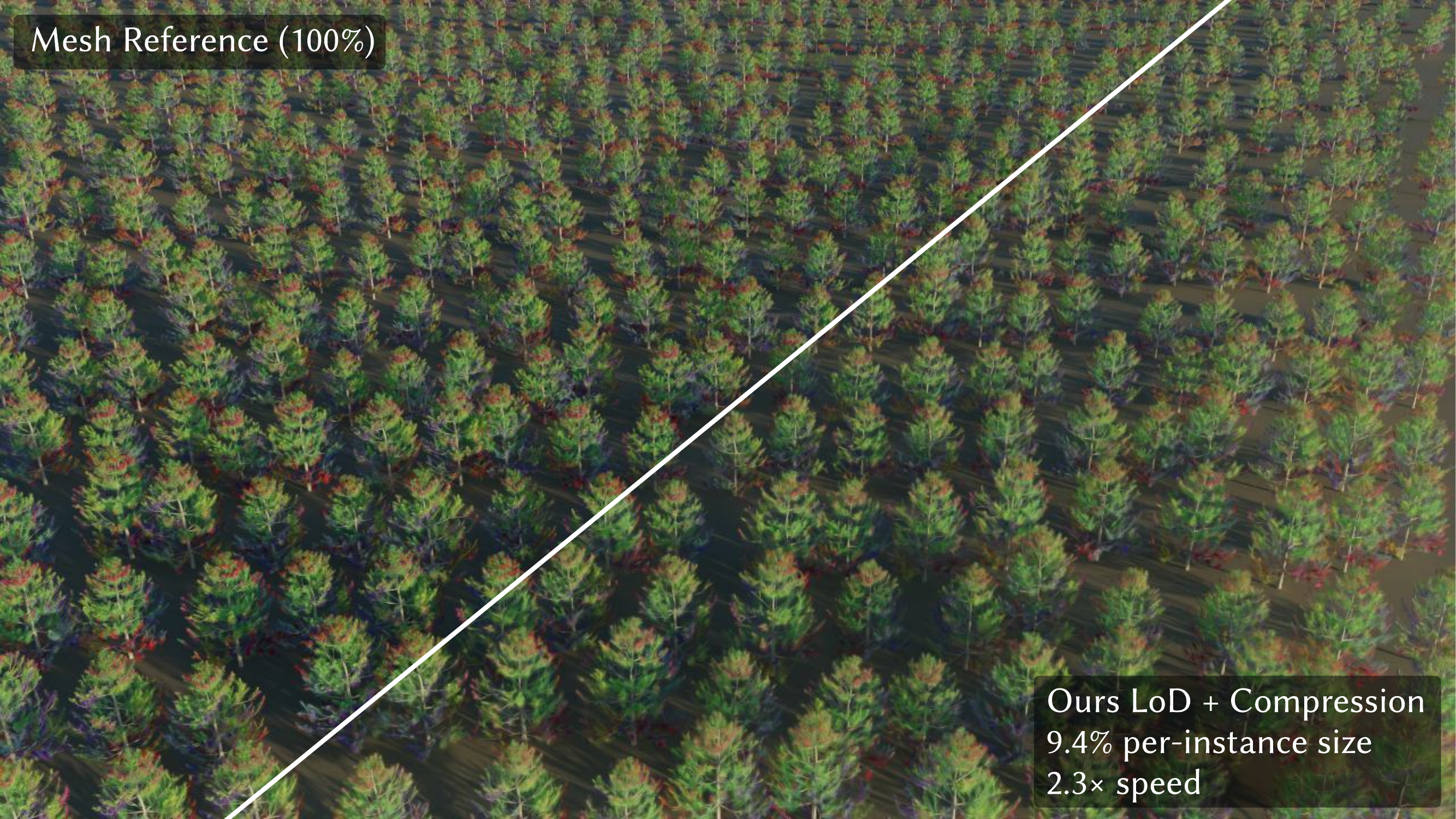}
    \end{tabular}
    \caption{\label{fig:lod_fan}
       Large outdoor scene rendering comparison. By utilizing discrete LoDs and codebook-based compression, our method 
       achieves significant memory reduction and speedup with minimal quality loss. PSNR: 27.71, SSIM: 0.878, LPIPS: 0.097, 
       \FLIP: 0.151.
    }
\end{figure}
Indeed, we demonstrate that our representation can be combined with both a discrete level-of-detail (LoD) scheme \textbf{and} 
compression to yield practical advantages. In \autoref{fig:lod_fan}, a total of 554 instances of the 
\emph{Color Tree} models are assembled in a large outdoor scene. We treat the 4 variants of the model with different primitive 
counts ($100\%$, $50\%$, $10\%$, and $1\%$) as a discrete LoD list of the original model. We select the LoD level of 
every instance based on its screen-space projection size: finer LoD near the camera, and coarser LoD toward 
the far horizon. This simple strategy allows us to aggressively reduce the primitive count with minimal quality loss: 
only ${\sim}10\%$ of instances keep the full variant, and overall we cut the primitive count to $25\%$ of the original.
Note that it is not possible to do so with na\"ive mesh simplification without significant quality degradation, as 
shown in \autoref{fig:mesh_comp}.
For each LoD variant, we further adopt a straightforward codebook-based quantization method~\citep{pacmcgit/PapantonakisKKLD24}, 
which further gives a $3.4\times$ size reduction on average. This compression method fully supports random access and 
has negligible decoding overhead. Combining both strategies, our per-instance RAM size is only $9.4\%$ of that of the original 
textured mesh. The reduced primitive count also boosts our rendering speed to be $2.3\times$ of that of the mesh reference scene.
Please refer to the figure for visual comparison and the caption for quantitative metrics.

\begin{table*}[tb]
	\centering
	\caption{
        Forward path tracing scene specifications and timings. Both total rendering times and per-spp times (in parentheses) are included. Table entries are sorted by CPU per-spp time. 
        We report uncompressed memory consumption.
    }
	\begin{tabular}{llllllllll}
		\Xhline{1pt}
		\textbf{Figure}                                         & \textbf{\#Prim.}  & \textbf{Mem.}    & \textbf{Res.}   & \textbf{Spp}   & \textbf{Bounces}   & \textbf{CPU Time} & \textbf{GPU Time} & \textbf{GPU Speedup}\\
        \hline
        \autoref{fig:material_edit}, \small{\emph{Dragon}}      & 500K    & 55.3 MB      & \small{1280$\times$720}   & 2048  & 2         & 2m 10s (63.4 ms) & 51.4s (25.1 ms) & $2.5\times$\\
        \autoref{fig:material_edit}, \small{\emph{Blanket}}     & 131K    & 14.5 MB      & \small{1280$\times$720}   & 2048  & 2         & 10m 56s (320.3 ms) & 4m 53s (143.1 ms) & $2.2\times$\\
        \autoref{fig:voxel_compare}, \small{\emph{Plant}}       & 220K    & 24.3 MB      & \small{1024$\times$1024}  & 256   & 1         & 2m 16s (531.2 ms) & 28.4s (111.0 ms) & $4.8\times$\\        
        \autoref{fig:voxel_compare}, \small{\emph{Color Tree}}  & 750K    & 83.0 MB      & \small{1024$\times$1024}  & 256   & 1         & 5m 16s (1234.3 ms) & 1m 40s (393.7 ms) & $3.1\times$\\
        \autoref{fig:teaser} \small{(\emph{night})}             & 5.5M    & 608.4 MB     & \small{2000$\times$1200}  & 4096  & 8         & 6h 58m (6122.5 ms) & 1h 30m (1319.5 ms) & $4.6\times$\\        
        \autoref{fig:gs_lit}, \small{\emph{Garden}}             & 5.8M    & 641.6 MB     & \small{1920$\times$1080}  & 2048  & 2         & 4h 20m (7617.2 ms) & 29m 56s (877.4 ms) & $8.7\times$\\
        \autoref{fig:gs_lit}, \small{\emph{Stump}}              & 4.9M    & 542.1 MB     & \small{1920$\times$1080}  & 2048  & 2         & 4h 50m (8496.1 ms) & 21m 45s (637.6 ms) & $13.3\times$\\
		\Xhline{1pt}
	\end{tabular}
	\label{table:fwd_timings} 
\end{table*}

Next, we discuss performance related to radiance fields, with timings provided in \autoref{table:rf_timings}. As radiance fields enjoy a simpler, analytic 
solution (\autoref{sec:radiance_field}), they are much faster to render, achieving interactive framerates. Qualitatively speaking, our rendering speed is favorable compared 
to NeRF-based approaches~\citep{barron2022mip, muller2022instant}, but falls short of approaches specialized for particle-based radiance field rendering~\citep{kerbl20233d,moenne20243d}. 
While both are general-purpose particle-based ray tracing approaches, our method can directly produce \emph{noise-free} renders for radiance fields (\autoref{eq:rf}), 
whereas \citet{tog/CondorSBBGDJ25} still rely on stochastic sampling. This gives us a significant performance advantage over their method.

For particle-based representations, a common cause of slow performance is excessive overlapping, which leads to either more overdraw or ray intersection tests per pixel.
As a positive side effect, the combination of our linear transmittance model and loss design mitigates overdraw, as visualized in \autoref{fig:rf_overdraw}. 
For the same scene, we compare a radiance field optimized by our approach versus converted from 3DGS (\autoref{subsec:conversion}) with similar primitive counts, 
and observe a 4-5$\times$ speedup in our renderer. We expect the benefit could be generalized to other techniques with different rasterization or ray tracing strategies.

Our optimization for a scene in general takes several hours to finish. This puts us in a similar relative position among other radiance field methods 
(with the exception of Instant-NGP which excels at fast optimization).
It is worth noting that our optimization performance is again limited by the lack of hardware-accelerated 
kd-tree construction as each iteration requires a kd-tree rebuild. The rebuilding operation easily dominates the optimization process, accounting for more 
than $90\%$ of the total time. In our current implementation, we employ a straightforward hybrid approach for SAH kd-tree construction by serially building upper levels 
before parallelizing over lower subtrees. This could be improved by either more sophisticated parallel kd-tree construction (updating) algorithms~\citep{tog/ZhouHWG08} 
or potential hardware advancement.

\begin{table}[tb]
	\centering
	\caption{
        Scene specifications and timings of the NeRF Blender dataset. All scenes take 30K optimization iterations. 
        Rendering timings (GPU) are averaged across all testing set views ($800\times800$ resolution).
    }
	\begin{tabular}{llllll}
		\Xhline{1pt}
		\textbf{Scene} & \textbf{\#Prims} & \textbf{Optim. Time} & \textbf{Avg. Render Time} \\
        \hline
        \emph{Chair}      & 453K          & 10h 30m   & 33.8 ms \\
        \emph{Drums}      & 333K          & 9h 7m   & 28.2 ms \\
        \emph{Ficus}      & 218K          & 4h 8m   & 23.8 ms \\
        \emph{Hotdog}     & 258K          & 7h 39m   & 30.2 ms \\
        \emph{Lego}       & 314K          & 6h 37m   & 21.3 ms  \\
        \emph{Materials}  & 266K          & 8h 58m   & 22.2 ms \\
        \emph{Mic}        & 273K          & 7h 8m   & 33.2 ms \\
        \emph{Ship}       & 460K          & 12h 47m   & 29.8 ms \\
		\Xhline{1pt}
	\end{tabular}
	\label{table:rf_timings} 
\end{table}
\section{Conclusion} \label{sec:conclusion}
In this work, we have presented a novel volumetric rendering primitive for unified scene representation. By combining anisotropic 3D Gaussian distribution and 
a non-exponential, 
linear transmittance model, our primitives can adapt to hard surfaces, thin structures, and aggregated elements. The primitive appearance is defined by a 
flexible phase function that incorporates both the NDF of an aggregation and the base BSDF of each aggregated element. Our representation provides efficient 
Monte Carlo operations to enable Monte Carlo path tracing for global illumination. We have demonstrated the generality and quality of our representation with 
various rendering applications and provided methods to acquire data from other existing representations. 
Furthermore, we have demonstrated the applicability of our method to differentiable rendering applications, including transmittance optimization and 
image-based radiance field reconstruction.

Our method has several limitations that could serve as fruitful topics for future research. Our transmittance model shares the common 
limitation with the model by \citet{vicini2021non} that it does not conform to certain physical constraints, such as the weak reciprocity proposed by 
\citet{d2018reciprocal}. Developing a reciprocal formulation for general heterogeneous non-exponential transport remains an open problem. 
Additionally, our 
method currently does not support refraction inside a solid. Rendering refraction requires tracking the change of index of refraction when a ray enters or exits 
a medium boundary. When a scene is entirely modeled by our volumetric Gaussian primitives, there is no clear definition of medium boundaries or mechanism to 
separate interior and exterior parts. 
However, translucency is possible using a thin-surface BSDF as the base BSDF of our phase function (\autoref{eq:phase_function}). 
\autoref{fig:leaf} shows a simple example.
Our method currently assumes that scattering in a primitive can be modeled accurately without an exitant positional distribution,
as considered in subsurface scattering~\citep{jensen2001bssrdf} or shell transport~\citep{moon2007rendering}. 
This is justified by the typical use case where a single primitive covers a small surface patch or a small cluster of oriented elements.
However, a more rigorous treatment of our appearance model is necessary, as the impact of the exitant positional distribution becomes more pronounced when a 
primitive covers a large volume in level-of-detail applications.
The experiments in \autoref{fig:mesh_comp} and \autoref{fig:lod_fan} serve as a proof-of-concept 
for our method as a multi-scale representation. A promising next step is to extend this simple 
discrete LoD scheme into a continuous, view-dependent LoD system~\citep{kerbl2024hierarchical}.
While out of the scope of this work, 
a full path-space differentiable rendering formulation for our representation, similar to that for exponential volumes \citep{zhang2021path}, will enable more 
powerful inverse rendering applications. 
Finally, both our CPU and GPU implementations offer significant potential for further optimization. Our current method prioritizes accurate computation of ray-primitive 
integrals and precise sampling of the free-flight distribution for overlapping primitives. Incorporating suitable approximations, such as the maximum 
response approximation~\citep{moenne20243d}, could yield notable performance gains.

Overall, we believe our work provides novel contributions toward a practical unified scene representation that encompasses both surfaces and volumes.
Such unification could offer benefits to both forward and inverse rendering techniques, as well as numerous downstream graphics applications.
\section*{Acknowledgments}
We would like to thank Bing Xu for numerous discussions and helpful suggestions, and Zhao Dong for proofreading the initial version of the paper.
In addition, we thank the authors of \citet{mildenhall2021nerf}, \citet{barron2022mip}, and \citet{jin2023tensoir} for releasing
the datasets accompanying their papers.
Finally, we would also like to thank the Editors-in-Chief, Eitan Grinspun and Carol O'Sullivan, for their gracious support and 
coordination throughout the revision cycle, as well as the anonymous reviewers for their valuable feedback.
The original \emph{Dragon} model is downloaded from \href{https://graphics.stanford.edu/data/3Dscanrep/}{The Stanford 3D Scanning Repository}.
The original leaf model in \autoref{fig:leaf} is created by Sketchfab user \textsf{crowinhand} (\href{https://creativecommons.org/licenses/by/4.0/}{CC-BY 4.0}).
The \emph{Sir Frog} model is created by Sketchfab user \textsf{Adrian Carter} (\href{https://creativecommons.org/licenses/by/4.0/}{CC-BY 4.0}).
Scenes and models in \autoref{fig:teaser}, \autoref{fig:diff}, and \autoref{fig:linear_vs_exp} are created using assets purchased from KitBash3D.
The environment map used in \autoref{fig:material_sphere} is created by \href{https://dativ.at/lightprobes/index.html}{Bernhard Vogl}.
All other environment maps are downloaded from Poly Haven (\href{https://creativecommons.org/publicdomain/zero/1.0/}{CC0}).

\bibliographystyle{ACM-Reference-Format}
\bibliography{main}

\appendix
\section{Ray Integral Derivation} \label{sec:ray_integral_derivation}
Without loss of generality, assume $\mu = 0$. Expanding the integral term in \autoref{eq:ray_integral} gives us
\begin{align} \label{eq:ray_integral_2}
    \int_{t_0}^{t_1} G(x_t) \,\D{t} 
    &= \int_{t_0}^{t_1} \exp( -\frac{1}{2} \, x_t\trans \Sigma^{-1} x_t ) \,\D{t} \nonumber \\ 
    &= \int_{t_0}^{t_1} \exp(-\frac{1}{2} \, (x+t\omega)\trans \Sigma^{-1} (x+t\omega)) \,\D{t} \nonumber \\ 
    &= \int_{t_0}^{t_1} \exp(-\frac{1}{2} \, (At^2 + Bt + C))  \,\D{t} \nonumber \\ 
    &= \int_{t_0}^{t_1} \exp \Big(-\frac{1}{2} ( (A(t+\frac{B}{2A}))^2 + (C- \frac{B^2}{4A}) ) \Big) \,\D{t} \nonumber \\
    &= \exp(\frac{B^2}{8A} - \frac{C}{2}) \int_{t_0}^{t_1} \exp (-(\sqrt{\frac{A}{2}}t + \frac{B}{2\sqrt{2A}})^2) \,\D{t}, \nonumber \\
    A &= \omega\trans\Sigma^{-1}\omega, \nonumber \\
    B &= \omega\trans\Sigma^{-1}x + x\trans\Sigma^{-1}\omega, \nonumber \\
    C &= x\trans\Sigma^{-1}x. 
\end{align}
To continue, Let $\tau = \sqrt{A/2} \,t + B/(2\sqrt{2A})$ and perform a change of variable:
\begin{align} \label{eq:ray_integral_3}
    &\int_{t_0}^{t_1} \exp (-(\sqrt{\frac{A}{2}}t + \frac{B}{2\sqrt{2A}})^2) \,\D{t} 
    = \sqrt{\frac{2}{A}}\int_{\tau_0}^{\tau_1} \exp (-\tau^2) \, \D{\tau} \nonumber \\
    &= \sqrt{\frac{\pi}{2A}} \Big( \erf(\sqrt{\frac{A}{2}}t_1 + \frac{B}{2\sqrt{2A}}) - \erf(\sqrt{\frac{A}{2}}t_0 + \frac{B}{2\sqrt{2A}}) \Big).
\end{align}
The final expression can then be obtained by substituting \autoref{eq:ray_integral_2} and \autoref{eq:ray_integral_3} into \autoref{eq:ray_integral}. If the 
ray is infinite ($t_0 = -\infty$ and $t_1 = \infty$), then \autoref{eq:ray_integral_3} simply evaluates to $\sqrt{2\pi / A}$.



\section{MIS PDF Approximation} \label{sec:approx_mis_pdf}
For the specular component of the Disney BSDF, the standard sampling procedure is to sample a half vector from the VNDF of the microfacet distribution 
\citep{heitz2018sampling}. We approximate the overall PDF by a roughened SGGX VNDF:
\begin{align} \label{eq:approx_phase_function_pdf_specular} 
    &\int_{\mathbb{S}^2}  D_{\omega_o}(n) \mathrm{pdf}_{\mathrm{spec}}(\omega_i | \omega_o; n) \,\D{n} \nonumber \\
    = &\int_{\mathbb{S}^2}  D_{\omega_o}(n) \frac{D_m(n\cdot\omega_h) G_1(n \cdot \omega_o) \langle \omega_h \cdot \omega_o \rangle}{|n\cdot\omega_o|} \,\D{n} \nonumber \\
    \approx &\int_{\mathbb{S}^2}  D_{\omega_o}(n) D_m(n\cdot\omega_h) \,\D{n} \nonumber \\
    \approx &\; \tilde{D}_{\omega_o}(\omega_h),
\end{align}
where we first drop the low-frequency components ($G_1(n \cdot \omega_o)$, $\langle \omega_h \cdot \omega_o \rangle$, and $|n\cdot\omega_o|$). We then 
utilize the fact that a SGGX distribution is equivalent to a double-sided GGX \citep{heitz2015sggx}. Therefore, the remaining integral becomes similar to a 
convolution between two SGGXs, which we further approximate by a roughened SGGX VNDF. Let $(\sigma_1^2, \sigma_2^2, \sigma_3^2)$ be the eigenvalues 
of original SGGX sorted in ascending order, and $\alpha$ be the isotropic roughness of the microfacet GGX distribution. The roughened SGGX has the adjusted 
eigenvalues $(\sigma_1^2 + \alpha^2\sigma_3^2, \sigma_2^2 + \alpha^2\sigma_3^2, \sigma_3^2)$ and the same eigenvectors as the original one. 

For the diffuse component of the Disney BSDF, the standard sampling procedure is usually just cosine-weighted hemisphere sampling. We simply approximate the overall 
PDF as a cosine-weighted hemisphere of the half vector:
\begin{equation} \label{eq:approx_phase_function_pdf_diffuse} 
    \int_{\mathbb{S}^2}  D_{\omega_o}(n) \mathrm{pdf}_{\mathrm{diff}}(\omega_i | \omega_o; n) \,\D{n}
    \approx \frac{1}{\pi} \langle \omega_h \cdot \omega_o \rangle.
\end{equation}
We then combine both components (with optional weighting by metallic and base color luminance) to get the overall approximation to 
\autoref{eq:phase_function_pdf}.
The approximate PDF 
admittedly lowers the effectiveness of MIS compared to the exact PDF, but overall still provides great variance reduction compared to not having MIS at all.
\section{Magnitude Remapping from 3DGS} \label{sec:3dgs_conversion}
3DGS renders a Gaussian by projecting it to the 2D screen space and evaluating the PDF of the projected 2D Gaussian. This is in fact similar to our ray 
integral for an infinite ray. To see it, consider an infinite ray with origin $x$ and direction $\omega$. Let $(u, v, \omega)$ be an orthonormal basis. 
The local-to-world transform is
\begin{equation*}
    M = \begin{pmatrix}
        R & x \\
        \bm{0} & \bm{1}
        \end{pmatrix}, \quad
    R = (u, v, \omega).
\end{equation*}
Following \citet{zwicker2001ewa}, the parameters of the projected 2D Gaussian are\footnote[2]{Note that
we omit the perspective projection as it does not affect normalization.}
\begin{align*}
    \hat{\mu} &= (M^{-1}\mu)_{2\times2}, \\ 
    \hat{\Sigma} &= (R^{-1}\Sigma R)_{2\times2},
\end{align*}
where $(\cdot)_{2\times2}$ denotes taking the $2\times2$ submatrix by deleting the third row and the third column. The projected 2D Gaussian is then evaluated 
\emph{without normalization}, and multiplied by the opacity $\alpha$ as the weight in the blending process:
\begin{equation*} 
    I_{\mathrm{3DGS}} = \alpha \exp \Big(-\frac{1}{2}\hat{\mu}\trans \hat{\Sigma}^{-1} \hat{\mu} \Big).
\end{equation*}
Mathematically, evaluating the PDF of the normalized projected 2D Gaussian is equal to integrating the original 3D Gaussian over an infinite ray 
defined by $x$ and $\omega$:
\begin{align*} 
    &\frac{1}{2\pi |\hat{\Sigma}|^{\frac{1}{2}}}  \exp \Big(-\frac{1}{2}\hat{\mu}\trans \hat{\Sigma}^{-1} \hat{\mu} \Big) = \\
    &\frac{1}{(2\pi )^{\frac{3}{2}} |\Sigma|^{\frac{1}{2}}}  \int_{t_0}^{t_1} \exp \Big(-\frac{1}{2}(x_t-\mu)\trans \Sigma^{-1} (x_t-\mu) \Big) \,\D{t}.
\end{align*}
Therefore, it is easy to see that by letting
\begin{equation}
    c = 2\pi |\hat{\Sigma}|^{\frac{1}{2}} \alpha,
\end{equation}
our infinite ray integral (\autoref{eq:ray_integral}) is equal to the above blending weight used by 3DGS
\begin{equation*}
    I(-\infty, \infty) = I_{\mathrm{3DGS}}.
\end{equation*}

\section{Sampling Exponential Transmittance Gaussians via Analog Decomposition Tracking} \label{sec:decomp_track}
\citet{kutz2017spectral} show that by decomposing an exponential medium into components, a free-flight sample from the original medium can be obtained by 
taking the minimum of independent free-flight samples from the components. This technique can thus be adopted to sample the exponential variants of scenes in 
\autoref{fig:linear_vs_exp}. Each is a heterogeneous exponential medium where the extinction coefficient is defined by a mixture of 3D Gaussians: 
$\sigma_{\mathrm{t}}(x) = \sum_k G_k(x)$. For each ray, we simply iterate all its intersecting Gaussians and sample each of them independently. 
Let $u \in [0, 1)$ be a random number and $x_t = x + t\omega, t \in [t_0, t_1]$ be a ray. Sampling the exponential free-flight distance from one 
Gaussian is straightforward by inverting \autoref{eq:exp_ff} and \autoref{eq:ray_integral}:
\begin{equation}
    u = 1 - \exp(-I(t_0, t)) \Leftrightarrow t = I^{-1}(-\ln(1-u); t_0).
\end{equation}
Finally, we take the nearest sample among samples from all Gaussians. 

It is important to recognize the difference between this approach and our approach with the linear transmittance Gaussian primitives.
Such approach falls under the conventional category of specifying a volume using \emph{per-point} phase function, as discussed in \autoref{sec:primitives}.
While the minimum operator in analog decomposition tracking conceptually resembles ``selecting a primitive'', it still samples an infinitesimal collision point 
based on the continuous free-flight PDF. Multiple scattering within the same primitive may happen in subsequent bounces. This is different from our approach 
where we sample an entire primitive and never explicitly account for sub-primitive multiple scattering (\autoref{fig:per_prim_illus}).

\end{document}